\documentclass[INR,biber]{nowfnt} %

\usepackage[utf8]{inputenc}

\title{User Simulation for Evaluating Information Access Systems}

\maintitleauthorlist{
Krisztian Balog \\
University of Stavanger \\
krisztian.balog@uis.no
\and
ChengXiang Zhai \\
University of Illinois at Urbana-Champaign \\
czhai@illinois.edu
}

\issuesetup
{%
 copyrightowner={K.~Balog and C.~Zhai},
 volume        = xx,
 issue         = xx,
 pubyear       = 2024,
 isbn          = xxx-x-xxxxx-xxx-x,
 eisbn         = xxx-x-xxxxx-xxx-x,
 doi           = 10.1561/XXXXXXXXX,
 firstpage     = 1, %
 lastpage      = 242
 }

\author[1]{Balog,Krisztian}
\author[2]{Zhai,ChengXiang}

\affil[1]{University of Stavanger; krisztian.balog@uis.no}
\affil[2]{University of Illinois at Urbana-Champaign; czhai@illinois.edu}

\articledatabox{} %

\usepackage{booktabs}
\usepackage{multirow}
\usepackage{mdframed}
\usepackage{stix}  %
\usepackage[shortlabels,inline]{enumitem}
\usepackage{hhline}
\usepackage{colortbl}
\usepackage{mathtools}

\definecolor{amber}{rgb}{1.0, 0.75, 0.0}
\definecolor{asparagus}{rgb}{0.53, 0.66, 0.42}
\definecolor{bluegray}{rgb}{0.4, 0.6, 0.8}
\definecolor{brinkpink}{rgb}{0.98, 0.38, 0.5}

\newcommand{\parheading}[1]{\noindent\emph{\textbf{#1}}}

\newlist{inlinelist}{enumerate*}{1}
\setlist[inlinelist]{label=(\arabic*)}

\def\prob{P}
\def\mR{\mathcal{R}}
\def\mC{\mathcal{C}}

\newcommand\mycircle[1]{\begin{picture}(1,1)
\ifnum0=#1\put(.5,.35){\circle{1}}\else
\ifnum10=#1\put(.5,.35){\circle*{1}}\else
\put(.5,.35){\circle{1}}\put(.5,.35){\circle*{.#1}}
\fi\fi\end{picture}}

\newcommand\flowstart{$\mycircle{10}$}
\newcommand\flowend{$\mycircle{7}$}
\newcommand\flowaction{$\mdlgwhtsquare$}
\newcommand\flowdecision{$\mdlgwhtdiamond$}

\usepackage{amsmath}

\begin{document}

\makeabstracttitle

\begin{abstract}
    Information access systems, such as search engines, recommender systems, and conversational assistants, have become integral to our daily lives as they help us satisfy our information needs.
    However, evaluating the effectiveness of these systems presents a long-standing and complex scientific challenge.
    This challenge is rooted in the difficulty of assessing a system's overall effectiveness in assisting users to complete tasks through interactive support, and further exacerbated by the substantial variation in user behaviour and preferences.
    To address this challenge, user simulation emerges as a promising solution.

    This book focuses on providing a thorough understanding of user simulation techniques designed specifically for evaluation purposes.
    We begin with a background of information access system evaluation and explore the diverse applications of user simulation.
    Subsequently, we systematically review the major research progress in user simulation, covering both general frameworks for designing user simulators, utilizing user simulation for evaluation, and specific models and algorithms for simulating user interactions with search engines, recommender systems, and conversational assistants.
    Realizing that user simulation is an interdisciplinary research topic, whenever possible, we attempt to establish connections with related fields, including machine learning, dialogue systems, user modeling, and economics.
    We end the book with a broad discussion of important future research directions, many of which extend beyond the evaluation of information access systems and are expected to have broader impact on how to evaluate interactive intelligent systems in general.
\end{abstract}

\chapter{Introduction}

Information access systems, such as search engines, recommender systems, and conversational assistants, have become increasingly intelligent in understanding users' intents, supporting their tasks, and interacting with them using natural language dialogue, thanks to recent progress in research in artificial intelligence (AI), especially in machine learning and natural language processing.
These information access systems are used by millions on a daily basis to perform a wide range of tasks where humans need help to find information relevant to a task.
In general, the interactions with these systems involve a user entering information needs or preferences (by typing queries, rating items, or asking natural language questions) and interacting with information objects (by clicking, typing, or speaking) that are presented by the system on some device (e.g., desktop, laptop, tablet, smart phone, or smart speaker) in some modality or combination of modalities (e.g,. text, rich snippets, voice).  With intelligent home devices becoming available, information access systems may also be used to power a wide range of intelligent agent systems, such as Google Home or Alexa, which can go beyond supporting information access to also support other user tasks (e.g., controlling home appliances, making appointments, or placing orders).

\vspace{\baselineskip}

Although information access systems have already become useful products for people, how to appropriately evaluate those systems remains an open scientific challenge.  It is especially challenging to evaluate a system's overall effectiveness in helping a user finish a task via interactive support.  The fact that users vary significantly in terms of their behaviour and preferences makes evaluation even more difficult.  As a promising strategy for evaluating information access systems using reproducible experiments, user simulation has attracted much attention recently. In this book, we review the recent progress in this area with a focus on user simulation for evaluating information access systems.

In the rest of this chapter, we first describe the spectrum of information access tasks.  Next, we briefly discuss the goals of evaluation and general methodologies of evaluation.  We then highlight the challenges involved in evaluating information access systems and how user simulation can help address those challenges.  Finally, we describe the aims and scope of this book.

\section{Information Access Tasks}
\label{sec:intro:tasks}

Information access refers to the ability to identify, retrieve, and use information effectively.\footnote{\url{https://www.encyclopedia.com/computing/news-wires-white-papers-and-books/information-access}} Access to the right information at the right time plays an important role in everyone's life and is vital to business operations. At a high-level, information access can happen in two modes~\citep{Zhai:2016:MC}: (1) \emph{pull mode}, where the user takes the initiative and uses a search engine to find needed information (``pull" relevant information to the user), and (2) \emph{push mode}, where the system takes the initiative and recommends relevant information to a user (``push" relevant information to the user). The two modes can be naturally mixed in conversational AI systems, which are increasingly common due to the emergence of large language models (LLMs)~\citep{McTear:2024:book}.

Search engines and recommender systems are the two most common widely used applications for information access. The two modes of information access are complementary and often supported simultaneously using a single system. For example, a search engine can not only support querying (pull mode) but also recommend related information to the user (push mode). Similarly, a recommender system may also recommend information in the form of a ranked list to enable a user to further interact with the recommended information and potentially enter a query to further explore the information space. Indeed, search and recommendation have been suggested as ``two sides of the same coin'' in~\citep{Belkin:1992:CACM}. As such, it is not surprising that search engines and recommender systems share many common technical challenges (e.g., modeling a user's information need and preferences, matching an information item with a user's interest, ranking items accurately, learning from a user's feedback information, evaluating a ranked list to assess its utility to a user), and tend to benefit from using similar techniques, including user simulation techniques. For this reason, we intend to cover the topic of user simulation in the broad context of information access with the understanding that most discussions are relevant to both search engines and recommender systems, even though the actual research work that we discuss may have been done for either just search engines or recommender systems.

Recently, conversational assistants have attracted much attention~\citep{McTear:2021:book}. They generally support mixed-initiative interaction via natural language to facilitate both search and recommendation in the same information access session~\citep{Zamani:2023:FnTIR}. Compared with traditional search engines and recommender systems, where the actions a user could potentially take are well specified by the user interface of the system, conversational assistants have more open-ended functions in the sense that a user can potentially ask questions, provide clarifications, and explore related topics using unrestricted natural language, thus adding complexity to user simulation. %
We note that conversational assistants, casually referred to as ``conversational AI,'' can cater to diverse user goals, including, e.g., social chatting. However, our primary focus in this book is on systems that are designed to support information access, i.e., tasks where there is an underlying information need and the system returns information objects (which may be documents, entities, answers, utterances, etc.). This focus increases the commonality between conversational assistants,  search engines, and recommender systems in terms of user simulation.

\section{Evaluation Methodologies}
\label{sec:intro:evaluation}

There are three widely-used evaluation methodologies for information access systems: reusable test collections~\citep{Sanderson:2010:FnTIR}, user studies~\citep{Kelly:2009:FnTIR}, and online evaluation~\citep{Hofmann:2016:FnTIR}.

\emph{Reusable test collections} (a.k.a. \emph{offline evaluation}) facilitate large-scale automatic evaluation and have been invaluable for comparing different algorithms and improving the state of the art.  They ensure repeatability and enable comparison between different approaches and study of effectiveness of individual components within complex methods.  However, they are static and are severely limited in their ability to capture many aspects of users and interactions adequately.  They measure system performance on an abstraction  of a given process (e.g., search or recommendation), where the user is also abstracted away. Specifically, they are based on simplified models of the information access process and user behaviour. Examples include the assumption that a system always presents a ranked list of results to a user or that a user can always recognize whether a document is relevant in a search result list. This has so far been the standard evaluation methodology for making relative comparisons between two systems in a repeatable and reproducible manner.  However, it is generally not so useful for the purpose of evaluating the actual utility of a system due to the significant deviation between the evaluation environment and the real-world application and the very limited inclusion of users. It is also generally hard or impossible to develop test collections for evaluating an {\em interactive} system, a limitation that can be addressed by user simulation.  

\emph{User studies} provide the highest fidelity in terms of capturing real users' interactions with an actual system in a controlled setting.  However, experiments that involve real users are costly to run.  Further, if multiple rounds of experimentation are performed, new users need to be involved in order to avoid misinterpretation of the findings due to fatigue and learning effects.  Also, experiments that involve an actual service have a bandwidth limit, which is set by the amount of users and their activity using that service.  User studies are useful for assessing the actual utility of a system, but they suffer from several limitations. First, the result is generally not reproducible (even the same user would not behave in the same way when repeating an experiment due to learning effects).  Thus, they have limited value for making relative comparisons between systems, especially when new systems---to be developed in the future---need to be included in the comparison. Second, the cost of recruitment is often high, and it is a challenge to recruit enough people from the right population. Major industry labs often have to invest significantly in recruiting users and conducting user studies, which smaller companies typically cannot afford. In academia, only a few examples of user studies of reasonable scale involve participants beyond university students~\citep{Brennan:2014:IIIX}. 

\emph{Online evaluation} (a.k.a. log-based studies) is based on the idea of observing real users of a fully operational system and assessing the system's performance by analyzing the recorded user behaviour. For instance, A/B tests can be used to evaluate different versions of a system. Online evaluation is widely used by companies that deploy real-world applications or services, and is regarded as the most direct and reliable measurement of quality and user experience. Like user studies, online evaluation enables measuring the actual utility of a system and comparing systems with real users, but at a much larger scale in terms of the number of users used for evaluation.  Unlike user studies, however, it generally does not provide control over the users, making it harder to interpret the results. As in the case of user studies, online evaluation suffers from being not reproducible and thus cannot be ``reused'' to compare different systems or analyze the effectiveness of various components. 
Another limitation of online evaluation is that the user interactions to be evaluated are limited to natural interactions with the system, thus it cannot accommodate counterfactual evaluation, i.e., where the potential outcomes of a hypothetical system (e.g., one with a new algorithm) and being compared to those of an existing system.
This hypothetical system cannot be used by real users, making online evaluation infeasible. Additionally, there is a risk of leaving a negative impression on users about a production system's performance if a system to be evaluated online turns out to perform poorly. 

\section{Challenges in Evaluating Information Access Systems and Simulation-Based Evaluation}
\label{sec:intro:evaluation_challenges}

In general, all the three methodologies discussed earlier can be applied to, and indeed have been regularly used for, evaluating information access systems.  However, none of those methodologies can be used to compare multiple interactive information access systems (in terms of their overall effectiveness in supporting users) using reproducible experiments due to the static nature of the test collection-based approach and the lack of reproducibility when real users are involved. These limitations can be addressed by using user simulation:  simulated users can be controlled and thus enable reproducible experiments. 

Note that the existing test collection-based evaluation methodology~\citep{Sanderson:2010:FnTIR} can be viewed as a simple form of user simulation. 
This means we are already utilizing user simulation, albeit implicitly, without explicitly articulating what kind of users are being simulated.
A key advantage of evaluation based on user simulation is to make assumptions about simulated users and their behaviour more explicit, while modeling a broader range of user actions than current measures consider (see Section~\ref{sec:frameworks:general} for more discussion on this).

We want to highlight the importance of evaluating the \emph{overall effectiveness} of a system (not just various components of a system), especially in the case of information access systems.  This is because users are likely to benefit the most from intelligent assistance in case of complex tasks.  Commonly, complex tasks are decomposed into a series of smaller and simpler components.  The decomposition process generally involves close collaboration between a user and a system in an interactive way, in which a user would iteratively direct the system to perform specific component functions and the results from multiple steps can be synthesized to generate solutions to a complex problem.
Most component-level tasks can be abstracted, studied and addressed in isolation, and evaluated using the reusable test collection methodology with reproducible experiments. While this is clearly necessary to allow for systematic progress to be made, evaluation of individual components alone is insufficient.
It is arguably more important to view them as components of a larger information access system and study how to evaluate the \emph{whole} system from a user's perspective.  Indeed, the ultimate goal of evaluation is to measure how well the user is aided in achieving their end goal.  It is vitally important to do this evaluation correctly; if not done appropriately, it would mislead us to draw wrong conclusions or deploy an inferior application system that negatively impacts the user experience.

\section{User Simulation}
\label{sec:intro:simulation}

What do we exactly mean by user simulation?  Informally, user simulation is to have an intelligent agent simulate how a user interacts with a system.  The agent can be built based on models/algorithms/rules and any knowledge we have about the user (their behaviour, knowledge, etc.).  The agent can also have parameters that can be varied in a meaningful way to simulate variations of users.  Once a user simulator is constructed, it can then be used to interact with any system that needs to be evaluated. In turn, we can measure the system's utility based on the observed behaviour of the agent while interacting with the system.
Simulation thus has the potential to enable repeatable and reproducible evaluations at a low cost, without using invaluable user time (human assessor time or online experimentation bandwidth).  Further, simulation can augment traditional evaluation methodologies by offering possibilities to gain insights into how system performance changes under different conditions and user behaviour.

\subsection{Problem Definition}
\label{sec:intro:simulation_definition}

User simulation is the process of modeling a user's behaviour and decision-making patterns within an interactive system, specifically designed to mimic and predict how a user will act in various interaction contexts or scenarios related to completing a task.
To effectively simulate a user's behaviour within an interactive system, configuration variables that influence this behaviour must be defined:
\begin{itemize}
	\item Task ($T$): A user's behaviour varies according to nature of the user's task. Tasks vary in complexity, and different tasks require different types and levels of interaction, decision-making processes, and completion strategies.
	\item System ($S$): A user's behaviour depends on the system they interact with. This includes the system's functionality, user interface, and overall usability and support for task goals. It is the system that dictates the types of possible actions, denoted as $\mathcal{A}$, that a user can perform at any given point during their interactions.
	\item User information ($U$): Different users may behave differently when completing the same task using the same system. Simulations must account for variations in individual user characteristics such as age, technical proficiency, preferences, and cognitive styles.
\end{itemize}
With these variables defined, the task of user simulation can be stated as the following computational problem:
\begin{quote}
    Given the variables $T$, $S$, and $U$, the goal is to create an agent that can simulate every action that user $U$ may take when attempting to complete task $T$ using system $S$.
\end{quote}
This problem involves developing a computational model that can dynamically generate user actions, reflecting the behavioural patterns and decision-making processes of a user, based on a specific task context. 
Formally, we define the computational model as $\pi: \mathcal{S} \rightarrow A$, where $\mathcal{S}=(T,U,S,H)$ represents the current state, encompassing information about the task $T$, system $S$, user $U$, as well as the history of previous interactions $H$ (including the actions taken by the user, the responses provided by the system, and any other relevant events that have occurred during the interaction), and $A 
\in \mathcal{A}$ is the action taken by the (simulated) user.
The choice of computational model (e.g., rule-based, probabilistic, or machine-learned algorithm) is influenced by the nature of the task, system, and user information.

\subsection{Scope}
\label{sec:intro:simulation_scope}

User simulation encompasses a wide spectrum, ranging from predicting single actions to modeling complex behaviour across multiple tasks. In our formulation, this scope is primarily determined by how the task information ($T$) is defined. At one end of the spectrum, $T$ might represent a very specific interaction context, such as predicting whether a user would click on a particular search result snippet. Here, the focus is on simulating a single, isolated action. Moving along the spectrum, $T$ could encompass a sequence of actions within a given context, such as reformulating search queries within a search session, requiring the model to consider dependencies between actions. Further expanding the scope, $T$ might represent an entire task, such as finding information on a particular topic or completing a purchase, where the simulation would involve multiple sequences of interactions. Finally, at the broadest level, $T$ could encompass a user's general preferences and behaviour across various tasks, necessitating models that capture long-term patterns and adapt to different contexts. Thus, by varying the granularity and breadth of $T$, our formulation allows for user simulations in a wide range of application scenarios at different levels of complexity.
Table~\ref{tbl:simulation_examples} lists specific examples of user simulation for various information access tasks.

\begin{table}[t]
    \centering
    \footnotesize
    \caption{Examples of user simulation, ranging from single actions to more complex behaviours.}
    \label{tbl:simulation_examples}
	\centering
	\begin{tabular}{p{2.5cm}p{2.9cm}p{2.9cm}p{1.8cm}}
		\toprule
		\textbf{Task ($T$)}                                                           & \textbf{System ($S$)} & \textbf{User information ($S$)} & \textbf{Actions ($\mathcal{A}$)} \\
		\midrule
            Rating a product to express satisfaction & E-commerce website with product pages and rating features & User's purchase history, browsing behaviour, and demographic information & Browsing, Rating \\
		\midrule
            Refining a search query to find specific information & Search engine with a search box, query suggestions, and navigable search result lists & User's initial query, search history, and click behaviour & Reformulating, Clicking \\
		\midrule
		Collecting as many relevant information items as possible & Search engine with a query box and navigable search result lists & University researcher conducting a comprehensive literature review on a topic & Querying, Clicking \\
		\midrule
		Finding a movie to watch & Recommender system with slates of items &
		Previous watch history & Clicking, Watching \\
		\midrule
            Seeking assistance with a technical issue & Conversational assistant with natural language chat interface & User's description of the problem, technical expertise, and previous interactions & Prompting \\
		\bottomrule
	\end{tabular}
\end{table}

\subsection{Approaches}
\label{sec:intro:simulation_approaches}

Approaches to user simulation can be classified based on the specific types of actions that they attempt to simulate. For example, some approaches may simulate how a user generates a query while others might simulate how a user responds to a search result list (e.g., simulating when a user might click on or skip a result).
Approaches simulating different types of actions can also be combined to simulate a whole session of actions of a user. As will be elaborated later in the book, most existing work tends to focus on simulating each type of action separately with significantly less work on simulating a whole user session.

The problem of simulating a user's action can often be framed as a classification problem when there is a relatively small set of actions to choose from; for example, simulation of a user's clicking action may be framed as a binary classification problem, where the algorithm would predict whether the simulated user would click on a result or not after examining a snippet. When there are potentially infinitely many actions to choose from (e.g., when formulating a query, any valid query would be potentially an option), in practice, we often make assumptions to restrict the number of actions to be considered when simulating those actions (e.g., limit the length of a query to be considered).

With the problem framed as a classification problem, different approaches generally vary in how they perform the classification (equivalently prediction) task. At a high level, we can distinguish two broad approaches: model-based and data-driven.

\begin{itemize}
	\item \textbf{\emph{Model-based}} approaches may be based on rules designed with knowledge about how users behave or on interpretable probablistic models that can more flexibly capture uncertainties using interpretable parameters.
	The parameters of such models may be set heuristically or empirically derived from observed user data. By varying those parameters, different types of users can be simulated.
	\item \emph{\textbf{Data-driven} (or machine-learned)} approaches emphasize maximizing accuracy of fitting any observed real user data, without necessarily imposing interpretability. Almost all such approaches are based on supervised machine learning, notably using deep neural networks which can learn effective, but non-interpretable representations from the data for predictive modeling.
\end{itemize}
\noindent
These two families of approaches may also be combined, e.g., by utilizing model-based techniques to compute effective features for data-driven approaches or employing machine-learned models in specific components of model-based approaches.
However, interpretability is desirable when building user simulators for evaluation to ensure that evaluation results are meaningful and to allow for the testing of verifiable hypotheses. Hence, this book primarily focuses on interpretable model-based approaches.

\subsection{Uses of Simulation}
\label{sec:intro:simulation_uses}

In general, user simulation has many uses, including at least the following:
\begin{itemize}
	\item Performing large-scale automatic evaluation of interactive systems (i.e., without the involvement of real users).
	\item Gaining insight into user behaviour to inform the design of systems and evaluation measures.
	\item Analyzing system performance under various conditions and user behaviours (answering \emph{what-if} questions, such as ``What is the influence of X on Y?'').
	\item Augmenting data with human feedback and generating synthetic data with the purpose of training machine learning models and addressing data scarcity or privacy concerns. More broadly, user simulation can facilitate machine learning approaches that require human input (interactive learning, reinforcement learning, or human-in-the-loop systems).
\end{itemize}
We note that all these uses require similar techniques, but our focus is on evaluation. 

\subsection{Requirements and Desiderata}
\label{sec:intro:simulation_requirements}

When utilizing user simulation for system evaluation, it is critical that simulators provide reliable and insightful assessments.
Two essential properties that ensure this are \emph{validity} and \emph{interpretability}.
\begin{itemize}
    \item \textbf{Validity}: Simulated users must exhibit behaviours that align with empirical observations of real user behaviour in similar contexts. This includes both high-level strategies (e.g., information seeking patterns) and low-level actions (e.g., clicking behaviour). Without validity, the insights gained from simulation cannot be trusted.
    \item \textbf{Interpretability}: While not strictly a requirement, interpretability is a highly desirable property. Interpretability means that the simulated behaviour can be understood and adjusted through controllable parameters. This allows researchers to (1) understand why the simulator produced certain behaviours and (2) investigate how changes in specific parameters  influence the behaviour of users. In general, as user behaviour and preferences vary significantly across users, interpretability is needed to facilitate interpretation of the evaluation results generated by user simulation, i.e., to understand what kind of real world users can be expected to produce similar results.
\end{itemize}
However, while striving for high validity is important, simulation does not need to be perfect in order to be useful.
For example, although the relevance judgments in almost all the test collections for information retrieval evaluation are incomplete, the conclusions about relative performance of different retrieval systems tend not be affected much by the approximation made in a test collection~\citep{Voorhees:2000:IPM}.
In fact, creating a ``perfect'' user simulator, i.e., one that flawlessly replicates human behaviour across all possible tasks and contexts, is likely an AI-complete problem, on par with achieving Artificial General Intelligence (AGI, cf. Section~\ref{sec:broader:agi}).
When evaluating systems, we are often interested in a \emph{relative comparison} between them with regards to some measure of utility, which is a weaker requirement than quantifying the \emph{actual utility} of technology (in terms of some measurable impact, such as enhanced productivity or user satisfaction).
Nevertheless, the practical utility of user simulation for relative comparisons lies in its \emph{sensitivity}: the better an evaluation can distinguish between systems, the more practically useful it is.  

While both validity and interpretability are desirable, there often exists a trade-off between the two. Data-driven (machine-learned) simulators, trained on large datasets of real user behaviour, can often achieve high predictive accuracy, capturing complex patterns and nuances in user actions. However, this predictive power comes at the cost of reduced interpretability. The internal workings of these models, often involving complex neural networks or ensemble methods, can be opaque and difficult to understand. This makes it challenging to pinpoint the specific reasons behind a simulated user's behaviour or to adjust the model's parameters in a controlled manner. 

Beyond the essential requirements of validity and interpretability, there are several other desirable properties that can enhance the realism of user simulation.
\begin{itemize}
    \item \textbf{Cognitive plausibility}: The decision-making processes underlying simulated user behaviour should be grounded in theories or models of human cognition, ensuring that the simulated actions are not arbitrary or random.
    \item \textbf{Variation}: While reflecting general user behaviour patterns, simulated users should also exhibit variability and occasional outliers, ``not replicating average behaviour completely''~\citep{Bignold:2021:Biomimetics}. That is, simulation should reflect the unpredictable nature of real human interactions.
    \item \textbf{Adaptability}: Simulated users should be able to learn from their interactions with the system, update their expectations about the system and adjust their behaviour accordingly~\citep{Balog:2021:DESIRES}. 
\end{itemize}
By incorporating these desirable properties, user simulators can achieve a higher level of realism and sophistication, enabling more accurate predictions of user behaviour and more insightful evaluations of interactive systems.

\section{Aims and Organization}

With the emergence of various information access systems exhibiting increasing complexity, there is a critical need for sound and scalable means of automatic evaluation.  Simulation has the potential to offer a solution here.  It has attracted attention from multiple angles and much progress has been made in the last decade.  Relevant research work, however, has been scattered in multiple research communities, including information retrieval, recommender systems, dialogue systems, and user modeling.  This book aims to synthesize that research into a coherent framework.  Given the substantial amount of work performed within the context of information access systems, this is where our main focus will lie.

Specifically, our main objective is to discuss how simulation may be employed to undertake evaluation of information access systems in order to (1) estimate how well they will perform under various circumstances, and (2) analyze how performance changes under different conditions and user behaviours.
However, we attempt to make our discussions as generic as possible, such that those working on other types of interactive systems, or applications of assistive AI, would also find it useful. Specifically, whenever possible, we would attempt to lay out general conceptual frameworks, discuss general challenges, and extract general ideas from specific research work, which we hope to be broadly useful to more readers as well as provide a stable meaningful high-level structure where future research work can be naturally incorporated and discussed. 
Our emphasis on the generality of discussion, however, means that our treatment of any specific research work is inevitably brief. Detailed information can be conveniently found in the numerous research papers cited throughout this book. When selecting specific work to elaborate, we have also chosen to focus more on representative work that is useful for illustrating major ideas instead of having an even coverage of all the work. Due to the interdisciplinary nature of the topic and quick growth of research, the references cited in our book are inevitably incomplete. Considering this limitation and anticipating the rapid progress of research in this area in the future, we have created a website (\url{https://usersim.ai/}) for an envisioned broad interdisciplinary research community on user simulation. This platform aims to foster collaboration among researchers from diverse fields, enabling them to collectively maintain a repository of up-to-date and relevant references over time.
 
The main intended audience of this book includes both researchers, who wish to further the research and development of simulation-based evaluation methods, as well as industry practitioners, who are interested in employing these techniques in operational settings.
Considering the fact that user simulation is broadly connected with multiple fields (see a more detailed discussion of this in Chapter~\ref{ch:broader}), we also hope this book will be broadly useful to an audience beyond those interested in using user simulation for evaluation. 

The rest of this book is organized as follows.
We start in Chapter~\ref{ch:background} by providing a background on the development of simulation techniques within different research communities.
Following this, Chapter~\ref{ch:overview} gives an overview of how user simulation has been employed in the past. %
Chapter~\ref{ch:frameworks} introduces a conceptual framework for generally modeling interactions between a user and a system and evaluating any interactive system using user simulation. 
Next, in Chapter~\ref{ch:decisions}, we discuss decision-making and cognitive processes of users, followed by the mathematical framework we will employ in subsequent chapters to model these.
We present simulation techniques for search engines and recommender systems in Chapter~\ref{ch:sim_search} and for conversational assistants in Chapter~\ref{ch:sim_conv}.
Chapter~\ref{ch:practical} focuses on applying user simulation in practice, covering issues related to configuring, validating, and building simulators.
Chapter~\ref{ch:broader} broadens the perspective on user simulation as an interdisciplinary research area intersecting with various fields beyond computer science, ultimately contributing to the progress towards AGI.
Finally, Chapter~\ref{ch:concl} concludes the book by highlighting open issues and possible future research directions.
\chapter{Background}
\label{ch:background}

This chapter provides a historical account on evaluation methodology and simulation techniques, from the perspectives of four research communities: information retrieval (IR), recommender systems (RecSys), dialogue systems, and user modeling.
We highlight how these fields have focused on different but complementary areas of evaluation and user simulation.

\section{Information Retrieval}
\label{sec:background:ir_recsys}

Due to the empirical nature of information retrieval, researchers in the IR field realized the importance of reproducible quantitative evaluation of an IR system through reusable test collections as early as in 1960s when the Cranfield evaluation methodology was developed~\citep{Cleverdon:1968:Cranfield}. 
However, given the amount of interactive support today's search engines provide, the static nature of test collections makes it challenging to apply this methodology in such highly \emph{interactive} settings.
Recognizing this, industrial research labs often use A/B tests~\citep{Hofmann:2016:FnTIR}, where the idea is to interleave the results from alternative methods or algorithms, and observations made on real users' behaviour in their natural environment, if interpreted correctly, could then be analyzed to assess the overall effectiveness of the systems or methods to be compared.  However, a major limitation is that online experiments cannot be reproduced.
User simulation techniques address these limitations, and generalize the Cranfield evaluation methodology to enable evaluation of not only interactive search systems, but also recommender systems, and assistive AI systems in general.

Below, we briefly introduce test collection-based evaluation (Section~\ref{sec:background:test_collection}), followed by research on interactive IR systems (Section~\ref{sec:background:iir}) and conversational information access systems (Section~\ref{sec:background:cia}).
A high-level historical account on the use of simulation in IR is presented in Section~\ref{sec:background:sim_ir}.

\subsection{Test Collection-Based Evaluation}
\label{sec:background:test_collection}

Test collection-based evaluation~\citep{Sanderson:2010:FnTIR} is the primary evaluation methodology used for assessing the utility of information access systems. The methodology was originally developed by information retrieval researchers in the 1960s to quantitatively evaluate information retrieval systems~\citep{Cleverdon:1968:Cranfield}. The methodology is often referred to as the Cranfield evaluation methodology.\footnote{https://en.wikipedia.org/wiki/Cranfield\_experiments} A test collection includes three components: (1) a set of documents, (2) a set of queries, and (3) relevance judgments of documents in the collection with respect to each query made by human assessors. A retrieval system would be evaluated by running the system to process each query on the document set to produce retrieval results. The relevance judgments could then be used to compute various measures for assessment of the quality of the retrieval results. 
For example, Precision and Recall are two basic measures proposed for evaluating a set of retrieved documents, with Precision indicating the percent of retrieved documents that are relevant according to relevance judgments, whereas Recall the percent of all the relevant documents in the collection that have been retrieved~\citep{Cleverdon:1968:Cranfield}. Other measures such as F1~\citep{VanRijsbergen:1979:book} can be used to give a single number summary of precision and recall (harmonic mean of Precision and Recall). A ranked list can be evaluated by using Average Precision~\citep{VanRijsbergen:1979:book,Harman:1995:IPM} and Normalized Cumulative Discounted Gain (NDCG)~\citep{Jarvelin:2002:TOIS}. The test collection-based evaluation methodology has since become the standard and been widely used in not only in the IR community but also many other communities for evaluation of virtually all kinds of empirical tasks, notably those in natural language processing, computer vision, and machine learning, and measures such as Precision, Recall, F1, Average Precision, and NDCG remain the main measures used today for evaluating IR systems, as well as many other intelligent systems in general~\citep{Burnell:2023:Science}.

The test collection-based evaluation can be regarded a simple form of evaluation using simulation where the simulated users are implicitly specified with many unrealistic assumptions made about the users. Despite that those strong assumptions made are not valid, the methodology has been shown to be reasonable and even reliable for relative comparison of retrieval systems based on studies in the context of TREC~\citep{Voorhees:2005:book}.  Most of the discussions in our book are about how to generalize the test collection-based evaluation method by explicitly specifying the users to be simulated, simulating various kinds of user actions, and making more realistic user simulation in general.  Another dimension of extension is to introduce general frameworks for defining user simulation and evaluation measures that can be applied across a range of interactive information access systems, including search engines, recommender systems, and conversational assistants.

\subsection{Interactive Information Retrieval}
\label{sec:background:iir}

While IR tends to have a strong system focus, \emph{interactive information retrieval} (IIR) focuses more on users and how they interact with the retrieval system~\citep{Ruthven:2008:ARIS}. 
Already at the dawn of IR as a research field, \citet{Cleverdon:1968:Cranfield} pointed out physical and intellectual user effort as an important factor. \citet{Salton:1970:ISR} identified user effort measures, including the attitudes and perceptions of users, as important components of IR evaluation.
Early measures for IIR can be categorized around relevance, efficiency, utility, user satisfaction, and success~\citep{Su:1992:IPM}.
A comprehensive overview of IIR research methods and measures is presented in~\citep{Kelly:2009:FnTIR}.

Some of the most important research findings in IIR relate to the discrepancy between interactive and non-interactive evaluation results.
For example, systems that were shown superior to others in (non-interactive) batch evaluation failed to show improvements in interactive situations when used by real users~\citep{Hersh:2000:SIGIR,Turpin:2001:SIGIR}.
\citet{Turpin:2006:SIGIR} found no significant relationship between the effectiveness of a search engine, measured by Mean Average Precision, and real user success in a precision-oriented task.
\citet{Smith:2008:SIGIR} demonstrated that users can adapt their behaviour and can be just as successful with a degraded search system than with a standard one.

Over the years, several notable advancements have been made in formal models for IIR, including the probability ranking principle for IIR~\citep{Fuhr:2008:IRJ},
economic models~\citep{Azzopardi:2011:SIGIR,Azzopardi:2014:SIGIR},
the interface card model~\citep{Zhang:2015:SIGIR},
and the game theoretic framework for IR~\citep{Zhai:2016:MC}.

Compared to the great deal of work on static test collections~\citep{Sanderson:2010:FnTIR}, there has been limited research on test collections that model dynamic user interactions, which might include, e.g., a user's query clickthrough, query reformulation, and multiple queries in a whole session. However, several attempts have been made to go beyond static test collections, including the TREC Interactive track, where interactive query formulation was studied~\citep{Over:2001:IPM},
the TREC Query track, with multiple queries for the same topic~\citep{Buckley:1999:TREC}, the TREC Web track, with multi-faceted relevance judgments~\citep{Clarke:2011:TREC}, and the TREC Session track, with actual user interactions in a session history~\citep{Carterette:2016:SIGIR}. In addition to TREC, evaluation of interactive retrieval tasks has also been studied in the context of CLEF~\citep{Ferro:2019:Book} and NTCIR~\citep{Sakai:2021:Book}.

\subsection{Conversational Information Access}
\label{sec:background:cia}

Even though interactions between a user and a traditional search/recommender system are a form of conversation, the users of those systems have access to only a limited set of actions allowed by a system's user interface (querying, clicking, rating, etc.).
In contrast, in a truly conversational setting, the user can freely express their request and feedback in \emph{natural language}.
Since user intents are no longer captured as explicit actions, such as querying or clicks, they need to be inferred from free text, which requires more advanced natural language understanding capabilities.
Even in seemingly straightforward cases, such as yes/no questions, humans might answer indirectly.  For example, when asked about their interest in an evening activity, the user might respond with ``I’d prefer to go to bed'' instead of simply saying ``no''~\citep{Louis:2020:EMNLP}.
On the other hand, there is a much richer possibility for describing intent and providing feedback to the system via natural language.
Overall, conversational systems can reduce effort for the user from a cognitive perspective, but there is also more potential for frustration~\citep{Luger:2016:CHI}, as the user has less direct control over the behaviour of the system.

Conversational search has been recognized as a frontier of strategic importance within IR~\citep{Anand:2020:Dagstuhl,Culpepper:2018:SIGIRForum}.
Similarly, in the RecSys community, conversational interfaces to recommendations have gained research attention~\citep{Jannach:2021:CSUR,Gao:2021:AI}.
Given the close relationship between search and recommendation, the delineation of the two tasks in a conversational setting is even more challenging. Therefore, the term \emph{conversational information access}~\citep{Balog:2021:DESIRES} or \emph{conversational information seeking}~\citep{Zamani:2023:FnTIR} is used to represent a broader scope that encompasses the different conversational goals or subdomains (search, recommendation, and also question answering).

Another crucial factor that sets conversational information access apart from traditional search and recommender systems is the notion of \emph{initiative}.
The interaction in traditional search systems is driven by the user, i.e., the user issues a request (query) and the system responds with a ranked list of results, also known as the query-response paradigm.
In recommender systems, conversely, the system takes initiative by offering suggestions to the user based on their history.
\emph{Mixed-initiative} refers to a conversational paradigm where the user and system both actively participate in addressing the user's information need.
Mixed-initiative has been identified as a critical component of conversational information access systems~\citep{Radlinski:2017:CHIIR,Trippas:2018:CHIIR,Wadhwa:2021:DESIRES}.
It enables systems to engage with users in natural language, for example, to ask clarification questions, solicit feedback, or to learn about the user's preferences in specific contexts.
Despite the great potential of such system-initiated questions, evaluation challenges are a bottleneck to progress: since the next system response is conditioned on the user's answer, and the user is not constrained to a predefined set of possible answers, it is not possible to create reusable offline test collections for end-to-end evaluation~\citep{Balog:2021:DESIRES}.

The advent of conversational information access systems accelerated the need for offline evaluation methods that consider user interactions and has triggered interest in user simulation techniques~\citep{Balog:2021:DESIRES,Lipani:2021:TOIS,Salle:2021:ECIR,Zhang:2020:KDD,Salle:2021:ECIR,Sekulic:2022:WSDM,Owoicho:2023:SIGIR}.

\subsection{Simulation in Information Retrieval}
\label{sec:background:sim_ir}

The general idea of simulation was explored in IR many decades ago. For example, \citet{Blunt:1965:techreport} presented a simulation approach to measure the effect of various system configurations on cost and on response time. Interestingly, costs were not only associated with equipment but also with personnel.  \citet{Cooper:1973:ISR} used simulation 
to generate synthetic queries and documents in order to analyze the effect of changes in query characteristics on the number of documents retrieved, but the relevance of documents was not considered, a main limitations acknowledged in the study.
\citet{Griffiths:1976:PhDThesis} introduced a general simulation framework for IR systems and proposed to estimate its parameters from empirical data.
\citet{Tague:1980:SIGIR} employed user simulation in bibliographic retrieval systems.
\citet{SparkJones:1979:JD} and \citet{Harman:1992:SIGIR} studied the effectiveness of relevance feedback with the help of simulation.
\citet{Gordon:1990:JASIS} measured the effectiveness of document retrieval systems using of simulated queries.

It was in the first decade of the 2000s when simulation gained significant traction, mostly within researchers working on interactive information retrieval.
Simulation has been used for studying relevance feedback~\citep{Leuski:2000:CIKM,White:2004:ECIR,White:2005:TOIS,Keskustalo:2008:IR}, information filtering~\citep{Mostafa:2003:IR}, and online vertical selection~\citep{Diaz:2009:SIGIR}.
Another line of work has focused synthetic query generation~\citep{Jordan:2006:JCDL,Azzopardi:2006:SIGIR},
The SIGIR 2010 workshop on the Simulation of Interaction (SimInt)~\citep{Azzopardi:2011:SIGIRForum} represents a milestone event.
Some of the most influential work on simulation around this time focused on query formulation~\citep{Baskaya:2012:SIGIR}, scanning/examination/stopping behaviour~\citep{Turpin:2009:SIGIR,Baskaya:2013:CIKM,Maxwell:2015:CIKM}, and the relationship between cost and effort~\citep{Azzopardi:2009:SIGIR,Azzopardi:2011:SIGIR,Baskaya:2012:SIGIR}.
This was followed by a relatively quiet period, until a renewed interest in simulation has been triggered, in a large part due to recent research on conversational information access systems (cf. Section~\ref{sec:background:cia}). This renewed interest was also marked by the Simulation for IR Evaluation (Sim4IR) workshop at SIGIR'21~\citep{Balog:2022:SIGIRForum}.

It is important to realize that elements of user simulation have been prevalent in the field of IR, even if they have usually not been called as such---virtually all evaluation measures assume some sort of user model, and thus may be viewed as naive simulators (cf. Section~\ref{sec:frameworks:naive}). 
While not treated in this book, a broader view of simulation could go beyond simulating the users themselves to simulate their whole task environment, particularly the simulation of test collections~\citep{Hawking:2020:Book}.

\section{Recommender Systems}

Both search and recommendation address the problem of providing users with items that are estimated to be relevant to the user's information need, preferences, and/or context, often presented as a ranked list.

Early work on recommender systems focused on the problem of rating prediction: predicting a given user's rating for a given item based~\citep{Steck:2013:RecSys}. 
It was then realized that this is an oversimplified abstraction of the actual user task. Users are presented with a list of recommended items and must then decide which items to explore further, taking into account various factors such as personal preferences, contextual information, and item characteristics.
This top-$k$ recommendation task has since become the standard, adopting IR evaluation measures~\citep{Cremonesi:2010:RecSys,Bellogin:2011:RecSys}.
However, a major difference from IR, from an evaluation perspective, is that users provide ratings themselves; there is no need to collect relevance assessments by third-party raters.
Also, unlike in search, where users have the initiative, recommendations are driven by system initiative.  What items are shown to users as suggestions influences what choices they can make. These can lead to various biases.

Over the years, different types of interactive recommender systems have been proposed, including critiquing-based systems~\citep{Chen:2012:UMUAI}, user-controlled recommenders~\citep{Jannach:2017:ECWeb}, visual recommenders~\citep{Kunkel:2017:IUI,Verbert:2016:TOIS}, and conversational recommenders~\citep{Gao:2021:AI,Jannach:2021:CSUR}.
The evaluation of such interactive systems necessitates some form of human involvement, i.e., either real or simulated users.

One of the earliest works of using simulation for evaluation is by \citet{Dzyabura:2013:RecSys}, who propose mechanisms to combine search and recommendations and demonstrate the benefits of such approach via simulations.
Recently, simulation has been used for measuring potential long-term effects of interactions between users and the recommender system~\citep{McInerney:2021:RecSys,Yao:2021:arXiv,Stavinova:2022:arXiv,Hazrati:2022:IPM} as well as for developing evaluation measures specific to recommender systems that employ carousel-based result presentation~\citep{Felicioni:2021:UMAP,Dacrema:2022:FBD,Rahdari:2024:TORS}.
The SimuRec workshop held at RecSys'21~\citep{Ekstrand:2021:RecSys} is further evidence that the need for simulation techniques has been recognized within the RecSys community.
Overall, ``although the volume of simulation-based research in the recommender system context is gradually increasing, simulations are most frequently used to explore the impact of recommender systems on various aspects of user behaviour rather than to assess their performance and effectiveness in a comparative way''~\citep{Rahdari:2024:TORS}.

\section{Dialogue Systems}
\label{sec:background:ds}

The goal of task-based dialogue systems is to help the user accomplish some task, such as make a restaurant reservation or buy a product.
All modern task-based dialogue systems are based around \emph{frames}: a knowledge structure representing the intents of the user, along with a set of slots and values that specify what the system needs to know in a given domain to accomplish the task~\citep{Jurafsky:2023:book}. In the travel domain, for example, slots include origin and destination cities, departure date and time, airline, etc. The system's goal then is to collect enough information, i.e., fill the required slots, to perform the task intended by the user.
In a \emph{dialogue-state} (or \emph{belief-state}) architecture, the \emph{dialogue state tracker} component maintains the current state of the dialogue (the constraints expressed so far), while the \emph{dialogue policy} defines the conversational behaviour of the system, i.e., what the system should say or do next based on the entire dialogue state.
Manually designing robust dialogue strategies is very time-consuming and expensive, especially as systems move towards greater levels of complexity.
Hence, an alternative is to hand-written rules is to employ machine learning approaches to automatically learn an optimal strategy.

One of the most important and fundamental ideas from dialogue systems research is modeling human-computer dialogue formally as a Markov Decision Process (MDP)~\citep{Levin:2000:TSAP,Young:1999:PTRS}, which we will further discuss as a general formal framework for user simulation in the book (see Section~\ref{sec:decisions:math}).
Crucially, mapping dialogue onto the MDP model requires a well-defined representation of the state space and action set.
For slot-filling dialogues, a common approach is to base the states on the various slots that need to be filled and the actions on the intents of the participants.
Dialogue policy learning has been approached using two different forms of reinforcement learning: model-based and simulation-based~\citep{Schatzmann:2006:KER}.  The model-based approach ignores actual user responses and considers only system actions to estimate transition probabilities directly from training data~\citep{Singh:2022:JAIR,Walker:1998:COLING}.
However, the vast space of possible dialogue states and strategies cannot be sufficiently explored from a fixed corpus to obtain reliable estimates; also, new strategies cannot be explored as they are not guaranteed to be present in the training data~\citep{Schatzmann:2006:KER}.
Instead, a dynamic environment is needed that allows for a large number of ``trial-and-error'' interactions.
According to the simulation-based approach, a user model is trained based on the available dialogue corpus and typically using some form of supervised learning.  Then, reinforcement learning techniques can be used for finding an optimal dialogue policy based on interactions with a simulated user.
This way, any number of training episodes can be generated in order to explore the space of possible strategies, including those that are not present in the original training data.
Consequently, simulation-based learning has become the predominant form of learning dialogue strategies~\citep{Schatzmann:2006:KER,Young:2010:CSL}.

Our focus in this book is on using user simulation to evaluate the utility of information access systems. While simulation has also been used to evaluate different aspects of a dialogue system~\citep{Griol:2013:AAI}, conversely to information retrieval, this aspect has been much less studied.  Instead, task-based dialogue systems are most commonly evaluated in terms of absolute task success; this, however, is only possible if the task is unambiguous.
Compared to conversational information access systems, which support a range of search and recommendation tasks as well as exploratory information gathering, the slot-filling nature of dialogue systems is rather restrictive.
Conversational information access requires a more flexible and interactive approach to addressing user information needs, which are often unclear or ambiguous and cannot be conveyed in terms of a few predefined slots.
Furthermore, the evaluation of task-oriented dialogue systems, in general, and the development of automatic evaluation measures that correlate well with human satisfaction, in particular, represent an open area of research~\citep{Deriu:2021:AIR}. Consequently, simulation-based evaluation has so far employed only simplistic measures, such as success rate.
Nevertheless, there is a solid body of existing work in dialogue systems related to user simulation to build on, including fundamental modeling, system architectures, and modeling dialogue strategies (cf. Chapter~\ref{ch:sim_conv}).

\section{User Modeling}
\label{sec:background:user_modeling}

Conceptually, user modeling is needed in any system that attempts to serve a user, especially if it is desirable to make the service personalized to suit the needs of each individual user or optimize its interactions with an individual user. As interactive AI applications become increasingly feasible, user modeling is also attracting much attention with new research results regularly reported in multiple conferences such as ACM UMAP,\footnote{https://www.um.org/conferences} ACM IUI,\footnote{https://iui.acm.org/} ACM CHI,\footnote{https://chi.acm.org/chi-series/} and the IR-focused CHIIR.\footnote{https://chiir.org/} From the perspective of user modeling, user simulation can be regarded as developing a complete and operational user model, which makes work on user modeling highly relevant to the discussion of user simulation. 

One way of categorizing models is to differentiate between descriptive and formal models.
\emph{Descriptive models}~\citep{Bates:1989:OR,Ellis:1989:JD,Ingwersen:2005:book,Kuhlthau:1988:RQ,Pirolli:1999:PR} provide reasoning and (post-hoc) explanation behind user behaviour. These models can help gain insights into the information seeking process and inform high-level design; we discuss them in more detail later, in Section~\ref{sec:decisions:conceptual}.
They, however, lack predictive capabilities; ``we can neither use these models to investigate what is likely to occur in different instantiations of a retrieval system; nor can we use them for simulating user behaviour''~\citep{Camara:2022:ECIR}.
\emph{Formal models}~\citep{Azzopardi:2011:SIGIR,Baskaya:2013:CIKM,Carterette:2011:CIKM,Fuhr:2008:IRJ} are expressed mathematically and have predictive power about why users behave in a certain way.
Such models have been explored in the field of Interactive IR (see Section~\ref{sec:background:iir}) and can be used as the basis for simulating user interactions.

Within the broader field of user modeling, the most relevant lines of work for include the following:
\begin{inlinelist}
    \item \emph{Search tasks and intent}, including, e.g., typology of search tasks~\citep{Marchionini:2006:ACM}, taxonomy of Web search~\citep{Broder:2002:SIGIRForum}, and taxonomy of e-commerce queries~\citep{Sondhi:2018:SIGIR}.  These studies provide a basis for studying simulation of different kinds of search tasks and intent.
    \item \emph{Information seeking models}, including, e.g., exploratory information seeking models~\citep{Ellis:1989:JD,Meho:2003:JASIST}.  Such models provide guidance on high-level workflow of a simulated user when performing an information seeking task.
    \item \emph{Cognitive models of users}, which attempt to model the knowledge state of a user; for example, the ASK theory~\citep{Belkin:1982:JD} and \citeauthor{Ingwersen:1996:JD}'s cognitive IR model~\citep{Ingwersen:1996:JD}.
    \item \emph{Contextual user modeling}~\citep{Ingwersen:2005:book}, which takes a broader view of the context that may affect a user's information seeking behaviour.
    \item \emph{Click models}~\citep{Chuklin:2015:Book} that can be directly used in a user simulator to simulate a user's clicking behaviour.
    \item \emph{Economic IR models}~\citep{Azzopardi:2011:SIGIR,Azzopardi:2014:SIGIR}, which provide guidance on how users make decisions during an interactive retrieval process and may also be directly used in a user simulator. 
    \item \emph{Choice modeling}~\citep{Hazrati:2024:UMUAI}, which leverage interpretable user simulators to simulate and analyze the overall behaviour of a community of users when interacting with a recommender system over a long time period. 
\end{inlinelist}

As user simulation is essentially to define a computational model for a user, the discussion of user modeling will appear throughout the book. However, as a distinct research topic, user modeling plays an increasingly crucial role as we strive to optimize the collaboration between AI systems and human users. This is because the objective function to be optimized by an AI system in assistive applications must formally describe the user with whom the AI system collaborates. So far, however, the user simulation techniques only utilize a limited portion of research results on user modeling, leaving much room for future work on incorporating more user modeling research results into user simulation.

User modeling has also often been used as a way to study dynamic behaviour of a whole community of users interacting with a system. For example, the overall impact of a recommender system on a community of users under certain assumptions about user behaviour has been studied in some work~\citep{Yao:2023:ICML,Hazrati:2024:UMUAI}. In such studies, interpretable (often simplified) user models are often used to enable the simulation and analysis of how a system interacts with a community of users over time. This can be viewed as modeling the collective behaviour of a user community within a social environment, based on the individual models of each user. Our book is focused on discussing simulation of individual users in the context of their interactions with an interactive information access system.
\chapter{Overview of User Simulation}
\label{ch:overview}

In this chapter, we provide a high-level broad overview of research on user simulation with a discussion of the different motivations and applications of how user simulation has been employed in the past.
Specifically, we identify three different purposes of studying user simulation:
(1) \emph{analysis} of user behaviour in specific applications,
(2) automatic \emph{evaluation} of systems,
(3) synthetic \emph{data generation} for training models (e.g., using reinforcement learning).
While our main interest in this book lies in (2), it is worth pointing out that the first two categories share a lot of similarities in terms of modeling user behaviour in an interpretable way; in contrast, the interpretability aspect may be less critical for the third category of application.
Moreover, as the research in the first two categories has historically been done mostly in the context of search engine applications, our discussion in the rest of this chapter will also be mostly about search engines, though we will also include a brief discussion of recommender systems and conversational assistants, which have attracted much attention recently.
Specifically, in the rest of the chapter, we will first discuss the different motivations for studying user simulation, roughly corresponding to different benefits of user simulation. We then provide an overview of the research done so far in the context of three representative information access systems, i.e., search engines, recommender systems, and conversational assistants. We end the chapter with a discussion of the different applications of user simulation, including particularly the use of simulation for evaluation, which is the main focus of this book.

\section{Motivations for User Simulation}

User simulation has been studied from different perspectives, often with a somewhat different motivation. \\

\parheading{Analysis of user behaviour.}
The study of user simulation has often been done in the context of studying user behaviour, where user simulation can provide multiple benefits, such as formally defining a hypothesis about a user's behaviour, modeling variations of user factors and their influences on user behaviour, and counterfactual analysis of impact of system components on user behaviour.
In general, studying how various factors influence user interaction is important for both understanding users and optimizing the design of user interfaces and algorithms to serve users. For this reason, the study of user behaviour in interactive systems has been the subject of decades of user-oriented IR research (cf. Section~\ref{sec:background:user_modeling}).
However, a major challenge in user studies is that the experimental conditions can introduce variables that are difficult to control (e.g., users' background, domain knowledge, or perseverance), but they can affect how users interact with an information access system. Such uncontrolled variables can be explicitly modeled in user simulations to enable more accurate modeling of a user's real behaviour.

Another benefit of using user simulation to analyze user behaviour is that a simulator articulates an explicit testable hypothesis about the behaviour in its user model. Such hypotheses can be tested by comparing data generated by the simulator against real data, e.g., user interaction logs.
The main outcomes of such analyses are insights regarding how users interact with a system.
These insights can help to develop more accurate user models, which in turn can inform the design of more realistic evaluation measures.

Moreover, simulation also allows for examining the effects of changes made to various system components on anticipated user behaviour.
It facilitates a more comprehensive assessment than what a live environment would typically allow (e.g., estimating long-term engagement/satisfaction of different user populations that can be controlled or varied in a counterfactual manner).
These insight-gaining type of simulations can answer questions like: \emph{What has the greatest influence on X?} or \emph{How will X and Y interact?}.

Research on user simulation from the perspective of user studies has led to many useful findings about users that provide insights about how to design a user simulator. Sometimes, a specific finding (e.g., typical lengths of user queries) may be directly used in a user simulation algorithm (e.g., query generation), but in many cases, it remains a technical challenge how to turn those findings into effective formal models or algorithms for user simulation. \\

\parheading{Evaluation of interactive systems.} 
The study of user simulation was also motivated by the need for using reproducible experiments %
to evaluate an interactive information access system. 
While it is always desirable to use real users for evaluating an interactive system, it is hard to conduct reproducible experiments to fairly compare two systems or evaluate the impact of any component in a system. It is especially difficult to control for the fact that once a user interacts with a system to perform a task, the user's cognitive state changes. The next time the same user might not interact the same way with that system, or with any other system, when performing the same or related tasks. It is also costly to use many real users to conduct large scale experiments. Building user simulators that can realistically simulate real users offers an attractive solution to this problem, since a simulated user can be ``reused'' again and again in a controlled manner to ensure the reproducibility of experiments without any bias. Moreover, when a simulated user has a meaningful user parameter, it would also allow for varying the parameter to simulate user variations. %
Simulations with the explicit focus of \emph{evaluating systems} enable exploration of changes in system performance under different conditions or behavioural patterns (``what-if'' analyses).

Research on user simulation in the evaluation context naturally emphasizes the development of algorithms for generating synthetic user behaviour; in contrast, the research of user simulation from the perspective of understanding user behaviour is not restricted by the requirement of describing a user's behaviour formally.\\

\parheading{Implicit feedback (for machine learning).} Research on user simulation may also be motivated by the need to learn from user feedback to optimize a ranking or recommendation algorithm. For example, an effective ranking/recommendation algorithm would essentially attempt to accurately estimate the likelihood that a user would click on an information item. If the estimated clicking probability is high, the item should be ranked high on the result/recommendation list. This connection means that optimizing an algorithm and accurately modeling a user's behaviour are in some sense the two sides of the same coin. As such, the models developed for optimizing ranking or recommendation can often be viewed as models for modeling a user's clicking behaviour. Research on user simulation from this perspective generally uses sophisticated machine learning algorithms to combine multiple factors (predictors) that can all potentially help improve the accuracy prediction, which is appropriate for the purpose of optimizing an algorithm in a system, since the system does have access to all the information. However, from the perspective of simulating a real user, many of those factors cannot be assumed to be available to a user when choosing an action to interact with a system. Therefore, a click model optimized for a system may not be an optimal model for simulating a real user, but many techniques developed for optimizing ranking may be adapted for modeling a user's clicking behaviour.\\

\parheading{Privacy protection.}
Study of user simulation may also address confidentiality concerns. Specifically, simulation enables us to experiment with, and evaluate, systems where the data and/or interactions are sensitive or confidential~\citep{Hawking:2020:Book}. There are many examples of applications where privacy needs to be protected, including, e.g., medical record search, email search, and enterprise search. In particular, simulation of Electronic Medical Records (EMRs)~\citep{Moniz:2009:JMIR} has been suggested and used in practice (see, e.g., MDClone\footnote{https://www.mdclone.com/}).  
If we have faith that a simulation accurately represents the salient aspects of some confidential setting, we can use that simulation for evaluation without ever seeing real data. Of course the accuracy of simulation still needs to be validated using the real data, but that validation can be performed by people who have access to the data. \\

\parheading{User interface design.}
Finally, user simulation can also benefit user interface (UI) design as simulations would enable experimentation with changes that we do not want real users to see (yet), such as changes to the user interface of a running system. It is hard to make large-scale UI changes, since people are generally reluctant to learn new tools or even new interaction patterns, and the cost of increasing the cognitive burden on the real users due to change of an interface may be as high as encouraging the user to switch to a different system. Interface designers might also be reluctant to expose a new idea prematurely to real people, which may be another reason why we do not want real users to test a new interface feature. 
Simulation could potentially help address these challenges. For example, a carefully designed simulation may help assess the expected utility of a new interface, especially with the use of interpretable user simulators~\citep{Azzopardi:2011:SIGIR,Chi:2003:CHI,Fu:2007:HCI}.

\section{Applications of User Simulation}

As discussed briefly earlier in this chapter, user simulation has been studied from different perspectives and with somewhat different motivations. In this section, we discuss those different applications of user simulation, including particularly the application of user simulation in evaluation, which is the focus of our book. We include a broad discussion of all those different applications for both completeness of treating the topic of user simulation and the connection of the work motivated from different application perspectives. Specifically, many techniques and frameworks that we will discuss in later chapters of the book are also relevant to other applications of user simulation.

\subsection{User Behaviour Analysis}

Since a user simulator must be mathematically well defined, it serves as a formally specified hypothesis about a user's behaviour. The benefit of a formally specified hypothesis is that the hypothesis can then be tested rigorously to examine its validity. This can be done by (1) designing focused user studies to examine to what extent the generated user behaviour of a formal user simulator match those of a real user or (2) fitting a formal user model to the observed real user data such as a search log to obtain estimates of the interpretable parameters in the formal model.
While the user behaviour that can be studied in this way is limited to what can be modeled formally, this kind of research of user behaviour, which may be called ``computational user modeling,'' should presumably be our long-term goal in fully and precisely understanding a user. In this sense, advancement of user simulation generally should lead to improved understanding of users.

An additional benefit of using user simulators to study users is that we can introduce meaningful user parameters to model latent states of users. For example, in the case of search engine interactions, a user's knowledge background or patience in interacting with results can all be potentially modeled with interpretable parameters that can be varied to model natural variations of users. The variation of a model also makes it possible to formally characterize the variations of real users as reflected in the observed user data such as search logs; for example, by fitting such a parameterized model to a search log, we may be able to learn the optimal parameter values that can best explain our data. Those learned parameter values can then be interpreted as a quantitative description of user variations. The interpretability of parameters in user simulation models can also directly facilitate the study of any hypothesis regarding a user behaviour related to the parameter. That is, the parameterized model can be used as a predictive model to predict a user's behaviour in a particular context, which can then be compared with the real user behaviour that we can obtain by designing appropriate user studies with the assumed context.

Lastly, as a valuable tool for user studies, a formal user simulator can facilitate various comparative analyses in association with other relevant user variables. For instance, by partitioning a large user population based on interesting variables (e.g., age, occupation, or geolocations), user simulation models can be employed to conduct comparative analyses among different subgroups to understand their commonality and differences. Similarly, any variable related to information access systems can also be varied to examine how a user's behaviour may vary depending on system-related variables; for example, a user's behaviour may be different depending on the collection of information items available for search (e.g., product search may be different from library search).

We should note that while we emphasize the benefit of user simulation for user studies here, it is equally important to recognize the inherent dependence of user simulation on user studies. Specifically, only when we base a user simulation model on sound findings of user studies would we be able to develop realistic interpretable simulation models. In this sense, those machine learning-based user models tend to have limited benefit for user studies despite the higher accuracy that they may show in mimicing a user's behaviour. Interpretability is an important goal that we must strive to achieve in user simulation in order to facilitate user studies.

It is also interesting to note that the emergence of large language models (LLMs) opens up new opportunities of creating user simulators that can exhibit human-like behaviours especially in natural language conversations~\citep{Aher:2023:ICML}. While such LLMs are not interpretable, they have been shown to replicate some population behaviour based on some Turing Experiments~\citep{Aher:2023:ICML}. There is great potential to leverage LLMs for user simulation (see Chapter~\ref{ch:broader}) with applications beyond evaluation (e.g., simulating complex social systems~\citep{Gao:2024:WWW}). 

\subsection{Evaluation of Interactive Systems}

Since the benefit of user simulation for evaluating information access system is a central topic of the book, we will be brief here in discussing this direction of application of user simulation and will just highlight a few major points.

First, it is our view that user simulation may be the only way to enable reproducible experiments for evaluating any interactive system that attempts to support a user in finishing a task. This is because the overall utility of such a system needs to be assessed based on a sequence of interactions in a whole task session, hence there must be a user involved in the loop for evaluation. However, a real user cannot be ``controlled,'' making it impossible to reproduce any experiments with humans. Our only option thus appears to be using a simulated user that we can control. The evaluation of any system can then be done by using a simulated user to interact with a system and measuring the overall utility based on the interaction data in a whole session. This view also makes us believe that although our focus is on information access systems, most of the discussion in the book may also be relevant to evaluation of any interactive intelligent systems (i.e., broadly interactive assistive AI systems).

Second, virtually all the existing test collections (for both IR and recommender systems) can be regarded as an evaluation method based on some kind of simplified user simulators (see Chapter~\ref{ch:frameworks} for more discussion of this). A major limitation of those existing collections is that they cannot be used to evaluate an {\em interactive} system. Given the research progress already made in user simulation, we have the opportunity to address this limitation by systematically converting existing test collections into new collections with a wide range of user simulators trained based on the existing user data, thus enabling automatic evaluation of interactive search and recommender systems (see Section~\ref{sec:concl:embracing}). 

Finally, the application of user simulation for evaluation is closely related to its application to user studies in that in both cases interpretable user models are required in order for the applications to be successful. The requirement of interpretability in the case of user studies is obvious, but it is also required for evaluation, otherwise the evaluation results would not be so meaningful. With intrepretability, user simulation for evaluation would not only be meaningful but also be powerful in generating simulated users with different characteristics to allow for evaluation with different kinds of users. The integration of research of user studies and the development of formal interpretable user models is therefore highly desirable and may reinforce each other in accelerating research in both areas. Indeed, we envision a healthy ecosystem where user studies would lead to more realistic user simulators, which further enable the development of more effective interactive systems (e.g., with new functions and new interface features), leading to new opportunities of studying user behaviour in a new system environment (with new functions or new interfaces), and new generations of user simulators being developed.

\subsection{Synthetic Data Generation}

In addition to analysis of user behaviour and evaluation of systems, user simulation also has applications in synthetic data generation to create/augment training data for supervised machine learning, which tends to require large quantities of labeled examples of training data. Because manually annotating a data set is labor-intensive and costly, using user simulation to generate synthetic training data is appealing since it does not involve manual effort. Compared with other approaches to synthetic data generation, user simulation enables the generation of more interpretable synthetic data when the parameters used in the simulators are interpretable. Due to the scarcity of interaction data available (e.g., because of issues such as the privacy), it is especially beneficial to leverage user simulation to generate interaction data, which can then be used for training reinforcement learning algorithms for learning an effective interaction policy.
Work on generation of synthetic data includes, but not limited to
\begin{inlinelist}
	\item modeling clicks and interactions between search results~\citep{Zhang:2019:CIKM},
	\item training spoken dialogue systems~\citep{Schatzmann:2006:KER,Schatzmann:2007:NAACL,Young:2010:CSL},
	\item generating synthetic conversations by leveraging large pre-trained language models~\citep{Mohapatra:2021:EMNLP,Kim:2021:ACL,Wan:2022:EMNLP},
    \item  generating a population of synthetic user profiles based on real user interaction data with items (simulation of tourists)~\citep{Merinov:2023:IIR},
	\item removing bias from RL-based training of recommender systems~\citep{Huang:2020:RecSys}, and
	\item deep reinforcement learning for IR~\citep{Zhang:2021:SIGIR} and RecSys~\citep{Chen:2021:arXiv}.
\end{inlinelist}

In this book, we focus on reviewing the second line of applications, i.e., using user simulation for evaluation. However, the techniques we will discuss for constructing user simulators are directly applicable for using user simulation to generate synthetic data.

\section{User Simulation for Search Engines}
\label{sec:overview:analyzing:search}

So far, most existing work on user simulation has been done in the context of search engines. In this section, we provide a detailed overview of this body of work. In the next two sections, we will provide a brief overview of user simulation in the context of recommender systems and conversational assistants, respectively, which reflects more what the recent research trend is.

The classic problem in information retrieval, \emph{ad hoc document retrieval}, is to return a ranked list of documents from a collection in decreasing order of their estimated relevance to a user-provided search query.
Typical system-oriented evaluation assumes a persistent scanning of a long list of search results, where the user reads each document before making a judgment about its relevance.
In practice, the ways in which users interact with text-based information retrieval systems when conducting searches are a lot more complex.
First, results are presented on a \emph{search engine results page} (SERP), with features designed to help people find information faster.
One of those features is \emph{snippets}, which provide a ``preview'' of each search result by including the title and URL of the website, and a textual excerpt from the document that highlights its relevance to the search query.
A searcher typically assesses the result snippet before clicking on them to read the full document.
When a user sequentially browses a document list, the probability of stopping at a lower position is generally higher than stopping at a higher position; in practice, few users scan past the second results page~\citep{Markey:2007:JASIST}.
It might also happen that the searcher abandons the SERP based on an initial impression, without inspecting any results in depth.
As a matter of fact, users often need to reformulate their queries to find what they are looking for.
Over the course of the search session, the user may learn by reading snippets and inspecting documents, which might shape their view of relevance.
The searcher will also at some point make a decision to stop their interaction with the search engine.
Stopping may occur for a variety of reasons, for example, the searcher has satisfied their information need, has been frustrated with the system, or has run out of time.

There is a rich body of research on aspects that can influence the behaviour and performance of searchers.
Below, we present a selection of studies around three main themes that have been extensively investigated in the past: formulating queries, examining search results, and modeling searcher effort.

\subsection{Formulating Queries}

Formulating effective queries is key to finding relevant information.
Searchers often need to modify their queries and try several variations before their information needs are fulfilled.
Query formulation behaviour is therefore studied on the session level, where a \emph{search session} is defined as ``a series of interactions by the user toward addressing a single information need''~\citep{Jansen:2007:JASIST}.
A search session typically lasts until either the information need is---at least partly---satisfied or users ``run out of time or ideas for a new query''~\citep{Baskaya:2013:CIKM}.
In simulations this is realized by assuming that the searcher is modifying queries up to a maximum number of times~\citep{Keskustalo:2009:AIRS} or up to an allowed time limit~\citep{Verberne:2015:ECIR}.
There is a cost associated with initial query formulation and subsequent reformulation as well as with examining search results.
These costs may affect what \emph{search strategy} a user would employ.
For example, they may choose to issue several short queries and assess only results at the top ranks (as typical of web search~\citep{Jansen:2000:IPM}) or rather carefully construct a single verbose query and examine the results at depth (as in Cranfield-style experiments performed at TREC~\citep{Voorhees:2005:book}).
For the above reason, query formulation cannot be meaningfully studied in isolation, without making some assumptions about how users would examine the results.
We shall discuss \emph{scanning strategies} in detail just in a bit, but in short, they determine the number of result snippets the user reads, before formulating the next query or ending the session.

Naturally, short queries are cheaper to formulate (i.e., users have to type less) and subsequent reformulations can be done in small increments at a minor cost.
To test whether web-like short queries could be effective for a TREC-type test collection, \citet{Keskustalo:2009:AIRS} compare four different session strategies.
The first three strategies involve multiple queries and users scanning the top-10 results.
Specifically: (S1) strictly single-word queries, (S2) incrementally longer queries, and (S3) variations of two core search terms plus a third term.
Under these three strategies, users are assumed to issue at most five queries and stop once a relevant document is found.
The fourth strategy, (S4), consists of a single verbose query, with the top-50 results examined.
Results show that multi-term query sessions (S2) and (S3) are successful, and are on par with or close to a single verbose query (S4), depending on the relevance criteria.
However, single-term query sessions (S1) are clearly inferior to the other strategies.

\citet{Azzopardi:2009:SIGIR} studies the relationship between user effort, in terms of query length, and utility, measured with standard retrieval metrics (e.g., average precision).
Under the Principle of Least Effort~\citep{Zipf:1949:Book}, it is assumed that the user aims to minimize their cost/effort while maximizing their expected utility.\footnote{\citet{Labhishetty:2022:ICTIR} propose and study a unified optimization framework for modeling both initial query formulation and subsequent reformulations based on this principle.}
An economic analysis of the productivity of querying is carried out using a set of simulated query-document pairs, which are produced using a generative probabilistic model (cf. Section~\ref{sec:sim_search:queries}).
The empirical results show that queries of length between two and five are the most effective---this closely matches the lengths of queries observed from real users. %

Assuming that a set of individual search terms are available for each query topic, \citet{Baskaya:2012:SIGIR} present five \emph{query modification strategies}, which define how to form a sequence of queries, based on term level changes that are grounded in observed real-world user behaviour.
These are:
(S1) single-word query with the search word repeatedly replaced ($w_1 \rightarrow w_2 \rightarrow w_3$),
(S2) an initial two-word query, followed by queries repeatedly changing the second word ($w_1, w_2 \rightarrow w_1,w_3 \rightarrow w_1,w_4$),
(S3) an initial three-word query, followed by queries repeatedly changing the third word ($w_1, w_2, w_3 \rightarrow w_1,w_2,w_4 \rightarrow w_1,w_2,w_5$),
(S4) an initial one-word query, followed by adding one word to each subsequent query ($w_1 \rightarrow w_1,w_2 \rightarrow w_1,w_2,w_3$),
(S5) an initial two-word query, followed by adding one word to each subsequent query ($w_1,w_2 \rightarrow w_1,w_2,w_3 \rightarrow w_1,w_2,w_3,w_4$).
Among these, S2 (two-word queries) and S3 (three-word queries) are found to be the most effective under strict time constraints; when more time is allocated, strategies with incremental extension (S4 and S5) catch up.

Query suggestion is a key functionality of modern search engines to assist users with a selection of possible follow-up queries.
\citet{Verberne:2015:ECIR} use simulation to decide which query the user would select from a list of suggestions.
Suggestion for terms to be added to the current query are extracted from documents that the user clicked on in the current search session.
The main finding of the study is that query suggestion is not universally beneficial for all user types; its effectiveness depends on how persevering the user is in examining results and how critical in selecting suggested queries.

\subsection{Examining Search Results}

When inspecting the returned result list, users need to make a decision based on the result snippet as to whether they are interested in viewing the full document.
If the snippet is deemed irrelevant, a relevant document may never get examined.
\citet{Turpin:2009:SIGIR} investigate the effects of including the snippet examination stage in the retrieval process and analyze how that might alter the relative ordering of different retrieval systems. %
It is observed that unless snippets are extremely accurate in how they reflect the content of the underlying document, they have a significant impact on system ordering.
\citet{Turpin:2009:SIGIR} therefore argue that snippets should be judged explicitly, in addition to documents, as part of the batch evaluation paradigm in order to model reality. However, there has been little work on actually modeling a user's interaction with snippets presumably due to the lack of such data sets. 

The \emph{scanning strategy}, also called a \emph{Browsing Model}~\citep{Carterette:2011:SIGIR, Moffat:2013:CIKM, Moffat:2022:SIGIR}, defines how the searcher interacts with the result list. Since the perceived utility of any search results is affected directly by how a user interacts with the results, it is not surprising that the browsing model has played an important role in defining evaluation metrics. Early work on evaluation measures has not explicitly specified a browsing model, making it unclear what assumptions have been made about users exactly. In the work ~\citep{Carterette:2011:SIGIR}, the authors suggested a framework for examining an IR measure that consists of three components, including a browsing model, a model of document utility, and a model of accumulation of utility, allowing for systamtic comparison of different measures. In a series of work by Moffat and coauthors~\citep{Moffat:2008:TOIS, Moffat:2013:CIKM, Moffat:2017:TOIS, Moffat:2022:SIGIR}, the browsing model is formally modeled with a probabilistic model, which enabled to derive a discounted co-efficient based on a browsing model when aggregating relevance information gain over search results. Specifically, with sequential browsing, the probability for a user to visit a result down on the list would generally be lower than the probability of visiting one ranked higher, leading to a natural discount of contribution of relevant information by a lowly ranked document as compared with a highly ranked one. The proposed frameworks, including the initial framework C/W/L~\citep{Moffat:2013:CIKM} and the extended one, C/W/L/A~\citep{Moffat:2022:SIGIR} provide a principled way to define evaluation measures based on an interpretable user model, including particularly an explicitly specified browsing model.  

In terms of specific browsing model, a commonly assumed scanning behaviour is \emph{persistent scanning}, which means that the user examines all results in the specified range (rank position $k$), from top to down, in sequential order~\citep{Baskaya:2013:CIKM}.
Often, scanning behaviour is considered from the opposite perspective of \emph{stopping behaviour}: determining when the user would terminate the search~\citep{Kraft:1979:IPM,Maxwell:2019:PhDThesis}.
Persistent scanning corresponds to stopping at a \emph{fixed depth}, regardless of the relevance of results.
This stopping strategy is encoded in several IR evaluation measures, for example, Precision@k and Normalized Discounted Cumulative Gain (NDCG).
Clearly, fixed depth stopping is a simplified model of user behaviour.
Intuitively, the user would decide when to stop scanning results based on their relevance.
\citet{Maxwell:2015:CIKM} explore how the performance of ad hoc document retrieval changes under different stopping strategies.
For example, the searcher would stop once they have observed a given number of non-relevant snippets (\emph{total non-relevant}) or a given number of non-relevant snippets in a row (\emph{contiguous non-relevant}); cf. Section~\ref{sec:sim_search:stopping}.
A comparison of a total of six different stopping strategies in terms of mean Cumulative Gain reveals that the aforementioned two strategies perform best~\citep{Maxwell:2015:CIKM}.
However, somewhat surprisingly, the baseline stopping strategy (fixed depth, with a well-chosen depth threshold) performs remarkably well, and has not been significantly outperformed by more advanced methods.

In practice, users do not jump straight to examining result snippets as dictated by the scanning strategy.
Instead, they form an initial impression of the SERP, and, if dissatisfied, might abandon it without looking at any of the results in detail~\citep{Diriye:2012:CIKM}.
\citet{Maxwell:2018:ECIR} introduce an additional SERP-level decision point, based on the notion \emph{information scent}.
As per the Patch Model in Foraging Theory (cf. Section~\ref{sec:decisions:conceptual}), the searcher first observes the SERP to determine whether it ``smells good enough'' to be worthy of closer inspection.
A baseline \emph{always examine} strategy is compared with stochastic judgments simulating \emph{average}, \emph{savvy}, and \emph{na\"{i}ve} users, depending on how likely they can distinguish between high-scent and low-scent SERPs.
Overall, all three stochastic user strategies outperform the \emph{always examine} baseline, and the better users are able to discern low-scent SERPs from high-scent ones, the more effective they are at searching.
These findings suggest that ``work should be directed towards improving how SERPs are rendered to increase how well people can identify good SERPs from the bad''~\citep{Maxwell:2018:ECIR}.
The results also suggest that evaluation measures should consider abandonment.

\subsection{Modeling Search Strategies}

So far, we have reviewed research focusing on how users issue queries and examine the results.
These works typically assume a simplified search environment and an ideal user behaviour.
Real environments, however, are more complex and there are several factors that influence the decisions users make (implicitly or explicitly) when interacting with a search engine.
For example, searchers might adapt their behaviour depending on the device (desktop computer vs. smart phone) or on the availability of features (e.g., query suggestions or auto-completions).
Next, we present some of the work that aims to gain insights into various search strategies users might employ, from a cost-benefit perspective.
Most commonly, cost is measured in terms of time spent on different actions (entering queries, scanning snippets, assessing documents, etc.), which is a reasonable and pragmatic proxy.\footnote{Cost could also be measured, for example, in terms of cognitive load, such as effort in reading and understanding a document.  While these options have been considered for evaluation, see, e.g.,~\citep{Arvola:2010:IR,Zhang:2014:SIGIR,Zuccon:2016:ECIR}, they tend to be more difficult to define and quantify for simulation.}
Gain is typically measured in terms of Cumulated Gain (CG)~\citep{Jarvelin:2002:TOIS}: the sum of the relevance scores of all seen documents in the session---that is, NDCG without discounting and normalization, to enable comparison between sessions of different lengths~\citep{Baskaya:2012:SIGIR}.
CG is also conveniently compatible with the graded relevance assessments that test collections offer~\citep{Paakkonen:2017:IRJ}.

\citet{Azzopardi:2011:SIGIR} applies economic theory to study the trade-off between querying (number of queries issued) and assessing (number of results examined) in sessions.
It is found that a more advanced retrieval model (BM25) supports a greater variety of search strategies, in order to achieve the same gain, compared to simpler models (like TF-IDF or Boolean retrieval). %
\citet{Baskaya:2012:SIGIR} study the effectiveness of various query modification and result scanning strategies, given an overall time constraint.
Depending on the device used for searching, desktop PC vs. smart phone, the costs associated with query formulation differ---as it takes more time to enter queries using a smart phone than using a regular keyboard.
The authors show that the device and overall search time allocation profoundly affect what kind of interactive behaviour can be successful.
Nonetheless, the more time available, the less it matters how one searches.
The authors also show that traditional rank-based metrics that focus on the quality of ranking alone (such as MAP and NDCG) are insufficient or even misleading when input costs are taken into account.

Search engine switching has been studied empirically in \citep{White:2009:CIKM,Guo:2011:SIGIR}. \citet{Smucker:2016:CHIIR} consider the scenario of a searcher switching from one search engine to another, in an expectation of better results.
By examining the gain curves for a set of TREC queries, the authors identify four classes of switching behaviour: never switch (ranking A has consistently faster gain rate than ranking B), switch immediately (the opposite of never switch), moot switch (rankings A and B have similar performance, switching neither helps nor hurts), and complex switch (one ranking has a high rate of gain initially, but it is not sustained).
It should be noted that while it is possible to calculate optimal switching behaviour, modeling this user decision would require ``some user computable rate of gain and its comparison to an expected rate''~\citep{Smucker:2016:CHIIR}.

Simulations often assume an ideal behaviour of the searcher, which may lead to unrealistic findings.
\citet{Baskaya:2013:CIKM} analyze session-level effectiveness under \emph{fallible searcher behaviour}, where users might err when making a judgment about the relevance of documents based on result snippets: they may skip relevant documents or click on non-relevant ones.
Not surprisingly, moving from ideal to fallible searcher behaviour affects overall effectiveness negatively, but highly relevant documents are affected to a lesser extent (because of fewer errors in their assessments).
\citet{Paakkonen:2015:CLEF} show that there is a gap in performance (in terms of cumulated gain over time) between novice and expert users, representing ordinary and more experienced searchers, respectively.
They investigate which behavioural dimensions contribute most to the difference, including query formulation strategies, scanning and stopping behaviour, session time allocation, and relevance-related behaviour (clicking and judging probabilities).
The results show that gain and cost structures contribute more than action probabilities to performance, suggesting that helping less experienced searchers to reduce their search costs, e.g., by making make snippet and document assessment faster, would offer them the greatest increase in performance.

\subsubsection{Variance in User Populations}

It might vary when users decide to stop scanning results: some individuals might be more patient than others or searchers might exhibit more persistence for certain types of queries.
\citet{Carterette:2011:CIKM} showcase how simulation can bring systems-based evaluations closer to user studies by allowing for more variance in user behaviour.
Specifically, they differentiate between \emph{patience profiles} corresponding to two broad classes of queries: informational and navigational.
For the former class, users are a lot more willing to go beyond rank 1 than for the latter class (the parameters of the distributions are estimated based on log data from a commercial search engine).
A comparison of systems submitted to the TREC 2005 and 2006 Terrabyte tracks reveals that for informational (ad hoc) queries, systems that provide the best user experience for impatient users (in terms of RBP~\citep{Moffat:2008:TOIS}) would not rank better than third by any of the traditional TREC evaluation measures.
For navigational (named page) queries, on the other hand, a single system is found to perform best for the majority of users.

\section{User Simulation for Recommender Systems}

Compared with the study of user behaviour in search engines, much less work has been done on analyzing user behaviour in recommender systems.
However, since search engines and recommender systems are closely related (cf. Section~\ref{sec:background:ir_recsys}), the findings and techniques developed from research of user simulation in the context of search engines may also be applicable to recommender systems. For example, the way users interact with a recommended (ranked) list of items may be essentially the same as how they would interact with a search result list.

Most of the existing work on analyzing user behaviour in recommender systems has focused on predicting whether a user would click on a recommended item (i.e., click modeling) often by using a broad spectrum of features computed for a candidate user-item pair. While many models are not interpretable, some work has suggested that a user's decision on whether to click on a recommended item is influenced by the same user's past actions and may also be correlated with the actions of other similar users. %
Some work has attempted to model consumer behaviour in an e-commerce context~\citep{Fleder:2007:EC,Fleder:2009:MS,Zhang:2020:ISR,Chaney:2021:arXiv}, analyze news recommenders (with respect to long-tail novelty and unexpectedness)~\citep{Bountouridis:2019:FAT}, and measure gender bias in music recommenders~\citep{Ferraro:2021:CHIIR}.

Unlike the case of search engines, where simulation is typically concerned with the behaviour of individual users, simulations for recommender systems are usually performed with the goal of understanding how certain algorithmic choices impact a population of users and study different forms of bias, e.g., preference bias~\citep{Zhou:2021:SSRN} and popularity bias~\citep{Yao:2021:arXiv}, or long-term dynamics~\citep{Hazrati:2024:UMUAI, Yao:2023:ICML}.

These are only a few simulation studies on recommender systems with a focus on evaluating system performance in terms of accuracy or diversity.
\citet{Szlvik:2011:ICWSM} study different user choice models in terms of which users of a movie recommender system would choose from the list of recommendations. These include always accepting the recommendations (Yes-men), rating only movies that are globally the highest rated (Trend-followers), and making random selections (Randomizers).
Their results show that (1) if the recommender system offers a personalized list of movies, then diversity is expected to be higher but overall ratings will suffer; (2) if only popular movies are recommended, then ratings will be higher but diversity will drop; and (3) if random movies are recommended then ratings are likely to become low, but users will receive a wide range of recommendations.
\citet{Jannach:2015:UMUAI} assess the effect of different recommendation strategies on the diversity of recommendations and find that the concentration bias of some algorithms on popular items can lead to a possibly undesired ``blockbuster'' effect.
In recent work, \citet{Rahdari:2024:TORS} demonstrate with the help of simulation that ``users find their desired item more efficiently when recommendations are presented as a list of carousels compared to a simple ranked list.''

The problem of recommendation has so far been framed as mostly a single round of interaction (i.e., the system recommends information to a user), instead of an interactive session as in the case of a search system. Thus, there is a limited set of possible user actions to take, which makes a user's behaviour less complicated than in the case of a search engine in some sense. 
Simulations have in fact been used in interactive recommender systems for evaluating critiquing and comparison-based feedback approaches~\citep{McGinty:2011:bookchapter}. %
However, there appears to be no standard interface or interaction workflow assumed about a user's further interactions, making it challenging to study user behaviour. As future recommender systems become increasingly interactive, it is reasonable to anticipate more research on user simulation for recommender systems.

\section{User Simulation for Conversational Assistants}

Conversational assistants represent an emerging and actively researched topic~\citep{Zamani:2023:FnTIR}. There is already a clear trend of growth of research work in this area including work on user simulation, especially in the context of task-oriented dialogue systems~\citep{ElAsri:2016:Interspeech,Gur:2018:SLT,Gur:2018:SLT,Lin:2021:SIGDIAL,Kreyssig:2018:SIGDIAL,Lin:2022:SIGDIAL,Hu:2023:CIKM,Davidson:2023:arXiv}.

Building on research on task-oriented dialogue systems, \citet{Zhang:2020:KDD} develop a user simulator specifically for the evaluation of conversational recommender systems use it to experimentally compare three preexisting conversational recommender systems.
\citet{Zhang:2022b:SIGIR} study how humans behave and formulate their utterances when a conversational assistant fails to understand them in order to mimic this behaviour in a simulator.

Another line of work utilizes large language models given their ability to generate fluent, human-like responses to a variety of inputs in natural language.
\citet{Sekulic:2022:WSDM} and \citet{Owoicho:2023:SIGIR} use user simulation to facilitate the evaluation of mixed-initiative conversational search systems. Specifically, their simulators can answer clarification questions that a system may ask of its user as well as provide explicit feedback on the presented results.

Given that conversational assistants likely represent the next generation of information access systems that can potentially support both search and recommendation capabilities via a more natural form of interaction, we may see rapid growth of research on user simulation for conversational assistants in the future.

\chapter{Simulation-Based Evaluation Frameworks}
\label{ch:frameworks}

In this chapter, we discuss how user simulation can be used to evaluate information access systems. We begin with the observation that traditional evaluation measures used in information access may be viewed as naive user simulators.
Next, we highlight the limitations of these metrics, and the additional components that would need to be modeled in order to going beyond them.
We then introduce a general evaluation framework for evaluating interactive information access systems using user simulation that can accommodate general interactions between a user and a system. 
While our discussion 
is mostly done in the context of search engine applications,
the main ideas, frameworks, and methods are all generally applicable to recommender systems and conversational assistants as well.

\section{Traditional Evaluation Measures as Naive Simulators}
\label{sec:frameworks:naive}

In this section, we discuss how traditional IR evaluation metrics may be viewed as evaluation using naive user simulators. This would provide some intuition of the general idea of simulation-based evaluation as well as motivate the need for a more general framework for simulating more realistic/sophisticated users.

Early work on simulation in IR already touched evaluation (see, e.g., ~\citep{Cooper:1973:ISR,Gordon:1990:JASIS}); for example, the expected search length can be regarded as an early discussion of how a metric can be viewed as quantifying a simulated user's effort in achieving the goal of satisfying the information need~\citep{Cooper:1968:AD}. 
Interpretation of commonly used retrieval measures form user simulation perspective has been studied in ~\citep{Moffat:2008:TOIS}, where a measure explicitly defined based on user modeling, called Rank-Biased Precision was proposed. \citet{Zhang:2017:ICTIR} further show that the whole Cranfield evaluation methodology can be regarded as a special case of using user simulation for evaluation and all the major past IR measures can and should be interpreted in the framework of user simulation, where, e.g., Average Precision is shown to model the perceived utility by a group of users with variable recall requirements. 

In general, we could interpret a measure such as Precision or Recall as the perceived utility of a retrieval result from the perspective of a simulated user. We can imagine the simulated (naive) user would sequentially ``browse'' the ranked list of retrieval results returned from the system and view every relevant document while skipping every non-relevant document. Now, suppose the user has viewed the top-$n$ documents and found that $k$ of these are relevant. In such a case, the user has spent effort to examine $n$ documents and gained relevant information from $k$ of them. Thus the Precision at $n$ documents, which is $k/n$, can be easily seen as measuring the ratio of the relevant information gain to the effort.  When we fix $n$, Precision at $n$ documents can be interpreted measuring the gain of relevant information for a fixed budgeted effort. Similarly, Recall can be interpreted as measuring the progress toward completion of the task of finding all the relevant documents. \citet{Zhang:2017:ICTIR} provide a detailed discussion of how multiple standard metrics can all be interpreted with a user simulation framework.

A particularly interesting observation is that Average Precision can be interpreted as simulating a set of users with variable recall expectations. Another interesting observation is that the discounting coefficients in NDCG can be interpreted as the stopping probability at different positions, and the value of NDCG can be interpreted as the expected gain with consideration of the uncertainty of stopping at different positions. Here, we can see clearly the benefit of using a user simulation framework to examine a traditional measure. Consider when we use a measure such as NDCG@k for evaluation, from an user simulation perspective, the parameter $k$ reflects where a user is expected to stop on a ranked list of results. Generally, this should depend on the amount of relevant information that the use rexpects to find, which often depends on the information need behind the query. For example, a query with high recall (e.g., from a user doing a literature survey) may tend to be associated with a larger $k$, whereas known item search should presumably use a smaller $k$. This means that ideally we should be using a query-specific value for $k$ when we evaluate a system over a set of queries. The argument for using variable $k$ for NDCG was indeed made recently by~\citet{Karmaker:2020:CIKM}, but embracing user simulation for evaluation would naturally reveal such a need.

The time-based NDCG measure~\citep{Smucker:2012:SIGIR} can also be interpreted as modeling the probability of stopping more accurately based on the time spent on examining documents. In the next section, we will provide a more complete discussion of many measures that go beyond the traditional basic measures such as Precision and Recall; they generally incorporate more explicit modeling of user interactions and correspond to a more realistic model of user behaviour, and are thus even closer to evaluation using user simulation.

\section{Going Beyond Traditional Evaluation Measures}
\label{sec:frameworks:advanced}

We now discuss what additional metrics need to be considered (going beyond the current metrics), especially those related to interactions with users,  thereby motivating the various components that need to be modeled in a simulation framework, including modeling a user's behaviour, modeling a user's cost/effort, and modeling the reward (gain of relevant information). We will also discuss the need for modeling the variation of users as well as variable behaviours of the same user in different contexts.  This would naturally motivate the need for a more general simulation framework to be discussed in detail in the next section.

\subsection{Measures Based on Explicit Models of User Behaviour}
\label{sec:frameworks:measures_ub}

From the perspective of user simulation, virtually all measures attempt to quantify the performance of a search result based on a combination of four factors:
\begin{inlinelist}
    \item the assumed user task (e.g., high precision vs. high recall),
    \item the assumed user behaviour when interacting with the results,
    \item measurement of the reward a user would receive from examining the result, and
    \item measurement of the effort a user would need to make in order to receive the reward.
\end{inlinelist}
Many measures have been proposed to improve the simple traditional measures, such as Precision and Recall, in one or more of these four factors. We summarize them based on how they model each of these factors.

The Rank-Biased Precision (RBP) measure~\citep{Moffat:2008:TOIS} makes an explicit assumption about the user's behaviour, which is to continue examining the next document with a probability of $p$. This provides a way to estimate the probability that a document at a particular rank position $i$ would be examined by a user is $p^{i-1}$, with the assumption that the very first one would be always examined.
Rank-Biased Precision is then defined as to measure the expected utility to a user with this kind of behaviour:
\begin{equation}
    RBP(r_1, ..., r_d)=(1-p) \sum_{i=1}^d r_i \, p^{i-1} ~,
\end{equation}
where $r_i$ is the gain (reward) of relevance information as defined in the NDCG measure and can be flexibly defined based on an application need. There is no explicit modeling of effort, but it can be inferred as uniform cost in examining any document. Unlike the classic Precision measure, where the user is assumed to have a fixed budget for examining documents, RBP assumes that a user has a fixed level of persistence in examining the results. In this measure, the task is not explicitly defined, but it is related to the ``persistence parameter'' $p$ in the sense that a user who wants higher recall can be assumed to have a higher $p$.

The assumption of a fixed continuation probability $p$ is not realistic in that a user who has already seen a highly relevant document may stop browsing earlier than one who has not.  Expected Reciprocal Rank (ERR)~\citep{Chapelle:2009:CIKM} was proposed to relax this assumption by using a more realistic cascading user model, in which a user is assumed to stop as soon as they have seen a sufficiently relevant document. The user's assumed task is thus to find just a single most relevant document. The reward is assumed to be based on the degree of relevance of a document and uniform cost is assumed.

Expected Browsing Utility (EBU)~\citep{Yilmaz:2010:CIKM} further improves over measures such as ERR and RBP by using an even more realistic sequential browsing model, where three interpretable parameters $\alpha$, $\beta$, and $\gamma$ are assumed to provide a more detailed characterization of a user's browsing behaviour, capturing the probability of leaving a document, probability of clicking on a relevant document, and persistence. Such models can be potentially estimated using search log data to obtain a more realistic user model. The task assumed is implicitly captured by the estimated user model; by varying the parameters, the measure can be adapted to simulate different tasks. The assumed examination cost is uniform, while the reward is based on the degree of relevance of documents.

The Time-calibrated measure~\citep{Smucker:2012:SIGIR} improves existing measures, where the examination cost has been assumed to be uniform, by using a more accurate measure of a user's effort in examining a document, with the assumption that a long document would take more effort/time. The browsing behaviour can then be tied to the amount of time that a user has spent up to a particular rank in the result list; with more time being spent, the user would more likely stop. The mapping between the time and stopping probability can be calibrated based on observed user behaviour. It can also support a measure that respects a ``budget'' on the user's effort as measured by time.

The C/W/L framework provides a more general framework for modeling the manner in which users examine a ranked list under the assumption that each user processes results top-down~\citep{Moffat:2017:TOIS,Azzopardi:2021:ICTIR}. The ``C'' component in the framework, $C(i)$, is a model of the browsing behaviour of a user encoded in the form of the probability of continuing to browse after examining the item at rank $i$. The ``W'' component, $W(i)$, refers to the fraction of a user's attention to the item at rank $i$, with the constraint $\sum_i W(i)=1$. The ``L'' component, $L(i)$, models the probability that the item at rank $i$ is the last item viewed by the user. With these component models, an evaluation measure can be defined for a particular user as the expected rate of gain (ERG), i.e., a weighted sum of the gain (of relevant information) from interacting with each document in a ranked list, where the weight for an item at rank $i$ is $W(i)$. Thus, an item at a rank with more attention from the user, as reflected by a higher $W(i)$, would be weighted more highly. As one might expect, the item at rank 1 always has the highest $W(i)$, which makes sense. An alternative measure, called Expected Total Gain (ETG), can be defined based on weighting using $L(i)$ instead of $W(i)$.
The C/W/L framework can cover many other specific measures that are based on the simple sequential browsing model as special cases. One limitation of C/W/L is that the aggregation of the reward over all examined items is a simple sum, implying the independence of their individual rewards. The C/W/L framework was further extended to addresss this limitation by separating the actions taken by a user and the benefit received via an action, leading to the C/W/L/A framework~\citep{Moffat:2022:SIGIR}.
The separation adds more flexibility to the framework to allow many evaluation metrics to be more systematically categorized and to reveal more gaps in the current evaluation metrics, bringing the evaluation methodology closer to a more general methodology of evaluation based on sophisticated user simulators.

A more general high-level formal framework for model-based measures was proposed in~\citep{Carterette:2011:SIGIR}, where a measure is formalized as quantifying the accumulated utility over all the documents examined by a user who adopts a certain browsing model, thus introducing explicitly three component models: (1) a \emph{browsing model} that captures how a user interacts with the ranked list of results, (2) a \emph{document utility model} that captures the utility of a single document, and (3) a \emph{utility accumulation model} that captures how the utility of each individual document would be combined to capture the total utility of all the examined documents. The separation enables different ways of instantiating each component model and combination of those different instantiations to obtain many different specific measures. In~\citep{Carterette:2011:SIGIR}, it was shown that the framework can not only cover many existing measures and provide an interpretation of those measures to reveal the assumed component models used in each of them, but also suggest many interesting new measures that might also be meaningful. Although the modeling of cost was not explicitly included in this framework, conceptually, this framework has paved way for further generalization of model-based evaluation measures to the general user simulation-based evaluation methodology advocated in~\citep{Zhang:2017:ICTIR}, where the browsing model would be generalized to a more general user simulator that can simulate not only browsing behaviour over a ranked list, but also any user action, including, e.g., query formulation/reformulation. We will describe this general framework later in this chapter.

The early IR measures have generally defined reward in terms of relevance-based gains. Later, novelty and diversity of the search results were also considered as part of the reward in some extended measures that consider both relevance and novelty, e.g., Normalized Discounted Cumulative Utility (NDCU) \citep{Yang:2007:SIGIR} and $\alpha$-NDCG \citep{Clarke:2008:SIGIR}. From the perspective of user simulation, their main differences are in the improvement of the modeling of reward as perceived by the simulated user.
Specifically, NDCU measures gain not solely based on relevance, but also based on novelty as compared with the information already consumed by the user, and $\alpha$-NDCG evaluates search results based on topic-level gain, instead of document-level gain, thus redundant documents that cover similar topics would not be favoured. 

\subsection{Measures for Search Sessions}

The measures discussed above are all restricted to evaluation of the results of a single query. However, retrieval tasks often require a user to reformulate queries, which can be due to an ineffective initial query or an exploratory information need. In such cases, it is more important to evaluate the {\em overall} performance of a whole search session.
To this end, some measures defined on a single query have been generalized to measure the performance over a session that consists of multiple queries. For example, NDCG has been extended to session-based search, taking user effort implicitly into account, leading to the Session NDCG (SDCG) measure~\citep{Jarvelin:2008:ECIR}. The main idea is to discount the results returned from a later query in a session. Technically, the DCG discount vectors from all the queries can be concatenated by applying an additional discount coefficient based on the query position in the session, thus enabling the computation of NDCG for the whole session by simply pretending the concatenated results were a single ranked list of documents. From a user simulation perspective, the main difference from a single query-based measure is the generalization of the sequential browsing of a ranked list to sequential browsing of multiple ranked lists in a session. A more general formal framework for modeling the user browsing behaviour and computing an Expected Global Utility over a session has been proposed in~\citep{Yang:2009:ICTIR}, which enables modeling the uncertainty of a user's browsing behaviour and computes the expected utility w.r.t. the distribution of all possible user browsing behaviours. The application of the framework to derive a session-based NDCG measure is discussed in~\citep{Yang:2009:ICTIR}. Similar ideas have been explored in~\citep{Kanoulas:2011:SIGIR}, where the idea of modeling a user's browsing behaviour in a session using a ``path'' was introduced, which can be roughly interpreted as modeling the perceived ranking of all the documents a user has interacted with in a session as a single ranked list. Any measure can then be defined based on such a perceived ranked list for the whole session. Two strategies have been explored in depth for implementing the idea, leading to model-free session evaluation (no explicit modeling of user behaviour, using the best path among a set of possibilities) and model-based session evaluation (with assumptions of user behaviour, using the expectation according to the assumed user behaviour distribution).

Compared with single query-based evaluation measures,
those session-based evaluation measures can be regarded as evaluation using more realistic and more sophisticated user behaviour models (browsing through a whole session vs. through a single ranked list). However, the browsing behaviour has not been modeled in detail in such measures; for example, the clicking decision that drives the browsing behaviour has not been considered, even though click models have been separately studied~\citep{Chuklin:2015:Book}. The Interactive IR Probability Ranking Principle (IIR-PRP)~\citep{Fuhr:2008:IRJ} theoretically integrates a user's clicking decision with a measure of the overall utility of a ranked list with consideration of dependency of document utility in the list. While IIR-PRP was proposed for deriving an optimal ranking algorithm, it can also be regarded as defining an evaluation measure based on relevance gain and cost of examining documents, with explicit modeling of sequential browsing decisions. All these measures, however, are based on the assumption of sequential browsing, which is only applicable when the search results are displayed as a ranked list (or there is a natural way for a user to perform sequential interaction with the results). Modern search and recommender systems provide much more sophisticated user interfaces that can easily go beyond sequential browsing (e.g., results organized in slates). In the next section, we will discuss a more general simulation-based evaluation framework for evaluating retrieval systems that support more sophisticated user interactions, including those going beyond sequential browsing.

\section{A General Simulation-Based Evaluation Framework}
\label{sec:frameworks:general}

Generally, we may consider interactions in either a search engine or a recommender system as ``moves'' in an interactive game.
The user may be viewed as an ``agent'' that constantly makes decisions based on the current interaction environment.
A user simulator is an operational agent that can simulate how a user makes those decisions.
To evaluate a system, we can have the system interact with a simulated user and measure the system's performance by defining measures based on the interaction data generated by the user simulator.

A general simulation-based evaluation framework can thus be defined as consisting of the following elements:
\begin{enumerate}
    \item A collection of user simulators are constructed to approximate real users.
    \item  A collection of task simulators are constructed to approximate real tasks.
    \item  Both user simulators and task simulators can be parameterized to enable modeling of variation in users and tasks.
    \item  Evaluation of a system is done by first having a simulated user perform a simulated task by using (interacting with) the system and then
          computing various measures based on the entire interaction history of the whole ``task session.''
\end{enumerate}
In~\citep{Zhang:2017:ICTIR}, such a general evaluation framework is formally defined as follows. 
Let $S$ be a system, $U$ be a user, and $I$ be the whole process of the interaction of $U$ and $S$ to finish task $T$. We would be interested in measuring the system's performance based on $I$. From a user's perspective, we can measure the performance in two dimensions:
(1) Interaction Reward, $R(I,T, U,S)$, which measures the total reward the user has received via the interaction, and (2) Interaction Cost, $C(I,T,U,S)$, which measures the total cost of the interaction. In general, the more interaction actions the user makes, the more reward the user can potentially receive and the more cost the user would have to bear (since the user needs to make more effort).  If one single measure is needed, the reward and cost can be combined, which can be in many different forms. The need for modeling both reward and cost was already recognized in the Interactive IR Probability Ranking Principle~\citep{Fuhr:2008:IRJ}, where the two are integrated in a decision model simulating how a user makes browsing decisions. In~\citep{Zhang:2017:ICTIR}, the integration of the reward and cost is regarded as the foundation of an evaluation measure. This reflects the intuition that an ideal system should maximize reward to a user while minimizing the cost or effort required of the user. Besides these general frameworks, there are many specific models, notably economic IR models, where cost and gain are explicitly modeled (see, e.g., ~\citep{Azzopardi:2011:SIGIR} and many follow-up studies).

When the user $U$ is a simulated user, the interaction sequence $I$ may be uncertain or stochastic. In such a case, a more general measure of reward or cost can be defined as the expected Interaction Reward or Interaction Cost w.r.t. the distribution of all the possible interaction sequences that the simulated user $U$ may make with system $S$, i.e., $\prob(I|T, U,S)$. In this way, we may define the following: (1) Simulator Reward: $R(T, U,S)=\sum_{I} \prob(I|T, U,S) R(I, T, U,S)$. (2) Simulator Cost: $C(T, U,S)=\sum_{I} \prob(I|T,U,S)C(I, T, U,S)$.

In~\citep{Zhang:2017:ICTIR}, the general framework is further refined to assume that $I$ is a sequence of user actions taken in response to a sequence of \emph{interface cards}, generated by system $S$. Under such an assumption, the interaction sequence is $I = ((z_1, a_1,q_1), (z_2, a_2,q_2), . . . , (z_n, a_n,q_n))$, where $a_i$ is a user action, $q_i$ is an interface card (i.e., a dynamic user interface) generated by the system, and $z_i$ is a representation of the user's state during the interaction. The meaning of $(z_i,a_i,q_i)$ is that while in state $z_i$, the user has taken action $a_i$ (e.g., entering a query), which leads to the system generating an interface card $q_i$ (e.g., showing a list of search results). With this refinement, reward and cost can be further decomposed into action-level reward.

It is easy to see that this general framework is applicable to the evaluation of any interactive information access system, or even any assistive AI system. When the user actions are restricted to browsing of documents in a ranked list sequentially, the framework can naturally cover some specific frameworks such as C/W/L~\citep{Moffat:2017:TOIS}, C/W/L/A~\citep{Moffat:2022:SIGIR} and the model-based measure framework~\citep{Carterette:2011:SIGIR} as special cases, thus also covering virtually all the existing specific IR measures.

The analysis of existing measures within this general framework has revealed interesting interpretations, potentially guiding the choice of appropriate measures for specific evaluation purposes.
For example, Average Precision can be interpreted as the expectation of a retrieval system's utility for a set of users whose tasks have variable levels of recall~\citep{Zhang:2017:ICTIR}.

It was shown in~\citep{Zhang:2017:ICTIR} that the general framework, with assistance from the Interface Card Model~\citep{Zhang:2015:SIGIR}, can be used to evaluate an interactive IR system with a computationally generated dynamic browsing interface using user simulation.
Specifically, the task was to browse to a target relevant document in a collection (known-item browsing), and the goal was to compare a system which can generate a browsing interface dynamically in adaptation to both the display size and the system's confidence in the inference about the user's interest~\citep{Zhang:2015:SIGIR} with a baseline system that has an adaptive interface to display size that can choose to use at any time. Simple user simulators were created and the system performance is measured based on the effort required in order to reach the target item (the number of clicks required). The results using user simulation were found to be consistent with those from using real user studies~\citep{Zhang:2017:ICTIR}.

In general, with the simulation framework, we can potentially evaluate a search engine or a recommender system using any sophisticated user simulator. In Chapters~\ref{ch:sim_search} and~\ref{ch:sim_conv}, we will discuss how such sophisticated user simulators can be constructed.

\chapter{User Simulation and Human Decision-making}
\label{ch:decisions}

Construction of user simulators requires specifying the functions of the system that a user interacts with as well as modeling how a user makes a decision between multiple actions that could be potentially taken based on the functionality of the system. As we intend to cover broadly multiple types of information access systems, including 
particularly search engines, recommender systems, and conversational assistants, we would need to address the question of what exactly their functions are. 
There are many possible answers to this question. In this book, we view search engines as tools that help people find information, recommender systems as tools that aid people in making better choices, and conversational assistants as tools that support users in completing their tasks via natural language interactions (cf. Section~\ref{sec:intro:tasks}). In order to successfully simulate user behaviour when interacting with such systems, it is important to have a broad understanding of the cognitive processes involved, which play a crucial role in shaping users' behaviour and decision-making. In this chapter, we first provide a high-level overview of research on how people seek information (Section~\ref{sec:decisions:conceptual}) and discuss various conceptual models that can provide theoretical guidance for modeling processes and decisions from an individual's perspective. We then discuss how users make choices in the context of recommender systems (Section~\ref{sec:decisions:choices}).
Finally, in Section~\ref{sec:decisions:math}, we discuss how to model decision-making processes mathematically using Markov decision processes (MDP). The MDP framework provides a general formal framework for constructing user simulators, which we will use to discuss specific user simulation techniques in the next two chapters, simulating interactions with search and recommender systems (Chapter~\ref{ch:sim_search}) and interactions with conversational assistants  (Chapter~\ref{ch:sim_conv}), respectively.   

\section{Conceptual Models of Information Seeking}
\label{sec:decisions:conceptual}

\begin{figure}
    \centering
    \includegraphics[width=0.6\textwidth]{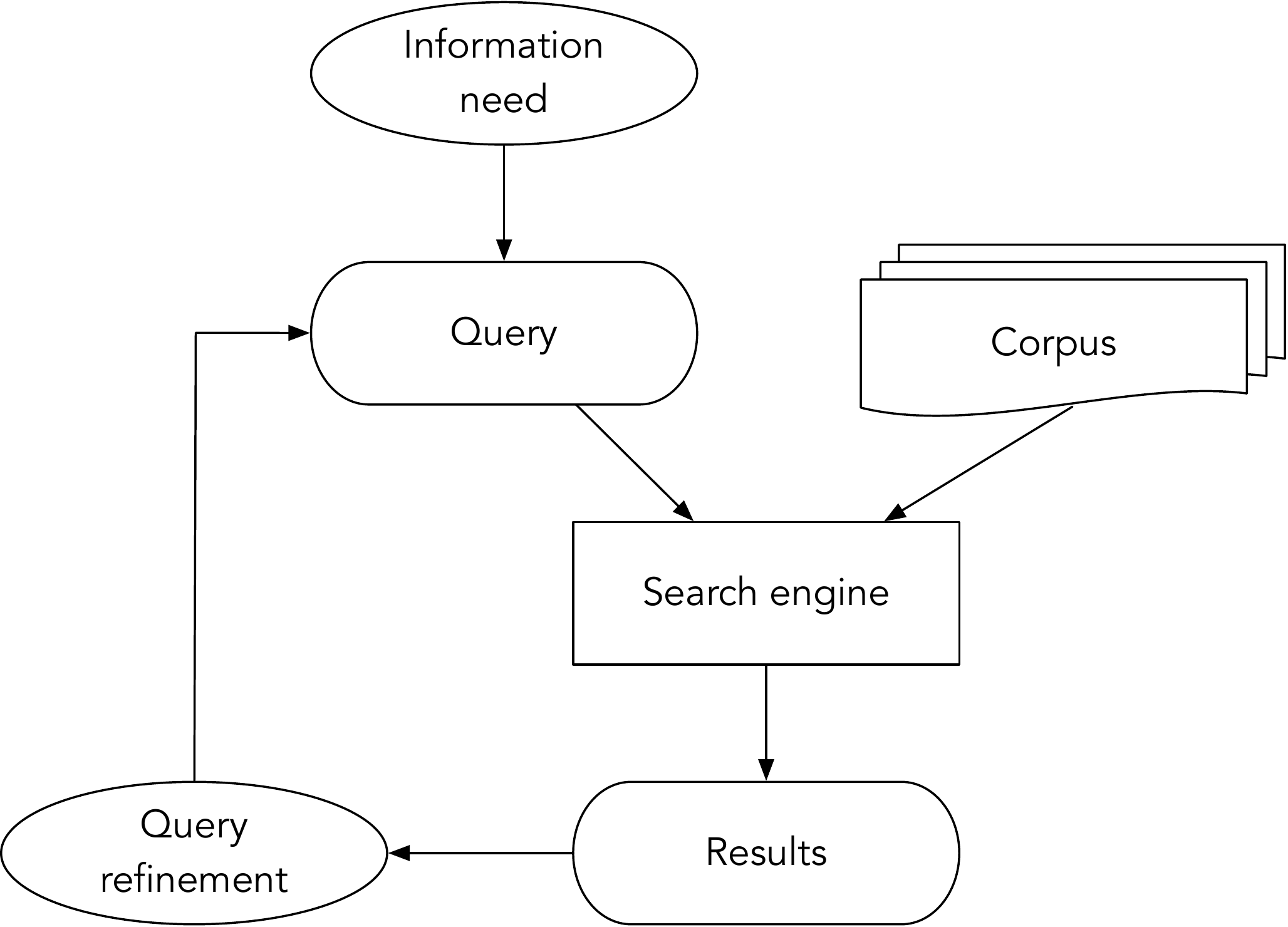}
    \caption{The classic model of information retrieval, adapted from~\citet{Broder:2002:SIGIRForum}.}
    \label{fig:classic_ir_model}
\end{figure}

Over the past decades, a great deal of research has been directed at understanding how people search for information, including their mental and strategic processes.
The classic model of IR (also referred to as the \emph{standard model of information seeking} in \citep{Hearst:2009:book}) assumes that the user has an information need, which is verbalized in the form of a query that is posed to the search engine.  The user examines the results, and if needed, reformulates the query.  These steps are repeated until a satisfactory result is found.  Figure~\ref{fig:classic_ir_model} illustrates the process.
While this model has promoted IR research in many ways (such as forming the basis of test collection-based evaluation), it is insufficient to capture how real humans interact with search systems.  Among others, it ignores how information needs might change dynamically as the user's interests evolve during the search process and also disregards the broader task context that motivated the search in the first place.

The field of library and information science has had an interest in developing models of information behaviour based on a cognitive viewpoint since the late 1970s~\citep{deMey:1977:CC}.
The central idea of the cognitive viewpoint is that information processing is human-oriented and thus draws attention to human perception, cognition, and structures of knowledge when interacting with information~\citep{Wilson:1984:SSIS}.
This stood in stark contrast with the predominantly system-focused IR research of that time, centered around the representation and matching of queries and documents (cf. Fig.~\ref{fig:classic_ir_model}).
Studies examining user's search behaviour began in the 1980s, focusing on intermediary-assisted search taking place in libraries~\citep{Pejtersen:1980:TAIR,Belkin:1982:JD,Ingwersen:1982:JD,Ellis:1989:JD}. 
The component of intermediary is later replaced by the construct of an interface/intermediary, allowing for interaction to occur between a user and an IR system~\citep{Ingwersen:1996:JD}.
During early 1990s, a cognitive turn took place in information seeking and retrieval, from focusing on the technological aspects only to a more human-oriented approach to information
interaction~\citep{Ingwersen:2005:book}.

In this section, we review models of search behaviour within three main categories:
(1) \emph{cognitive models}, focusing on the cognitive processes underlying the information-seeking activity,
(2) \emph{process models}, representing the different stages and activities during the search process, and
(3) \emph{strategic models}, describing tactics that users employ when searching for information.
Without aiming to be comprehensive, we discuss a few models within each category that are deemed particularly illustrative for the purposes of this book.
For a detailed account of models and frameworks for information seeking, we refer the reader to \citep{White:2016:book}[Chapter 3].

Before we get into the specifics, a general note on the limitations and relevance of these models.
As \citet{Wilson:1999:JD} explains, ``most models in the general field of information behaviour are [...] statements, often in the form of diagrams, that attempt to describe an information-seeking activity, the causes and consequences of that activity, or the relationships among stages in information-seeking behaviour. [...] The limitation of this kind of model, however, is that it does little more than provide a map of the area and draw attention to gaps in research.''
Furthermore, the validation of information seeking models based on large-scale behaviour data remains largely an open question~\citep{White:2016:book}.
That said, these pioneering models are still relevant because they describe ``relationships between information seekers, human intermediaries, information resources and information systems''~\citep{Savolainen:2018:JD}.
Conceptual models can provide theoretical guidance for building user models and inform the high-level design of user simulators.
These works also provide useful directions for future research to make simulators more realistic (cf. Section~\ref{sec:concl_future:realisticity}). For example, the representation of context and modeling of the user's knowledge state have been largely unexplored to date---this gap presents opportunities to link to research on ``searching as learning,'' where understanding how users acquire knowledge during search sessions is crucial~\citep{Vakkari:2016:JIS,Ghosh:2018:CHIIR}.

\subsection{Cognitive Models}

Cognitive models focus on the individual's internal representation of a problem situation and how that situation may be influenced by a number of contextual factors.
According to Belkin's Anomalous State of Knowledge (ASK) hypothesis, ``an information need arises from a recognized anomaly in the user's state of knowledge concerning some topic or situation and that, in general, the user is unable to specify precisely what is needed to resolve that anomaly''~\citep{Belkin:1982:JD}. The ASK hypothesis thus proposes a specific reason as to why people engage in an information-seeking behaviour.
Belkin assumes the presence of a human intermediary and proposes the ASK to be resolved via co-operative \emph{dialogue} between the user and the intermediary (as opposed to the user asking the IR system directly).

\begin{figure}
    \centering
    \includegraphics[width=0.9\textwidth]{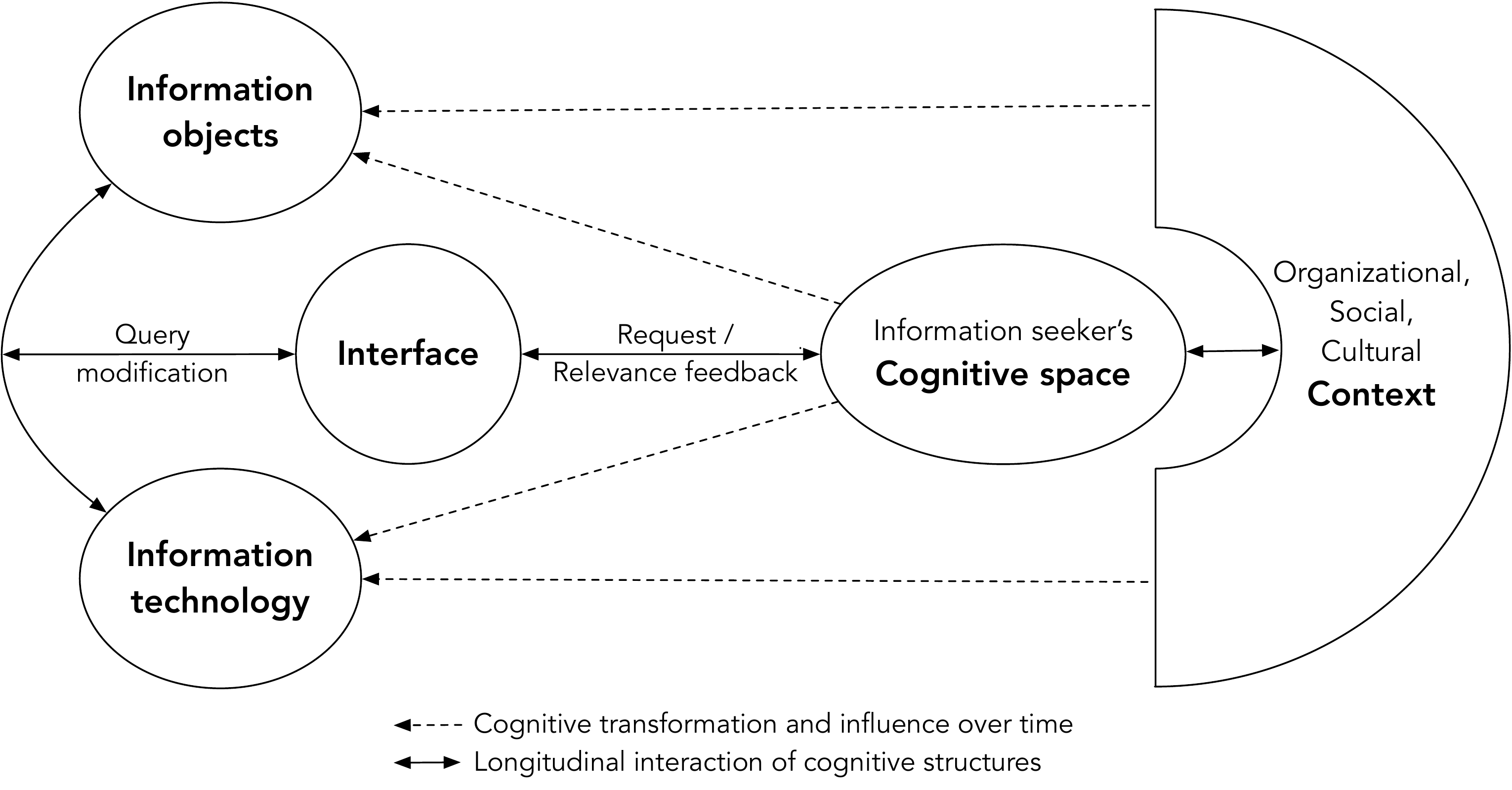}
    \caption{Integrated information seeking and retrieval (IS\&R) research framework, adapted from~\citet{Ingwersen:2005:book}.}
    \label{fig:ingwersen_isr_model}
\end{figure}

\citet{Ingwersen:1996:JD} specifies elements of a cognitive IR theory, combining both system-oriented and cognitive research.
IR elements include the representation of information objects, queries, and retrieval techniques.
Cognitive elements include the individual's work task, cognitive state, problem space, and information behaviour, as well as their socio-organizational environment.
The strength of the model lies in the detailed description of essential processes from both the user and system perspectives.  On the other hand, it paints a fairly complicated picture of information interactions and cognitive structures, while the model's applicability to system evaluation is limited~\citep{Savolainen:2018:JD}.
\citet{Ingwersen:2005:book} elaborate further on this model and present a generalized research framework for information seeking and retrieval (IS\&R), illustrated in Fig.~\ref{fig:ingwersen_isr_model}.  It consists of five main components: information objects, information technology, interface, information seeker, and context.
There are two possible types of relationships between the components: (1) cognitive transformation and influence over time, and (2) longitudinal interaction of cognitive structures.
The IS\&R model treats the information seeker as the central actor and emphasizes ``the \emph{interaction} between the information seeker(s) and the environment surrounding that individual, also over time''~\citep{Ingwersen:2005:book}.
The model further emphasizes the mutual dependencies between the different cognitive components, including intra-component structures.
While it represents the most systematic and holistic framework for information interaction to date, it remains at a very high level of conceptualization and does not describe the nature of interactions in greater detail~\citep{Savolainen:2018:JD}. Such a framework, however, provides useful guidance for the design of realistic user simulators.

\subsection{Process Models}

A line of research has examined the various stages of the information seeking process from the user's perspective.
\citet{Kuhlthau:1991:JASIST} identifies the following six stages:
(1) \emph{initiation}, recognizing a need for information;
(2) \emph{selection} of the general topic and approach that is expected to yield the best outcome;
(3) \emph{exploration} of the general topic in order to further personal understanding;
(4) \emph{formulation}, where a focused perspective on the topic emerges;
(5) \emph{collection} of the information related to the focused topic; and
(6) \emph{presentation}, which completes the search and prepares the results to be presented or used.
\citet{Kuhlthau:1991:JASIST} also describes the affective (feelings) and cognitive (thoughts) patterns associated with each of these steps.
It is to be noted that these stages characterize complex information needs and ``are not necessarily representative for more light-weight tasks''~\citep{Hearst:2009:book}.

\begin{figure}
    \centering
    \includegraphics[width=\textwidth]{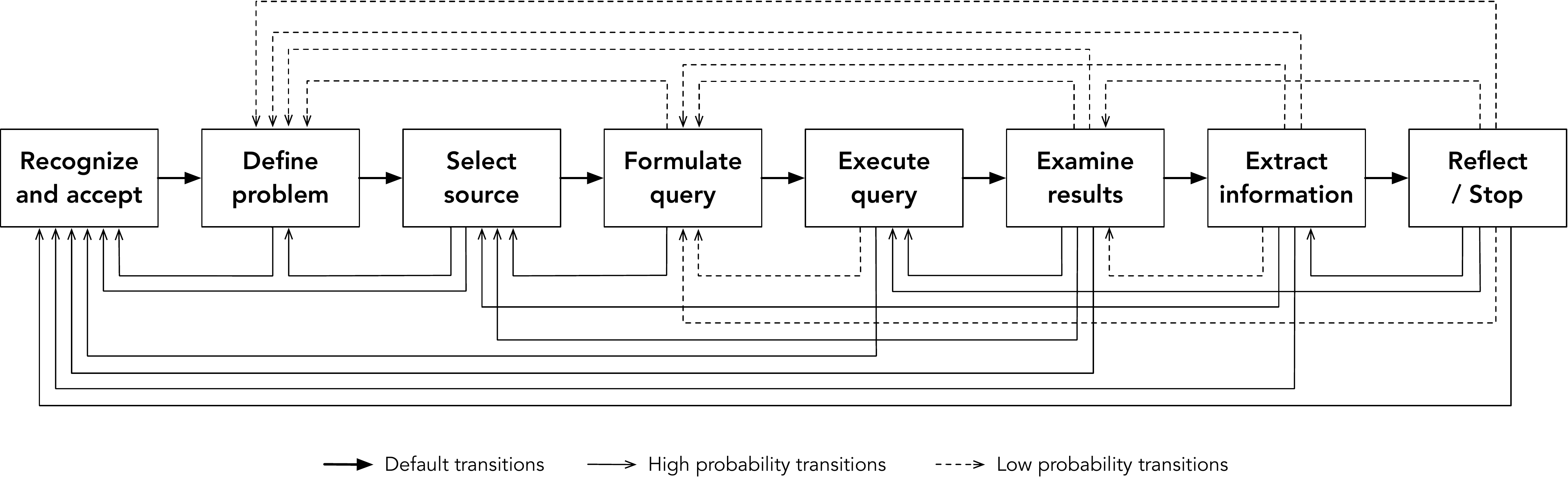}
    \caption{Information seeking process model, adapted from~\citet{Marchionini:1995:book}.}
    \label{fig:marchionini_process}
\end{figure}

\citet{Marchionini:1995:book} propose another model, where the information-seeking process is composed of eight sub-processes, as shown in Fig~\ref{fig:marchionini_process}:
(1) recognizing and accepting an information problem,
(2) defining and understanding the problem,
(3) choosing a search system,
(4) formulating a query,
(5) executing the search,
(6) examining the results,
(7) extracting information, and
(8) reflecting on the process and deciding whether to iterate further or stop.
Importantly, these sub-processes do not necessarily follow each other in a sequential order, but may develop in parallel and at different rates.  Thus, the main information seeking process can take advantage of opportunities arising from intermediate results.
\citet{Marchionini:1995:book} further categorizes the sub-processes into three classes: (1) understanding, (2) planning and execution, and (3) evaluation and use.  Understanding is mainly a mental activity, while the other two of these classes are both mental and behavioural activities.

Process models are useful for simulation as they can provide reliable scaffolds to model searcher behaviour and to situate the user at the correct stage within a complex information seeking process.  They can also be instrumental, for example, in simulating how users might take different paths, according to their skills and experience.

\subsection{Strategic Models}
\label{sec:construction:conceptual:strategic}

Next, we present two conceptual models that describe high level search strategies for exploratory search.
These models are different from the models discussed above in that they do not conceptualize the nature of information interaction or specify the stages in a search process, but rather present frameworks of search tactics using analogies from the physical world. 

\begin{figure}
    \centering
    \includegraphics[width=0.8\textwidth]{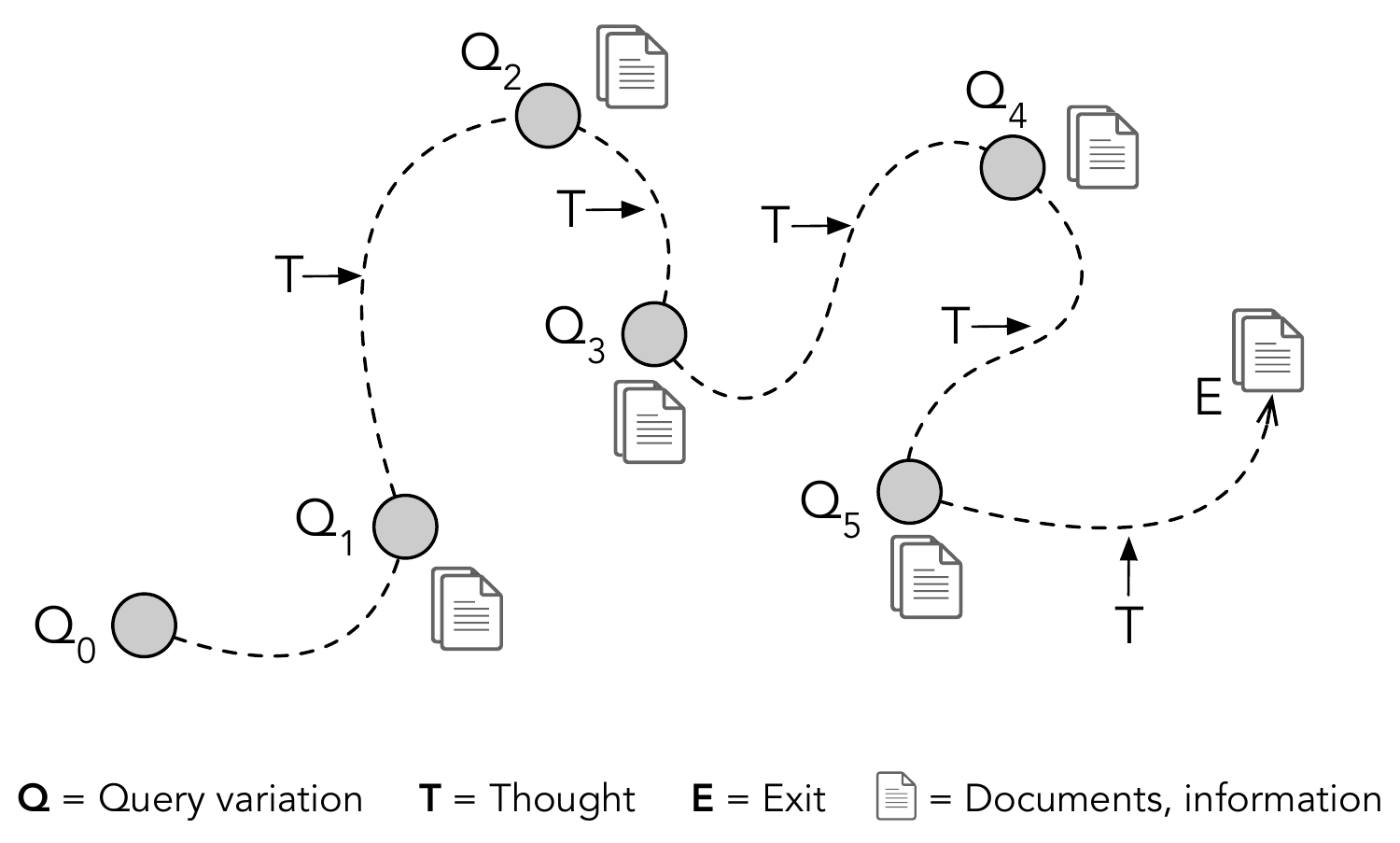}
    \caption{Search, as an evolving process, according to the berry-picking model~\citep{Bates:1989:OR} (illustration adapted from~\citet{White:2016:book}).}
    \label{fig:berrypicking}
\end{figure}

The \emph{berry-picking model}, proposed by \citet{Bates:1989:OR}, considers information seeking analogous to foragers looking for food.
One central point of the model is that searchers' needs are not satisfied by a single set of retrieved results (as assumed by the classic more of IR, cf. Fig.~\ref{fig:classic_ir_model}).  Rather, the information they seek is scattered like berries on bushes, and needs to be collected bit by bit.
Another main point of the model is that as searchers encounter new pieces of information and references along the way, those might give them new ideas and directions to follow.  Thus, search is an evolving process and as it progresses, not just the queries but the underlying information needs are also continually shifting, in part or whole; see Fig.~\ref{fig:berrypicking} for an illustration.
The berry-picking model is supported by observational studies~\citep{ODay:1993:CHI,Borgman:1996:JASIST} and the assumption that information needs are satisfied as a result of multiple queries are also captured in many of the session- or task-based models in Web search~\citep{White:2010:SIGIR,Eickhoff:2014:WSDM}.

\emph{Information foraging theory}~\citep{Pirolli:1999:PR} applies ideas from optimal foraging theory to understand how people search for information and interact with potential sources of information.  Optimal foraging theory explains how animals maximize their fitness while they search for food, i.e., gain the most energy for the lowest cost~\citep{Stephens:1986:book}.  Information foraging theory thus assumes that humans use the same innate foraging mechanisms and behave in a way to maximize their rate of gaining valuable information over time.
Central to the theory is the notion of a \emph{patch}---in the wild, this is an area where food can be acquired.  Foragers need to decide how long they want to stay in a patch before moving to the next patch.  Applied to the task of information seeking, staying on a patch corresponds to examining the results presented on a SERP, while moving between patches is akin to issuing a new query~\citep{Azzopardi:2015:ICTIR}.
Another central concept of information foraging theory is \emph{information scent}---similar to how animals can rely on scents to indicate their chances of finding prey, searchers can examine cues presented to them on web pages or SERPs to help locate information~\citep{Woodruff:2001:CHI,Olston:2003:TCHI,Pirolli:2007:book}.
There is empirical evidence showing that when information scent starts to decrease, searchers transition to other information sources~\citep{Card:2001:CHI} and that users examine the SERP to greater depths if it appears to contain many relevant results~\citep{Card:2001:CHI,Wu:2014:SIGIR}.
\citeauthor{Azzopardi:2018:SIGIR}'s information foraging metric~\citep{Azzopardi:2018:SIGIR} is a useful example of \citeauthor{Pirolli:1999:PR}'s work translated directly into a user model and a metric.

The difference between berry-picking and information foraging is that the former focuses on the dynamics of complex information needs and finding answers incrementally in stages, while the latter is concerned with the exploitation-exploration trade-offs when navigating between information sources.
Hence, information foraging theory also bears a connection to economic models of the search process, which try to model the cost-benefit trade-offs of various user actions~\citep{Birchler:2007:book,Azzopardi:2011:SIGIR}.

It should also be noted that all three kinds of models discussed in this section are interconnected, each offering a unique perspective.

\section{Choice and Decision Making in Recommender Systems}
\label{sec:decisions:choices}

Recommender systems are meant to help users make better choices through a collaborative process between a user and the system. In order to realistically simulate users' interactions with recommender systems, it is useful to have a broad understanding of how humans make choices. Indeed, a realistic user simulator should attempt to simulate how humans make such choices in various scenarios when interacting with the options provided by a system. 

\subsection{Choice Patterns}

The ASPECT model~\citep{Jameson:2014:FNTHCI} distinguishes six human \emph{choice patterns}. We briefly discuss these, along with possible ways a recommender system can support each pattern, following~\citet{Jameson:2015:bookchapter}. Notice that these patterns refer to ``options'' in general; in our context, these typically mean items from a collection (products, locations, recipes, etc.).

\begin{itemize}
    \item \textbf{Attribute-based choice}: The various options can be described in terms of attributes, some of which are considered more important for the chooser than others. Recommender systems often help with attribute-based choices by reducing the space of options (i.e., collection of all items) to a smaller \emph{consideration set}.

    \item \textbf{Consequence-based choice}: A different way of thinking about choice is to consider the consequences of choosing a particular option. The recommender system can help, for example, by determining (in a personalized way) when there is a choice to be made, identifying available options that the user was not aware of,  or highlighting (and possibly warning against) particular consequences that the user might not be familiar with.

    \item \textbf{Experience-based choice}: This pattern refers to the case when the person has past experience either with the given choice situation or with particular options. On a high level, case-based~\citep{Smyth:2007:bookchapter} and content-based~\citep{deGemmis:2015:bookchapter} recommender systems support experience-based choice by analyzing similar previous experiences to determine which of the available options to suggest.

    \item \textbf{Socially-based choice}: People often let their decisions influenced by the choices or advice of others. Collaborative filtering~\citep{Koren:2015:bookchapter} can be seen as a way of supporting this pattern, by considering choices of other people who are similar to the chooser in some respect. However, people that are considered similar by collaborative filtering algorithms are not always the most relevant class for the chooser~\citep{Jameson:2014:FNTHCI}. The social relationships between people are taken into account in some trust-based recommender systems~\citep{Victor:2011:bookchapter}.

    \item \textbf{Policy-based choice}: Choices can be made according to a specific policy, albeit this is more common in an organizational setting than for individuals. The choice process can be divided into two phases. The first is about finding a policy; recommender system can aid in that by helping the user find policies (e.g., a diet or exercise regime to follow).
          The second phase concerns applying that policy; recommender system can ask people to specify their preferences explicitly, then consider these preferences when presenting choices (e.g., ordering news stories based on expressed preferences).

    \item \textbf{Trial-and-error based choice}: In instances, and especially when none of the above patterns leads to a clear decision, a person may opt to randomly select an option to assess it. This does not necessarily lead to a full ``execution'' (purchase, consumption, etc.) but could also mean a closer examination (e.g., reading reviews). Recommender systems can support trial-and-error based choice by suggesting other options to try out next and aspects to pay attention to, and then learn from the user's experiences after having tried an option. For example, in critique-based recommenders~\citep{McGinty:2011:bookchapter} the user is presented with one or more items that they can give feedback on, then the system suggests further items; these steps are repeated until the user arrives at a satisfactory option. In some domains, such as music recommendation~\citep{Schedl:2015:bookchapter}, the user is often invited to try out something new, even if it is uncertain whether they will like it.

\end{itemize}
The above patterns are presented in a distilled form, but in practice they are often used in combination. For example, a simple attribute-based strategy is employed to narrow down the space of options, then another pattern is used to choose among those options.
While the six ASPECT choice patterns were originally formulated to explain individual choices, they can also be applied in ``scenarios in which there is a need to support the choice of a group of persons''~\citep{Jameson:2022:bookchapter}.

\subsection{Preferences}

A different conception, which is often used in economics, is that of choice based on \emph{preferences}~\citep{Hausman:2011:book}. Accordingly, the goal of the recommender system is to create a \emph{preference model}, based on information acquired about a person's preferences, that can be used to predict what that person will like.

A widely adopted method for soliciting preferences is in the form of \emph{ratings}, which are regarded as absolute preferences for specific options. Generally, a rating given to an item reflects the \emph{associations} the person has with that item, which includes beliefs, experiences, and affective responses~\citep{Jameson:2015:bookchapter}. These associations can also be organized according to the ASPECT choice patterns, e.g., ratings with respect to important attributes (attribute-based) or affective responses and stored evaluations of the item (experience-based).
However, it is generally not helpful to regard a rating as a single, well-defined variable that reflects the true degree of preference, but rather ``as the reflections of different \emph{perspectives} on the item which arise under different conditions''~\citep{Jameson:2015:bookchapter}. The perspective that the rater takes can depend on various factors, such as the specific rating questions asked, the rater's current mood, reference points that are provided~\citep{Nguyen:2013:RecSys}, or the time between consuming and rating an item~\citep{Bollen:2012:RecSys}.

\section{Mathematical Framework}
\label{sec:decisions:math}

In order to realistically simulate user behaviour, it is important to realize that the actions users take are not independent events. While they are often studied in isolation, these actions are part of a larger decision-making process. 
Earlier in this chapter we have explored the cognitive side of this process---how users make choices when seeking information or interacting with recommended items. 
Now, let us introduce a mathematical framework that helps us formally model decision-making from the user's point of view: the Markov Decision Process (MDP).

Imagine a user interacting with an information access system such as a search engine. At any given moment, they are in a specific \emph{state} ($s \in \mathcal{S}$, e.g., typing a query, scanning the result list, clicking on a result snippet).  They have choices, i.e., a set of \emph{actions} ($a \in \mathcal{A}$) they can take (e.g., submit a query, click on a result, leave the search engine). Each action leads them to a new state ($s'$), like following a path. An MDP helps us model this journey and is made up of:
\begin{itemize}
    \item \textbf{States} ($\mathcal{S}$): The different situations a user can be in.
    \item \textbf{Actions} ($\mathcal{A}$): The choices available to the user in each state.
    \item \textbf{Transition probabilities} ($P(s'|s,a)$): The likelihood of moving from one state ($s$) to another ($s'$) after taking a specific action ($a$). 
    \item \textbf{Rewards} ($R(a,s)$): The immediate benefit (or cost) the user gets from taking a particular action ($a$) in a given state ($s$). This could be, e.g., finding the information they seek or experiencing frustration. Sometimes, the reward is not immediate but can only be received at the end of a sequence of interactions. For example, successfully purchasing an item or abandoning the system after a series of failed attempts.
\end{itemize}
In an MDP, we take discrete time steps. The Markov property, a fundamental assumption of this framework, states that the next state ($s'$) the system transitions to depends solely on the current state ($s$) and the action taken ($a$). In other words, the system has no ``memory'' of its past states and actions. %
This simplifies modeling and reduces computational complexity. 

In the MDP framework, a \emph{policy} ($\pi:\mathcal{S} \rightarrow \mathcal{A}$) is a function that determines which action to take in each state ($\pi(s) \rightarrow a$).
That is, the policy guides the simulated user's behaviour, dictating which action to choose when faced with different scenarios within the system.

MDPs are commonly used for modeling systems that make sequences of decisions to maximize rewards, such as in robotics or game playing. In these applications, the MDP framework enables reinforcement learning (RL), where an agent (that controls the system's decision-making) learns to interact with an environment through trial and error. The agent takes actions, observes the resulting states and rewards, and adjusts its policy iteratively to achieve the highest possible cumulative reward over time.

Although both reinforcement learning and user simulation utilize MDPs, there are important differences arising from their distinct objectives.
In RL, the main focus revolves around finding an optimal policy~\citep{Singh:1999:NIPS}.
Designing effective reward functions is crucial to achieve that, while transition probabilities are often observed from an external environment, without necessarily modeling them explicitly, though in model-based RL, predictive models for state transitions and actions may be used~\citep{Moerland:2023:Survey}. 
Conversely, in user simulation, the reward function can be used to encapsulate the costs and rewards based on observed data (from logs or user studies), while transition probabilities are modeled explicitly based on some model of user behaviour.\footnote{Parameters of the user model may be hand-crafted and/or learned from data.} 
Even though a rational user behaviour may generally be assumed, the policy, governing the user's decision making, does not need to be optimal.  Indeed, humans are unlikely to behave optimally when exposed to a new information access system for the first time, as it takes time to learn how the system works and how it can be utilized most effectively.
In user simulation, the policy is also based on an explicit model of user behaviour.  Some decisions might be made based on deterministic rules while others vary across users, depending on their personal characteristics.
Crucially, for the purposes of simulation-based evaluation, the policy needs to be controllable by the system designer. 

Decisions may also be modeled in terms of costs and benefits from an economic perspective~\citep{Azzopardi:2011:SIGIR,Azzopardi:2014:SIGIR,Azzopardi:2016:ICTIR}.
Economic models can help better understand user interactions via testable hypotheses, which can in turn inform system design.
However, they may not be effective in predicting user decisions in real-world situations due to their assumptions of perfect information and rational decision-making, which may not hold true in practice.

One benefit of framing the user simulation problem as to create a user simulation agent based on MDP is the natural connection with various learning algorithms such as RL and imitation learning~\citep{Hussein:2017:CSUR,Labhishetty:2023:thesis}. It also allows systematic discussion of all components in a user simulator. We use the MDP framework to systematically organize the simulation of various components of search and recommender systems in Chapter~\ref{ch:sim_search}.
In the context of task-oriented dialogue systems, MDPs have been used for user simulation for a long time; we continue to build on this in Chapter~\ref{ch:sim_conv}.

\chapter{Simulating Interactions with Search and Recommender Systems}
\label{ch:sim_search}

This chapter focuses on simulating how a user interacts with search engines and recommender systems.
These represent the two main modes of information access: \emph{pull}, where search is initiated by the user, and \emph{push}, where the system takes initiative and recommends content to the user (cf. Section~\ref{sec:intro:tasks}).
There are several reasons for treating and discussing these two modes together.
First, it has been established early on that the two problems are essentially ``two sides of the same coin''~\citep{Belkin:1992:CACM} and it has also been reiterated more recently that there are theoretical and practical reasons to model the two tasks together~\citep{Zamani:2018:DESIRES}.
Indeed, the separation between the two tasks is increasingly blurred in modern information service platforms, where search results may be accompanied by related items as recommendations, and recommendations may be filtered based on search queries.
Second, there is a lot of common elements when it comes to how a user interacts with these systems: in both cases, the user needs to examine a ranked list of items, select which ones they find attractive enough to click on, and decide when to stop interacting with the system.
There are, of course, differences as well: query formulation is specific to search, while leaving explicit feedback (such as likes or ratings) is a signature characteristic of recommender systems.
Overall, there has been considerably more research attention given to user modeling and evaluation for search engines than for recommender systems, which will be reflected in this chapter.
For simplicity, we will be using standard terminology from information retrieval throughout most of the chapter. %
That is, we will be talking about \emph{result lists} served on search engine result pages (SERPs), consisting of a ranked list of \emph{results} or \emph{snippets} that link to \emph{documents}. However, snippets may be replaced with \emph{previews} and documents may be replaced with arbitrary \emph{information objects} (e.g., entities, products, images, videos).

We start in Section~\ref{sec:sim_search:workflows} by presenting models that describe interaction workflows, specifying the space of user actions and system responses, and possible transitions between them.
From the perspective of the cooperative game framework~\citep{Zhai:2016:IEEE} (cf. Section~\ref{sec:frameworks:general}), these models define the ``moves'' that a user and a system can each make. In terms of a Markov Decision Process (MDP) framework (cf. Section~\ref{sec:decisions:math}), these define the space of actions and the possible transitions between them.
Then, we discuss specific user actions: query formulation in Section~\ref{sec:sim_search:queries}, scanning behaviour in Section~\ref{sec:sim_search:scanning}, clicks in Section~\ref{sec:sim_search:clicks}, effort involved in processing documents in Section~\ref{sec:sim_search:processing}, %
and stopping in Section~\ref{sec:sim_search:stopping}.
The length and depth of discussion in each of these sections are proportional to the amount of research effort invested in each respective area. %
We close this chapter in Section~\ref{sec:sim_search:summary} with a summary, discussing how the various component-level models fit together and identifying where gaps and limitations remain.

\section{Workflow Models}
\label{sec:sim_search:workflows}

Simulation relies on simplified models (of workflows and user behaviour), which allows for ``unnecessary complications'' to be abstracted away.
These models are ``neither complete nor even accurate with respect to reality, which is simulated and investigated''~\citep{Baskaya:2014:PhDThesis}.
With the aim of arriving at more accurate models, the main research challenge is determining what elements of human behaviour to capture in these abstractions, while keeping the models as simple as possible.
We look at workflow models that attempt to capture the cognitive processes of searchers, by considering specific actions and decision points.
These will provide us with the necessary scaffolding to develop user simulators.

\subsection{Search Workflows}

\begin{figure}[t]
    \centering
    \vspace*{-0.5\baselineskip}
    \includegraphics[width=0.8\textwidth]{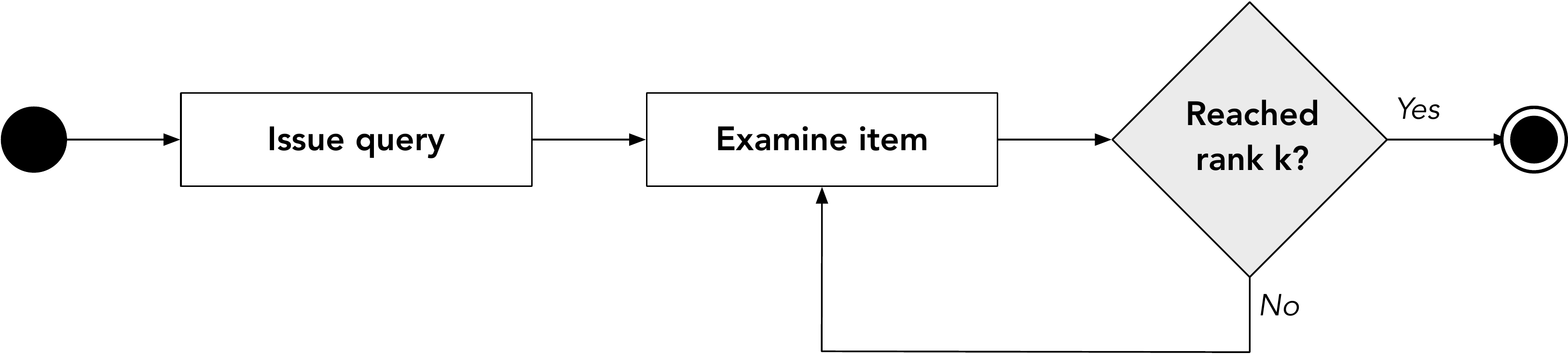}
    \caption{Flowchart of a naive searcher model, corresponding to highly abstracted user (illustration adapted from~\citep{Maxwell:2019:PhDThesis}).}
    \label{fig:naive_searcher}
\end{figure}

To begin, consider the simple searcher model shown in Fig.~\ref{fig:naive_searcher}, which describes how a user is assumed to behave according to traditional evaluation measures, such as MAP or NDCG@k (cf. Section~\ref{sec:frameworks:naive}).
In this flowchart, following the notation in \citep{Maxwell:2019:PhDThesis}, \flowstart~and \flowend~signify the start and the end of the search process, respectively, arrows $\rightarrow$ show the direction of the flow, actions are denoted by unfilled rectangles \flowaction, and decision points are represented as rhombuses \flowdecision.
The naive searcher model corresponds to a highly abstracted user, who would examine each individual result in linear order up to a fixed ranked depth of $k$, then stop.

\begin{figure}[t]
    \centering
    \includegraphics[width=0.8\textwidth]{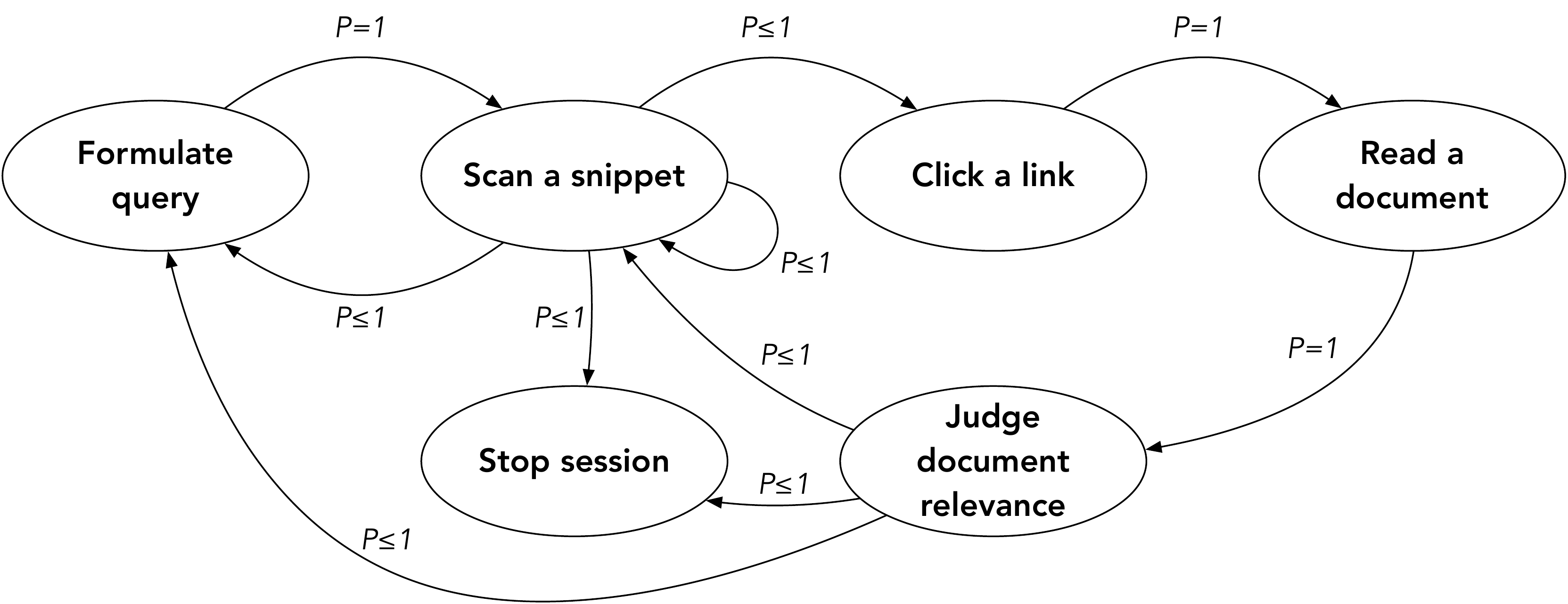}
    \caption{Automaton expressing the subtasks performed during a search session, according to \citet{Baskaya:2013:CIKM} (illustration adapted from \citep{Baskaya:2013:CIKM}).}
    \label{fig:search_session}
\end{figure}

\begin{figure}[t]
    \centering
    \includegraphics[width=0.9\textwidth]{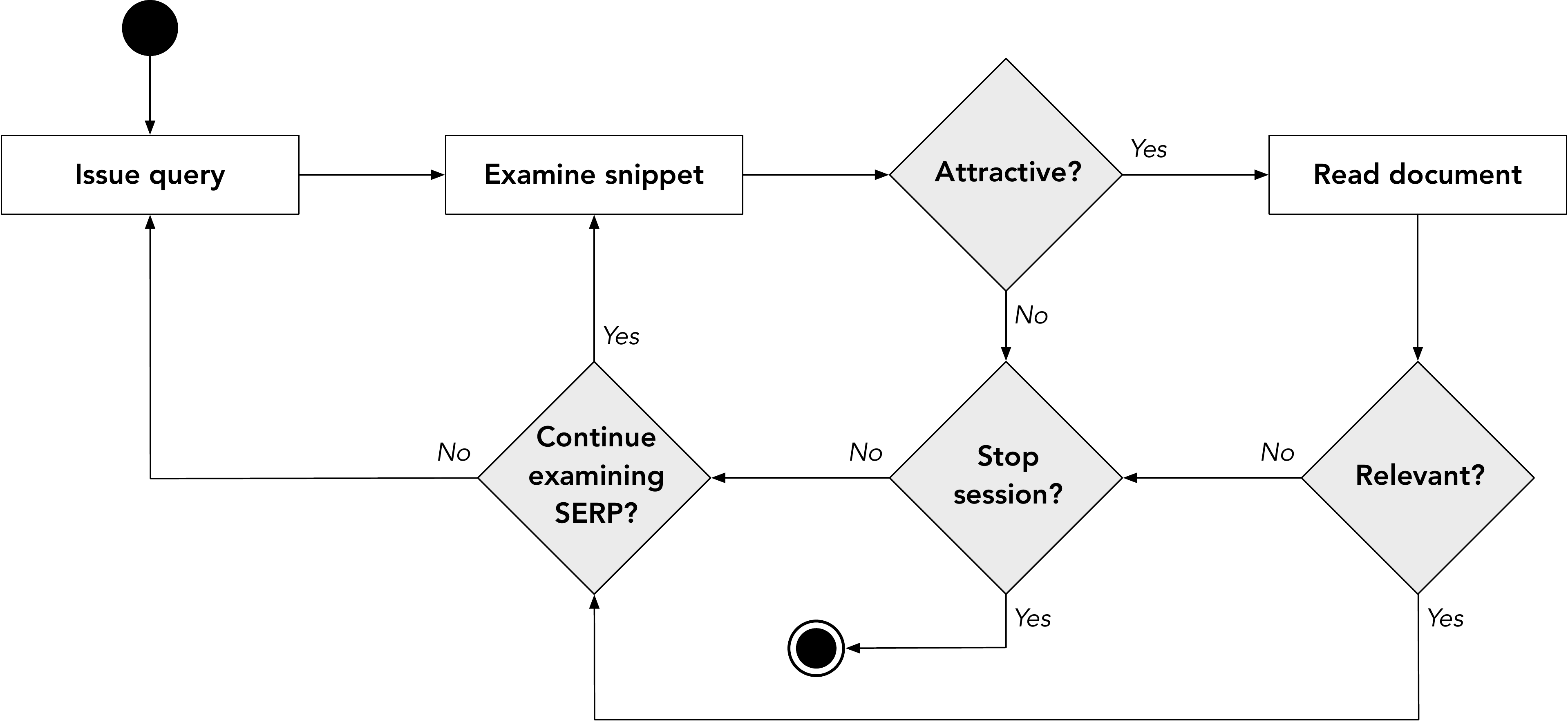}
    \caption{Searcher model by \citet{Baskaya:2013:CIKM}, visualized as a flowchart (illustration adapted from~\citet{Maxwell:2019:PhDThesis}).}
    \label{fig:searcher_model}
\end{figure}

In reality, search is a lot more interactive process that involves more decision points and a much broader space of actions, including (1) query (re)formulation, (2) snippet scanning and assessment, (3) results clicking, (4) document examination (reading), (5) document assessment, and (6) session stopping.
A state-based stochastic model of the user with the above actions is presented in \citep{Baskaya:2013:CIKM} and is shown in Fig.~\ref{fig:search_session}.
Here, search is modeled as a Markov process, which considers the different actions as states, with transition probabilities that define the probability of the user performing an action at a given state (which represents the previous action taken by the user).
There are more detailed, and thereby more realistic, models of the search process, which we will look at shortly.
Nevertheless, the most important user actions are captured in this model.  Figure~\ref{fig:search_session} can thus serve as a guide to structuring our section on the modeling of various user decisions.
Figure~\ref{fig:searcher_model} shows \citeauthor{Baskaya:2013:CIKM}'s searcher model as a flowchart, as derived by \citet{Maxwell:2019:PhDThesis}.
While the underlying model is the same as in Fig.~\ref{fig:search_session}, the flowchart notation emphasizes the differences between the user making a decision  (\flowdecision) and taking an action (\flowaction).\footnote{Making a decision could also be regarded as taking an ``implicit action'' that is not observed.}
Contrasting with the naive searcher model (Fig.~\ref{fig:naive_searcher}), improvements to make it more realistic include (1) considering multiple queries within a search session, (2) separating the examination of result snippets from the actual documents, and (3) permitting the user to stop at any point.

\begin{figure}[t]
    \centering
    \includegraphics[width=0.9\textwidth]{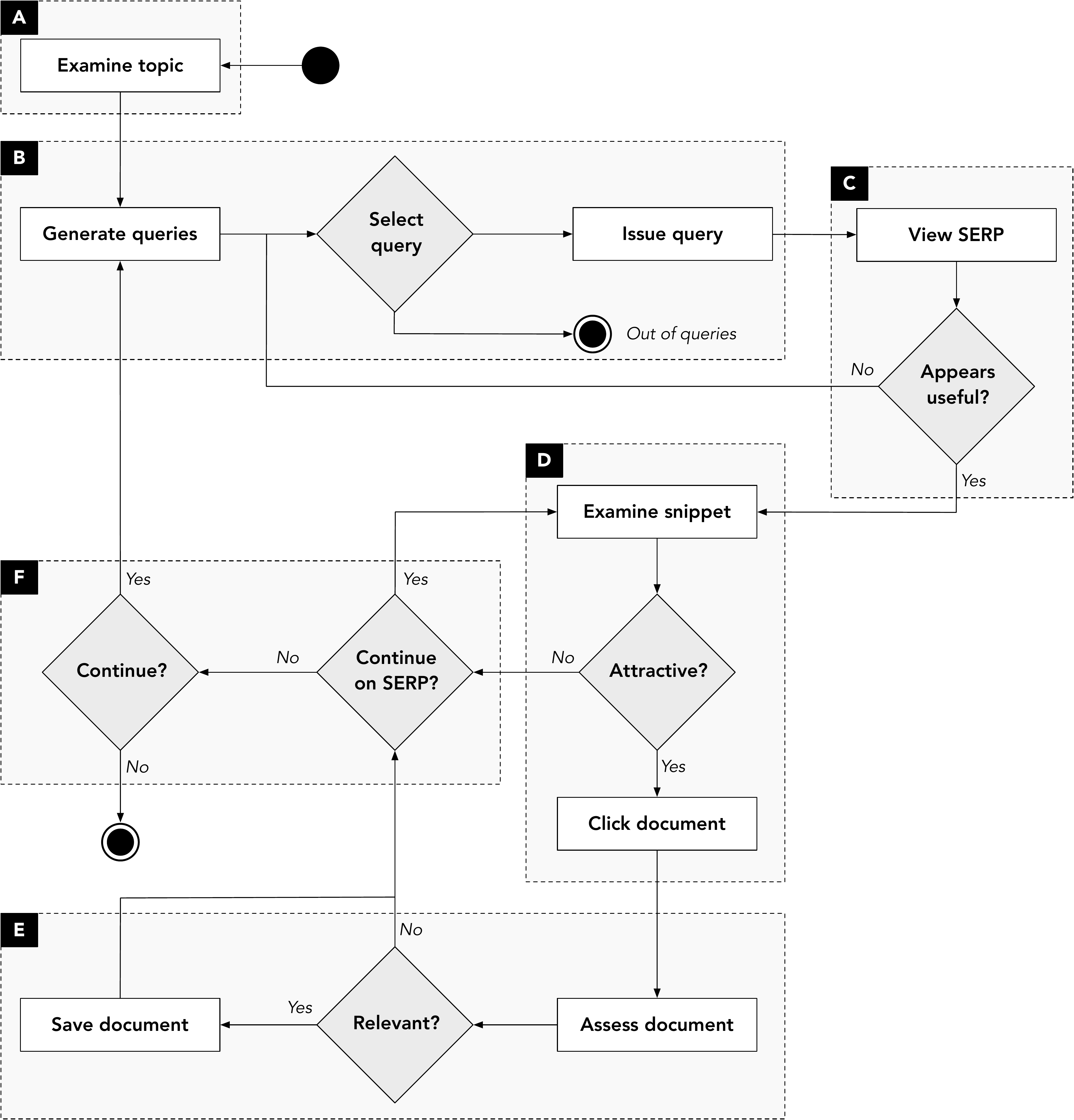}
    \caption{Flowchart of the Complex Searcher Model (illustration adapted from~\citet{Maxwell:2019:PhDThesis}). The main components are: (A) topic examination, (B) querying, (C) SERP examination, (D) result summary examination, (E) document examination, and (F) deciding to stop.}
    \label{fig:complex_searcher_model}
\end{figure}

The Complex Searcher Model, proposed by \citet{Maxwell:2015:CIKM} and then further updated in \citep{Maxwell:2018:ECIR}, combines actions from prior searcher models~\citep{Baskaya:2013:CIKM,Thomas:2014:IIiX} and additionally introduces a SERP-level examination decision; see Fig.~\ref{fig:complex_searcher_model}.
The boxes (A)--(F) on this figure denote sets of interactions:
\begin{enumerate*}[(A)]
    \item topic examination, which assumes a TREC-like topic description outlining the information need that motivates the search;\footnote{For ordinary users, without TREC-like topic descriptions, this step corresponds to reflecting on their information need.}
    \item querying, i.e., the process of generating potential queries to address the underlying information need and deciding which ones to issue;
    \item SERP examination, where the searcher obtains an initial impression of the SERP and decides whether it appears promising enough to examine the results in detail or rather it should be abandoned;
    \item result summary examination, that is, looking at the result snippets and determining if they look attractive enough to be clicked for further examination;
    \item document examination, which involves reading (parts of the) document to assess whether it is relevant to the information need, and ``saving'' it, if it is;
    \item deciding to stop, i.e., making decisions about whether or not to stay on the current SERP and, if not, whether to continue with the search session (and issue further queries) or terminate the session.
\end{enumerate*}

There exist various extensions to the Complex Searcher Model.  For example, the User State Model~\citep{Maxwell:2016:CIKM} extends it with modeling the cognitive state of the user, which maintain what the user knows, has done and seen, and considers relevant.
The Subtopic Aware Complex Searcher Model~\citep{Camara:2022:ECIR} considers multiple aspects of a complex information need as subtopics and models the subtopic selection and subtopic switching steps in the search process, as well as the user's cognitive state for each subtopic.

All the above models assume that the user is interacting with a single search engine.  The selection of the search service itself could also be made as a decision point in the process, as suggested in \citep{Thomas:2014:IIiX}.

In Sections~\ref{sec:sim_search:queries}--\ref{sec:sim_search:processing} and \ref{sec:sim_search:stopping}, we will look at each of the main user actions depicted in Fig.~\ref{fig:search_session} in detail and discuss approaches to simulate respective user behaviour.

\subsection{Recommendation Workflow}

\begin{figure}[t]
    \centering
    \vspace*{-0.5\baselineskip}
    \includegraphics[width=0.58\textwidth]{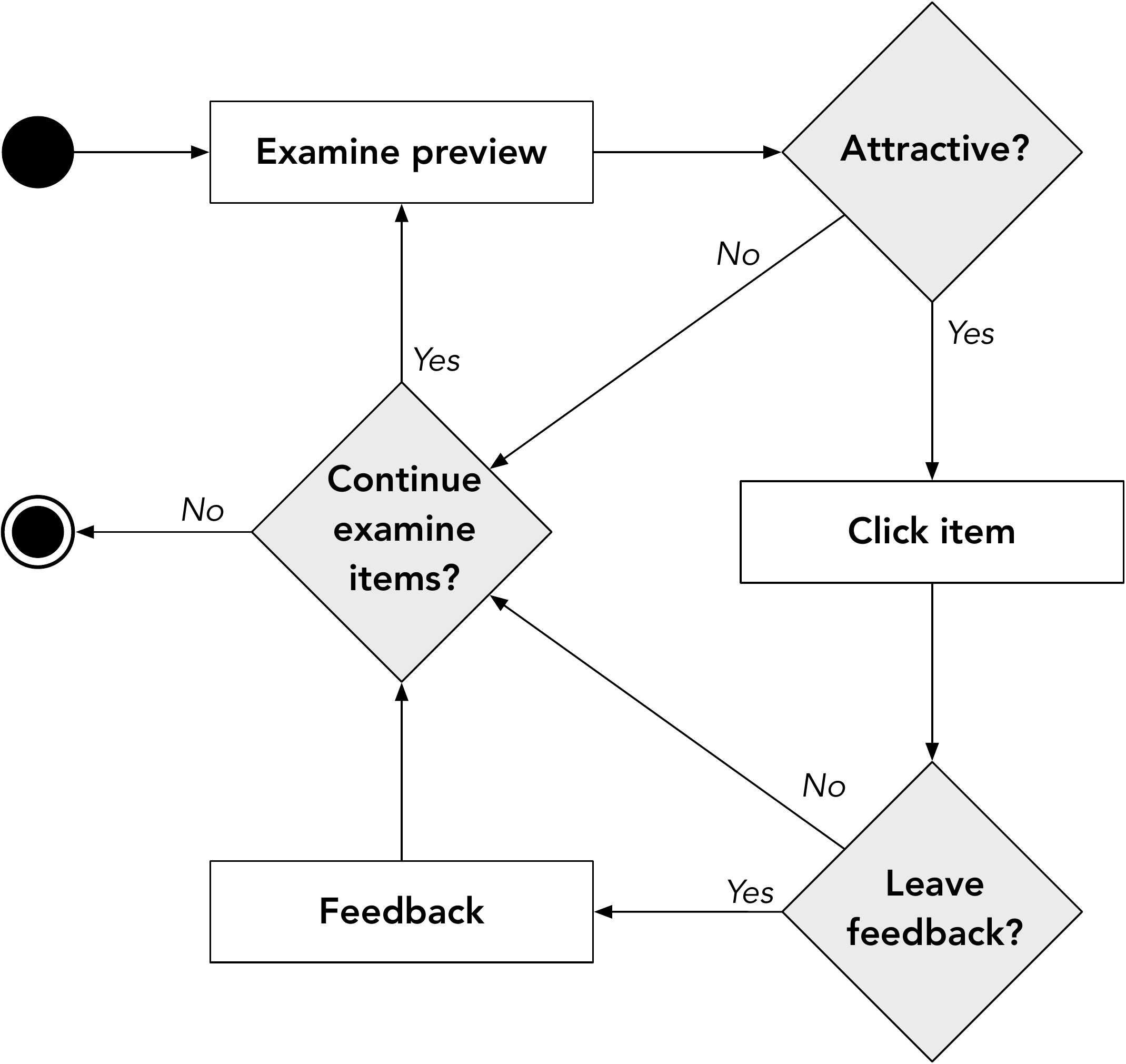}
    \caption{Flowchart of a basic user model for recommender systems that present recommendations as a ranked list.}
    \label{fig:recsys_model}
\end{figure}

\begin{figure}[t]
    \centering
    \includegraphics[width=0.8\textwidth]{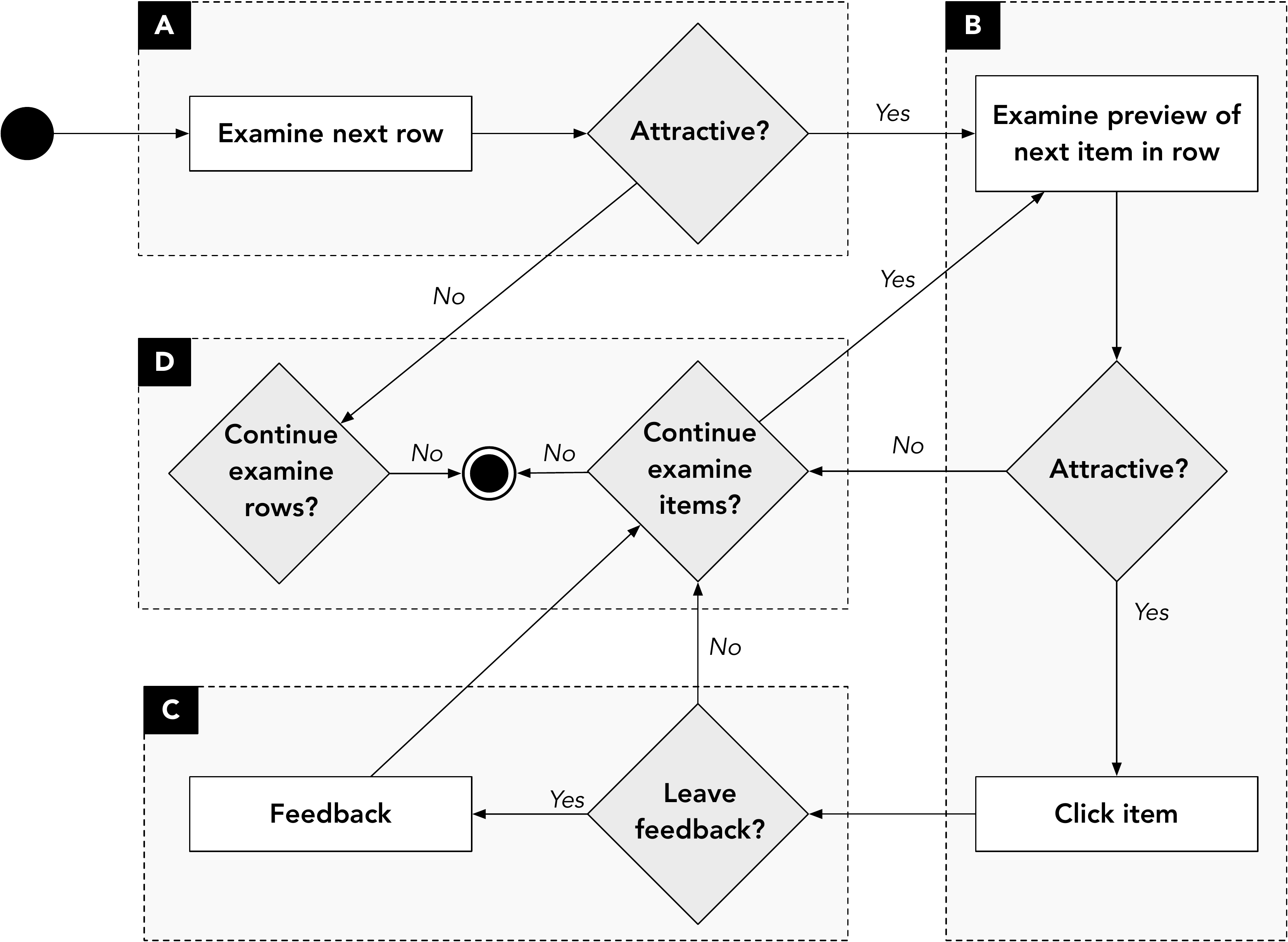}
    \caption{Flowchart of an advanced user model for recommender systems, corresponding to carousel-based interfaces with multiple ranked lists (rows). The main components are: (A) row examination, (B) item examination, (C) feedback, and (D) stopping decisions.}
    \label{fig:recsys_model_advanced}
\end{figure}

We are not aware of any workflow models being proposed specifically for recommendation, though elements of a workflow pertaining to users' scanning and clicking behaviour in carousel-based user interfaces are described in~\citep{Rahdari:2024:TORS}.
In the most basic setting, where recommendations are presented as a ranked list, we can imagine that a user would interact with a recommender system according to the following simplified workflow, depicted in Fig.~\ref{fig:recsys_model}.  The user examines previews of items in a result list, and they decide for each if they find it attractive enough to click on it.  Once they clicked on it (and consumed the item, which is not an action we consider in this simple model), they decide if they want to leave feedback on it.  Then, they can either go back to the result list (which may or may not be updated based on feedback left during the session) and continue examining items or stop interacting with the recommender.

Clearly, the above workflow is largely incomplete as it does not consider how modern recommender systems typically present results using carousel-based interfaces (cf. Section~\ref{sec:sim_search:scanning_complex}).
There, the results are not presented as a single ranked list but rather as several ranked lists in multiple rows, each labeled with a category or topic.
Inspired by the work of~\citet{Rahdari:2024:TORS}, we present an advanced workflow in Fig.~\ref{fig:recsys_model_advanced} corresponding to carousel-based interfaces. Each carousel contains a ranked list of items, which we refer to simply as a \emph{row}. It is assumed that the user examines rows in sequential order. If they find a row attractive, they will examine the items contained in the row, also in sequential order, and can click on them and leave feedback the same way as in the basic model. They can decide to stop examining items within a row and, then, to continue to the next row or stop.

This workflow might be further extended based on additional functionalities offered by modern recommender systems, such as exploratory browsing by showing additional recommendations based on the current item being viewed. These additional recommendations can appear in various forms, such as related items that are similar or complementary (e.g., in an e-commerce setting, under ``similar products'' and ``others also bought'' sections), encouraging a deeper exploration of available items.
However, this is beyond the scope of this book, and the development of more complex workflows that correspond to modern recommender system UIs remains an open area for future research.

The various user actions depicted in Figs.~\ref{fig:recsys_model} and~\ref{fig:recsys_model_advanced} will be discussed in Sections~\ref{sec:sim_search:scanning}--\ref{sec:sim_search:stopping}.

\subsection{Multiple Workflows}

It is worth noting that multiple workflows, including both homogeneous workflows (e.g., multiple search engine workflows or multiple recommendation workflows) and heterogeneous workflows (e.g., search workflow combined with recommendation workflow), can generally be combined in a mixed-mode information access system, potentially generating infinitely many different variations of workflows. From a user simulation perspective, it means that for different systems with different workflows, we generally need to model user actions differently since the user actions could vary significantly. Another factor is the context or state in the MDP framework used for modeling a user, which clearly depends on the specific workflow. A general way of modeling interactions with any interface is to use the Interface Card Model~\citep{Zhang:2015:SIGIR}, where a user is modeled with three component probabilistic models, i.e., probability of actions to be taken on any given interface, the cost of performing each action, and the utility of performing each action. How this model can be used to support a general simulation-based evaluation framework was discussed earlier in Chapter~\ref{ch:frameworks}.
\section{Simulating Queries}
\label{sec:sim_search:queries}

Formulating a query is generally the very first action that a user takes in search-based information seeking. A user also often reformulates the query later in a session either because the previous queries are not effective or because they want to explore a topic more broadly.
Before presenting respective simulation techniques, we discuss how these actions may be modeled within the MDP framework.

Simulating a user's initial query can be treated as designing a policy to decide the action that a user would take in the initial state of information seeking. Since there are infinitely many possible queries that could potentially be issued, existing work has made assumptions to constrain the query space and make simulation tractable, e.g., the query length does not exceed three words.
The initial state, which may include the user's knowledge state, information need, as well as the search system state (e.g., the collection), would also potentially influence the query to be generated by the user. It is especially interesting to note that the ASK theory of information need~\citep{Belkin:1982:JD} naturally suggests a representation of the user's information need using the MDP framework as some kind of ``hole" in the user's knowledge space. %
In a regular search scenario, the query would generally be composed of keywords that the user believe might match with relevant content that can fill the hole.
It is thus easy to see that the quality of a user's query would highly depend on the user's knowledge about the relevant content to be found. Consequently, query simulation requires an assumption to be made about the user's knowledge. We may identify two extremes: In one extreme case, the user may actually know everything about the missing information (no ``hole" in the knowledge space), and can thus formulate a ``perfect'' query with the right keywords. In such a case, the user's search goal is generally to find some ``known items" rather than relevant information. %
This scenario is referred to as \emph{known item search}. In the other extreme case, the user has little or no knowledge about the relevant content; this happens, e.g., in the case of learning about a completely new concept or topic (e.g., in a medical search scenario). In those instances, the user generally faces difficult challenges in formulating an effective query due to a gap in their vocabulary. Overall, the state in the MDP framework for simulating an initial query represents multi-dimensional information, including especially the user's information need and search task.

Query reformulations happen in a different state than initial querying, though the process of generating a query may be similar. The main difference is that in the initial stage, the user has no choice of other kinds of actions (e.g., browsing results)---the only choice is between the different queries that could possibly be issued. However, in the case of reformulating a query, in addition to choosing between different queries, the user generally also has the option to continue browsing the current search result list. %
The user's state when reformulating a query is also in general different from the initial state due to learning during the search process; the user generally would have gained more information about relevant and/or non-relevant content, which may help them to come up with a potentially more effective query reformulation.

Regardless of whether it is to simulate initial query formulation or subsequent query reformulation, the state in the MDP would make a similar transition to a new state, where the user would be exposed to the search results returned by the search engine in response to the query. In the new state, the user would then face new decisions on different actions (i.e., different ways to scan the results, which will be discussed later). \\

\begin{table}[t]
    \centering
    \footnotesize
    \caption{Overview of query generation approaches discussed in Section~\ref{sec:sim_search:queries}.
        Notation: $q$ query, $d$ document, $T$ topic, $q_0$ seed query, $R$ set of relevant documents, $s$ topic description, $Q$ set of queries, $S_{1..i-1}$ session history. The method can be (Prob)abilistic or (Det)erministic and (Dyn)amic or (Stat)ic.}
    \label{tab:query_simulation}
    \begin{tabular}{@{~}llll@{~~~}l@{~}}
        \toprule
        \textbf{Generation} & \textbf{Reference}            & \textbf{Input} $\Rightarrow$ \textbf{Output}              & \multicolumn{2}{l}{\textbf{Method}}         \\
        \midrule
        Individual queries
                            & \citep{Azzopardi:2007:SIGIR}  & $\emptyset \Rightarrow (q,d)$                             & Prob.                               & Stat. \\
                            & \citep{Azzopardi:2009:SIGIR}  & $T=(q_0,R) \Rightarrow q$                                 & Prob.                               & Stat. \\
        \midrule
        Controlled query sets
                            & \citep{Jordan:2006:JCDL}      & $R \Rightarrow \langle q_1,.., q_n \rangle$               & Det.                                & Stat. \\
        \midrule
        Query reformulations
                            & \citep{Baskaya:2012:SIGIR}    & $\{t_1, .., t_m\} \Rightarrow \langle q_1,.. q_n \rangle$ & Det.                                & Stat. \\
                            & \citep{Carterette:2015:ICTIR} & $T=(s,Q)$, $S_{1..i-1} \Rightarrow q_i$                   & Prob.                               & Dyn.  \\
        \bottomrule
    \end{tabular}
\end{table}

\noindent
The problem of simulating queries has attracted considerable attention in IR. We discuss seminal work organized into three main groups: the generation of individual queries, the generation of query sequences based on discriminative power, and query reformulations; see Table~\ref{tab:query_simulation} for an overview.
Generally, most methods work by sampling terms from documents/topic language models and often make use of existing test collections, i.e., TREC-like topic definitions (see Fig.~\ref{fig:trec_topic} for an example) and relevance judgments.  The generation process may be probabilistic or deterministic in nature.  Most methods are \emph{static} in a sense that they are not affected by what the system does; \emph{dynamic} query generation, on the other hand, considers previous queries and results returned for those queries.
It is worth emphasizing that simulated queries ``are synthetic in nature and are not necessarily representative of real user queries''~\citep{Jordan:2006:JCDL}. 
For example, \citet{Breuer:2022:ECIR} show that is is possible to reproduce statistical properties of real queries and preserve the relative system orderings using simulated queries, but the query terms are very different from those used in real queries.

\begin{figure}[t]
    \footnotesize
    \fbox{
        \parbox{\textwidth}{
            <num> Number: 303\\
            <title> Hubble Telescope Achievements\\

            <desc> Description: \\
            Identify positive accomplishments of the Hubble telescope since it
            was launched in 1991.\\

            <narr> Narrative:\\
            Documents are relevant that show the Hubble telescope has produced
            new data, better quality data than previously available, data that
            has increased human knowledge of the universe, or data that has led
            to disproving previously existing theories or hypotheses.  Documents
            limited to the shortcomings of the telescope would be irrelevant.
            Details of repairs or modifications to the telescope without
            reference to positive achievements would not be relevant.
        }}
    \caption{Example TREC topic definition (from Robust 2003 track).  The terms present in such topic definitions are often used as the basis of query generation.}
    \label{fig:trec_topic}
\end{figure}

\subsection{Generating Individual Queries}

\citet{Azzopardi:2007:SIGIR} propose generative probabilistic models that generate queries for \emph{known item search}.  It is assumed that the user has a particular, previously seen document in mind, the ``known item,'' that they want to retrieve.  The user thus tries to construct a query by recalling terms of phrases from that document.  
The process involves sampling a document $d$ to be the known item with probability $P(d)$, selecting a query length $l$ with probability $P(l)$, and then creating a query by sampling terms $t_1,\dots,t_l$ from the (unigram) language model of document $d$ with probability $\prob(t_i|\theta_d)$. The probability distributions $\prob(d)$, $\prob(l)$, and $\prob(t_i|d)$ characterize the type and style of queries that are generated.

Notice that this process generates not only queries but query-document pairs; since there is a single relevant document (the known item), it allows for automatic evaluation with a query set of any size, without needing to collect relevance judgments. A user may also be assumed to recognize the relevant document when the search result contains it and thus click on such a document (see the discussion of click modeling later).

\citet{Azzopardi:2009:SIGIR} extend the above process for the generation of \emph{ad hoc queries}.
Here, it is assumed that a topic $T=(q_0, R)$ is given as input, which is defined by a seed query $q_0$ (i.e., TREC topic titles) and the corresponding set of relevant documents $R$.
Queries of a given length are generated by repeatedly sampling terms according to a two-component mixture query language model,  where the first term ($\prob(t_i|\theta_T)$) corresponds to the selection of terms that are on topic, while the second term (\prob$(t_i|C)$) corresponds to sampling from the background collection (i.e., noise).
To estimate the topic model, possible strategies include selecting frequent or discriminative terms from the set $R$ of relevant documents, or conditioning term selection on the seed query $q_0$.

\subsection{Controlled Generation of Query Sequences}

Next, we move from generating single queries to creating a set of queries with the same underlying search intent.
The Controlled Query Generation method~\citep{Jordan:2006:JCDL} takes a set $R$ of documents as input and generates a sequence of queries $\langle q_1, \dots, q_n \rangle$ targeting these documents.
This deterministic process identifies the most discriminating terms in the target documents using relative entropy, with the first query ($q_1$) theoretically being the best and the last query $(q_n)$ the worst in identifying the source document set.
A language model is generated from the documents in $R$, and terms are sorted based on their contribution to relative entropy. Query generation is initiated with the most discriminating term, and iteratively, the next most discriminating term not present in previous queries is identified to formulate the next query. The process terminates when no further terms remain or the highest term score falls below a predefined threshold. Three types of queries are considered: (a) single-term queries, (b) two-term queries, and (c) queries of varying length.

\citet{Jordan:2006:JCDL} construct source document sets ($R$) by utilizing metadata annotations of documents (i.e., topic and location tags).  These document sets are categorically related, but could be too artificial; for example, it is recognized that some sets may be very broad topically.  Nevertheless, a large number of sets can be created cheaply this way, as the process does not require any human involvement.  Alternatively, relevant document sets of existing test collections could also be used as $R$~\citep{Azzopardi:2009:SIGIR}.

\subsection{Generating Query Reformulations}

Queries within a search session are not randomly sampled from some predefined set of possible queries, but evolve in some way.  For example, they are reformulated according to some structural patterns or based on the results returned for previous queries in the session.

Building on and extending the work of \citet{Keskustalo:2009:AIRS}, \citet{Baskaya:2012:SIGIR} present five prototypical query modification strategies based on observed user behaviour.
It is assumed that a fixed set of terms ($t_1,\dots,t_m$) is available for each topic, from which queries may be constructed.  
The strategies are based on term level changes that users may employ and include:
(S1) repeated replacement of the initial term ($t_1 \rightarrow t_2 \rightarrow t_3 \rightarrow \dots$), 
(S2) variation of the second term in a two-term query ($t_1, t_2 \rightarrow t_1, t_3 \rightarrow t_1, t_4 \rightarrow \dots$),
(S3) variation of the third term in a three-term query ($t_1, t_2, t_3 \rightarrow t_1, t_2, t_4 \rightarrow t_1, t_2, t_5 \rightarrow \dots $),
(S4) extension of a single-term query with new terms ($t_1 \rightarrow t_1, t_2 \rightarrow t_1, t_2, t_3 \rightarrow \dots$), and 
(S5) extension of a two-term query with new terms ($t_1, t_2 \rightarrow t_1, t_2, t_3 \rightarrow t_1, t_2, t_3, t_4 \rightarrow \dots$).

These strategies assume that query terms are given in a predefined order and therefore the generation of query sequences is deterministic.
To implement these approaches in practice, query terms may be selected and ordered by human annotators~\citep{Keskustalo:2009:AIRS}, taken from the topic description by maintaining the original word order~\citep{Verberne:2015:ECIR}, or sampled and ranked using a language model of the topic description~\citep{Maxwell:2016:CIKM}.

\citet{Carterette:2015:ICTIR} generate queries dynamically, considering the history (queries and corresponding results) up to a given point $i$ in the session.
It is assumed that topics $T=(s,Q)$ come with a textual description $s$ and a set of queries $Q$; such data is available, for example, in the TREC Session track~\citep{Carterette:2016:SIGIR}.
One key component of the model is the conditioning of the query length $l$ on the topic, $\prob(l|T)$.  This is motivated by the fact that some topics naturally lend themselves to longer queries.
Another key element is that the language model from which query terms are sampled, $\prob(t|T,l,i)$, is continuously updated based on the results the user has seen for previous queries in the session.  
This is a simple way of modeling how users learn during the search process, which is disregarded in other query generation approaches.
Instead of sampling queries from a traditional multinomial (unigram) language model, a two-stage process is used that yields more realistic queries.
In the first step, multiple candidate queries are generated by sampling terms according to the binomial model $\prob(t|T,l,i)$. In the second step, one query is selected based on its discriminative power among topics, given by $\prob(q|T)$.

\subsection{Query Generation as an Optimization Problem}

Query generation has also been framed as an optimization problem in recent work.
In \citep{Labhishetty:2022:ICTIR}, an optimization framework called PRE (Precision-Recall-Effort) is proposed to simulate query generation, based on two assumptions about the user's behaviour. The first is that when a user generates a query (either an initial query or a reformulated query), they would attempt to choose a query that maximizes both the expected Precision and expected Recall, while simultaneously attempting to minimize the effort to be made. The second assumption is that a user's estimate of Precision and Recall depends on the user's knowledge state, which would be updated during a search session based on any new knowledge/information that the user could gain in the search process. The explicit dependency of query formulation on the user's knowledge state makes it possible to model the initial query formulation and subsequent reformulations uniformly using the same optimization framework with the same objectives. Part of the user's knowledge is the knowledge about how a search system works, and the user is also assumed to learn more about how that system works over time and adjust query formulation strategies accordingly.

Specifically, the expected Recall and Precision are both modeled with probabilistic models that are conditioned on the user's knowledge state, which would be updated over time to reflect the incremental acquisition of knowledge.
Formally, the expected Recall of a query $q$ is modeled as
\begin{equation}
    \mathbf{Rec}(q,\mR) = \prod_{d \in \mR} \prob(\mathit{Match}=1|q,d) ~,
\end{equation}
where $\mathit{Match} \in \{0,1\}$ is a binary variable that denotes whether there is a match ($\mathit{Match}=1$) or not ($\mathit{Match}=0$) between a query $q$ and a document $d$ in the collection $\mC$, and $\mR \subset \mC$ denotes the subset of relevant documents. The conjunctive expression here encodes the objective of finding a query that can match every relevant document in $\mR$, thus capturing the desire of maximizing Recall.

The expected precision of query $q$ is modeled as
\begin{equation}
    \mathbf{Prec}(q,\mR) = \prod_{d \in \mC-\mR}(1-\prob(\mathit{Match}=1|q,d)) ~,
\end{equation}
where $\mC-\mR$ is the set of non-relevant documents and
$1-\prob(\mathit{Match}=1|q,d)$ is the probability that $q$ does not match $d$. The conjunctive relation here captures the goal of not matching any of the non-relevant documents, which intuitively captures the goal of maximizing Precision. Note that by avoiding matching with non-relevant documents, the Precision objective does not necessarily ensure matching relevant ones, which, however, is achieved via the Recall objective. Thus, the Recall and Precision objectives should be used together. 

Adding a model $\mathbf{E}(q)$ of the effort needed to formulate query $q$, the PRE framework states that the query formulation process is a multi-objective optimization process involving three potentially conflicting objectives: (1) model of Recall: $\mathbf{Rec}(q,\mR)$, (2) model of Precision: $\mathbf{Prec}(q,\mR)$, and (3) user effort: $\mathbf{E}(q)$, formally defined as follows:
\begin{equation}
    q^* = \arg\max_q \alpha \log \mathbf{Rec}(q, \mR) + (1-\alpha)\log \mathbf{Prec}(q,\mR)-\lambda \mathbf{E}(q) ~,
\end{equation}
where $\alpha$ and $\lambda$ are parameters to control the trade-off between multiple objectives.

The PRE framework serves as a general computational simulation framework for query formulation with an explicitly specified interpretable objective function to be optimized. Most other specific query formulation simulation models can be regarded as special cases of the framework. In particular, $\prob(\mathit{Match}=1|q,d)$, which is the basis for estimating both Recall and Precision,  can be instantiated in many different ways to model different notions of ``matching'' that a user can potentially take when estimating the Recall or Precision of a query. A user's knowledge about terms often used in relevant or non-relevant documents can be naturally involved in defining $\prob(\mathit{Match}=1|q,d)$, thus allowing for modeling query reformulation after a user learns more about the content of relevant and non-relevant documents via interacting with search results.  Some basic strategies to instantiate each component of PRE have been presented and studied in~\citep{Labhishetty:2022:ICTIR}, where the results show that the need for optimizing both Recall and Precision when formulating a query is indeed empirically supported.

\subsection{User Knowledge and Cognitive State}
\label{sec:sim_search:cognitive_state}

Users with different backgrounds exhibit notable variations in their topical knowledge and prior experience with search engines. These factors influence the queries they would issue. Also, during the search session, users gain more knowledge about the topic as well as about how the search system works, both in general and with respect to their current information need, which influences future interactions.
Currently, these factors are either completely ignored or modeled in a simple fashion, by representing users' knowledge as term distributions~\citep{Maxwell:2016:CIKM,Camara:2022:ECIR}.

\citet{Labhishetty:2020:SDMWorkshop} propose a cognitive state user model (CSUM) for e-commerce search, modeling the information need and knowledge state of the user. %
In particular, it attempts to model to what extent a user's information need is exploratory as opposed to a fixed information need, which is particularly meaningful in e-commerce search since a user may sometimes want to purchase the same (consumable) product regularly (fixed information need) or purchase a product for a purpose such as using it as a Christmas gift without necessarily having a specific product in mind (exploratory information need).  CSUM also includes components to model all the major search behaviour of a user, including query formulation, clicking on results, and query reformulation, and thus provides a relatively complete model for users of e-commerce.

\section{Simulating Scanning Behaviour}
\label{sec:sim_search:scanning}
\vspace*{-0.25\baselineskip}

Scanning behaviour is concerned with how the user processes the list of results presented to them in response to their search query.
While the results are often displayed as a ranked list of snippets, more complex interfaces are also possible. %
From the perspective of modeling in the MDP framework, the user is in state $S_S$ when interacting with the presented search results.
In this state, the actions that a user could potentially take $A_S$ depend on the interface features provided by the search system.
(To be consistent in notations, we tend to use variable $S$ to denote a state and variable $A$ to denote an action. The subscript $s$ here stands for ``scanning.")
For example, in a typical ranked list of snippets, a user can selectively examine some of the displayed snippets.
The task of simulating scanning behaviour is to model the policy of a user for choosing which action to take based on the state $S_S$.
Scanning behaviour is closely related to stopping behaviour as when viewed from decision perspective, modeling scanning behaviour includes modeling a user's decision regarding whether to continue scanning or stop scanning. However, while the behaviour of continuing to scan a result list is well defined, the stopping behaviour is more complicated as there may be multiple reasons for stopping, including, e.g., when a user's information need is already satisfied, when a user has given up search, or when a user has decided to reformulate the query. For this reason, we have decided to discuss the stopping behaviour in a separate section later (see Section~\ref{sec:sim_search:scanning}).

In most of the existing work, \emph{sequential browsing} is assumed (a.k.a. \emph{linear traversal hypothesis}~\citep{Guo:2009:WWW}), where the user examines the search results in a ranked list sequentially from the top to the bottom. With this assumption, the probability of examimining a result snippet ranked at position $i$ is generally dependent on what the user has seen/done when interacting with the results ranked above position $i$. Eye tracking experiments confirm that users indeed scan results in rank order and that they are less likely to look at lower-ranked results~\citep{Joachims:2007:TOIS,Dumais:2010:IIIX}. This hypothesis is also used in most evaluation measures, in which we assume a simple user model, where the user starts at the top of the result list and works their way down the items, then eventually stop. The probability of the user examining a result snippet decreases as they go further down the list, or, inversely, the probability of stopping increases.  This can be accounted for in evaluation measures by discounting the relevance/utility of results with their position in the ranking~\citep{Zhang:2010:IRJ}.

Under the assumption of sequential browsing, in the state of interacting with a search result at a particular ranking position $i$ (denoted by $S_S(i)$), a user would first decide whether to examine the snippet at the position or not. Once this decision is made, the state of the MDP would transition to a new state $S_S(i)'$, and the user would need to make additional decisions. For example, if the user decided to examine the snippet, the user would further need to decide whether to click on the snippet to view the result in detail or skip the snippet. The modeling of this clicking decision will be discussed in the next section. Regardless of whether the user has examined the snippet, after the user interacts with such a result, the user would make another binary decision, i.e., whether to stop or continue. If the user stops (quit scanning the results), they would further need to decide whether to continue the search session or quit the whole search session completely; this decision reflects the two different reasons for stopping, i.e., either the user's information need is satisfied or the user has given up (due to the impossibility of finding the needed information).  If the user continues, the user would likely reformulate the query, so the action of ``continuing to search'' is essentially the same as reformulating the query.  We thus see that the issue of simulating scanning behaviour is complex and closely related to the simulation of other behaviours such as query formulation and stopping.

\subsection{Models for Simulating Scanning Behaviour}

Many models for simulating a user's scanning behaviour have been proposed based on the sequential browsing hypothesis. The C/W/L framework~\citep{Moffat:2013:CIKM,Moffat:2017:TOIS} and its extension, the C/W/L/A framework~\citep{Moffat:2022:SIGIR} are particularly interesting because a variable $C(i)$ is introduced to explicitly model the probability of continuing to scan after examining an item at position $i$, referred to as a browsing model in \citep{Moffat:2022:SIGIR}. This enables instantiating the evaluation framework with any specific browsing model, which can be regarded as the same as a scanning model. Similarly, in the work~\citep{Carterette:2011:SIGIR}, a general conceptual framework also explicitly identified browsing model as a separate component (enabling different instantiations), along with two other components, i.e., model of document utility, and utility accumulation model.

Specific scanning/browsing models have often been developed as part of a clicking model for convenience of evaluation since it is hard to obtain the ground truth of a user's scanning behaviour (which requires using devices such as eye-tracking), whereas the clicking actions of a user are easily observed and generally available in search log data. Since a clicking action can only happen after a user examines a snippet, when a result has been clicked, it implies that the user had examined that result (before clicking it). Under the sequential browsing assumption, one may also assume that any skipped results above a clicked result have also been examined by the user, but the user has just decided not to click on it (due to an unattractive snippet). However, whether a user has also examined a result ranked below the last clicked result would generally be unknown from the search log.

In their book, \citet{Chuklin:2015:Book} provide a comprehensive review of many click models for modeling clicks in a search system, where the authors distinguished \emph{basic models}, which cover the major assumptions made about user behaviour in the click models but use limited information about user or content, from \emph{advanced models}, which extend basic models to model a user's clicking behaviour more accurately often by relaxing some simplification assumptions made in a basic model and incorporating additional latent variables or content-based features.  In many cases, the scanning behaviour and clicking actions are not distinguished explicitly, but such models can generally be regarded as mostly modeling scanning behaviour with the assumption that the user would always click on a result that is relevant and skip one that is non-relevant. Under such a perfect snippet assumption, the probability of clicking is equal to the probability of examining a relevant result. In reality, however, a user may not always recognize a relevant result just based on the limited information in the snippet. Even though some models have explicitly separated the probability of examination from that of clicking, the probability of clicking (i.e., attractiveness) is often a parameter adjusted for the purpose of fitting the observed clicking actions of users in search log data. In general, there is a lack of models that can be used to predict the clicking action based on snippets, probably because of the unavailability of data with realistic snippets.

Many basic click models vary in their way of defining the examination model $p(E=1|Rank=i)$, particularly in the specific dependency relation assumed between the examination of a results at position $i$, the examination probabilities for the results ranked above position $i$, and whether the examined result at a previous position is satisfactory to a user (relevant or not).
Examination of a result at position $i$ depends on the extent with which a user is satisfied with the results previously examined. While not explicitly mentioned in any of these models, this assumption touches an important user variable, i.e., task completion, or user satisfaction in general, which presumably plays an important role in modeling clicks. For example, for high-recall tasks, a user may be assumed to be more patient and thus would tend to have a higher probability of continuing the browsing. Similarly, for a high-stake search task (e.g., medical search), even if the search results may be poor, the user may still have the tendency of having a higher probability of continuing the browsing. Theoretically speaking, this decision mechanism can be better modeled based on more information about the user and the search context. This issue can be addressed in a more principled way by modeling these variables as part of the user state in the MDP framework.

A particularly interesting model is the \emph{cascade model} introduced by \citet{Craswell:2008:WSDM}, which assumes a linear traversal through the ranking.  The user examines each result and decides whether the snippet is deemed relevant enough to warrant a click, before moving to the next.  The cascade model also assumes that snippets below a clicked result are not examined. That is, the user would stop after having found a relevant result.

Another interesting model is the \emph{user browsing model}~\citep{Dupret:2008:SIGIR}, which assumes sequential scanning behaviour. At each rank position, the user first decides whether to look at the snippet or not; if they find it attractive enough, then they click on it.  Whether the result gets clicked or not, they will then resume the scan of the result list from the next rank position.
Unlike the cascade model, this model does not assume that the user examines all result snippets up to a click.  Instead, it models the event that user \emph{examines} the snippet and, independently from it, whether they find the snippet \emph{attractive}.
Formally, the probability of result at rank position $i$ being examined is expressed as $\prob(E=1|R_i,C_1,...,C_{i-1})$, where $E$ is a binary random variable denoting the event of examination, $R_i$ is the result snippet at rank position $i$, and $C_j$ are a binary random variables indicating whether the result snippets at earlier rank positions got clicked.\footnote{In \citep{Dupret:2008:SIGIR} only the distance from the last clicked result is considered.} The probability of result $R_i$ being attractive is expressed as $\prob(A=1|R_i)$.
\citet{Dupret:2008:SIGIR} model both examination and attractiveness as Bernoulli variables.  Neither $E$ nor $A$ are directly observed,
hence the model parameters are estimated from clicks, which are modeled as a composition of these two events.  The parameters of the examination and attractiveness models are estimated using the Expectation–Maximization (EM) algorithm.
Their empirical results are consistent with the findings of the eye tracking studies of \citet{Joachims:2007:TOIS}, namely: (1) the first two snippets receive substantially more user attention than later snippets, (2) the probability of examination or click decreases as the distance to the last click increases, and (3) users almost always look at the the snippet below a clicked snippet (referred to as ``Skip Next'' strategy in \citep{Joachims:2007:TOIS}).

The advanced models reviewed in~\citep{Chuklin:2015:Book} generally extend the basic models by introducing additional parameters to characterize a user's behaviour in more detail, or using more data (particularly content-based features), or incorporating additional
knowledge about user behaviour, but they have mostly stayed with the same assumptions about the user behaviour discussed above.  They include models for modeling clicks in aggregated search (federated search) scenarios where the interface is more complicated than a simple ranked list~\citep{Chen:2012:WSDM}, models that attempt to model behaviours in a whole query session or multiple sessions~\citep{Zhang:2011:KDD}, models that use additional signals such as eye-tracking and mouse activity-tracking~\citep{Huang:2012:SIGIR}, models that attempt to leverage relevance labels~\citep{Hofmann:2011:CIKM}, models that capture non-linear examination behaviour of users when interacting with results~\citep{Wang:2013:SIGIR}, and models that leverage supervised machine learning which allows for flexible incorporation of various features~\citep{Zhu:2010:WSDM}.

Note that the discussion above indicates the existence of complex dependencies between scanning behaviour and what kind of relevant information and how much relevant information the user has been able to gain in the process of interacting with the search results. This complexity coupled with the challenge in observing the actual scanning behaviour of a user makes the construction of an accurate model for simulating scanning behaviour quite challenging. Since it is impossible to observe the latent decision/preference behaviour of a user while scanning through the results, we have not yet seen results to examine to what extent a scanning behaviour model can capture a real user's scanning behaviour. As one solution, in the advanced models, a user's scanning behaviour can be modeled via a latent variable parameter, which can then be estimated by fitting the model to the observed log data. While such an indirect way of modeling a user's scanning behaviour could bypass the challenge to some extent, more user studies that target at directly studying a user's scanning behaviour are clearly needed to make more substantial progress in simulating scanning behaviour.

\subsection{Variations of Scanning Strategies}

While users generally scan results in similar ways, individuals can exhibit different behavioural patterns regarding the order in which and the depth to which they examine results. This further increases the complexity of simulating scanning behaviour.

\citet{Klockner:2004:CHI} observe that most users (52-65\%) employ a \emph{depth-first strategy}, where they click immediately on potentially relevant items as they encounter them. However, a non-negligible minority (11-15\%) uses an \emph{breadth-first strategy} and look through the entire result list, before clicking on any of the results. The remaining users (20-37\%) apply a mixed \emph{partially breadth-first} strategy, where they sometimes look ahead at the next few entries before selecting an item to click.
\citet{Aula:2005:INTERACT} identify two groups: \emph{economic} (46\%) and \emph{exhaustive} (54\%) evaluators.
Economic evaluators tend to scan less than half of the results visible on the screen before their first action. Exhaustive evaluators tend to examine over half of the results visible on the screen, or even scroll down, before performing their first action. These groups resemble the depth-first and breadth-first strategies in \citep{Klockner:2004:CHI}.
\citet{Smucker:2011:HCIR} distinguish between ``fast and liberal'' and ``slow and neutral'' searchers. Users in the former group are fast in examining snippets and liberally click on them, but rather spend more time evaluating the full documents. The latter group follows a reverse strategy: they take over twice as long to make a decision about which snippet to click to then speed up the evaluation of full documents.
\citet{Dumais:2010:IIIX} examine searcher behaviour when the SERP contains both organic results and advertisements.  They identify three main clusters of users: those who explore the SERP broadly (\emph{exhaustive}, 32\%) and more narrowly (\emph{economic}), with the latter category further broken down based on the attention given to advertisements (\emph{economic-results}, 39\% vs. \emph{economic-ads}, 29\%).
Overall, two broad categories emerge from these studies: searchers who examine results exhaustively before making their first click and those who examine results more narrowly and click more liberally.

The examination and clicking decisions are strongly interconnected and therefore can more accurately be modeled together; we discuss these approaches in the next section.

\subsection{Complex Presentation Layouts}
\label{sec:sim_search:scanning_complex}

\begin{figure}[t]
  \centering
  \begin{tabular}{cc}
    \includegraphics[width=0.48\textwidth]{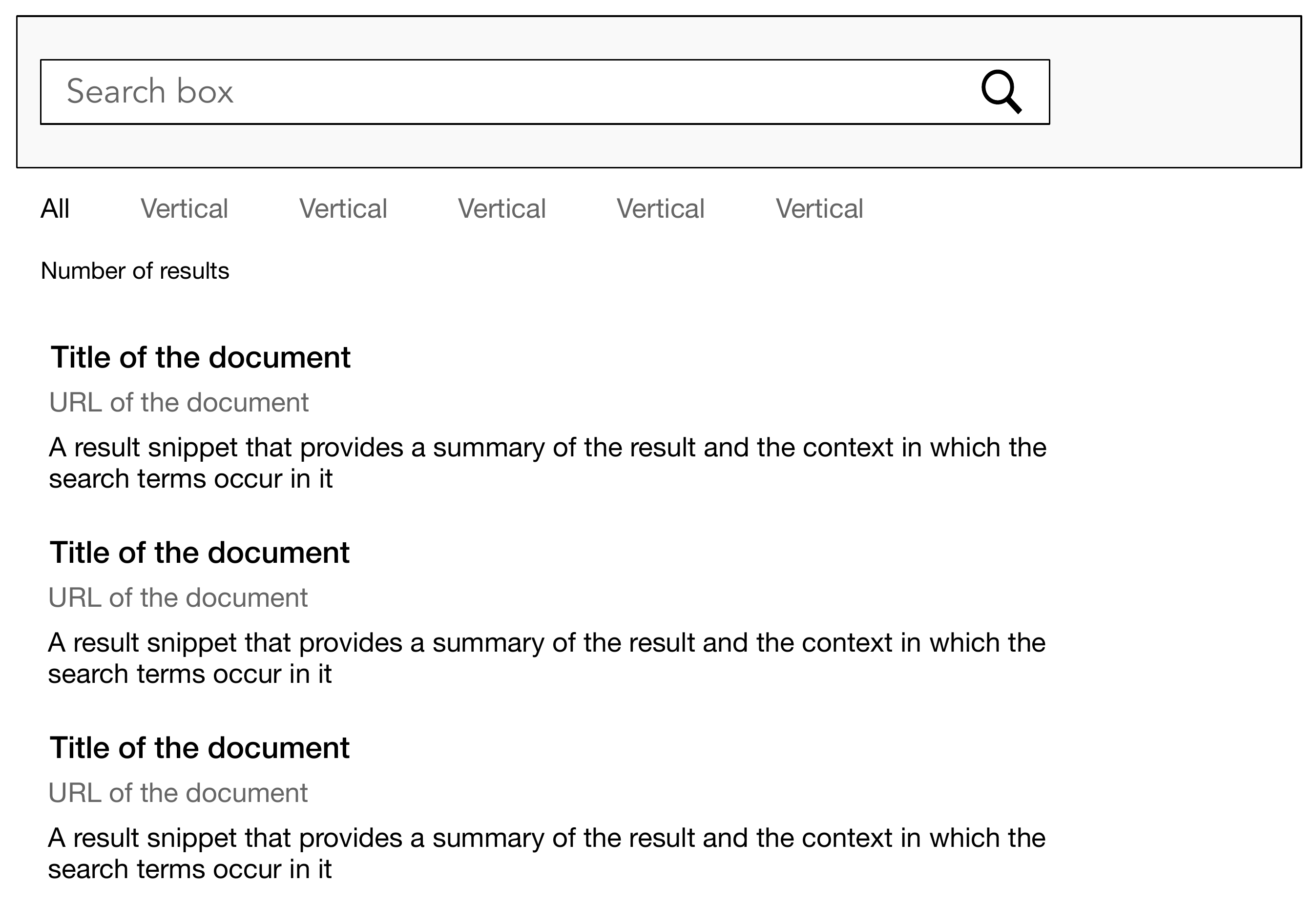} &
    \includegraphics[width=0.48\textwidth]{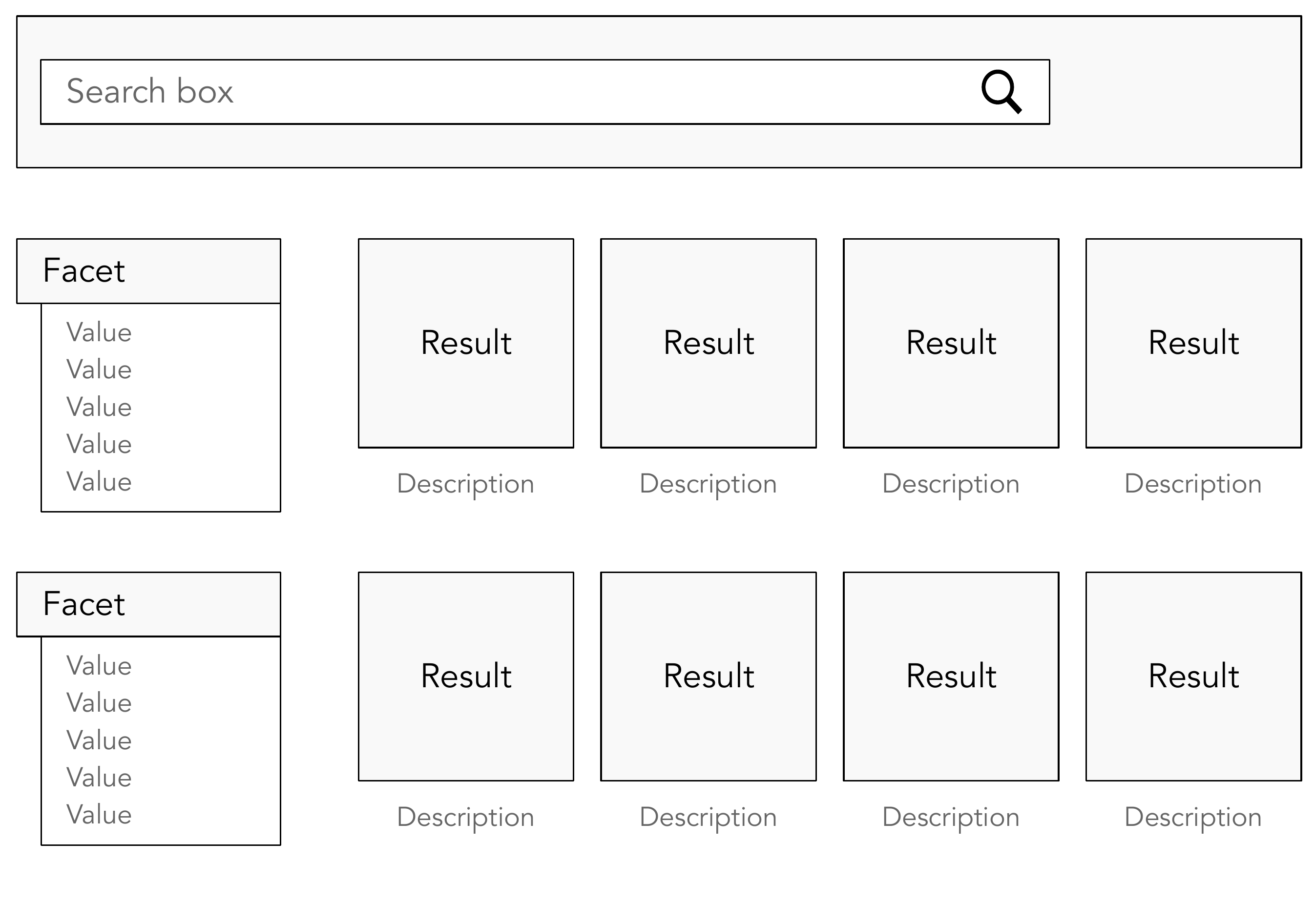}
    \\
    \footnotesize{(a) A traditional ``ten blue links'' layout.}       &
    \footnotesize{(b) A product search layout.}
    \\
                                                                      & \\
    \includegraphics[width=0.48\textwidth]{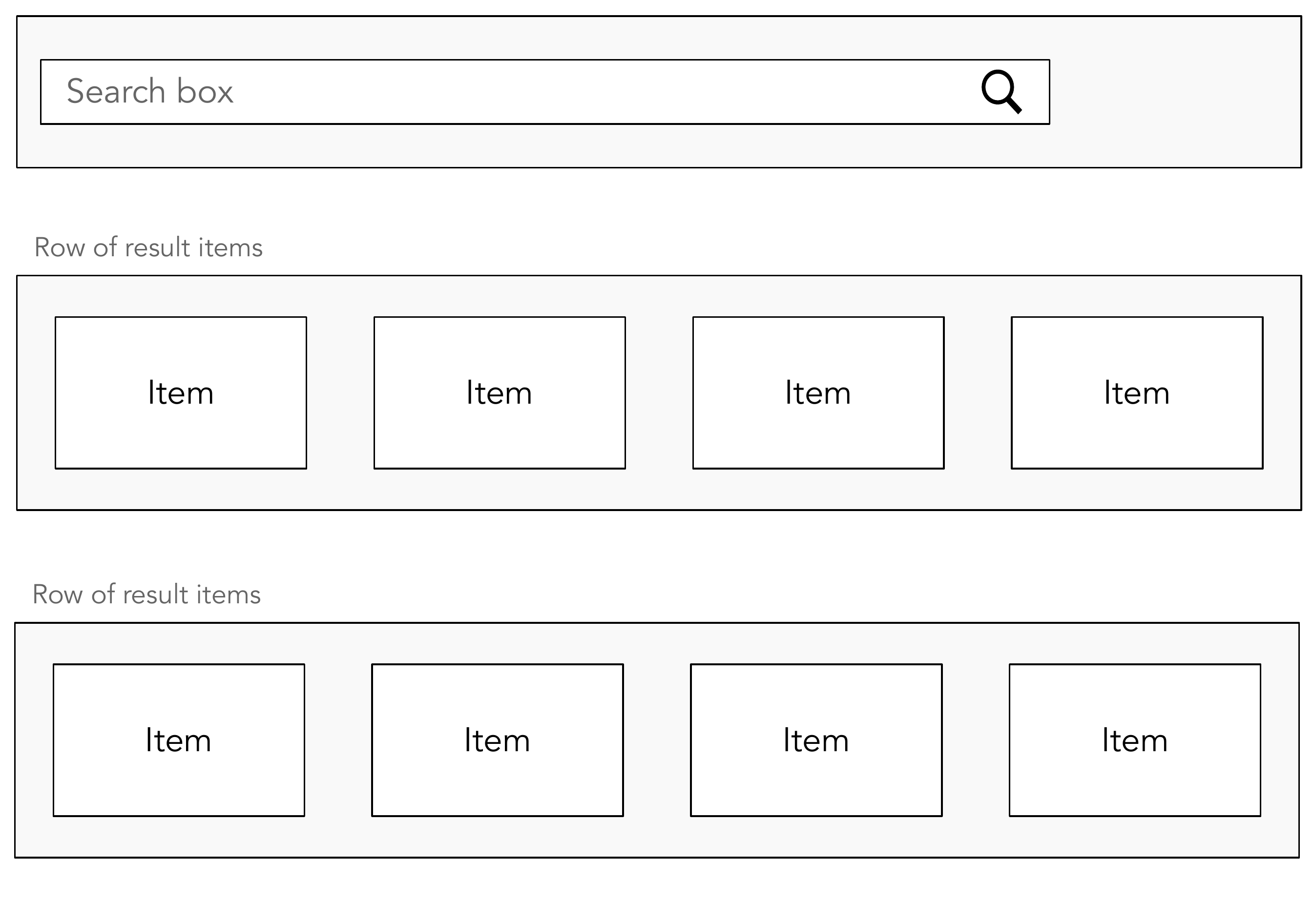} &
    \includegraphics[width=0.48\textwidth]{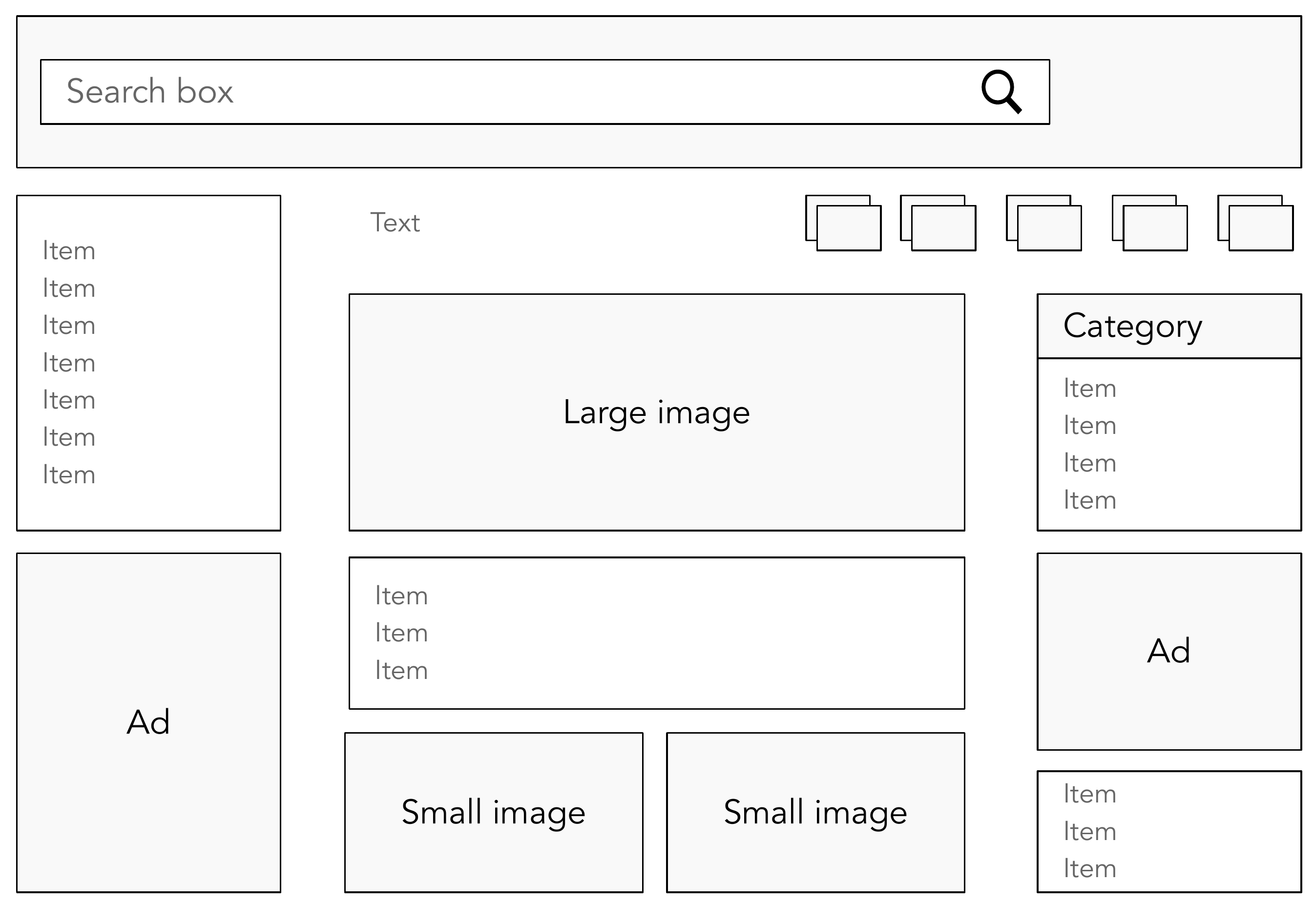}
    \\
    \footnotesize{(c) A video recommendation layout.}                 &
    \footnotesize{(4) An advertisement layout.}
    \\
  \end{tabular}
  \caption{Examples of different result presentation layouts (illustration adapted from~\citet{Oosterhuis:2018:SIGIR}).}
  \label{fig:serp_layouts}
\end{figure}

A major limitation of the existing work on scanning behaviour, and in user simulation in general, is that they have rarely considered modern SERPs and alternative presentation layouts.
Indeed, the vast majority of prior work on user simulation has been conducted in the context of web search, focusing on how a user interacts with a list of search results, presented as snippets.  Search engine result pages, however, have evolved significantly over the last two decades from the traditional ``ten blue links.'' Besides the list of organic results, modern SERPs may contain direct answers, advertisements, entity cards, related questions, as well as results from various verticals depending on the intent (shopping, news, maps, images, etc).
This directly challenges the top-down traversal assumption~\citep{Diaz:2013:CIKM} and necessitates whole-page optimization~\citep{Wang:2016:WSDM}.
It has been shown that heterogeneous content in a SERP changes how users interact with the results, e.g., in the presence of vertical results~\citep{Liu:2015:SIGIR}, adverisements~\citep{Dumais:2010:IIIX}, entity cards~\citep{Bota:2016:CHIIR}, rich information panels~\citep{Navalpakkam:2013:WWW}, or right-rail results~\citep{Shao:2022:ACI}.
Users not only pay more attention to certain region of the page (\emph{position bias}) but also to specific type of items (\emph{vertical bias})~\citep{Chen:2012:WSDM,Lagun:2014:SIGIR,Liu:2015:SIGIR}.

It is also important to remember that web search, while widely used, represents just one flavor of search.
There are application domains where different result presentation formats are employed.  A prime example is vertical search and recommender systems that focus on specific types of entities, such as products or videos.
Vertical search often employs a grid layout to display search results, while many recommender systems employ a carousel-based presentation format~\citep{Bendada:2020:RecSys,Rahdari:2022:HT}.
See Fig.~\ref{fig:serp_layouts} for examples of various result presentation layouts; in many of these, the top-down scanning assumption is misguided~\citep{Oosterhuis:2018:SIGIR}.
Simulation of user behaviour enabled by complex SERPs with heterogeneous content and/or nonlinear page layouts requires the modeling of additional scanning, browsing, and clicking behaviours.
How to formally model such complex browsing and interaction actions of a user is an important future direction.
There is ongoing work in this area, for example, on user interaction models for carousel-based layouts for recommender systems~\citep{Felicioni:2021:UMAP,Rahdari:2024:TORS}.
However, given the increasing complexity of result presentation interfaces, model-based approaches might have inherent limitations compared to data-driven approaches~\citep{Azzopardi:2020:CHIIR}, which goes back to the more fundamental question of model-based vs. data-driven simulation (cf. Section~\ref{sec:intro:simulation_approaches}). %

\section{Simulating Clicks}
\label{sec:sim_search:clicks}

The task of simulating a user's clicks is to mimic a user's decision on whether to click on a search result (to view it in detail) after being exposed to a search result often in the form of a snippet in the context of a ranked list. Using the MDP framework, the simulation of clicks can be associated with a formal specification of part of the policy on choosing from two candidate actions
$A_C \in \{0,1\}$ based on the current state $S_C$ which includes all the relevant context information to this decision, including, e.g., a user's current query $Q$, the snippet $R_i$ at the current position $i$, the whole ranked list of results, $R_1, R_2, ..., R_k$, any other useful information about the user $U$, and any (historical) context information that might affect a user's decision on whether to click on a result $H$ (e.g., historical interactions of the user $U$ or other similar users). The clicking policy generates a value for $A_C$ based on $S_C$: $A_C=\pi_C(S_C)$. For example, in an extremely simple case, the policy may only use the current ranking position to determine whether to click a result. In this case, $\pi_C(S_C) \approx \pi_C(i)$, leading to a stochastic clicking policy specified based on a position-specific clicking probability. Intuitively, a higher ranking position (i.e., a smaller $i$) would have a higher probability of clicking. As a more specific example, the clicking policy can be defined as $\pi_C(i)=1/\log_2 (i+1)$, which would give us an interpretation of the discounting coefficients used in the nDCG evaluation measure as a naive clicking policy.

Note that here we make the assumption that the user has already examined the snippet $R_i$. That is, the clicking policy would only be used in simulating a user when the simulated scanning strategy has predicted examination of the result $R_i$. Thus, there are multiple scenarios of interaction with each search result in general as shown in Table~\ref{tab:clicking}. Note that the interpretation of viewing a snippet without clicking on the result (to further examine the content of the document) is complicated since there is a possibility that the snippet might have already provided the relevant information that the user is searching for (e.g., a telephone number). In such a case, we cannot assume the snippet to be non-relevant; it is indeed the opposite case. From a user simulation perspective, this adds complexity to the space of actions since there is a non-observable action of ``good abandonment''~\citep{Li:2009:SIGIR}. With enriched user interfaces, where relevant information can be easily shown to users directly, becoming more common, such special unobservable actions may also become more common. This action is more related to examination of document content than clicking on a document (see Section~\ref{sec:sim_search:processing}) and how to incorporate such actions into an evaluation measure may be complicated, but in general, it needs to be modeled with consideration of the overall utility of the system as suggested in~\citep{Machmouchi:2017:CIKM}. The framework we presented in Chapter~\ref{ch:frameworks} can be used to define such an overall utility based on an entire interaction sequence of user actions.

\begin{table}
    \caption{User interaction with search results: examination vs. clicking.}
    \label{tab:clicking}
    \footnotesize
    \centering
    \begin{tabular}{cccl}
        \toprule
        \textbf{Shown}    & \textbf{Examined} & \textbf{Clicked}  & \multirow{2}{*}{\textbf{Status of result}}     \\
        \textbf{to user?} & \textbf{by user?} & \textbf{by user?} &                                                \\
        \midrule
        No                & N/A               & N/A               & Unexposed result                               \\
        Yes               & No                & N/A               & Ignored result (affected by stopping strategy) \\
        Yes               & Yes               & No                & Skipped result (negative feedback)             \\
        Yes               & Yes               & Yes               & Clicked result                                 \\
        \bottomrule
    \end{tabular}
    \vspace*{-0.5\baselineskip}
\end{table}

As discussed in the previous section, the modeling of clicking actions has often been integrated with modeling of scanning behaviour, and the book \citep{Chuklin:2015:Book} provides a comprehensive review of many such (combined) click models.
From the perspective of user simulation, those click models have made the assumption that clicking is the result of examining an attractive result (snippet) (i.e., the \emph{examination hypothesis}~\citep{Guo:2009:WWW}): a user is assumed to click on a result if they (1) have examined the result snippet, and (2) have found the snippet attractive. This assumption enables separation of a click model into two component models: the \emph{examination model}, $\prob(E=1|Rank=i)$, which models the probability that a user would examine the result at position $i$, and the \emph{attractiveness model}, $\prob(A=1|R_i)$, which models how attractive the snippet at position $i$. This also enables separation of modeling scanning behaviour (discussed in the previous section) and modeling clicking behaviour discussed in this section. Under this assumption, $\prob(C=1|Rank=i, R_1, R_2, ..., R_k, Q, U, H)=p(E=1|Rank=i) \; \prob(A=1|R_i)$. Many click models are designed to estimate the relevance of clicked results and the separation naturally addresses the problem of position bias when using the clickthroughs to estimate the relevance of a clicked result~\citep{Joachims:2007:TOIS}, since the position bias would be captured by the examination model, thus allowing the attractiveness model to be unbiased.

Without necessarily separating the scanning behaviour from the clicking decision after examining a snippet, the different click models proposed in the literature generally vary in their assumptions about user behaviour in browsing the results and to what extent the content information is used. It is worth distinguishing two high-level categories of click models: position-based and content-based simulation.

In \emph{position-based simulation}, the estimate of the clicking probability is assumed to only depend on the rank positions. That is, $\prob(Click=1|Rank=i, R_1, R_2, ..., R_k) \approx \prob(Click=1|Rank=i)$.   Such a position-based model is generally overly simplified for the purpose of simulating users for evaluation. When only positions are considered, a \emph{perfect snippet assumption} is often made implicitly, that is, a user is assumed to be able to tell whether a result is relevant based on the snippet and would always click on a result if it is relevant.  This is not a realistic assumption and could only facilitate construction of a user simulator when we have available the relevance judgments for a query (e.g., when using existing test collection to create user simulators).
Despite the simplicity of position-based models, they have the advantage of easily ``generalizing'' to new search scenarios and can thus be conveniently used to simulate any user to generate interaction data even though the simulation is not accurate and is based on oversimplified user behaviour.

In contrast with the position-based simulation strategy, the \emph{content-based simulation} strategy uses the snippet content to model the probability of clicking with the intuition that a relevant snippet would likely be clicked, whereas a non-relevant one would likely be skipped. These models are intuitively more accurate, but they suffer from the overfitting problem in that the model may not generalize well to simulate clicking in a new scenario. For example, if the parameters of a click model are tied to specific results, the model can only be used to simulate clicks on the same results (seen in the training data) and cannot be used to predict clicking probability on an unseen result. Using query-document (or query-item) features in a supervised learning framework helps alleviate the problem, but such a model may lose interpretability and thus cannot be easily varied to simulate variations of users. The click models designed based on the relevance status of results can also be used to estimate the relevance of a clicked item (which can help obtain more accurate training data for training a ranking algorithm); for example, in~\citep{Wang:2013:WWW}, a Bayesian Sequential State Model is proposed to model the clicking action with consideration of content and dependency between information items.

In general, the parameters of a click model can be estimated based on search log data. When the click model involves only parameters representing observable events, the parameters can be easily estimated by normalizing the counts of the corresponding events of those parameters based on the analytical form of the Maximum Likelihood (ML) estimator. When a model involves a latent variable that is not observable, which are clearly needed to model the scanning behaviour of a user, there is generally no analytical solution to the ML-estimation problem, and an iterative numerical algorithm such as the EM algorithm is often used to compute the ML estimate and estimate all the parameter values~\citep{Chuklin:2015:Book}.

\subsection{Modeling Clicks in Recommender Systems}

Much work has been done recently on modeling clicks and predicting clickthrough rates (CTRs) in the context of recommender systems~\citep{Zhou:2018:SIGKDD,Zhang:2022:SIGIR} and online advertisement~\citep{Yang:2022:IPM}. In this line of work, the goal is to predict how likely a user would click on a recommended item or an advertised item based on both information about the user and information about the items. The intended application is to use the CTR prediction model to optimize the ranking of items for recommendation or advertising instead of simulating user behaviour, but these two goals are essentially the same in the sense that an item predicted to have high CTRs should be recommended to the user. That is, an effective ranking algorithm for recommendation or advertisement applications needs to be able to predict well whether a user would click on an item if it was recommended to them. In this sense, such models could technically be regarded as modeling user clicks; indeed, they are generally evaluated based on how accurate their prediction of user clicks is.
However, because of the emphasis on the use of such predictive models as recommendation algorithms rather than user simulators, these models generally will attempt to use as many relevant features as input to the prediction as possible and are often based on deep neural networks. Those models are commonly used to score an item (with or without a user) and rank all the items with the goal of ranking the items with high CTRs on the top. The performance of those models is typically measured using a ranking-based measure such as Area Under Curve (AUC). Popular models include the Deep Interest Network~\citep{Zhou:2018:SIGKDD}, Wide and Deep Network~\citep{Cheng:2016:Workshop}, and Deep Factorization Machines~\citep{Guo:2017:IJCAI}. These models often attempt to use historical context information of users, which is typically available in the context of recommender system applications, to improve prediction accuracy~\citep{Zhang:2022:SIGIR}. Compared with the models reviewed in~\citep{Chuklin:2015:Book}, these newer models use richer user information and more sophisticated supervised learning models. The review~\citep{Yang:2022:IPM} provides an excellent detailed review and comparison of the major models in this direction.

While most existing work can be regarded as non-interpretable approaches as discussed above, interpretable models have also been recently explored.
For example, \citep{Rahdari:2024:TORS} develop a carousel click model for modeling clicking behaviour on carousel-based recommender interfaces, which generalizes the traditional cascade model \citep{Craswell:2008:WSDM}.
Another interesting recent line of work on choice models \citep{Hazrati:2022:IPM,Hazrati:2024:UMUAI} has taken a more interpretable approach to simulate clicking in recommender systems. For example, \citep{Hazrati:2022:IPM,Hazrati:2024:UMUAI} use three choice models, Age-CM, Popularity-CM, and Rating-CM, each modeling a somewhat different clicking behaviour. Specifically, \citet{Hazrati:2024:UMUAI} describe them as: ``Age-CM models users that tend to choose newer items among the recommended ones. Popularity-CM instead models users that tend to choose more popular items. Finally, Rating-CM models users that choose items with higher ratings.'' A baseline choice model, Base-CM, is also considered where the simulated user would faithfully accept recommended results (thus potentially containing position bias). The authors employ these models in simulation experiments involving a population of simulated users repeatedly interacting with a simulated recommender system. Such a simulation experiment enables the analysis of many questions about the community-wide behaviour of users and systems over a long period with some interesting findings about the complicated interactions between users' choices and the recommender systems. From evaluation perspective, this line of work demonstrates the feasibility of using user simulation to evaluate the overall utility and impact of a recommender system if deployed to a community of users. Although the perspective of evaluation we have taken in the book is more on comparing two interactive systems, such a broader perspective of user simulation for analysis of a complex interactive ecosystem of users and systems is no doubt a very interesting future research direction, where user simulation also plays a critical role.

\subsection{On Using Click Models in User Simulators}

From the perspective of user simulation, the existing click models can all potentially be used to simulate a user's clicking activities, but there are two important considerations.
First, there is trade-off between click prediction accuracy and interpretability.
While more sophisticated models, notably those relying on deep learning, have been shown to be more accurate in predicting clicks, they are deficient in their interpretability for the purpose of simulating users, thus making it hard to vary such models to simulate variations of users.

Second, some models may not be realistic due to their lack of consideration of the actual context information a user would be exposed to when deciding whether to click on an result. For example, a user's click decision in a search engine is generally made based on the information shown in the result snippet of a result without having access to the whole document. This means that a click model based on the content of a document instead of the snippet of the document is unrealistic.
However, the user's prior background knowledge about the query topic is also relevant. For example, an expert user may be able to recognize a relevant document based on just a short snippet, where a novice user might not.
Therefore, the click models used in a user simulator must also attempt to model the user's information need and cognitive state.
Recently, clicking simulation has also been studied with further introduction of a user cognitive state~\citep{Labhishetty:2020:SDMWorkshop}, cf. Section~\ref{sec:sim_search:cognitive_state}.

The following are some examples of specific click models that can be used for user simulation.

\begin{itemize}
    \item \citet{Maxwell:2015:CIKM} set the probability of the user clicking on relevant and non-relevant snippets, based on a user study, as:
          \begin{equation}
              \prob(C=1|D) = \begin{cases}
                  0.21 & \text{relevance} = 0 \\
                  0.36 & \text{relevance} = 1 \\
              \end{cases}
          \end{equation}

    \item \citet{Baskaya:2013:CIKM} base click probabilities on the relevance level of the underlying document (from TREC qrels):
          \begin{equation}
              \prob(C=1|D) = \begin{cases}
                  0.27 & \text{relevance} \leq 1 \\
                  0.34 & \text{relevance} = 2    \\
                  0.61 & \text{relevance} = 3    \\
              \end{cases}
          \end{equation}
    \item \citet{Carterette:2015:ICTIR} base clicks primarily on features of document titles, URLs, and snippets (features derived from LETOR), which are available to users.
    \item \citet{Maxwell:2016:CIKM} predict click decisions based on probabilistic language models, comparing the language model representing the user's background knowledge with a language model created from the snippet's content.
\end{itemize}
From the perspective of the MDP framework, the models for scanning behaviour and clicking actions together can be regarded as defining the policy used by the simulated user agent when interacting with search results. However, to completely plug in the scanning and clicking models into the MDP framework, we need to further study how to model the user state and the search result context in detail as well as to model a user's decision process when deciding whether to click on a document with consideration of alternatives if the user would not click on a result. For example, a user may also reformulate a query as an alternative way to continue exploring new documents instead of continuing to click on the current results. The decision at least partially depends on a user's knowledge about the search topic; if the user is an expert, it would be easier to reformulate a query and thus the user may have the tendency of clicking on fewer results than a novice user who might perceive the cost of reformulating a query as higher than simply continuing the examination of additional results. Thus, in general, modeling clicks needs to consider a user's cognitive state.

\section{Simulating Document Processing}
\label{sec:sim_search:processing}

Upon clicking on a selected search result, the user processes the corresponding full document.
Processing, i.e., reading and understanding, a document requires an \emph{effort} from the user and yields some \emph{utility} to them. Viewed in the MDP framework, the state represents the context of having already anticipated that the document may be relevant when clicking on the result. Thus, the user may be looking for the larger context of the relevant part of the snippet with query words in mind. The exact actions that a user can take naturally depend on the specific user interface provided by the system, but allowing a user to scroll up/down on a screen to view all the content of a document is generally supported. Such observable scrolling actions and the observable dwelling time on a document are usually determined by a latent decision whether to continue processing a document based on what has already been seen so far.
In economic terms, we can talk about the \emph{costs} and \emph{benefits} associated with such a latent decision of a user when assessing a document~\citep{Azzopardi:2016:ICTIR}.

\subsection{Simulating Effort}
\label{sec:sim_search:effort}

The effort or cost involved in processing a document is difficult to define, quantify, and measure.  Most commonly, time is used as a proxy.
\emph{Dwell time} refers to the amount of time the user spends on looking at the full document after clicking on a result in the SERP.
Most evaluation measures implicitly assume that as users work their way down the ranked list, they process documents at a constant rate.
This is clearly not the case. For example, one could reasonably expect that shorter documents take less time to read than longer documents.
The time-biased gain metric~\citep{Smucker:2012:SIGIR} estimates the time (in seconds) needed to process a document of length $l$, measured in words, as
\begin{equation}
    T_D(l) = al + b ~,
\end{equation}
where the user is reading at a rate of $a$ seconds per word, and then uses a constant amount of $b$ seconds to make an assessment about the document's relevance.  \citet{Smucker:2012:SIGIR} estimate $a=0.018$ and $b=7.8$ based on a user study.
\citet{Arvola:2010:IR} also measure reading effort, but on the character level.

\citet{Carterette:2015:ICTIR} present models for estimating dwell times based on (1) features of that document, (2) features of other results on the SERP (e.g., URLs, snippets, titles), and (3) the clicks the user made on other ranked results.
Document features are derived from those used in the LETOR datasets for learning to rank~\citep{Qin:2010:IRJ}, which include retrieval model scores between query and document URL/title/snippet, statistics based on query term frequencies, URL length and depth, number of inlinks and outlinks, and spam score.
The time the user spends on reading a document would also depend on the difficulty of the text of that document.  This is modeled by computing various reading level scores, such as SMOG~\citep{McLaughlin:1969:JR}, Flesch-Kincaid~\citep{Kincaid:1975:TechReport}, and Coleman-Liau~\citep{Coleman:1975:JAP}.
Dwell time is a continuous number as opposed to a categorical value, therefore prediction is modeled as a regression problem. Viewed within the MDP framework, the two actions are modeled, continue vs. stop, and the policy is captured by the predictive model based on various features, including the accumulated effort (dwell time) at any point.

Alternatively, the effort involved in processing a document may also be modeled in terms of cognitive load, e.g., understandability.  Commonly, reading difficulty is used as a proxy, which can be approximated by measures of readability, like the ones listed above.
We note that while understandability can be incorporated into estimates document processing effort (as done, e.g., by \citet{Carterette:2015:ICTIR} above), it is more often considered when modeling document utility, as we shall discuss next.

\subsection{Simulating Document Utility}
\label{sec:sim_search:utility}

Next, we present ways of modeling the utility or benefit that processing a selected (clicked) document provides to the user.

Already back in the early days of IR, \citet{Cooper:1973:JASIST} argued that retrieval evaluation should ultimately be based on \emph{utility}, i.e., the value it brings to the user.  Utility is meant to be a broader concept than topical relevance---it includes, among others, quality, novelty, importance, and credibility. It encompasses everything that the user values, for instance, a witty or engaging writing style. \citet{Cooper:1973:JASIST} further notes that a document might also have a nonutilitarian value, e.g., entertainment.
Utility, however, is elusive to measure, therefore in practice document relevance is used commonly as a proxy.

The question to answer, then, is: Given a document retrieved for a query, how would users estimate its relevance? An evaluation measure is generally meant to quantify the overall utility of a list of results based on the estimated utility of each individual document a user has been exposed to, so this question is also the main question behind the design of an evaluation measure. A more accurate model of how a user estimates relevance would better capture the {\em perceived} utility of a document; it would improves both user simulation and design of evaluation measures.

In this existing work, this step is commonly identified in the searcher workflow as the task of judging document relevance (cf. Fig.~\ref{fig:search_session}) or is sometimes associated with the action of ``saving''~\citep{Smucker:2011:HCIR} or ``marking'' a document~\citep{Maxwell:2015:CIKM}.
Notice that the naive searcher model (cf. Fig~\ref{fig:naive_searcher}) has no recourse to relevance---the user would blindly consider all documents as relevant~\citep{Maxwell:2016:CIKM}.

\subsubsection{Leveraging Ground Truth Relevance Assessments}

Most existing work assumes the presence of relevance assessments for query-document pairs, e.g., by utilizing TREC corpora. This can be understood as assuming such information is included in the state variable of the MDP framework when simulating a user. This is appropriate for simulating scenarios where it is reasonable to assume that the user has already had excellent knowledge about what constitutes relevant content for the given query, e.g., an expert user or a known item search task.

For clarity, we shall refer to the ground truth labels in the test collection as \emph{relevance degree}, $r_{q,d}$, and the ability of the simulated user to correctly judge whether the document is relevant or not as \emph{relevance estimate}, $\prob(Rel=1|q,d)$.
If the user correctly estimates relevance, they are rewarded the corresponding relevance degree $r_{q,d}$ as gain.
There is a number of implicit underlying assumptions here: (1) the ground truth relevance labels reflect the true value of the document, (2) this value is the same for all users, and (3) the relevance of a document is independent of other documents encountered before during the course of interaction with the search engine.

The simplest option is to deterministically decide whether a document is relevant or not, based on a predefined relevance threshold $\mu_R$:
$\prob(Rel=1|q,d) = 1$ if $r_{q,d}\geq \mu_R$~\citep{Keskustalo:2006:ECIR}.
A high $\mu_R$ value indicates that ``the user is capable and willing to recognize and accept only highly relevant documents''~\citep{Keskustalo:2006:ECIR}, whereas a low $\mu_R$ value corresponds to the setting where the user liberally accepts even marginally relevant documents.
This approach, however, is unrealistic in that it assumes that users are perfectly capable of determining the ``true'' (i.e., ground truth) relevance of documents.

Therefore, it is more common to stochastically determine whether the user would find a document relevant based on the ground truth relevance labels.  For example, distinguishing only between relevant and non-relevant documents, \citet{Maxwell:2015:CIKM} set the following probabilities based on a user study:
\begin{equation}
    \prob(Rel=1|q,d) = \begin{cases}
        0.53 & r_{q,d} = 0       \\
        0.71 & r_{q,d} \geq 1 ~. \\
    \end{cases}
\end{equation}
Considering multiple degrees of relevance, \citet{Baskaya:2013:CIKM} set the probability of the user judging a document relevant as follows:
\begin{equation}
    \prob(Rel=1|q,d) = \begin{cases}
        0.20 & r_{q,d} = 0    \\
        0.88 & r_{q,d} = 1    \\
        0.95 & r_{q,d} = 2    \\
        0.97 & r_{q,d} = 3 ~. \\
    \end{cases}
\end{equation}
Further refinements are possible by defining different click distributions for different groups of users, e.g., based on expertise~\citep{Paakkonen:2015:CLEF}.

The vast majority of existing simulation work is indeed based upon existing (TREC) test collections. %
However, reliance on existing test collections is limiting both in terms of scale, i.e., there is a fixed set of information needs to experiment with, and in terms of modeling personal preferences, i.e., it varies across users and contexts what result they would find relevant, while test collections have a single ground truth relevance label associated with each document.
Finally, there are also aspect beyond topical relevance that would represent utility for the user, which are not captured here. %

\subsubsection{Without Ground Truth Relevance Assessments}

Next, we consider the case where ground truth relevance assessments are unavailable, i.e., when the state variable of the MDP framework does not include this information. This is appropriate, e.g., when simulating a user who has a significant vocabulary gap.
Consequently, the probability of the user judging a document relevant for the input query, $\prob(Rel=1|q,d)$, can only be estimated based on the content and/or features of the document. %
However, setting this probability in a deterministic manner or without considering additional user-specific information, it would amount to the exact same relevance scoring task that a retrieval system has to perform, therefore providing limited utility.

\citet{Maxwell:2016:CIKM} use unigram language models to predict whether the user would find the document relevant, thereby conditioning the relevance estimation on the individual user that is being simulated, $\prob(Rel=1|q,d,u)$.
Users are assumed to have some topical background knowledge and, accordingly, expect certain other terms or concepts to be present in the document.
This background knowledge is modeled as a term distribution, $\prob(t|\theta_B)$, and built from words that are associated with topic terms, utilizing word embeddings (word2vec).\footnote{\citet{Maxwell:2016:CIKM} make use of TREC topic and title descriptions, which provide details on what constitutes as a relevant document, and generate queries from these topic descriptions.}
This term distribution is then combined with language models created from the topic description, $\prob(t|\theta_T)$, and from relevant documents found by the user during the course of the session, $\prob(t|\theta_{RD})$, and is smoothed with a collection language model, $\prob(t|\theta_C)$, to estimate a relevance language model, $\prob(t|\theta_R)$:
\begin{equation}
    \prob(t|\theta_R) = \lambda \Big (
    \frac{w_T}{z}\prob(t|\theta_T)
    + \frac{w_I}{z}\prob(t|\theta_{RD})
    + \frac{w_B}{z} \prob(t|\theta_B)  \Big)
    + (1-\lambda) \prob(t|\theta_C) ~.
\end{equation}
The weights $w_T$, $w_I$, and $w_B$ control how much emphasis is placed upon the topic, interaction, and background user knowledge, respectively, with $z$ being a normalizing constant.

Each observed document $d$ is then scored against the relevance model using the normalized log likelihood method~\citep{Meij:2009:CIKM}:
\begin{equation}
    O(d|\theta_R) = \frac{1}{|d|}\sum_t \prob(t|d)\frac{\prob(t|\theta_R)}{\prob(t|\theta_C)} ~,
\end{equation}
where $|d|$ is the length of the document and $\prob(t|d)$ is the maximum likelihood estimate of term $t$ appearing in the document.
If this likelihood is above a predefined relevance threshold $\mu_R$, then the document is considered relevant.  In their experiments, \citet{Maxwell:2016:CIKM} consider $\mu_R$ values between -0.4 (liberal) and 0.4 (strict).
The probability distribution $\prob(t|\theta_{RD})$ is updated each time a document is judged as relevant by the user (by counting the number of times each term appears in these documents, then normalizing to obtain a maximum likelihood estimate).  Thus, $\theta_{RD}$ serves as the representation of the user's knowledge state and captures how that evolves based on the documents encountered.

It is worth pointing out the connection to recent work on employing large language models for estimating relevance judgments, which may be considered a data-driven and non-interpretable simulation approach~\citep{Faggioli:2023:ICTIR}. However, due to the large amount of information ``seen" by LLMs during the pre-training stage, they might be more suitable for simulating ``expert users'' who possess substantial knowledge on the topic, rather than those who are completely new to it (see Section~\ref{sec:broader:llms} for further discussion on the use of LLMs for simulation).

\subsubsection{Beyond Topical Relevance}

There are other factors beyond topical relevance that could be used to determine document utility.
It has been known for decades that even relevance, which is a narrower notion than utility, is a multidimensional concept; topicality is one of its essential constituents, but alone is insufficient~\citep{Boyce:1982:IPM,Cosijn:2000:IPM}.
For example, taking a cognitive approach, \citet{Xu:2006:JASIST} identify four additional criteria users employ in making relevance judgment beyond topicality: novelty, reliability, understandability, and scope.
\citet{Taylor:2007:IPM} further show that users' relevance criteria change depending on their stage in the search process.
Of the different facets of relevance, novelty and diversity have been addressed by fragmenting the information need into subtopics (or nuggets) and performing relevance assessments explicitly for each of the subtopics of the query~\citep{Zhai:2003:SIGIR,Clarke:2008:SIGIR,Sakai:2011:SIGIR}.
One of the main challenges in incorporating additional aspects of relevance lies in the difficulty of obtaining ground truth assessments.
Notably, automatic readability measures have been employed as estimators of understandability~\citep{Zuccon:2016:ECIR}.
The multi-dimensional view of utility that goes beyond topical relevance should clearly be adopted for both designing more accurate evaluation measures and developing more accurate user simulators.

\subsection{Search Goals and Context}

A user's search goal and context may significantly affect how a user processes results.
The underlying user goal in the majority of studies is to find as much relevant material as possible withing a predefined amount of time~\citep{Paakkonen:2015:CLEF,Baskaya:2013:CIKM,Verberne:2015:ECIR,Maxwell:2016:CIKM} or until some other stopping condition is met~\citep{Carterette:2015:ICTIR,Maxwell:2015:CIKM}.  Progress is then measured in terms of the number of relevant documents found or accumulated information gain.
Ultimately, users want answers, not documents, therefore modeling satisfaction should extend beyond the unit of documents, for example, by considering key information nuggets.
Further, what constitutes a satisfactory answer might depend on the user's context.  Consider the example of a user searching for the ``fastest way to get to the airport.''  If the person is performing the search at home, with time to plan their journey and access to a desktop computer or laptop, a satisfactory answer would likely include detailed information about different transportation options, such as driving, public transportation, or ride-sharing services. It may also include information on the estimated time it would take to reach the airport using each option, as well as any potential delays or traffic issues to be aware of. However, if the person is already on the move, using a mobile device, and needs an immediate answer, a satisfactory answer may be more focused on quick and easy directions, such as a map with the fastest route highlighted or step-by-step instructions.
To date, there has been limited work on considering different search contexts, even though it has been shown that interactions on the SERP are influenced not only by the presence of heterogeneous content and layout, but also by task complexity~\citep{Roy:2022:SIGIR}.
For example, when facing complex tasks, users tend to interact more with various components of the SERP~\citep{Arguello:2012:SIGIR}, examine the result list deeper and faster~\citep{Thomas:2013:AIRS}.
\citet{Wu:2020:CIKM} find that the presence of direct answers affects users' interaction patterns differently for factoid vs. complex question types, resulting in fewer clicks for factoid questions.
\section{Simulating Stopping Behaviour}
\label{sec:sim_search:stopping}

During the search process, users make decisions regarding when to stop examining a result list (\emph{query-level stopping}) and when to stop issuing queries (\emph{session-level stopping}), cf. Section~\ref{sec:sim_search:workflows}. The difference between the two lies in the granularity at which the information seeking process that would be stopped. Specifically, query-level stopping is an action taken by a user during the sub-process of viewing results from a particular query. %
Session-level stopping may be regarded as a stopping action taken by a user during the higher-level process of a whole information seeking session. Conceptually, stopping at the session level may be regarded as an implicit action taken by the user which triggers the transition of the (mental) state from ``during-information-seeking'' to ``information-need-satisfied'' or ``abandon.''
The modeling of user decisions around stopping may be complex.
For example, imagine a user that has made a query-level stopping decision, but their information need has not been satisfied.
In such a case, the user may attempt to formulate a new query, but if they are unable to come up with an appropriate query then they may decide to stop the whole session. In general, from the perspective of an MDP, stopping can be defined as a special action to stop a current information seeking process (e.g., examining a list). Once the current process is stopped, the state would transition to a different state, in which the user then faces additional actions to take, including again whether to continue or stop the whole process. A state may be characterized by multiple variables denoting different granularity of process and a stopping action can be applied to any of such processes. A consecutive sequence of stoppings may be possible if they are applied to processes of different granularity. Those processes are closely related to tracking task completion and cost accumulation as well as environmental or contextual constraints.

Stopping behaviour has been studied both from a \emph{user-centric} perspective, trying to understand the cognitive and decision-making aspects of it, as well as from a \emph{system-centric} perspective, with a focus on incorporating models of stopping behaviour into evaluation measures.
While early retrieval measures (e.g., Precision@k or MAP) assume that the user is examining the results up to a fixed depth $k$ (cf. Fig.~\ref{fig:naive_searcher}), in contemporary measures (e.g., nDCG and RBP) the likelihood of continuation decreases the deeper the result list is traversed, i.e., searcher persistence is directly related to stopping behaviour.

Several user studies have used interviews to understand \emph{why} people decide to stop, e.g.,~\citep{Zach:2005:JASIST,Berryman:2006:IR,Wu:2014:SIGIR}.
The main conclusion from these studies is that users do not apply predetermined criteria, but rather base stopping decisions on the feeling of ``good enough.''  Factors influencing this assessment include time constraints, diminishing returns of further information seeking, and increasing redundancy of information encountered.  However, these factors vary from person to person and as such are intrinsically difficult to capture effectively~\citep{Maxwell:2019:PhDThesis}.
Next, we present different heuristic rules that have been proposed in the literature in an attempt to quantitatively characterize the sense of ``good enough.''

\subsection{Stopping Heuristics}

Modeling search as an iterative decision making process, stopping heuristics may be divided into two broad classes of approaches: judgment-based and reasoning-based rules~\citep{Nickles:1995:PhDThesis}. \\

\parheading{Judgement-based rules} assume that the user sets and maintains some mental threshold along a given dimension and keeps track of a measure relative to this dimension~\citep{Gettys:1979:OBHR}.  Stopping happens when the measure exceeds the threshold.

\begin{itemize}
    \item \textbf{Satisfaction point}: The searcher would stop as soon as they have found enough relevant material, i.e., the desired number of relevant documents, based on their subjective notion of what constitutes a relevant result~\citep{Cooper:1973:JASIST}. This is also known as the \emph{satiation rule}~\citep{Kraft:1979:IPM}.
    \item \textbf{Frustration point}: Conversely to the previous rule, this heuristic considers the searcher's tolerance to non-relevant information: they stop as soon as they become frustrated with the results, i.e., having examined too many irrelevant documents~\citep{Cooper:1973:JASIST}. This is also known as the \emph{disgust rule}~\citep{Kraft:1979:IPM}.
    \item \textbf{Satisfaction or frustration}: The above two simple rules can be combined, allowing the user to stop once they have found enough relevant information or become frustrated by too many irrelevant results---whichever of the two conditions is met first~\citep{Kraft:1979:IPM}.
    \item \textbf{Magnitude threshold}: This rule considers the searcher's belief that once the cumulative impact of the information found reaches their internal threshold, it constitutes sufficient evidence in support of a particular conclusion~\citep{Nickles:1995:PhDThesis}.  This is similar to the satisfaction point heuristic, but ``enough'' is defined in terms of cumulative amount of information collected as opposed to number of relevant documents found.
    \item \textbf{Difference threshold}: The user assesses the value of the new result found (according to some dimension) with respect to the previously examined content.  They stop when the absolute difference between the two falls below some threshold (as it means that nothing new is being learnt)~\citep{Nickles:1995:PhDThesis}.
    \item \textbf{Single criterion}: This heuristic assumes that the user is searching for information related to a single criterion, which is typically the most important to them, and then stop as soon as they have found enough information about that criterion (for example, interest rates on mortgages offered)~\citep{Browne:2005:HICSS}.
\end{itemize}

\parheading{Reasoning-based rules} are dominated by the individual's informal reasoning processes as they develop a mental representation of the topic during the course of search~\citep{Yates:1990:Book}.

\begin{itemize}
    \item \textbf{Representational stability}: As a person searches for information, their mental model, or representation, of the underlying information need shifts and develops.  Once their internal representation of the problem stabilizes, they stop accessing additional information~\citep{Nickles:1995:PhDThesis}.
    \item \textbf{Propositional stability}: Similarly to the previous heuristic, this one is also concerned with achieving stability.  Here, the focus is on the conclusions that the searcher may tentatively form while accessing information.  At some point, these conclusions do not change anymore (i.e., stability is achieved), which prompts the user to stop~\citep{Nickles:1995:PhDThesis}.
    \item \textbf{The mental list}: The searcher is assumed to develop (or recall from long-term memory) a mental list of elements that they believe are required to be completed.  They will stop once they have collected information on the requisite elements~\citep{Nickles:1995:PhDThesis}.
\end{itemize}

\noindent
It should be noted that the different heuristics are likely to behave differently under different contexts~\citep{Maxwell:2019:PhDThesis}.  For example, the mental list rule works well when a person is searching for information concerning the purchase of a home; their mental list would comprise of specific properties, such as the number of bedrooms, parking facilities, neighborhood etc.  Similarly, one could easily construct a list of requirements for a movie that a family with kids could watch on a Sunday afternoon.
However, the construction of a mental list might be impossible if the person is unfamiliar with the topic and has no prior knowledge of the criteria that would need to be met.
\citet{Browne:2007:MISQ} hypothesize and present empirical evidence that for tasks that are more structured and lend themselves naturally to decomposition, people tend to use the single criterion and mental list rules. On the other hand, for poorly structured, holistic tasks the threshold- or stability-based rules are more often employed.

The dependency of stopping on the task of a user means that the state variable in the MDP for modeling stopping should also include information regarding (1) the progress toward completion of the task, (2) importance of finishing the task; searching for medical information may be modeled as a more important task than checking out today's news; due to the higher utility of the former task, the user would be willing to spend more effort on search, thus leading to delayed stopping as compared with the latter case, where a user could easily give up on search, especially if there are environmental constraints, (3) any constraints from the search context or environment (e.g., downloading and viewing a long document on a mobile device might not be possible if the Internet connection is slow or a user does not have enough time to wait for it to be completed), (4) the perceived cost/effort of continuing the browsing or search (e.g., typing a query while driving is more costly (and may be impossible) but it has lower cost when done at a meeting, whereas using a speech query has lower cost during driving than at a meeting).

Stopping heuristics may also be formulated based on theoretical models of search.  For example, in case of \emph{information foraging theory} (cf. Section~\ref{sec:construction:conceptual:strategic}), foragers need to decide how long to stay on a given patch (i.e., examine the SERP) before moving on to a new patch (i.e., issue a new query).  The optimal behaviour for the forager, according to the \emph{marginal value theorem}~\citep{Charnov:1976:TPB}, would be to stay on the same patch until the rate of gain outperforms the average rate of gain expected within the patch.  After that point, there are diminishing returns on foraging there (i.e., examining content on the SERP).
Other heuristics developed as part of \emph{optimal foraging theory}~\citep{Stephens:1986:book} are based on time; a forager would stop after spending a given amount of time on a patch~\citep{Krebs:1973:bookchapter} or if a certain amount of time has elapsed since the last prey found~\citep{Kerbs:1974:AB}.
For a more detailed account of stopping heuristics, including their relationship to theoretical models and further examination via user studies, the interested reader is referred to \citet{Maxwell:2019:PhDThesis}[Chapter 3].

\subsection{Modeling Query-level Stopping}

While the above stopping heuristics offer plausible approaches to characterizing why and when searchers would stop, these ideas are not necessarily straightforward to operationalize.
Next, we present various ways of modeling query-level stopping behaviour; these can be thought of as programmable means of implementing some of these above heuristics.
Specifically, we model the user's decision to stop examining the result list after having examined a result snippet (or, more generally: item).

\begin{itemize}
    \item \textbf{Fixed depth}: The searcher is assumed to examine snippets in the result list until they reach a fixed depth.  This assumption corresponds to the naive searcher model (cf. Fig.~\ref{fig:naive_searcher}) and is widely held across many IR measures.  On average, this is not an unreasonable approach, however, users in practice would employ different thresholds for different queries.
    \item \textbf{Searcher frustration}: \citet{Maxwell:2015:CIKM} propose to operationalize the frustration point rule~\citep{Cooper:1973:JASIST,Kraft:1979:IPM} by counting the number of non-relevant snippets observed by the user in two different ways: within the search session in total (\emph{total non-relevant}) and contiguously in a row (\emph{contiguous non-relevant}).
    \item \textbf{Satisfaction}: Following the satisfaction point heuristic~\citep{Cooper:1973:JASIST,Kraft:1979:IPM}, the searcher will stop after encountering a pre-defined number of relevant snippets.
    \item \textbf{Satisfaction of frustration}: The searcher would employ both the satisfaction and the frustration stopping heuristics, and stops as soon as one of the two conditions is met.
    \item \textbf{Difference}: \citet{Maxwell:2015:CIKM} present two separate stopping strategies to operationalize the difference threshold heuristic~\citep{Nickles:1995:PhDThesis}.  The current snippet is compared against all snippets previously examined by the user. If its similarity falls below a predefined threshold, that is, the snippet is considered too similar to previously seen content, then the user stops. Similarity may be measured in terms of term overlap or Kullback–Leibler divergence.
    \item \textbf{Rate of gain}: Building on information foraging theory and assuming that the searcher have some idea of the of the average rate of gain, the user will stop if the rate of gain from the documents examined so far drops below a threshold. \citet{Maxwell:2015:CIKM} determine the rate gain at the current snippet based on the DCG scores of documents observed up to that point and the total time taken. Additionally, they require searchers to examine a minimum number of documents, before making a decision based on the rate of gain.
    \item \textbf{Time-based}: Another set of heuristics inspired by information foraging theory based stopping decisions on time. It can be the total amount of time spent on the SERP or time elapsed after the last relevant document found~\citep{Maxwell:2019:PhDThesis}. A time-based strategy may also be combined with the satisfaction heuristic: if the SERP yields a high volume of relevant content early on, then satisfaction-based stopping would be triggered, while if relevant items are at greater depths then the time-based heuristic is used instead~\citep{Maxwell:2019:PhDThesis}.
\end{itemize}

\noindent
Note that query-level stopping may be affected by whether the user can easily reformulate the query; intuitively, if the user can easily come up with a new, potentially more effective query, then they may tend to stop browsing the current search results earlier, but if the user has already reformulated a few queries, they may have no choice but to continue browsing the current list.

\subsection{Modeling Other Stopping Decision Points}

\begin{figure}
    \centering
    \includegraphics[width=\textwidth]{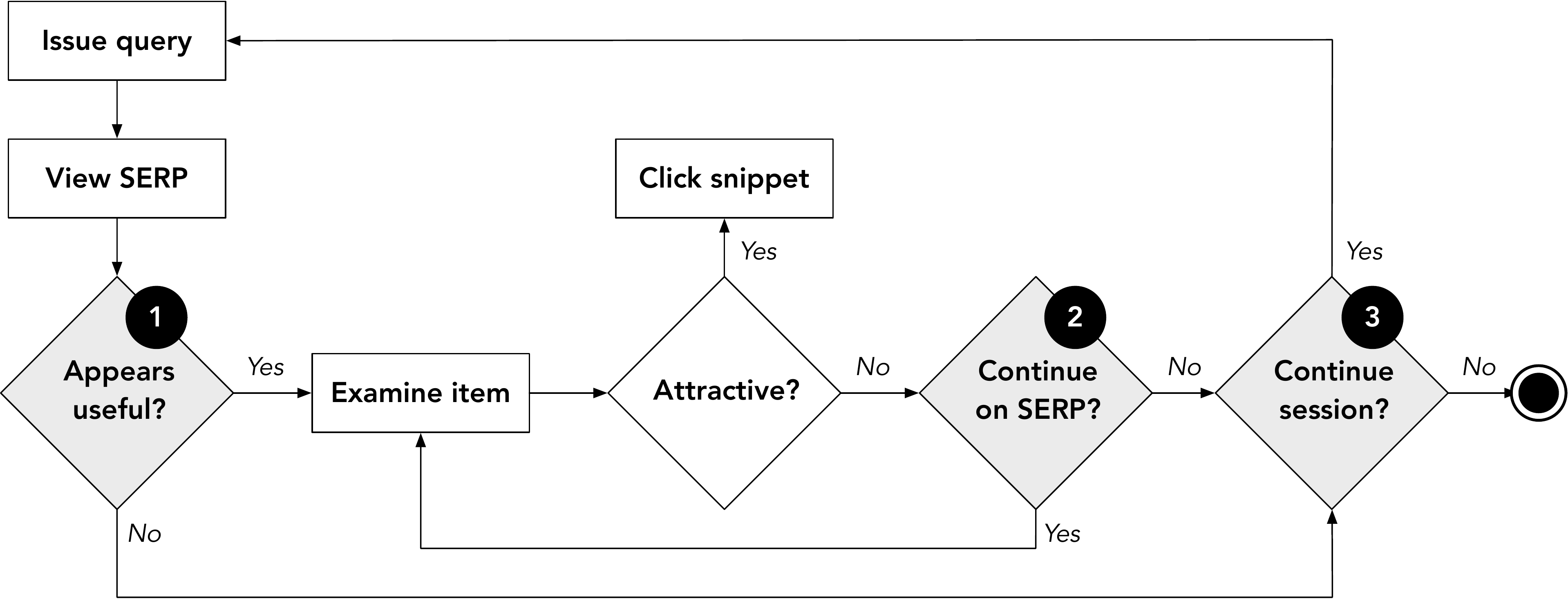}
    \caption{Excerpt from the updated Complex Searcher Model~\citep{Maxwell:2018:ECIR}, highlighting various stopping decision points: (1) SERP-level stopping, (2) query-level stopping, and (3) session-level stopping.}
    \label{fig:stopping_points}
    \vspace*{-0.5\baselineskip}
\end{figure}

The majority of work assumes that either the user will examine results up to a fixed depth or stop based on some criteria (see above).
However, it is more appropriate to think of stopping as a cascade of choices the searcher has to make, as illustrated in Fig.~\ref{fig:stopping_points}.
The first decision happens on the \emph{SERP level}, motivated by the notion of SERP abandonment~\citep{Li:2009:SIGIR,Diriye:2012:CIKM}.  That is, the user might leave the SERP without clicking on any of the results, because their information need has been satisfied (e.g., a direct answer was presented in a knowledge panel) or because none of the results seemed interesting enough to warrant further examination.
\citet{Maxwell:2018:ECIR} estimate the probabilities of examining SERPs with high information scent (i.e., good results) and low information scent (i.e., poor results) based on ground truth relevance judgments associated with the top ranked snippets on the SERP.
If the user decides to abandon the SERP, they are faced with two options: (a) formulate a new query or (b) abandon the search session altogether (or switch to a different search engine).

If the user chooses to examine the SERP, after each snippet they have to decide whether to continue on the SERP or not, i.e., make a \emph{query-level stopping} decision. Should they decide to not continue, they again have the following two choices: (a) formulate a new query or (b) abandon the entire search session.
With the exception of the fixed depth strategy, the query-level approaches discussed above can also be applied to modeling this \emph{session-level stopping}. For example, the user would stop issuing queries once they become satisfied with the results or reach their frustration threshold.

Note that with more information-dense SERPs, there is also an increased possibility of ``good abandonment,'' i.e., that the user's information need is answered by the SERP without needing to click on any of the results~\citep{Li:2009:SIGIR,Chuklin:2012:WWW}.
Thus, there is a need for more advanced user browsing models that are able to capture such implicit action.
\citet{Thomas:2018:ADCS} present an extended user model that takes into account the full variety of results searchers see, based on the notion of \emph{cards}, which refer to various SERP elements, including static, clickable, and interactive ones.
\section{Summary and Future Challenges}
\label{sec:sim_search:summary}
\label{sec:sim_search:summary_mdp}

It is evident that there is a plethora of research on individual components, spanning from rudimentary to sophisticated techniques.
At the same time, there is a scarcity of research that integrates these component-level models into a complete end-to-end framework.
There exist studies that consider a reasonably complete set of user actions and states, for example, \citet{Carterette:2015:ICTIR} simulate query reformulations, clicks, dwell times, and session abandonment for evaluating retrieval systems that make use of session history, \citet{Maxwell:2016:SIGIR} instantiate all components of the Complex Searcher Model (cf. Fig.~\ref{fig:complex_searcher_model}), and \citet{Labhishetty:2020:SDMWorkshop} attempt to model and update a user's cognitive (knowledge) state and ``exploratoriness'' of information need.
However, there are many simplifying assumptions made regarding how users interact with a search system as well as in terms of what functionality a (modern) search engine provides.

Throughout the chapter, we have discussed how the various decisions a user has to make when interacting with search and recommender systems can be modeled in a general MDP framework in terms of actions, states, rewards, and policy. A component user simulation model (e.g., a click model) can be viewed as modeling the actions that a user can potentially take in a particular state of interaction with a system (e.g., a state of interacting with the search results) and the (local) policy that the user may use to decide which action(s) to take in that state (e.g., whether to click a result). Thus, the MDP framework can serve as a powerful and mathematically sound framework for studying user simulation.
Different user simulation models and techniques can all be viewed as specific instantiations of the general framework under various assumptions, and machine learning algorithms, particularly model-based reinforcement learning~\citep{Moerland:2023:Survey} and imitation learning~\citep{Hussein:2017:CSUR,Labhishetty:2023:thesis}, can be naturally used to train an interpretable user simulator based on user data.

However, current approaches generally only model a user superficially with many overly simplified assumptions. They also fail to holistically consider the user's overall objectives in a particular context and optimize a global policy by making cost-benefit trade-offs; instead, user decisions are currently based on local (component-level) policies.
Below, we summarize existing techniques for constructing user simulators using the MDP framework, and outline areas that require further investigation in order to make simulators more realistic and human-like.\\

\parheading{State}:
The state variable is key to modularizing the construction of a user simulator. At a high level, it can be explicitly defined based on the workflow of an information seeking session. For example, the basic search workflow dictates that there are various high-level states that we need to model, including the initial query formulation, interaction with search results, which can be decomposed into two separate states, i.e., scanning and clicking, and reformulation of queries. For each of these explicit states, various simulation methods have been proposed in the literature. Within each high-level state, we may further distinguish fine-grained (low-level) states, which are often based on a user's latent cognitive state. For example, in the case of modeling scanning behaviour, we may further model a user's cognitive state including capturing how a user might have gained increasingly more relevant information while interacting with more search results, meaning that we can define a separate (low-level) state after a user interacts with each single snippet.
In general, the state should include both the interface provided by the information access system and the user's mental/cognitive state. The latter further consists of a user's task, goal, intent, preferences, and cognitive and emotional states.

Existing work has generally only focused on the explicit high-level states, with some work modeling the cognitive state of users during the process of interaction with a simple list of results. More detailed and broad modeling of the state is needed in the future, which requires
(1) modeling a user's task deeply and refining task workflows; 
(2) modeling complex user interfaces, including different layouts as well as conversational interfaces;
(3) modeling the context/environment of the search (e.g., time and location); and
(4) better understanding of complex processes such as cognition and emotional reactions (such as frustration or impatience). \\

\parheading{Actions}:
In each state, a user would have multiple options of taking actions. As in the case of state, we may also distinguish explicit and implicit actions that a user might take. The explicit actions that are available for a user to take are generally determined by the features/functions provided by a system. At a high-level, there are some common functions supported by all search systems, including, e.g., query formulation/reformulation, scanning the results, clicking on results, etc. Many of these common actions have been studied extensively in prior work and specific approaches have been proposed for simulating a user taking them.
Modern user interfaces of search and recommender systems, however, have rich interaction features, requiring a more detailed modeling of a broader scope of user actions than what has currently been studied.

There are also implicit actions that we cannot (easily) observe. For example, eye movement can only be tracked by using a special device.
Therefore, instead of treating them as actions, these observations are often incorporated in the form of specific assumptions, such as the linear traversal hypothesis~\citep{Guo:2009:WWW}. The change of a user's cognitive state can further be viewed as a result of a user taking an implicit action (e.g., updating the information need and task status as the user gains more relevant information). It is impossible to observe such implicit actions at this point, so they can only be modeled via latent variables.
The modeling of implicit actions is particularly challenging as those may not be experimentally verifiable in the near future. Nevertheless, a realistic user simulator should still attempt to use latent variables to model cognitive and emotional states of users and model the implicit actions (which are often related to decisions) that can be taken by a user. \\

\parheading{State Transitions}:
According to an MDP, a user's action would trigger a transition from one state to another. This kind of state transition is straightforward when we consider only explicit states and explicit actions, as the transitions are mostly determined by how a system responds to the user's action. However, transitions of implicit states are much more complicated and may depend on the user's prior background knowledge, physical environment, and context of information seeking, as well as their personality and cognitive skills. So far, such transitions have only been studied in a limited manner with overly simplified assumptions. Since the modeling of cognitive state in existing work has mostly focused on representing the user's information need, it is not surprising that the transition of implicit state has been mostly studied in terms of updating a user's knowledge about relevant and non-relevant information.
A detailed modeling of implicit state transitions requires a better understanding of human cognition and emotional reactions, and a contextualizaion of such human behaviour in different kinds of physical environment and task context.  \\

\parheading{Reward (and Cost)}:
The reward in an MDP models a user's objective of information seeking, which is closely related to the user's task. A related factor is the effort a user must make in order to achieve the goal (i.e., the cost).
When it comes to decision making, a user would likely consider the cost associated with a given action and the potential increase of reward resulting from that action.  However, it is important to consider not only immediate but also accumulated costs, as certain decisions, such as abandoning a search session, might be mostly based on accumulated cost. The accumulated cost could be captured as part of the state in the MDP.
Analogously, we may distinguish the global reward, which is the overall accomplishment of a task, from the local reward that a user could perceive after taking a specific action. A user may be assumed to mainly optimize the global (long-term) reward, but the short-term reward affects their perceived cost/effort and more directly influences the decision on which action to take.

Since existing work on user simulation has not yet considered the comprehensive sequential decision model suggested by the MDP framework, the notion of global reward has not been well studied, though there has been work on studying user satisfaction/frustration. Further, it is often (implicitly) assumed in information seeking settings that reward equals finding topically relevant material and cost equals time/effort spent. However, reward should ultimately be based on the broader notion of utility
and users might not mind the extra cost, i.e., spending more time interacting with the system, if it offers an engaging user experience. \\

\parheading{Policy}:
The policy of an MDP determines how to choose an action in each state. This is at the core of computationally modeling a user's behaviour and connects all the other components in the MDP. Given the current state, which provides all the information for deciding an action, including, e.g., a user's current cognitive and emotional states, the current interface a user is interacting with, the policy would determine the best action for a user to take (which could also be an implicit action that a user takes mentally). The chosen action would then cause the state to transition to a new state and the process repeats in this way. Since  existing work has mostly focused on simulating some specific actions of a user, the models developed in existing work can only be used to define the policy function for some special states, such as the state in which a user would formulate an initial query, the state in which a user would scan results in a list, or click on a result, and the state in which a user would decide whether to reformulate a query. The models of policy can be either simple but interpretable models or machine-learned non-interpretable predictive models of user behaviour.

A major gap is the lack of work on integrating multiple component policies into a unified approach, with a systematic treatment of all the states and their transitions, thus completely defining the policy function. There have already been some attempts to address this by loosely connecting policy models of the common explicit states in a search system, thus integrating initial and subsequent query formulation/reformulation and click modeling. In the future, tighter integration of component models with full specification of states and state transitions is needed, so as to bring all the research results on individual components together in the MDP framework with the purpose of building more realistic user simulators.

In order to realistically model decision making, it is important to consider how humans tend to employ a high-level strategy to balance costs and benefits, and adopt these trade-offs based on their current context. Such higher-level strategies might be informed by conceptual models of information seeking and decision making (cf. Sections~\ref{sec:decisions:conceptual} and ~\ref{sec:decisions:choices}).\\

\noindent
We have shown that the MDP framework can be used as a general framework to synthesize both past and future work on simulating user interactions with search and recommender systems. It is worth noting that other mathematical frameworks could also be used, and the exploration of alternatives and the advantages they can offer is an important direction for future research.
Nevertheless, the fundamental principles and insights discussed in this chapter, such as modeling the space of actions and associated costs/rewards for decision making, are likely to remain relevant across different modeling frameworks. In particular, as we will further discuss in the next chapter, the MDP framework is also applicable to simulating user interactions with conversational assistants.

Finally, we would like to point out the close connection of user simulation and user studies. Many user studies have revealed useful findings about user characteristics and behaviours. In general, these findings have not yet made into formal design of user simulators, though some findings have clearly influenced the design of user simulation models. The topic of user simulation naturally connects user studies and formal model design, therefore we anticipate such integrated work to be critical to eventually develop realistic user simulators. Table~\ref{tab:user_characteristics} lists some specific user characteristics that have been considered in past work, but the challenge of incorporating them into policies in a generic manner remains to be addressed.

\chapter{Simulating Interactions with Conversational Assistants}
\label{ch:sim_conv}

\vspace*{-0.75\baselineskip}
Conversational AI broadly refers to systems that are capable of natural language understanding and responding in a way that mimics human dialogue, thereby providing an interactive and engaging experience for users.
Traditionally, these systems are categorized as being \emph{goal-driven} (or \emph{task-oriented}), aiming to assist users to complete some specific task, or \emph{non-goal-driven} (also known as \emph{chatbots}), aiming to carry on an extended conversation (``chit-chat''), usually with the purpose on entertainment~\citep{Chen:2017:SIGKDD,Jurafsky:2023:book}.
Our interest in this book lies in the former category, that is, conversational systems that have a clear goal: assisting the user in completing some task with an underlying information need, which can be satisfied through natural language dialogue.
This includes the tasks of search, recommendation, and question answering, collectively referred to as the problem of \emph{conversational information access} (CIA)~\citep{Balog:2021:DESIRES}.
CIA can be seen as a specific form of task-oriented dialogue (TOD), where the user's goal can, for instance, be to satisfy some ad-hoc information need, find an item matching their preferences, or learn about a topic.
What makes user simulation for conversational systems particularly challenging is that a simulator also needs to possess advanced natural language understanding and generation capabilities (cf. Section~\ref{sec:background:cia}).
Consequently, constructing a fully functional user simulator is akin to building an independent conversational agent.

We begin by establishing basic concepts in Section~\ref{sec:sim_conv:preliminaries}.
Simulator architectures, which closely mirror the architectural principles of TOD systems, are introduced in Section~\ref{sec:sim_conv:architectures}.
User simulation for TOD has a longstanding history in dialogue systems research and has recently gained increased interest within the natural language processing community, with a growing number of approaches utilizing pre-trained large language models (LLMs). We review simulation for TOD systems in Section~\ref{sec:sim_conv:tod}.
Although conversational information access can be seen as a form of task-oriented dialogue, it goes beyond the identification and filling of slot-value pairs, typical of traditional TOD systems (e.g., departure and destination city/airport, departure and return dates, etc. for travel booking). Therefore, in Section~\ref{sec:sim_conv:cia}, we discuss how CIA can be conceptualized in terms of intents and dialogue structure.
This is followed by the discussion of simulation approaches developed specifically for conversational search and conversational recommendation in Sections~\ref{sec:sim_conv:conv_search} and~\ref{sec:sim_conv:conv_rec}, respectively.
It is worth emphasizing that user simulation for CIA evaluation is very much an area of ongoing work, where notable advancements are being made in terms of conversational models as well as evaluation measures and methodologies.
We identify and discuss some of the future challenges in Section~\ref{sec:sim_conv:summary}.

\section{Preliminaries}
\label{sec:sim_conv:preliminaries}

We assume that a conversation happens between a user and a system, who take turns. Following existing research, we assume the modality to be text, either written or spoken (and transcribed), noting that other modalities are also receiving increased attention.  A dialogue, thus, is a sequence of turns, where each turn $t$ is a natural language utterance from either the user $x_t^u$ or the system $x_t^s$: $x_0^s \rightarrow x_0^u \rightarrow x_1^s \rightarrow x_1^u \rightarrow \dots \rightarrow x_{t}^s \rightarrow x_t^u$.

The notion of \emph{dialogue act} originates from (spoken) dialogue systems research and is meant to represent the function or high-level intention of an utterance in a dialogue, for example, whether it is a question, offer, suggestion, etc.~\citep{McTear:2016:book}.
Dialogue acts are typically represented as tuples, consisting of an \emph{intent} and, optionally, a number of slot-value pairs (e.g., \texttt{AFFIRM} or \texttt{INFORM(a=x,b=y,...)}).\footnote{We note that concept of \emph{dialogue act} is not used consistently in the literature; in some cases it refers to the intent, while in other cases it means the intent along with the associated slot-value pairs. For clarity, we use \emph{intent} in the former and \emph{dialogue act} in the latter case. When the distinction is not important or the sense is clear from the context, we may simply write \emph{action}.}
Formally, a dialogue act is a tuple $a=(I,V)$, where $I \in \mathcal{I}$ is an intent from the set of possible dialogue intents and $V$ is a list of slot-value pairs $V=\langle (slot_1,value_1), \dots, (slot_n,value_n) \rangle$.
Human-machine dialogue can then be represented semantically as a sequence dialogue acts by the system ($a_i^s$) and the user ($a_i^u$) as they take turns: $a_0^s \rightarrow a_0^u \rightarrow a_1^s \rightarrow a_1^u \rightarrow \dots \rightarrow a_{t}^s \rightarrow a_t^u$. 

The space of possible intents and slots need to be defined specific to the objectives of the dialogue application. Over the years, various taxonomies have been developed.
For example, the Dialogue State Tracking Challenge~\citep{Williams:2016:DD} has defined user and system actions for task-oriented dialogue systems (bus, restaurant, and tourist information).
The international standard ISO 24617-2 defines communication functions for general dialogue and also provides a formal language, the Dialogue Act Markup Language (DiAML), for expressing dialogue annotations~\citep{Bunt:2017:chapter,Bunt:2020:LREC}.

\section{Simulator Architectures}
\label{sec:sim_conv:architectures}

The main challenge in simulating users in a conversational setting is that the simulator needs to have advanced natural language understanding and generation capabilities, much like that of a conversational system.  Hence, adopting the architecture of task-oriented dialogue systems is a natural approach in the construction of user simulators.
Accordingly, two broad categories of simulator architectures can be distinguished: modular (or pipelined) and end-to-end systems.
The difference between these two is more fundamental than just an architectural choice, it is also a question of granularity: modular simulators model user responses semantically on the level of dialogue acts, and then generate the corresponding natural language utterances. End-to-end systems, on the other hand, operate on the utterance level, i.e., generate textual responses directly.  
While end-to-end systems might yield more fluent dialogues, they do not allow us to interpret user behaviour, and therefore are not that well suited for our purposes of evaluation.
Parameterization of such systems with meaningful user variables would be desirable in general, but this requires modeling a user beyond their surface-level utterances in order to capture latent user states using more interpretable models.

\subsection{Modular Systems}
\label{sec:sim_conv:architectures_modular}

\begin{figure}[t]
	\centering
	\includegraphics[width=0.8\textwidth]{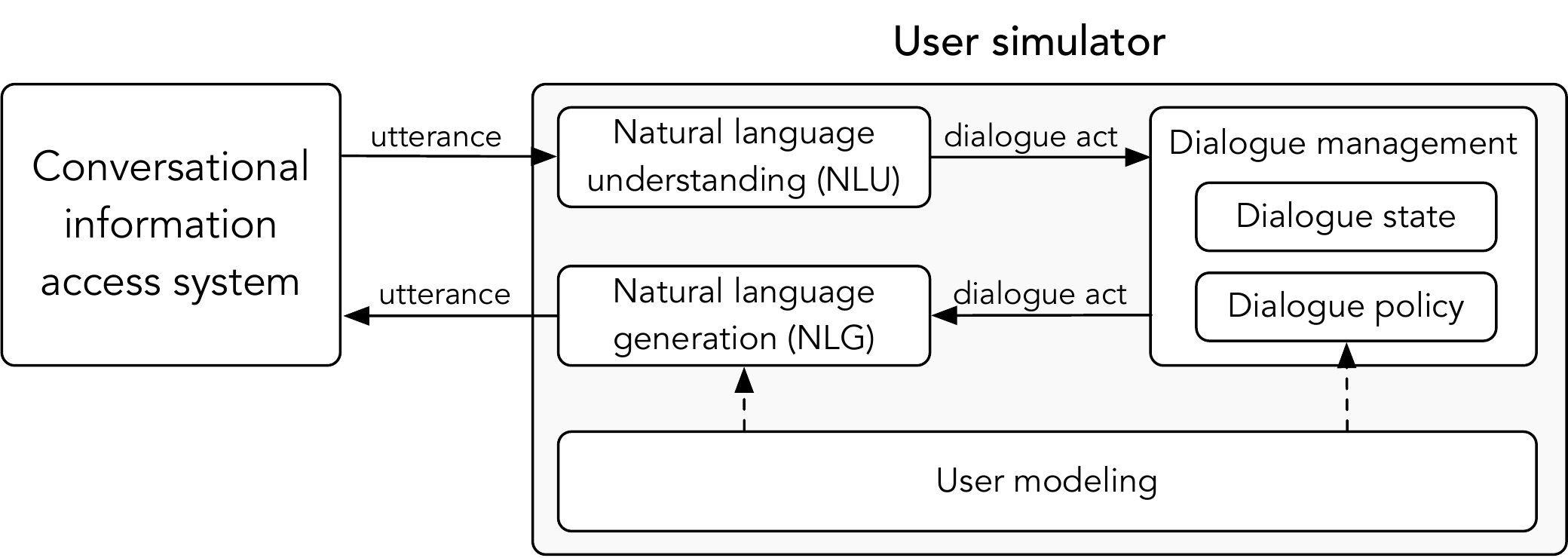}
	\caption{Modular user simulator architecture for conversational information access.}
	\label{fig:convsim_arch_modular}
\end{figure}

Following the architecture of traditional task-oriented dialogue systems, user simulation can be decomposed into the sequentially dependent modules of natural language understanding, dialogue management, and natural language generation. Additionally, we separate the modeling of individual characteristics into a dedicated user modeling component, following~\citet{Afzali:2023:WSDM}. \\

\parheading{Natural language understanding (NLU)} is the task of converting the (raw) system utterance into an internal semantic representation, that is, to a dialogue act, which consist of an intent and its parameters (slot-value pairs), referred to as the task of \emph{slot filling}. \emph{Intent detection} is naturally approached as a classification task, while slot filling is a sequence labelling problem~\citep{Chen:2017:SIGKDD}. Traditionally the two tasks have been addressed independently, however, in more recent work the two tasks are modeled jointly, leveraging the strong relationship that exists between them~\citep{Weld:2022:CSUR}.
It may be argued that users rarely misunderstand the system, hence the NLU component may be ``bypassed'' and the simulator could receive the system utterance directly on a semantic level, i.e., as dialogue acts~\citep{ElAsri:2016:Interspeech,Kreyssig:2018:SIGDIAL,Lin:2021:SIGDIAL}.
While this assumption is not unreasonable to make, it requires access to the conversational system and does not work in cases where the system needs to be treated as a ``black box.'' \\

\parheading{Dialogue management} is responsible for maintaining the dialogue state and determining the next user action based on the system's utterance and the goals, knowledge, and preferences of the individual.
The \emph{dialogue state} in traditional task-oriented dialogue systems is based around the notion of a \emph{semantic frame}, which is a knowledge structure consisting of a collection of slots that together specify what the system needs to know to complete a given task~\citep{Jurafsky:2023:book}.\footnote{In spoken dialogue systems, a considerable research effort is centered on dealing with the uncertainty resulting from noise in automatic speech recognition. There, it is common for the dialogue system to maintain multiple hypotheses of the true dialog state~\citep{Williams:2016:DD}. However, for text-based interactions, this is typically not a concern.}
For user simulation, the dialogue state needs to capture dialogue context from the user's perspective.  This includes representing the current information need, the dialogue history, and the system's last action. The dialogue state  may be represented explicitly, in a directly observable and interpretable way, such as previous dialogue acts and the status of various slots, or as a latent representation in an embedding space.
The \emph{dialogue policy} determines how the user should respond, i.e., the user intent and associated parameters; this might be influenced by the individual characteristics of the user. To ensure that the simulator maintains a dialogue strategy, a commonly used approach is an \emph{agenda-based simulator}; it maintains an agenda or a stack of actions and determines which action to execute at each turn. Alternatively, the policy may be learned directly from data. We discuss these alternatives for TOD in more detail in Section~\ref{sec:sim_conv:tod}. \\

\parheading{Natural language generation (NLG)} is concerned with turning the generated response from a structured representation (dialogue act) into natural language. At its simplest form, this is based on templates~\citep{Gur:2018:SLT,Shi:2019:EMNLP}. Other alternatives include retrieval-based methods~\citep{Wu:2017:ACL,Hu:2014:NIPS}, neural text generation models~\citep{Wen:2015:EMNLP,Tseng:2021:ACL}, and hybrid solutions~\citep{Li:2017:arXiv,Sun:2023:TOIS}. Natural language generation can also consult the user modeling component to produce outputs that are customized based on individual characteristics. \\

\parheading{User modeling} focuses on capturing the characteristics of individuals that would influence how they interact with the system. These could include, among others, information about the user's goal, knowledge, preferences, personal characteristics (e.g., patience), and beliefs about the system (e.g., based on past interactions). In case of information access tasks, one can often identify commonalities (similar user state variables) between user modeling in a conversational system and traditional search/recommender systems. However, both the information available for inferring a user's state and the observable actions are different. In particular, for conversational systems, we would need to model how a user's latent state influences the natural language utterances generated by the user as well as infer a user's latent state based on the natural language system responses.

\subsection{End-to-End Systems}
\label{sec:sim_conv:architectures_e2e}

\begin{figure}[t]
	\centering
	\includegraphics[width=0.5\textwidth]{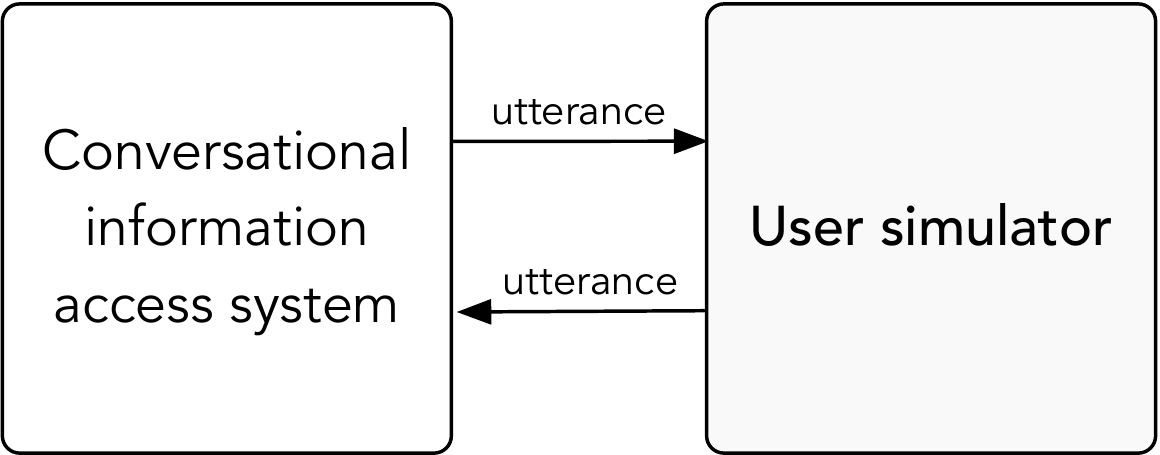}
	\caption{End-to-end user simulator.}
	\label{fig:convsim_arch_end2end}
\end{figure}

More recently, there have been numerous attempts to build end-to-end trainable task-oriented dialogue systems, leveraging advances in end-to-end neural generative models~\citep{Gao:2019:FnTIR,Ni:2022:AIR}. The main benefits over modular systems include the possibility of global optimization and easier adaptation to new domains~\citep{Chen:2017:SIGKDD}.
End-to-end architectures may also be used for constructing user simulators, by casting the generation of user responses as a sequence-to-sequence (or source-to-target transduction) problem: given a dialogue history up to the given turn as input, generate the user's natural language response as the output sequence.\footnote{As noted above for modular architectures, in end-to-end simulation the NLU component may also be omitted~\citep{Lin:2022:SIGDIAL}.  Instead, the dialogue act corresponding to the system utterance is taken as input, effectively equipping the simulated user with perfect language understanding capabilities.}
This problem can be effectively addressed by transformer-based natural language generation models~\citep{Tseng:2021:ACL,Lin:2022:SIGDIAL}.
However, as indicated earlier, such model-based end-to-end approaches are more suitable for other uses of simulation, such as RL-based learning~\citep{Lin:2022:SIGDIAL,Tseng:2021:ACL} or data augmentation~\citep{Wan:2022:EMNLP,Mohapatra:2021:EMNLP,Kim:2021:ACL}. For simulation-based evaluation, a more controllable generation of user responses is required, which (currently) cannot be guaranteed with end-to-end systems.

\section{User Simulation for Task-Oriented Dialogue}
\label{sec:sim_conv:tod}

In this section, we discuss work on simulating users in the context of traditional (``slot-filling'') task-oriented dialogue systems.
Such systems are built for specific domains, each characterized by a \emph{domain ontology} that describes the specific intents, slots, and entities that can be talked about. The supported conversational functionality is limited to the user specifying their constraints in terms of \emph{informable slots} ($C$) and requesting information on \emph{requestable slots} ($R$) to find matching entities in a knowledge base.
This is analogous to some scenarios considered within conversational information access, such as finding items where the user already has a reasonably concrete idea of attributes and the constraints on them, or looking up known items or their properties. However, CIA systems ultimately support a much broader array of information needs. This is discussed further in Section~\ref{sec:sim_conv:cia}.

Below, we present a selection of influential approaches, categorized into model-based and data-driven approaches, with respect to the modeling of user decisions. Within each category, the approaches are discussed in a roughly chronological order, highlighting the progression of methods and the ways in which subsequent work builds upon and is influenced by earlier research.

\subsection{Model-Based Approaches}

Model-based approaches are characteristic of modular simulator architectures, where the user's response is governed by a \emph{dialogue policy} (a.k.a. \emph{dialogue strategy}).\footnote{It is not to be confused by the state model and policy of the conversational agent; while task-oriented dialogue systems are often built on top MDPs, those are different for the system and the (simulated) user.}
The policy $\pi$ determines what action $a^u_{t+1}$ the user should take next, given the dialogue history up to turn $t$.
A main challenge here is that the simulator should give responses that are coherent across multiple conversational turns and are driven by some user goal. This can naturally be modeled in the MDP framework.
Statistical user modeling techniques ``exploit the Markov assumption that the next user action can be predicted based on some representation of the current state of the dialogue''~\citep{Schatzmann:2006:KER}.
As the user and system take turns, the simulator receives the corresponding dialogue acts ($a^u_{t}$ for user and $a^s_{t}$ for system), and updates the state:
\begin{equation}
	s_t \rightarrow a^u_t \rightarrow s_{t+1} \rightarrow a^s_{t+1} \rightarrow s_{t+2} \rightarrow \dots
\end{equation}
Assuming a Markovian state representation, user behaviour can be decomposed into three models: $\prob(a^u_{t+1}|s_t)$ for next user action selection (i.e., the dialogue policy $\pi(s_{t})$), $\prob(s_{t+1}|s_t,a^u_t)$ for transitioning to the next state based on a user action, and $\prob(s_{t+2}|s_{t+1},a^s_{t+1})$ for transitioning to the next state based on a system action.
The challenge lies in finding a state representation $s_t$ that is as compact as possible, while containing enough information about the dialogue so that the Markov property can be met.
The state representation typically includes the user goal, the previous system action, and some history or memory of the dialogue.

\subsubsection{N-grams Models}

One of the earliest data-driven approaches is proposed by \citet{Eckert:1997:workshop}, modeling the user's next response based on the dialogue history, resembling the estimation of language models (hence the name n-gram model):
\begin{equation}
	\pi(s_t)=\prob(a^u_{t+1}|a^s_t,a^u_t,a^s_{t-1},a^u_{t-1},\dots,a^u_0,a^s_0) ~.
\end{equation}
However, since these probabilities are difficult to obtain reliably due to data sparsity, a strong simplifying assumption is made to condition the next user action exclusively on the preceding system action:
\begin{equation}
	\pi(s_t)=\prob(a^u_{t+1}|a^s_t) ~,
\end{equation}
where the conditional probabilities are estimated based on an annotated corpus of human-machine dialogues.
This model does not have any information about the user's goal and does not place any constraints on the simulated user behaviour, therefore fails to produce realistic dialogues.
Subsequent work tried to correct for it by estimating a selected set of simpler probabilities corresponding to ``sensible'' pairs of system and user actions (specifically, responding to greetings, constraining questions, and relaxation prompts), while other are assumed to be zero, thereby setting some constraints on the flow of the dialogue~\citep{Levin:2000:TSAP}.\footnote{In principle, all the selected probabilities could be estimated from a dialogue corpus, but in practice, the probabilities corresponding to constraining or relaxing questions are set manually in~\citep{Levin:2000:TSAP} as they were not present in the corpus used.} While this yields somewhat more realistic dialogues, the consistency between user responses across the dialogue is still not guaranteed. Also, simulated dialogues might run infinitely long if the user keeps changing their answer to the same question.

\subsubsection{Goal-directed User Model with Memory}

To overcome these issues, i.e., to make the simulated user goal-directed and behave consistently according to its goal, \citet{Pietquin:2004:PhDThesis} extend the work of \citet{Levin:2000:TSAP} by conditioning their set of selected probabilities on an explicit representation of the user goal and dialogue history. The user's goal is represented as a sequence of slot-value pairs randomly generated once per dialog: $G=\langle (slot_1,value_1,prior_1), \dots, (slot_n,value_n,prior_n) \rangle$. Each slot-value pair in the goal is also assigned a priority; when the user is prompted for the relaxation of some attribute, slot-value pairs with a higher priority are less likely to be relaxed.
The dialogue history at time $t$ is represented as a vector $h_t = \langle c_1, \dots, c_n \rangle$, where $c_i$ is the count of the occurrences a value is provided for the corresponding $slot_i$. This enables the simulator to disclose new information to the system if mixed initiative is supported.
In addition to the set of probabilities considered in~\citep{Levin:2000:TSAP}, the probability to terminate the session early is also introduced, which prevents dialogues running infinitely long. Early termination could happen, for example, because of unsatisfactory behaviour of the system (triggered by $c_i$ exceeding some threshold).
Since goals are specified as slot-value pairs, this model allows for automatic evaluation in terms of full or partial task completion.

\subsubsection{Agenda-based Simulator}

The agenda-based user simulator~\citep{Schatzmann:2007:NAACL} is a widely used approach for training task-oriented dialogue systems.
It factors the user state into an agenda and a goal, $s_t=(A_t,G_t)$, where the agenda $A_t$ is a stack-like structure representing the pending intentions the user aims to achieve, that is, a priority ordered sequence of user actions that need to be executed.
The goal is a tuple $G_t=(C_t,R_t)$, where $C_t$ is a set of domain-specific constraints the user wants to impose on the dialogue and $R_t$ specify requests, i.e., slots whose values are initially unknown to the user and will need to be filled out during the conversation.
For example, in the context of restaurant recommendation, looking for the name, address, and phone number of a centrally located bar serving beer would be represented as:
\begin{align*}
	C_0 =
	\begin{bmatrix*}[l]
		\mathrm{type} & = & \mathrm{bar} \\
		\mathrm{drinks} & = & \mathrm{beer} \\
		\mathrm{area} & = & \mathrm{central} \\
	\end{bmatrix*}
	 &  &
	R_0 =
	\begin{bmatrix*}[l]
		\mathrm{name} & = &  \\
		\mathrm{addr} & = & \\
		\mathrm{phone} & = &  \\
	\end{bmatrix*}
\end{align*}
In~\citep{Schatzmann:2007:NAACL}, the agenda is initialized with all goal constraints set to \texttt{INFORM} acts and all goal requests set to \texttt{REQUEST} acts; the \texttt{BYE} act is added at the bottom of the agenda to close the dialogue. For the above example, this corresponds to:
\begin{align*}
	A_0 =
	\begin{bmatrix*}[l]
		\mathrm{INFORM(type = bar)} \\
		\mathrm{INFORM(drinks = beer)} \\
		\mathrm{INFORM(area = central)} \\
		\mathrm{REQUEST(name)} \\
		\mathrm{REQUEST(addr)} \\
		\mathrm{REQUEST(phone)} \\
		\mathrm{BYE} \\
	\end{bmatrix*}
\end{align*}
As the conversation progresses, the agenda and goal are dynamically updated.
At each user turn, determining the next user action simplifies to popping items from the top of the agenda.
Specifically, let the agenda be of length $N_t$, where $A_t[N_t]$ denotes the top and $A_t[1]$ denotes the bottom item.
Using $A_t[:-n]$ as a Python-style shorthand notation for the top $n$ items on the agenda ($A_t[:-n]=A_t[N_t-n+1..N_t]$), action selection can be modeled as a Dirac delta function:
\begin{equation}
	\prob(a^u_{t+1}|s_t) = \prob(a^u_{t+1}|A_t,G_t) = \delta(a^u_{t+1},A_t[:-n]) ~,
\end{equation}
where $n$ is a binary random variable that depends on the degree of user initiative, but is typically a small value.\footnote{We note that since each user utterance is represented as a single dialogue act, popping $n>1$ items from the agenda (1) only makes sense there are multiple dialogue acts with the same intent among the top $n$, such that multiple slot-value pairs be merged together under a single dialogue act with that intent (e.g., \texttt{INFORM(type=bar,drinks=beer)}) or (2) requires the definition of compound acts.}
The new agenda after executing $a_t^u$ becomes $A_{t+1}[i]=A_t[i] \; \forall i \in [1..N_t-n]$. Continuing with our running example, popping $n=2$ items off of the agenda above would generate the user dialogue act \texttt{INFORM(type=bar,drinks=beer)} and the following updated agenda:
\begin{align*}
	A_1 =
	\begin{bmatrix*}[l]
		\mathrm{INFORM(area = central)} \\
		\mathrm{REQUEST(name)} \\
		\mathrm{REQUEST(addr)} \\
		\mathrm{REQUEST(phone)} \\
		\mathrm{BYE} \\
	\end{bmatrix*}
\end{align*}
The state transition models incorporate the agenda and goal as follows.
When the user executes $a^u_t$, we can assume that the goal remains constant and the state transition thus becomes:
\begin{align}
	\prob(s_{t+1}|s_t,a_t^u) & = \prob(A_{t+1},G_{t+1}|A_t,G_t,a_t^u)                             \\
	                         & = \delta(A_{t+1},A_t[1..N_t]) \: \delta(G_{t+1}, G_t) ~. \nonumber
\end{align}
The second type of state transition, after executing system action $a^s_t$, is decomposed into agenda and goal updates:
\begin{equation}
	\prob(s_{t+1}|s_t,a^s_t) = \prob(A_{t+1}|A_t,G_{t+1},a^s_t) \: \prob(G_{t+1}|G_t,a^s_t) ~.
\end{equation}
In \citep{Schatzmann:2007:NAACL}, the agenda update ($A_t \rightarrow A_{t+1}$) is viewed as a sequence of push operations, where dialogue acts get added on top of the agenda.
This is typically performed using handcrafted heuristics. For example, if a $slot_i=value_i$ provided in $a^s_t$ violates one of the goal constraints in $G_t$, then one of the following may be pushed onto $A_{t+1}$: \texttt{NEGATE}, \texttt{INFORM($slot_i=value_j$)}, or \texttt{DENY($slot_i=value_i,slot_i=value_j$)}.
The push operations are also followed by a deterministic ``clean up'' step, where duplicated or unnecessary dialogue acts (e.g., requests for already filled goals) are removed.

The goal update ($G_t \rightarrow G_{t+1}$) can be decomposed into changes in goal constraints $C_t$ and requests $R_t$ according to the following model:
\begin{equation}
	\prob(G_{t+1}|G_t,a_t^s) = P(R_{t+1}|R_t,C_{t+1},a_t^s) P(C_{t+1}|R_t,C_t,a_t^s) ~.
\end{equation}
To simplify the estimation of component probabilities, \citet{Schatzmann:2007:NAACL} make a number of simplifying assumptions.
Requests slots are assumed to be independent; each slot in $R_{t+1}$ is either filled based on a value provided in $a^s_t$ and given the corresponding goal constraints in $C_{t+1}$, or is left unchanged.
Goal constraints $C_{t+1}$ can be updated by adding a new constraint or changing an existing constraint slot in $C_t$ to a different value, or left unchanged.

We observe that the agenda-based simulator is commonly categorized incorrectly in the literature as a rule-based approach. To be precise, the agenda-based approach provides a probabilistic framework for instantiating user simulators. The probabilities corresponding to the maintenance of the agenda may be set using expert-defined rules, as done in \citep{Schatzmann:2007:NAACL}, but could also be learned directly from training data~\citep{Schatzmann:2009:TASLP}.
With careful designing of the state and action space, and factoring of component probabilities, the need for manual handcrafting can be minimized.

\subsection{Data-Driven Approaches}

Next, we turn to approaches to create user simulators that can be learned in a fully data-driven manner from dialogue corpora.
These models can operate on the level of dialogue acts (typical of modular architectures) or directly generate natural language utterances (common in end-to-end architectures), noting that hybrid solutions also exist.
It is worth pointing out that the majority of the works discussed here use simulation as an interactive environment for the training of dialogue systems, and not for evaluation, which might limit their realism~\citep{Sun:2023:TOIS}.

\subsubsection{Sequence-to-Sequence Models}

Sequence-to-sequence models, originally developed for machine translation, have been successfully applied to simulate user behaviour in TOD. This is a natural fit, as both tasks involve generating a sequence of tokens in response to an input sequence. Here, the input sequence is some representation of the dialogue history, and the output sequence is the next user action and/or utterance; see Table~\ref{tab:s2s_sim} for an overview.

\begin{table}[t]
	\caption{Overview of user simulators using a sequence-to-sequence model structure. The output may be a dialogue act (act), utterance (utt), user satisfaction score (sat), or some combination of these.}
	\label{tab:s2s_sim}
	\vspace*{-0.25\baselineskip}
	\footnotesize
	\begin{tabular}{@{~}l@{~~}l@{~~}l@{~~~}l@{~}c@{~}c@{~}}
		\toprule
		\multirow{2}{*}{\textbf{Reference}} & \multirow{2}{*}{\textbf{Architecture}} & \multirow{2}{*}{\textbf{Input}} & \multirow{2}{*}{\textbf{Output}} & \textbf{Modeling} & \textbf{Multi-}  \\
		                                    &                                        &                                 &                                  & \textbf{goal?}    & \textbf{domain?} \\
		\midrule
		\citep{ElAsri:2016:Interspeech}     & RNN-LSTM                               & feature vect.                   & act                              & Y                 & N                \\
		\citep{Crook:2017:Interspeech}      & RNN-GRU/LSTM                           & utt                             & utt                              & N                 & N                \\
		\citep{Kreyssig:2018:SIGDIAL}       & RNN-LSTM                               & feature vect.                   & utt                              & Y                 & N                \\
		\citep{Gur:2018:SLT}                & RNN-GRU                                & act                             & act                              & Y                 & N                \\
		\citep{Lin:2021:SIGDIAL}            & Transformer                            & feature vect.                   & act                              & Y                 & Y                \\
		\citep{Sun:2021:SIGIR}              & Transformer                            & context                         & act+sat                          & Y                 & Y                \\
		\citep{Lin:2022:SIGDIAL}            & Transformer                            & context                         & act+utt                          & Y                 & Y                \\
		\citep{Kim:2022:SIGIR}              & Transformer                            & context                         & act+utt+sat                      & Y                 & Y                \\
		\citep{Sun:2023:TOIS}               & Transformer                            & context                         & act+sat                          & Y                 & Y                \\
		\bottomrule
	\end{tabular}
	\vspace*{-0.75\baselineskip}
\end{table}

\citet{ElAsri:2016:Interspeech} propose an LSTM-based approach that takes as input a sequence of dialogue contexts and outputs a sequence of user actions.
At the beginning of each dialogue, a user goal with constraints and requests, $G = (C,R)$, is randomly generated, similar to the agenda-based approach, which remain fixed during the dialogue.
In each turn $t$, the simulated user takes as input the conversation contexts observed so far, $\langle c_1, \dots, c_t \rangle$.
Specifically, $c_t$ consists of four components (all represented as binary vectors): the most recent machine action, the inconsistency between machine information and user goal (i.e., slots that have been misunderstood by the system so that these may be corrected), the constraint status (to inform the system about preferences), and the request status (to keep track of requests that have not yet been fulfilled).
This is passed to an RNN, which encodes this dialogue history into a single vector $v_t$. Then, $v_t$ is used to initialize the decoder RNN at each time step, which outputs a sequence of $l$ user intents $\langle I_1,\dots,I_l \rangle$. These intents are turned to dialogue acts $a_{t+1}^u$ by looking at the current user goal and uniformly drawing among the constraints/requests left.
Note that the model here can be considered partially interpretable, as the factors influencing the user's next action are made explicit. The actual decision-making, however, is governed by a machine-learned model.

Unlike the above model that works on the semantic level of dialogue acts, sequence-to-sequence models have also been used to generate the text of user utterances directly.
\citet{Crook:2017:Interspeech} extend a basic encoder-decoder neural network with an additional dense layer between the encoder and decoder for better representation of context, following a similar architecture that has been proposed for non-goal-driven dialogue systems in \citep{Serban:2016:AAAI}.

\citet{Kreyssig:2018:SIGDIAL} present the Neural User Simulator (NUS), which follows a very similar architecture and context representation as \citet{ElAsri:2016:Interspeech}, but the output is a sequence of words.
NUS allows for goal changes (when the system indicates that the goal is unachievable) and is able to generate a diverse set of utterances in the same dialogue context.

\citet{Gur:2018:SLT} propose an end-to-end hierarchical user simulator (HUS), which encodes the entire dialogue history based on the user goal and system dialogue act, without requiring feature extraction or a supervised signal for state tracking.
A pivotal aspect of their approach is to represent system dialogue acts on a more coarse level, by replacing specific slot values with one of the following:
\texttt{Requested}, if the value is requested by the system,
\texttt{ValueInGoal}, if the value appears in the user goal,
\texttt{ValueContradictsGoal}, if the value contradicts the user goal,
\texttt{DontCare}, if the value in the user goal is flexible,
or \texttt{Other} otherwise.
The actual values can then be substituted either from the user goal, system dialogue act, or from the knowledge base.
A modified version of their model, the variational hierarchical user simulator (VHUS), introduces a variational sampling step before decoding the user turn to generate more diverse responses.
\citet{Gur:2018:SLT} also develop a goal regularization approach to enforce a more direct correspondence and avoid generating too long dialogues when the user diverges from the initial goal.

Transformer-based models~\citep{Vaswani:2017:NIPS}, renowned for their strong natural language understanding capabilities, have spurred a new generation of user simulation approaches.
\citet{Lin:2021:SIGDIAL} present a transformer-based user simulator (TUS), employing a feature-based input representation similar to \citet{ElAsri:2016:Interspeech}, but design it to be domain-independent. They use variable input lengths based on the slots mentioned by the system. Using a transformer model as the user policy network, input sequences of arbitrary lengths are handled and the relationships between the different slots can be captured by self-attention.
The output includes the domain that should be selected for the given turn and a list of one-hot vectors to determine the values of the slots, using a coarse-grained representation similar to \citet{Gur:2018:SLT}. (Note that the user intent is not explicitly predicted, but since only two possible user intents are considered, \texttt{INFORM} and \texttt{REQUEST}, those can be inferred based on how slot-values are represented.)

\citet{Sun:2021:SIGIR} propose to simulate not only the next user action but also predict user satisfaction. The dialogue context, simply a concatenation of all utterances, is encoded into a latent representation using BERT~\citep{Devlin:2019:NAACL}. Two independent models are trained for two tasks, such that ``action prediction takes the predicted output of satisfaction prediction model as the input''~\citep{Sun:2021:SIGIR}.

\citet{Lin:2022:SIGDIAL} present a generative transformer-based simulator (GenTUS) that generates semantic as well as natural language representations of the user action, that is, both a dialogue act and an utterance.
Specifically, GenTUS takes the last system dialogue act, the previous three user dialogue acts, the user goal, and the turn number $t$ as input (encoded as a JSON-formatted string). The decoder generates the next user dialogue act and the corresponding language response as output, leveraging the ontology to constrain the generation process to prevent generating illegal semantic actions.

\citet{Kim:2022:SIGIR} fine-tune a T5 model~\citep{Raffel:2020:JMLR} in a multi-task learning setting to create a user simulator that predicts user satisfaction and actions, and also generates utterances.
Through an ablation study, the authors show that the three tasks help each other to better simulate users.

\citet{Sun:2023:TOIS} propose a metaphorical user simulator, MetaSim, which aims to improve the realism of simulation by mimicking the user's analogical thinking in interactions with the system.
It relies on a database of training dialogue records, from which it retrieves the top-$k$ most similar utterances based on the current dialogue state, to serve as the \emph{metaphor}. The metaphor is then taken as additional input to the policy, to determine the next user action, along with a prediction of user satisfaction. Both the metaphor module and the dialogue policy are based on T5.

\emph{\textbf{Summary.}} Looking at Table~\ref{tab:s2s_sim}, we can observe a trend of moving from manual feature engineering to progressively adopting end-to-end approaches. Further, approaches have become not only multi-domain, but also multi-task, simultaneously training models to perform multiple tasks relevant to user simulation, such as dialogue act prediction and utterance generation or user satisfaction estimation.
It is interesting to observe that while early work was done exclusively in the field of dialogue systems~\citep{Schatzmann:2006:KER}, more recent work is being increasingly carried out within the field of natural language processing and, to a smaller extent, within information retrieval.
As models become more end-to-end, their interpretability diminishes, while offering limited control over the behaviour of the simulated user.
Effectively, dialog planning can only be controlled indirectly, through the input data that is supplied for training, which is problematic for several reasons.
First, it requires a sufficient amount of training data in the target domains, which makes these approaches less effective in low-resource scenarios.
Second, if some actions or conversation paths are not present in the training corpus then those will not be captured in the simulator.
Third, data-driven simulators may suffer from bias present in the corpus.
Fourth, if users with various personalities are to be modeled (e.g., based on the level of cooperativeness), these characteristics need to be provided as supervised training signals along with each training example~\citep{Gur:2018:SLT}. Different experiments might require additional data annotation/collection effort, which would contradict the primary goal of using simulation as a means of automatic evaluation without human involvement.
Generally, basing user dialogue policy on sequence-to-sequence models suffers from the same fundamental issues as end-to-end simulators; cf. our earlier discussion in Section~\ref{sec:sim_conv:architectures_e2e}.

\subsubsection{In-Context Learning}

Recently, the emergence of large pre-trained language models capable of following instructions have been utilized for creating user simulators through a technique known as in-context learning. This approach involves providing the LLM with a prompt that describes the task, optionally accompanied by a few examples for demonstrating the desired behaviour, to guide its responses, thus eliminating the need for fine-tuning on historical dialogues.
This is a rapidly evolving area of research, with new approaches and techniques continuing to emerge.
Rather than attempting an exhaustive review, we highlight some key research themes and discuss some of the open challenges.

In-context learning has been shown to be effective in simulating user utterances~\citep{Davidson:2023:arXiv,Terragni:2023:arXiv} and feedback~\citep{Hu:2023:CIKM}. It has also been successfully applied to simulate both the user and system sides of a conversation, generating annotated dialogues~\citep{Li:2022:EMNLP}.
Within this line of research, several avenues have been explored, including investigating strategies for sampling dialogue examples as ``shots'' for the model, with techniques ranging from random sampling~\citep{Li:2022:EMNLP} to similarity-based selection according to the user goal~\citep{Terragni:2023:arXiv}. Additionally, researchers have experimented with various prompt design strategies, such as varying the number of examples and comparing structured (bullet point) formats with descriptive ones~\citep{Terragni:2023:arXiv}, as well as utilizing different LLMs for the task~\citep{Terragni:2023:arXiv,Davidson:2023:arXiv}.

These approaches are able to generate linguistically diverse and human-like utterances, but tend to be sensitive to the prompt used~\citep{Davidson:2023:arXiv}.
Compared to sequence-to-sequence approaches, there is even less control over the data generation process as there are no guarantees that the LLM would follow the instructions in the prompt.
For example, \citet{Terragni:2023:arXiv} find that ``in contrast to an agenda-based simulator, a prompt-based user simulator may easily give up after a dialog breakdown, deviating from the requirement to fulfill all the objectives described in the prompt.''
It is also common for LLMs to hallucinate, e.g., ``the user arbitrarily expresses disinterest in Chinese food, despite it not being mentioned in their original goal''~\citep{Terragni:2023:arXiv}.
Additionally, these types of simulators are prone to getting stuck in loops with the TOD system, wherein both keep repeating their last utterance~\citep{Davidson:2023:arXiv,Terragni:2023:arXiv}.
\citet{Terragni:2023:arXiv} further observe cases where the user simulator disrupts the dialog by assuming the role of an assistant and conclude that ``the issue likely arises from the training data for LLMs, which typically emphasizes emulating assistant behaviour rather than accurately representing user behaviour.''
Mitigating these deficiencies of LLMs is an active area of ongoing work. For example, \citet{Li:2022:EMNLP} propose a method to generate a belief state at each turn, to check whether it is consistent with the user utterance.
Further improvements proposed to prompt design include
``prompting the user simulator to repeat the sub-goal on which it is grounding its output'' to provide additional grounding and ``recasting the dialogue as a chain-of-thought reasoning problem''~\citep{Davidson:2023:arXiv}.

\section{From Task-Oriented Dialogue to Conversational Information Access}
\label{sec:sim_conv:cia}

Next, we turn our attention to conversational information access, which represents a particular kind of task-oriented dialogue. It is a broad task that encompasses the goals of conversational search, recommendation, and question answering.
CIA systems are expected to (1) support multiple user goals, including search, recommendation and exploratory information gathering, (2) learn user preferences and personalize responses accordingly, and (3) be capable of taking initiative~\citep{Balog:2021:DESIRES}.
\citet{Radlinski:2017:CHIIR} identify desirable properties of conversational search system, which include the following: (1) user revealment (enabling the user to express/discover their information need), (2) system revealment (enabling the system to reveal its capabilities and corpus), (3) mixed initiative (allowing both the user and system to take initiative), (4) memory (the ability to reference and incorporate past statements), and (5) set retrieval (considering utility over sets of complementary items).
It is important to note that while the goals that CIA systems should support are clear, much of the envisioned functionality, as of today, mostly exists only on a conceptual level and approaches that support multiple conversational goals (search, recommendation, and QA) are yet to be developed.
It is only very recently that datasets that support multiple conversational goals have started to become available~\citep{Bernard:2023:SIGIR}.
In existing work, simulation approaches are developed in a goal-specific manner, therefore we will discuss them along this distinction: conversational search (Section~\ref{sec:sim_conv:conv_search}) and conversational recommendation (Section~\ref{sec:sim_conv:conv_rec}).\footnote{Conversational QA is not covered as there has not been simulation work on that until very recently~\citep{Abbasiantaeb:2024:WSDM}.}

Nevertheless, it is helpful to conceptualize the process of information access as a type of task-oriented dialogue, and abstract out both low-level actions and decision points (Section~\ref{sec:sim_conv:actions}) as well as higher-level dialogue patterns (Section~\ref{sec:sim_conv:dialogue}) that manifest during conversational interactions.

\subsection{Action Space}
\label{sec:sim_conv:actions}
\vspace*{-0.25\baselineskip}

\renewcommand{\arraystretch}{1.15}
\begin{table}[!t]
  \centering
  \caption{Taxonomy of user and system actions in~\citep{Azzopardi:2018:CAIR}. Fn. (light grey) refers to conversational functionality according to~\citep{Radlinski:2017:CHIIR} and Pr. (dark grey) refers to search process in~\citep{Trippas:2018:CHIIR}.}
  \small
  \label{tab:dialogue_acts}
  \begin{tabular}{l@{~~}llllll@{~~}l}
    \toprule
    \textbf{Fn.}                                                                       &  & \textbf{Pr.}                                                                           & \textbf{User actions} & \textbf{System actions} & \multicolumn{3}{c}{\textbf{Fn.}}                                                                                                                                      \\
    \midrule
                                                                                       &  & \cellcolor[gray]{0.75}                                                                 & \textbf{Reveal}       & \textbf{Inquire}        & \cellcolor[gray]{0.9}                                                                 &  & \cellcolor[gray]{0.9}                                                      \\
                                                                                       &  & \cellcolor[gray]{0.75}                                                                 & -~Disclose            & -~Extract               & \cellcolor[gray]{0.9}                                                                 &  & \cellcolor[gray]{0.9}                                                      \\
                                                                                       &  & \cellcolor[gray]{0.75}                                                                 & -~Non-disclose        & -~Elicit                & \cellcolor[gray]{0.9}                                                                 &  & \cellcolor[gray]{0.9}                                                      \\
                                                                                       &  & \cellcolor[gray]{0.75}                                                                 & -~Revise              & -~Clarify               & \cellcolor[gray]{0.9}                                                                 &  & \cellcolor[gray]{0.9}                                                      \\
                                                                                       &  & \cellcolor[gray]{0.75}                                                                 & -~Refine              &                         & \cellcolor[gray]{0.9}                                                                 &  & \cellcolor[gray]{0.9}                                                      \\
                                                                                       &  & \cellcolor[gray]{0.75} \multirow{-6}{*}{\rotatebox[origin=c]{90}{Query formulation}}   & -~Expand              &                         & \cellcolor[gray]{0.9} \multirow{-6}{*}{\rotatebox[origin=c]{-90}{User revealment}}    &  & \cellcolor[gray]{0.9}                                                      \\

                                                                                       &  &                                                                                        &                       &                         &                                                                                       &  & \cellcolor[gray]{0.9}                                                      \\ [-0.9em]

    \cellcolor[gray]{0.9}                                                              &  & \cellcolor[gray]{0.75}                                                                 & \textbf{Inquire}      & \textbf{Reveal}         & \cellcolor[gray]{0.9}                                                                 &  & \cellcolor[gray]{0.9}                                                      \\
    \cellcolor[gray]{0.9}                                                              &  & \cellcolor[gray]{0.75}                                                                 & -~List                & -~List                  & \cellcolor[gray]{0.9}                                                                 &  & \cellcolor[gray]{0.9}                                                      \\
    \cellcolor[gray]{0.9}                                                              &  & \cellcolor[gray]{0.75}                                                                 & -~Summarize           & -~Summarize             & \cellcolor[gray]{0.9}                                                                 &  & \cellcolor[gray]{0.9}                                                      \\
    \cellcolor[gray]{0.9}                                                              &  & \cellcolor[gray]{0.75}                                                                 & -~Compare             & -~Compare               & \cellcolor[gray]{0.9}                                                                 &  & \cellcolor[gray]{0.9}                                                      \\
    \cellcolor[gray]{0.9}                                                              &  & \cellcolor[gray]{0.75}                                                                 & -~Subset              & -~Subset                & \cellcolor[gray]{0.9}                                                                 &  & \cellcolor[gray]{0.9}                                                      \\
    \cellcolor[gray]{0.9}                                                              &  & \cellcolor[gray]{0.75}                                                                 & -~Similar             & -~Similar               & \cellcolor[gray]{0.9}                                                                 &  & \cellcolor[gray]{0.9}                                                      \\

    \cellcolor[gray]{0.9}                                                              &  & \cellcolor[gray]{0.75}                                                                 & \textbf{Navigate}     & \textbf{Traverse}       & \cellcolor[gray]{0.9}                                                                 &  & \cellcolor[gray]{0.9}                                                      \\
    \cellcolor[gray]{0.9}                                                              &  & \cellcolor[gray]{0.75}                                                                 & -~Repeat              & -~Repeat                & \cellcolor[gray]{0.9}                                                                 &  & \cellcolor[gray]{0.9}                                                      \\
    \cellcolor[gray]{0.9}                                                              &  & \cellcolor[gray]{0.75}                                                                 & -~Back                & -~Back                  & \cellcolor[gray]{0.9}                                                                 &  & \cellcolor[gray]{0.9}                                                      \\
    \cellcolor[gray]{0.9}                                                              &  & \cellcolor[gray]{0.75}                                                                 & -~More                & -~More                  & \cellcolor[gray]{0.9}                                                                 &  & \cellcolor[gray]{0.9}                                                      \\
    \cellcolor[gray]{0.9}                                                              &  & \cellcolor[gray]{0.75}                                                                 & ...                   & ...                     & \cellcolor[gray]{0.9}                                                                 &  & \cellcolor[gray]{0.9}                                                      \\
    \cellcolor[gray]{0.9} \multirow{-12}{*}{\rotatebox[origin=c]{90}{Set retrieval}}   &  & \cellcolor[gray]{0.75} \multirow{-12}{*}{\rotatebox[origin=c]{90}{Result exploration}} & -~Note                & -~Record                & \cellcolor[gray]{0.9}                                                                 &  & \cellcolor[gray]{0.9}                                                      \\

                                                                                       &  &                                                                                        &                       &                         & \cellcolor[gray]{0.9}                                                                 &  & \cellcolor[gray]{0.9}                                                      \\ [-0.9em]

    \cellcolor[gray]{0.9}                                                              &  &                                                                                        & \textbf{Interrupt}    & \textbf{Suggest}        & \cellcolor[gray]{0.9}                                                                 &  & \cellcolor[gray]{0.9}                                                      \\
    \cellcolor[gray]{0.9}                                                              &  &                                                                                        & -~Interrupt           & -~Recommend             & \cellcolor[gray]{0.9}                                                                 &  & \cellcolor[gray]{0.9}                                                      \\
    \cellcolor[gray]{0.9}                                                              &  &                                                                                        &                       & -~Hypothesize           & \cellcolor[gray]{0.9}                                                                 &  & \cellcolor[gray]{0.9}                                                      \\
    \cellcolor[gray]{0.9}                                                              &  &                                                                                        & \textbf{Interrogate}  & \textbf{Explain}        & \cellcolor[gray]{0.9}                                                                 &  & \cellcolor[gray]{0.9}                                                      \\
    \cellcolor[gray]{0.9}                                                              &  &                                                                                        & -~Understand          & -~Report                & \cellcolor[gray]{0.9}                                                                 &  & \cellcolor[gray]{0.9}                                                      \\
    \cellcolor[gray]{0.9} \multirow{-6}{*}{\rotatebox[origin=c]{90}{Mixed initiative}} &  &                                                                                        & -~Explain             & -~Reason                & \cellcolor[gray]{0.9} \multirow{-18}{*}{\rotatebox[origin=c]{-90}{System revealment}} &  & \cellcolor[gray]{0.9} \multirow{-24}{*}{\rotatebox[origin=c]{-90}{Memory}} \\
    \bottomrule
  \end{tabular}
\end{table}
\renewcommand{\arraystretch}{1.0}

A taxonomy of actions designed specifically for conversational search is presented in
\citep{Azzopardi:2018:CAIR}, with reference to the high-level desiderata in \citep{Radlinski:2017:CHIIR} as well as the different phases of the search process in \citep{Trippas:2018:CHIIR}; see Table~\ref{tab:dialogue_acts}.
Overall, it is assumed that the conversational process revolves around the system trying to help the user resolve their information need. There are two main types of user actions, which either (1) change the state of the information need or (2) are related to the space of information objects.
The first type of actions (Reveal) corresponds to the user disclosing details regarding their information need, which may also evolve during the search. The second type of actions involves inquiring about the available options (Inquire), navigating them (Navigate), or asking the system how it understood the information need or why it presented particular suggestions (Interrogate). Additionally, there may be points in the conversation when the user wants to interrupt or stop the system (Interrupt).
\citet{Azzopardi:2018:CAIR} note that their list attempts to represent the main actions previously observed and discussed in the literature, but it is non-exhaustive and is meant to be taken as a starting point.
For example, \citet{Bernard:2023:SIGIR} expand a selected set of communicative functions from ISO 24617-2 with intents from \citep{Azzopardi:2018:CAIR} in order to characterize multi-goal conversations in an e-commerce setting.

\subsubsection{Dialogue Structure}
\label{sec:sim_conv:dialogue}
\vspace*{-0.5\baselineskip}

It is useful to characterize dialogues in terms of overall organization, sequencing, and components, which we collectively refer to as \emph{dialogue structure}. Dialogue structure serves as a blueprint for how participants should interact, and helps ensure that the conversation flows and that there is a natural progression towards task completion.

\begin{figure}[t]
  \centering
  \includegraphics[width=0.9\textwidth]{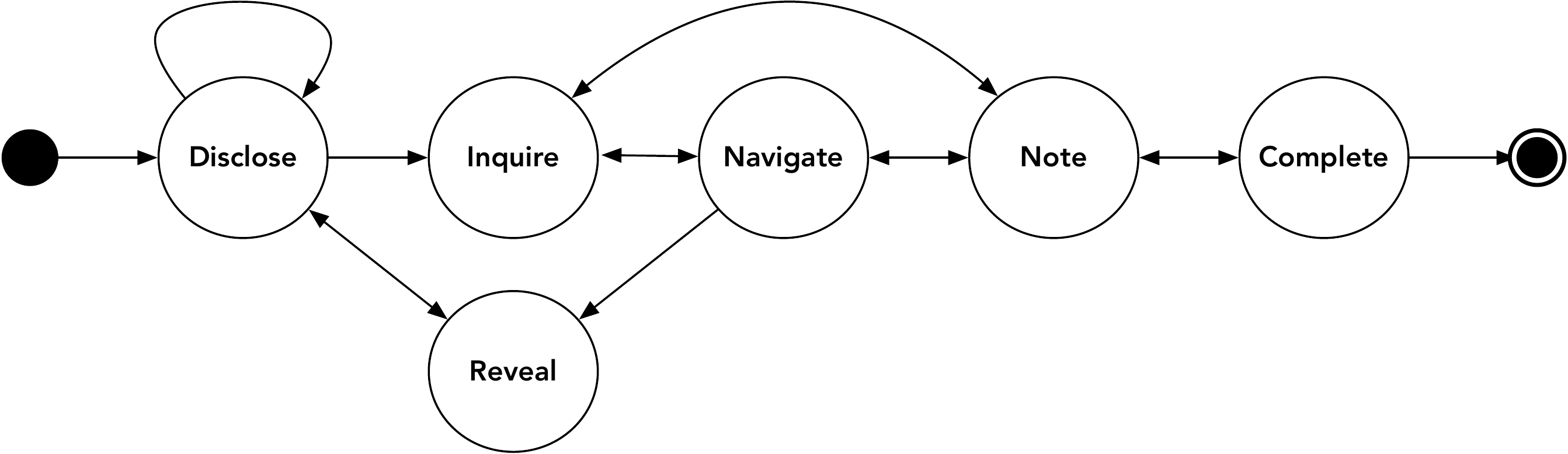}
  \caption{CIR6 dialogue workflow model, designed specifically for the task of conversational item recommendation (illustration adapted from \citet{Zhang:2020:KDD}).}
  \label{fig:cir6_model}
\end{figure}

Dialogue structure can be designed by envisioning the optimal sequencing of turns, identifying key dialogue acts and their dependencies, anticipating potential user intents and system responses, and incorporating strategies for effective information flow.
\citet{Trippas:2018:CHIIR} provide a different high-level formalization, distinguishing three stages of the conversational search process: (1) query formulation, (2) search results exploration, and (3) query re-formulation.
\citet{Zhang:2018:CIKM} suggest to classify interactions in an e-commerce conversational search setting into three stages: initiation, conversation, and display.
\citet{Aliannejadi:2021:CIKM} propose a user model for mixed-initiative conversational search that consists of three major phases: querying, feedback, and browsing.
\citet{Zhang:2020:KDD} propose a dialogue workflow model (CIR6) specific to the task of conversational recommendation by defining six key user intents and their dependencies; see Fig.~\ref{fig:cir6_model}.

\begin{figure}[t]
  \centering
  \includegraphics[width=0.4\textwidth]{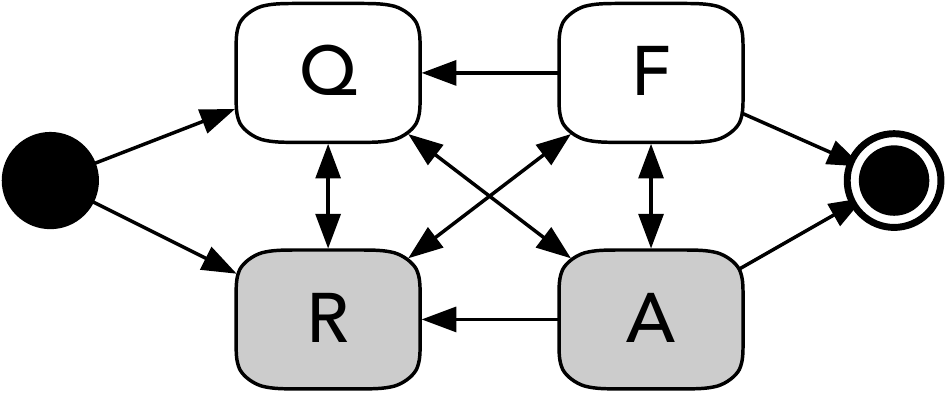}
  \caption{The QRFA model for conversational search; user actions are displayed with a white, system actions with a grey background (illustration adapted from \citet{Vakulenko:2019:ECIR}).}
  \label{fig:qrfa_model}
\end{figure}

Another way of identifying structural patterns in dialogues is through the annotation and analysis of transcripts of human-human conversations, which represent an ideal setting in terms of how people naturally interact.
\citet{Qu:2018:SIGIR} find that a frequent user intent transition pattern in information-seeking conversations is: START $\Rightarrow$ original question ($\Rightarrow$ potential answer $\Rightarrow$ further details)x3 $\Rightarrow$  potential answer $\Rightarrow$ positive feedback $\Rightarrow$ END.
\citet{Lyu:2021:WWW} analyze context-driven recommendation in the restaurant domain and break conversations down into three distinct stages, based on the recommendations made (0, 1, or 2+). They observe that conversations shift from preference elicitation and refinement in the first stage to inquiry and critiquing in subsequent stages, with additional comparisons in the last stage.
A similar pattern is identified in \citep{Bernard:2023:SIGIR}, analyzing multi-goal dialogues in an e-commerce setting: the first part of the conversations tends to focus on uncovering the user's needs and preferences to recommend some products (i.e., recommendation), while the second part is mostly concerned with product-related information seeking (i.e., search and QA), with users sometimes asking for other options (i.e., a second peak in recommendation).
\citet{Zhang:2022b:SIGIR} study how users reformulate their requests in case a conversational recommender system fails to understand them. They observe a typical pattern where a persistent user would first try to rephrase, then simplify before giving up. They also observe that rephrasing tends to happen earlier in dialogues than simplification.

On a higher level of abstraction, \citet{Vakulenko:2019:ECIR} present the QRFA (Query, Request, Feedback, and Accept) model, which is a generic model of conversational information seeking processes. The model has four basic classes: two for user (Query and Feedback) and two for system (Request and Answer), representing the role the utterance plays in the conversation (proactive or reactive); see Fig.~\ref{fig:qrfa_model}.
\citet{Trippas:2020:IPM} explore how people interact in an entirely voice-based conversational search setting. They develop a two-tiered model with two main themes: task-level, which is centered around the search task, and discourse-level, which surrounds it and is concerned with the more general communication mechanism.

Similar to the information-seeking workflows discussed earlier (see Section~\ref{sec:sim_search:workflows}), dialogue structures can provide a basis for defining the states within the MDP framework for user simulation. Modeling the transitions between states would be necessary for simulating a user's actions in the right context of the conversation
To enhance the realism of simulations, we can draw upon extensive past work from linguistics and psychology on how conversations are structured and their common patterns, ranging from immediate ``adjacency pairs''~\citep{Sacks:1974:L} to higher-level discourse patterns~\citep{Thomas:2021:TOIS}.

\section{User Simulation for Conversational Search}
\label{sec:sim_conv:conv_search}

Conversational search is a type of task-oriented dialogue, where the goal is to satisfy the user's information need.
We have discussed in Section~\ref{sec:sim_conv:actions} the space of user actions for information-seeking conversations. This, combined with an appropriate state representation, could allow for modeling the user's decision-making process as an MDP. 
However, a key challenge lies in creating a computational representation of the user's goal and modeling progress towards the completion of that goal as part of the user's state. 
When system responses are not limited to predefined information objects (e.g., passages or documents) but can be arbitrary generated text, relying on preexisting object-level relevance judgments is no longer feasible.
Consequently, current user simulation techniques often make strong assumptions about the dialogue structure and require responses to be documents/paragraphs with existing relevance judgments.
Given that mixed-initiative interaction is a defining characteristic of conversational information access, much of the research has focused on simulating responses to clarification questions.
Notably, current research is limited to simulating specific user actions within a conversational search dialogue, with no existing methods capable of fully automatic, end-to-end dialogue simulation.

In the following, we discuss existing approaches categorized according to being model-based or data-driven. 

\subsection{Model-Based Approaches}

In a conversational search setting, the user receives a single answer, as opposed to a list of results presented in a SERP. Therefore, ``it is likely that their next query will be dependent on what that answer is, whether it is to directly follow up with another query, to clarify what they wanted, to move to a different aspect of their need, or to stop the search altogether''~\citep{Lipani:2021:TOIS}.
This section focuses on model-based approaches for simulating such user queries.

\begin{figure}[t]
	\centering
	\includegraphics[width=0.5\textwidth]{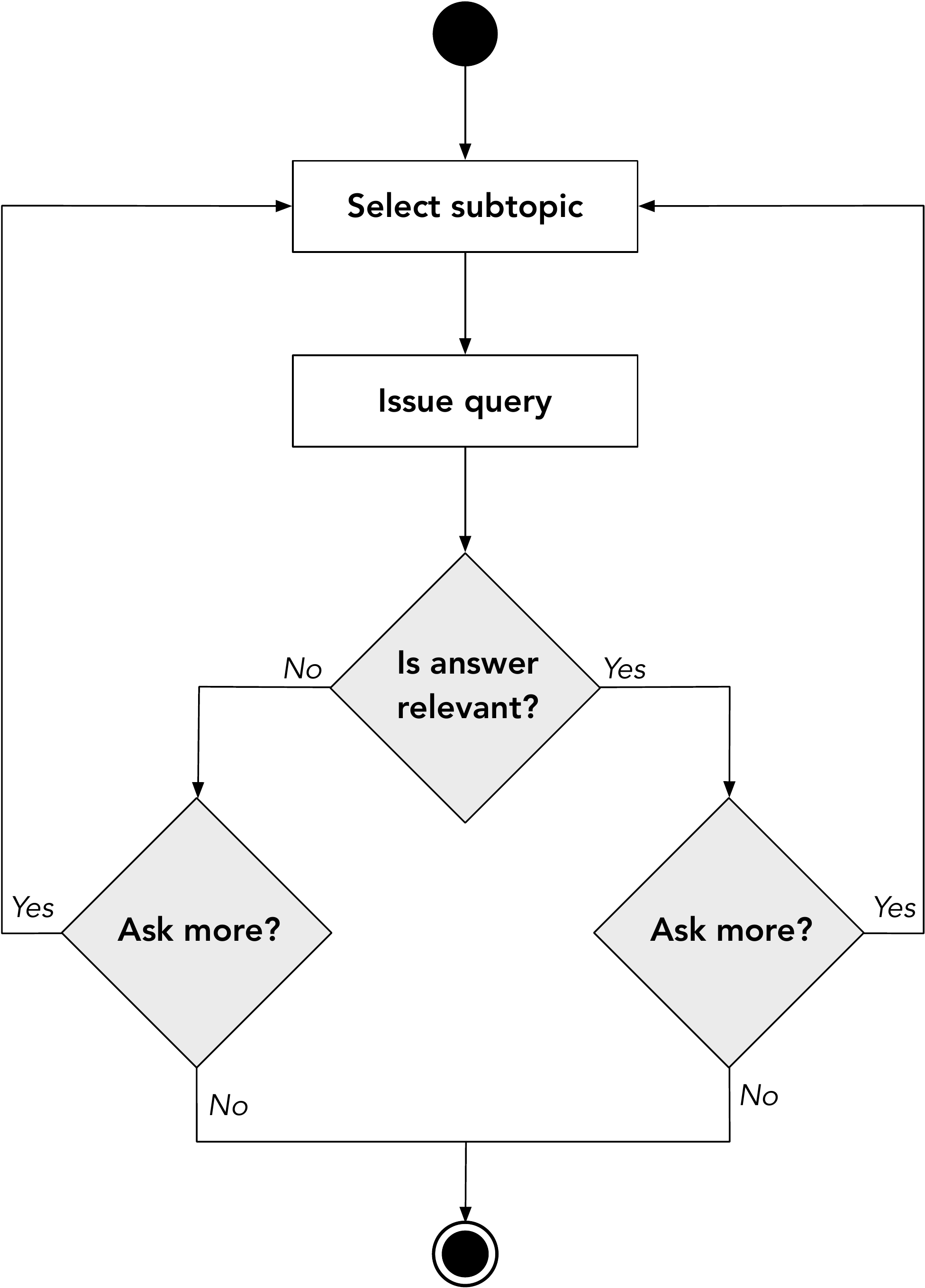}
	\caption{Flowchart of the user model in \citep{Lipani:2021:TOIS} for simulating user interactions with a conversational search system.}
	\label{fig:lipani_model}
\end{figure}

\citet{Lipani:2021:TOIS} propose to represent both user queries and system responses based on the idea of \emph{subtopics}, similar to those used in diversity and novelty evaluation~\citep{Clarke:2011:TREC}.
Specifically, it is assumed that the user's goal is to learn about a set of subtopics by interacting with the system, and at each dialogue turn the user asks about a particular subtopic. Based on the relevance of the system's response, the user will ask further questions (about the same subtopic or a different one) or stop querying; see Fig.~\ref{fig:lipani_model}.

Formally, a conversation is modeled as a sequence of query-answer pairs $\langle (q_1,a_1), \dots, (q_m,a_m) \rangle$.
The user's goal is to learn about a set of subtopics $\mathcal{S}=\{s_1,\dots,s_n\}$. The dialogue is modeled as an MDP, where each state corresponds to a subtopic. Additionally, $s_0=start$ and $s_{n+1}=end$ are two special states, which correspond to the start and end of the conversation.

\citet{Lipani:2021:TOIS} model the progression of the dialogue through subtopic-to-subtopic transitions, presenting two alternative approaches.
The \emph{relevance independent} approach assumes that the transitions are independent of the system's answers:
\begin{equation}
	\prob(S_{t+1}=s_j|S_t=s_i) = c_{i,j} ~,	\label{eq:lipani_state_ri}
\end{equation}
where $S_t$ is the user state at turn $t$ and $c_{i,j}$ is the normalized count of transitions $s_i \rightarrow s_j$ observed in a dialogue dataset ($i \in [0..n]$, $j \in [1..n+1]$).

The more realistic \emph{relevance dependent} approach conditions the transitions on the relevance of the answer given by the system:
\begin{equation}
	\prob(S_{t+1}=s_j|S_t=s_i,R_t) = \label{eq:lipani_state_rd}
	\begin{cases}
		\prob(S_{t+1}=s_j|S_t=s_i,R_t=r) ~       & \text{if}\; r          \\
		\prob(S_{t+1}=s_j|S_t=s_i,R_t=\bar{r}) ~ & \text{if}\; \bar{r} ~,
	\end{cases}
\end{equation}
where the random variable $R=\{r,\bar{r}\}$ indicates whether or not the the last system answer $a_t$ was relevant to query $q_t$. The probabilities are estimated as above, i.e., based on transitions observed in historical dialogue data, but are additionally conditioned on relevance.

The user dialogue policy is based on the notion of persistence in querying the system, depending on the relevance of the answer to the previous query.
This persistence model is formally expressed as:
\begin{eqnarray}
	\prob(Q_1=q) & = & 1 ~, \nonumber \\
	\prob(Q_{t+1}) & = & \prob(L_t|Q_t,R_t) \; P(Q_t) ~, \label{eq:lipani_probQ}
\end{eqnarray}
where the random variables indicate the act of querying or not, $Q=\{q,\bar{q}\}$, and the act of leaving or not, $L=\{l,\bar{l}\}$.
The first equation expresses that the user always starts the conversation with querying.
For any subsequent turn $t$, the user would leave with probability $\prob(L_t=l|Q_t=q,R_t=r)$ if the system response was relevant and with probability $\prob(L_t=l|Q_t=q,R_t=\bar{r})$ if the result was not relevant; both probabilities are estimated from user logs.

Putting everything together, the process of simulation is as follows (assuming the beginning of a new turn $t$ and the simulator being in state $S_t=s$):
\begin{itemize}
	\item If $s=end$ then leave the session; otherwise, sample a query $q_t$ from the set $\mathcal{Q}_s$ of queries that are associated with subtopic $s$.
	\item The query is sent to conversational system, which responds with $a_t$.
	\item The relevance of the answer to the query, $R_t$, is determined.
	\item The next action (continue querying or end the session) is determined based on the persistence model given in Eq.~\eqref{eq:lipani_probQ}.
	\item The state $s$ is updated according to Eq.~\eqref{eq:lipani_state_ri} or~\eqref{eq:lipani_state_rd} and the turn counter $t$ is incremented.
\end{itemize}
Note that this method is not concerned with the generation of natural language questions, but rather focuses on asking questions, from a predefined set, in a natural order.
Overall, the following data components are required: (1) a sample of information needs (i.e., topics), (2) for each topic, a predefined set of subtopics, (3) user queries corresponding to those subtopics, (4) subtopic-level relevance judgments, and (5) a dialogue dataset with subtopic annotations for the estimation of state transition probabilities.

\subsection{Data-Driven Approaches}

One of the most studied forms of system initiative in conversational search is clarifying questions, which aim to elucidate the user's information need~\citep{Aliannejadi:2019:SIGIR,Zamani:2020:WWW,Bi:2021:ICTIR}.
Data-driven approaches to simulating user responses to such questions leverage LLMs, either through specialized supervised training procedures or via in-context learning (i.e., prompting); see Table~\ref{tab:clarif_q_sim} for an overview.

\begin{table}[t]
  \caption{Overview of LLM-based user simulators for simulating responses to clarifying questions.}
  \label{tab:clarif_q_sim}
  \footnotesize
  \begin{tabular}{llll}
    \toprule
    \textbf{Reference} & \textbf{Learning paradigm} & \textbf{LLM} & \textbf{Behavioural parameters} \\
    \midrule
    \citep{Salle:2021:ECIR} & Supervised fine-tuning & BERT & cooperativeness, patience \\
    \citep{Sekulic:2022:WSDM} & Supervised fine-tuning & GPT-2 & - \\
    \citep{Owoicho:2023:SIGIR} & In-context learning & GPT-3 & patience \\
    \citep{Wang:2024:WWW} & Supervised fine-tuning & T5 & cooperativeness \\
    \bottomrule		
  \end{tabular}
\end{table}

\subsubsection{CoSearcher \citep{Salle:2021:ECIR}}

\citet{Salle:2021:ECIR} simulate how a user would respond to clarifying questions that are in the form: ``Are you looking for \emph{[facet]}?''
Their simulator, called CoSearcher, has two main components: a user intent model and a persona model.
The \emph{user intent model} represents the user's information need and returns a boolean value depending on whether the clarifying question matches the user's intent. This is implemented in \citep{Salle:2021:ECIR} by fine-tuning a BERT model~\citep{Devlin:2019:NAACL} for binary classification.
The \emph{persona model} (referred to as user parameters in the original work) specifies personal user characteristics, namely cooperativeness and patience. \emph{Cooperativeness} ($\in [0,1]$) represents the user's willingness to help the system by giving an informative answer (e.g., ``No, but you are close'' or ``No, I'm looking for \emph{[intent]}'') as opposed to a minimal response (``Yes'' or ``No'').
\emph{Patience} specifies the maximum effort, measured in the number of turns, the user is willing to spend interacting with the system.

\subsubsection{USi \citep{Sekulic:2022:WSDM}}

CoSearcher targets a specific type of clarifying questions, which ask directly about a single facet, and does not consider conversational context. Further, it assumes that there is a predefined set of facets for each information need that the system can ask about.
\citet{Sekulic:2022:WSDM} address these limitations in their conversational user simulator, called USi, by fine-tuning a transformer-based LLM for the task of answering clarifying questions.
Specifically, they employ DoubleHead GPT-2~\citep{Radford:2019:misc} with language modeling and classification losses, and utilize existing datasets: Qulac~\citep{Aliannejadi:2019:SIGIR} and ClariQ~\citep{Aliannejadi:2020:arXiv}. 
The first part of the training input is given to the model as the sequence \texttt{in[SEP]q[SEP]cq[bos]a[eos]}, where $in$ is the textual description of the user's information need that is underlying the user's query $q$, $cq$ is the clarifying question asked by the system, and $a$ is the answer given by the user; \texttt{[bos]} and \texttt{[eos]} are special tokens indicating the beginning and end of a sequence, respectively, and \texttt{[SEP]} is a separation token.
The second part of the training input for the DoubleHead model is given another input sequence with a distractor answer and a binary label indicating which of the answers is preferable. The distractor answers are sampled from the training dataset heuristically (e.g., if the answer starts with ``Yes'' then the distractor answer starts with ``No,'' and vice versa).
At inference time, the above input sequence is given without the answer segment, which will be generated by the LLM.
A conversation history-aware variant of this approach is also proposed, which is capable of generating answers to further follow-up clarification questions, by extending the input sequence given to the LLM with previous user and system utterances.

\subsubsection{ConvSim \citep{Owoicho:2023:SIGIR}}

\citet{Owoicho:2023:SIGIR} extend upon the work of \citet{Sekulic:2022:WSDM} and develop ConvSim, a user simulator that is capable of not only answering clarifying questions but also providing explicit feedback to system responses, by few-shot prompting a GPT-3 model~\citep{Brown:2020:NeurIPS}.
All generated utterances need to be in line with the underlying information need $in$. 
Specifically, at a given conversational turn $t$, it is assumed that there exists a turn-level description of the information need $in_t$---this is created manually by expert annotators in \citep{Owoicho:2023:SIGIR}.
Then, the LLM is prompted with the respective task description (i.e., whether to answer a clarifying question or provide feedback), the turn-level information need $in_t$, sample transcripts with the desired behaviour, and the conversation history up to turn $t$.
ConvSim is capable of simulating multi-turn interactions; following \citet{Salle:2021:ECIR}, the approach considers patience, i.e., the maximum number of turns of feedback the simulated user is willing to provide.

\subsubsection{Type+QA \citep{Wang:2024:WWW}} 

\citet{Wang:2024:WWW} present a user simulator by fine-tuning a pretrained T5 model~\citep{Raffel:2020:JMLR} for generating answers to clarifying questions, using the same datasets as~\citet{Sekulic:2022:WSDM}. Based on the analysis of low-scoring examples in the development set, they identify wrong answer types as the most common error, i.e., the model generates an answer with the wrong meaning, for example, ``no'' instead of ``yes''.
They propose a two-step model to address this issue. First, the answer type is predicted using a 4-way categorization (yes, no, open, irrelevant) by fine-tuning a pretrained text classification model (RoBERTa~\citep{Liu:2019:arXiv}). This prediction is turned into a decoding constraint at inference time: ``if the predicted answer type is `yes' or `no,' then the generation starts with `yes' or `no' correspondingly''~\citep{Wang:2024:WWW}.
Based on the realization that simulating user responses to clarifying questions is similar to the problem of QA, further gains are obtained by performing the fine-tuning for answer generation on top of a T5 checkpoint that has been trained on QA datasets~\citep{Khashabi:2020:EMNLP}.
Simulators are trained with two different levels of cooperativeness, by partitioning the training data into two subsets, cooperative and uncooperative, based on response length.

\subsubsection{Analysis of Simulated Responses}

Multiple recent studies have performed an analysis of simulated user utterances to gain insights and understand their limitations.
\citet{Sekulic:2024:TIST} compare two representatives of LLM-based user simulators: USi~\citep{Sekulic:2022:WSDM}, which follows a specific supervised training procedure and ConvSim~\citep{Owoicho:2023:SIGIR}, which utilizes in-context learning (i.e., prompting). The comparison is done using both automatic natural language generation (NLG) metrics and human judges.
While ConvSim performs worse than USi on automatic NLG metrics, this outcome is attributed to the known unreliability of such measures~\citep{Novikova:2017:EMNLP}. Conversely, in manual evaluation, ConvSim outperforms USi in both naturalness and usefulness, with the performance difference becoming more pronounced over multiple turns of interaction.
Qualitative analysis of the responses reveals a greater frequency of short answers in case of USi, while the majority of the answers by ConvSim aim to clarify and refine the underlying information need.
Additionally, it is worth noting that USi is based on GPT-2, while ConvSim is based on GPT-3. Consequently, some of the observed differences may be attributable to differences in the capabilities of the underlying LLMs.

\citet{Wang:2024:WWW} compare open LLMs (Llama2 and Flan) with a commercial LLM (GPT-4) and find that open LLMs in a zero-shot setting do not sufficiently understand the user simulation task. GPT-4, on the other hand, can generate seemingly perfect answers even in situations where humans fail to understand the meaning of the clarifying question. This is likely not the desired human-like behaviour.

\section{User Simulation for Conversational Recommendation}
\label{sec:sim_conv:conv_rec}

The main task of conversational recommender systems is to elicit user preferences using natural language interactions, point users to potential items of interest, and process feedback by users on the made suggestions~\citep{Jannach:2021:CSUR}.
Conversational recommender systems can naturally be framed in the traditional sense of task-oriented dialogue systems: the goal is to find items that satisfy the set of constraints expressed by the user, which can be represented in terms of slot-value pairs: $C=\langle (slot_1,value_1), \dots, (slot_n,value_n) \rangle$.
These constraints are based on the user's long-term and short-term preferences and may change depending on the context and the items that are available.

Below, we discuss work on user simulation for conversational recommender systems, categorized into model-based vs. data-driven approaches.

\subsection{Model-Based Approaches}

Given the close resemblance to task-oriented dialogue systems, it is natural to discuss techniques along the main components of a modular simulator architecture depicted in Fig.~\ref{fig:convsim_arch_modular}.

\subsubsection{Natural Language Understanding}

The NLU task for user simulation is exactly the same as for task-oriented dialogue: given a textual utterance as input, turn it into a semantic representation, i.e., dialogue act. Therefore, existing solutions for intent detection and slot filling can be utilized here (cf. Section~\ref{sec:sim_conv:architectures_modular}).
Automatic approaches are inherently error-prone and unlikely to produce perfect results.
\citet{Zhang:2020:KDD} utilize the fact that many conversational systems use a limited set of language expressions (often as a result of a template-based NLG) and train the NLU specific to the conversational system that is being evaluated, by collecting and annotating a small sample of dialogues from that system.

\subsubsection{Dialogue Management and User Modeling}

One of the main challenges from a user simulation perspective is to maintain consistency during the dialogue, both in terms of dialogue flow (i.e., collaborating with the system to progress towards the user goal) and preferences (i.e., not contradicting preferences that were disclosed earlier).

\citet{Zhang:2018:CIKM} assume a simple ``system ask user respond'' workflow, where the system keeps asking the user questions until it is confident in the results.
It starts with an initial user request, prompting the system to search for candidate items. If the system lacks sufficient confidence in the results, it generates a  question for the user and refines the results based on their answer (along with any previously collected question-answer pairs). The questioning repeats until the system is confident enough to present the final results to the user.
\citet{Zhang:2018:CIKM} utilize textual item reviews and transform each review into a question-answer sequence to simulate the conversation that resulted in the purchase of the given item. 
The initial request is constructed from the product category. The questions ask about specific product aspects, according to the order in which they are mentioned, and the simulated user answers these questions with the values mentioned in the review.

\citet{Zhang:2020:KDD} employ an agenda-based dialogue policy that is guided by an \emph{interaction model}. The interaction model specifies (1) the set of user actions, (2) the allowed transitions between them, and (3) the expected system response (action) for each user action.
The latter component allows the simulator to determine whether the system responds to the user with an appropriate action (i.e., ``understood'' the user); if not, then a replacement action can be pushed onto the agenda.
The interaction model can be a general-purpose one, i.e., the QRFA model in Fig.~\ref{fig:qrfa_model}, or created specifically for the problem of conversational recommendation, i.e., the CIR6 model in Fig.~\ref{fig:cir6_model}.

The ensure the consistency of preferences, \citet{Zhang:2020:KDD} introduce a \emph{preference model}, which may be seen as part of the user modeling component in Fig.~\ref{fig:convsim_arch_modular}.
The preference model maintains a knowledge structure with $(slot, value, pref)$ triples, which can be queried for preferences on a given slot (to answer questions like ``What are your favourite movie genres?'') and on specific slot-value pairs (to answer questions like ``Do you like action movies?'').
Note that the slot can also be a specific type of entity with the value being the unique identifier of the entity (which allows for answering questions like ``Have you seen Inception?'' and ``How did you like it?'').
The preference model may be empty at the beginning of the conversation or seeded with initial preferences. Whenever it is queried for slots or slot-value pairs, either an existing value is returned, or a new value is generated and saved for future requests.
To create a preference model grounded in actual user preferences, \citet{Zhang:2020:KDD} randomly sample a user from a dataset of historical user-item interactions, then subsample item ratings of that user (the rest of the ratings are used as held-out data for automatic evaluation).
To ensure the consistency of preferences, the notion of a personal knowledge graph (PKG) is utilized, which is a ``resource of structured information about entities personally related to its user''~\citep{Balog:2019:ICTIR}.
The PKG has two types of nodes: items and attributes. Preferences on items are taken from historical data; for attributes, preferences are inferred by aggregating the preferences of entities with that attribute.

\subsubsection{Natural Language Generation}

The simplest approach to NLG is to base it on predefined language patterns, i.e., templates~\citep{Zhang:2018:CIKM}. To increase language variety, for each intent a number of different articulations may be created~\citep{Zhang:2020:KDD}.
To improve the realism of simulators, \citet{Zhang:2022b:SIGIR} identify specific utterance reformulation patterns, based on how humans interact with conversational systems when the system fails to understand them.
Given an original utterance (e.g., ``I'm looking for something light-hearted'') a persistent user would often first try to rephrase (e.g., ``I want a funny comedy that is light hearted''), then simplify (e.g., ``funny comedy''), before giving up.
\citet{Zhang:2022b:SIGIR} present an approach based on transformer models that are guided by guided reformulation type, with an additional reading difficulty filter, to generate such utterances.
The integration of this solution into a simulator would also require changes to the dialogue policy, to consider how likely the user would try again, based on their persistence (which can be a global setting or a parameter in the user model that varies across individuals). Additionally, the dialogue manager needs to keep track of the number of attempts on the same dialogue act and pass this value down to the NLG component.

\subsection{Data-Driven Approaches}

Data-driven approaches have been employed to improve the realism of simulation in terms dialogue flow as well as the fluency and naturalness of utterances.

\citet{Wang:2023:KDD} propose a multi-stage generation process that first generates a \emph{conversation flow} and then realizes the dialogue based on the generated flow.
A conversation flow is characterized as a sequence of  entities mentioned in the conversation. In order to produce fluent, coherent conversations, frequent \emph{flow schemas}, which characterize conversations as a sequence of entity types, are identified from real datasets using frequent pattern mining algorithms. 
Conversation flows are generated in an autoregressive manner based on learned user and schema prompts.
The realization of dialogues follows a simple template-based approach, with templates collected from observed dialogues via delexicalization.

Recent work has investigated the capacity of LLMs as user simulators also in the context of conversational recommender systems.
\citet{Wang:2023:EMNLP} propose an interactive evaluation approach, iEvaLM, where each simulated user is given a persona based on preferences established from ground truth items from existing datasets. Simulated users can perform one of the following actions in each turn: (1) express preferences when prompted by the system, (2) provide feedback when presented with a list of recommendations, and (3) finish the conversation when a target item is found or a certain number of dialogue turns is reached.
At each conversational turn, ChatGPT~\citep{Ouyang:2022:NeurIPS} is zero-shot prompted to generate the user utterance, with the conversation up to that point provided as part of the input.
The simulator is evaluated in terms of naturalness and usefulness, and is reported to outperform real humans in both dimensions. 

\citet{Friedman:2023:arXiv} explore ways to guide the behaviour of an LLM to increase the realism of the simulation.  
In \emph{session-level control}, a single variable, such as the user profile, is defined at the beginning of the session.
In \emph{turn-level control}, a distinct variable, such as the intent, is defined at each conversational turn to condition the simulator for that turn.
The values of the variables may be sampled according to empirical distributions obtained from real conversations.
In practice, ``one way to execute the control is to translate the variable into text that can be included as part of the simulator's input along  with the rest of the conversation''~\citep{Friedman:2023:arXiv}.

\citet{Yoon:2024:arXiv} study to what extent LLM-based simulators can represent and reflect the characteristics of users in conversational recommendation. 
The authors design specific tasks for each key property that a user simulator should exhibit: 
(1) choosing items to talk about,
(2) expressing binary preference,
(3) expressing open-ended preference,
(4) requesting recommendations, and
(5) giving feedback.
For each task, a population of simulators is created and given a task-specific prompt. The outcomes are then compared to curated human data.
\citet{Yoon:2024:arXiv} ``observe that simulators tend to favor mentioning popular items, correlate little with human preferences, exhibit lack of personalization in requests, and occasionally give incoherent feedback.''
\section{Summary and Future Challenges}
\label{sec:sim_conv:summary}

Research on conversational information access systems is still in a relatively early phase. Much of the research has focused on individual components and highly goal-specific dialogue flows. There are user simulators available to support the automatic evaluation of these scenarios, but it is also clear that there is much headroom for improving the realism and robustness of these simulators.\\\\

An interesting observation is that while simulation approaches for traditional search/recommender systems tend to focus on specific actions, as discussed in the previous chapter, there is more work on simulating entire conversations.
This is likely due to the fact natural language dialogue primarily involves simulating a single type of low-level action (generating utterances) rather than workflows of diverse actions (like querying, browsing, and clicking).
Especially with LLMs, it is relatively easy to simulate plausible, human-like utterances in a given dialogue context. However, such LLM-based simulations can lack control and may produce unrealistic behaviours~\citep{Terragni:2023:arXiv,Davidson:2023:arXiv,Wang:2024:WWW,Yoon:2024:arXiv}.
Furthermore, there is a risk or removing desirable variations that are characteristic of human behaviour.
The results in \citep{Wang:2023:EMNLP} show that LLMs are generally ``stronger'' (more knowledgeable) than human users, so using them to simulate a user might end up simulating a ``superuser.'' With appropriate prompting, an LLM could be instructed to simulate a more realistic user, but it is unclear whether the LLM would `listen'' to the prompt and follow it strictly. Another concern with LLM is its reproducibility. Unless the same model checkpoint can be used, the behaviour may be different in future versions of an LLM.

One of the biggest challenges is that the evaluation of conversational systems itself is---still---an open question.
Even though a number of different evaluation measures have been proposed~\citep{Deriu:2021:AIR,Liu:2021:TOIS}, these fail to capture what makes a conversation ultimately successful.
As \citet{Zamani:2023:FnTIR} explain, ``the same conversation may be considered of high or of low quality depending on context: For example, if a user is in a rush or not, or if the user requires high confidence in the conclusion or not.''
Shorter time to task completion is not necessarily sufficient; a prolonged but engaging interactions may lead to a more satisfying overall experience.
Also, users might trade off long- and short-term utility~\citep{Radlinski:2017:CHIIR}.

Several of the challenges discussed in the context of search and recommender systems in the previous chapter are also applicable here, including the modeling of context, user characteristics, and evolving user goals (see Section~\ref{sec:sim_search:summary}).

Additionally, facilitating multi-modal interactions would be especially pertinent in a conversational setting.
While past research has primarily focused on uni-modal (i.e., natural language) interactions, other modalities as input (e.g., speech, pointing/clicking, body gestures) and output (e.g., speech and multimedia elements) have also begun to receive increasing attention~\citep{Liao:2021:SIGIR,Kostric:2022:RecSys,Deldjoo:2021:SIGIR,Hauptmann:2020:MM}.
Modeling the action space around these and selecting the appropriate modality for a given user action/context are interesting challenges to be addressed in future simulators.
A particularly interesting line of work within the context of dialog-based interactive image retrieval is relative captioning, to simulate users' feedback on images using natural language~\citep{Guo:2018:NeurIPS,Zhang:2019:NeurIPS,Wu:2021:CVPR,Wu:2021:RecSys,Wu:2022:RecSys,Wu:2023:RecSys,Vlachou:2024:arXiv}.

Progress on conversational information access systems that support and seamlessly integrate multiple users goals critically depends on appropriate evaluation methodology and resources~\citep{Balog:2021:DESIRES}.  User simulation techniques play a pivotal role in fostering research and development in this area. Increasingly more work has leveraged LLMs, which have great potential for simulating how users interact with conversational information access systems.

\chapter{User Simulation in Practice: From Theory to Application}
\label{ch:practical}

There are several important factors to consider when it comes to employing user simulation in practice, beyond the specific modeling approach used.
First, one needs to consider the system, task, and ground the parameters controlling the behaviour of simulated users in empirical observations of the behaviour of the real user(s) that we hope to simulate.  
We discuss the configuration of simulators in Section~\ref{sec:practical:instantiating}.
Second, another critical issue concerns the evaluation of simulators themselves. Such validation is essential in order to verify that the results obtained using simulation are reliable and can be trusted. We present assessment criteria and validation approaches in Section~\ref{sec:practical:validation}.
Third, 
while user simulation is broadly regarded as a more cost-effective evaluation methodology compared to approaches involving human participants, developing these simulators can require significant engineering effort and resources.
In Section~\ref{sec:practical:workflow}, we discuss the role of user simulation within the broader evaluation workflow, highlighting how it complements other evaluation methodologies.
Finally, Section~\ref{sec:practical:toolkits} reviews toolkits and resources that can serve as a starting point for building simulators or be utilized directly.

\section{Instantiating Simulators}
\label{sec:practical:instantiating}

As stated in Section~\ref{sec:intro:simulation_definition}, the goal of user simulation is to create an agent that can simulate the actions that a user $U$ would take when attempting complete task $T$ using system $S$.
Instantiating a simulator requires configuring the variables $S$, $T$, and $U$.

\subsection{System}

Chapters~\ref{ch:sim_search} and~\ref{ch:sim_conv} discussed techniques for building simulators for specific types of systems: search/recommender engines and conversational assistants, respectively.
For each type of system, functionality is closely tied to its UI elements, such as a search box and a result list for search engines and a text input field and chat history display for conversational assistants. Individual systems differ in the features they offer, e.g., some search engines have query auto-completion or suggestion functionality, while some conversational assistants provide interactive buttons or scrollable results.
For most systems it can be useful to make a further distinction based on the device that is used, for example, a mobile phone vs. a PC, as it can have a major influence on users' behaviour. From the perspective of user simulation, we are most interested in modeling the actions that a user can take when interacting with each possible interface of the system. 

\subsection{Task}

The tasks discussed in this book are generally concerned with accessing information, but the nature and complexity of these tasks can vary significantly. For example, some ad hoc information needs, such as finding the current time in Tokyo or looking up a recipe for lasagna, can usually be resolved in a single search session and require minimal interaction with the system beyond the initial query. In contrast, complex information needs, such as preparing a literature review on climate change or planning a multi-destination vacation, often necessitate multiple search sessions. These tasks involve gathering diverse sources, refining search strategies, and synthesizing information across sessions~\citep{Marchionini:1995:book}.
Similarly, the task of re-finding an email or document one has read before involves searching through a personal data collection with potentially known keywords~\citep{Dumais:2003:SIGIR}, which is quite different from the extensive and often iterative process of finding all documents that may be relevant to a legal case~\citep{Sansone:2022:IS}. 
Despite all the above tasks utilizing the same basic functionality of a search box and a ranked list of results, the depth, breadth, and methods of interaction with the system differ greatly, leading to variable user behaviour that a user simulator would have to model based on the specific task that a user is performing.

\subsection{User}

Due to the complexity of real user behaviour, simulations of user interactions with interactive systems necessarily rely on simplified models of user behaviour. These models incorporate various configurable parameters that aim to capture the diverse ways in which users approach and engage with the system when completing a given type of task. Below, we discuss several key parameters that can be adjusted within these models, illustrating the range of user characteristics and preferences that can be simulated (see Table~\ref{tab:user_characteristics}). 

\begin{table}[t]
    \centering
    \caption{Examples of user characteristics considered in simulation studies.}
    \footnotesize
    \label{tab:user_characteristics}
    \begin{tabular}{lp{5.5cm}l}
        \toprule
        \textbf{Characteristic} & \textbf{Description}                                              & \textbf{Reference}           \\
        \midrule
        Cooperativeness               & Willingness to provide relevance feedback                         & \citep{Keskustalo:2006:ECIR} \\
         & Willingness to provide clarification information & \citep{Salle:2021:ECIR} \\        
        Expertise               & Search experience                                                 & \citep{Paakkonen:2015:CLEF}  \\
        Exploration             & Willingness to explore different subtopics                        & \citep{Camara:2022:ECIR}     \\
        Learning speed          & Speed of the user incorporating novel terms into their vocabulary                & \citep{Camara:2022:ECIR}     \\
        Patience                & Patience to browse through the list of retrieved documents        & \citep{Keskustalo:2006:ECIR} \\
         & Willingness to engage in a long dialogue with the system & \citep{Salle:2021:ECIR} \\
        Persistence & Willingness to examining the next search result & \citep{Lipani:2021:TOIS} \\ 
        Scanning behaviour      & Time allocation between examining summaries and reading documents & \citep{Smucker:2011:HCIR}    \\
        Stopping strategy & When to stop examining results for a given query & \citep{Maxwell:2015:CIKM} \\
        Subtopic switching      & Strategy employed to navigate through subtopics                   & \citep{Camara:2022:ECIR}     \\                
        Tolerance               & Willingness to click on a search result snippet                   & \citep{Camara:2022:ECIR}     \\
        \bottomrule
    \end{tabular}
\end{table}

\begin{itemize}
    \item \textbf{Task-related factors}: These parameters pertain to how users interact with the system to achieve specific goals. Examples include query length~\citep{Azzopardi:2007:SIGIR} and session length, which can be measured either as the number of rounds of querying~\citep{Carterette:2015:ICTIR} or the total time spent interacting with the system~\citep{Smucker:2012:HCIR,Verberne:2015:ECIR}.
    \item \textbf{Individual user traits}: These reflect the diverse properties and characteristics that can influence user behaviour, including, e.g., 
    (1) \emph{demographic information} (age, gender, location, etc.),
    (2) \emph{cognitive and behavioural traits}, such as search strategy, decision-making process, persistence in examining search results, patience in interacting with the system, cooperativenes in providing clarification, learning speed, and willingness to explore (see Table~\ref{tab:user_characteristics}),  and 
    (3) \emph{historical interaction data} (past queries, clicks, etc.).
    \item \textbf{Interaction cost and reward}: These quantify the effort required for users to interact with the system and the perceived benefit or utility they gain from those interactions.
    Cost is often measured in terms of time needed to perform a specific type of action, for example, to formulate a query, undertake an initial SERP examination, examine an individual result snippet, examine a document, or provide feedback~\citep{Baskaya:2013:CIKM,Maxwell:2015:SIGIR}. A user's cognitive effort should also be considered~\citep{Labhishetty:2022:ICTIR}. The parameters capturing cost are often calibrated based on empirical data gathered from user studies. 
    Reward, on the other hand, is commonly estimated based on the gain users receive when provided with relevant and novel information~\citep{Azzopardi:2011:SIGIR,Smucker:2012:SIGIR}. When task information is available, the reward can be estimated based on the progress in task completion. 
\end{itemize}

\noindent
Since log data capture real user behaviours, they are very useful for building user simulators. However, log data may be biased in various ways.
For example, in recommender systems, it has been widely recognized that the ratings left by users are subjective and are not missing at random~\citep{Steck:2010:KDD}.
Therefore, one needs to take appropriate measures to avoid further reinforcing those biases in simulators~\citep{Huang:2020:RecSys}.
\section{Validating Simulators}
\label{sec:practical:validation}

For the results of simulations to be credible, it is essential to validate that the simulator imitates the behaviour of real users sufficiently well.
As \citet{Pietquin:2013:KER} emphasize, ``it is necessary to provide a clear idea of the purpose of a user simulation,'' since requirements differ for simulators that are used for training RL-based dialogue managers, expanding existing datasets, or facilitating automatic evaluation of systems. 

While the ultimate measure of a user simulator's success is its ability to predict system performance with real users, various validation approaches can be employed to assess different aspects of the simulator's behaviour. These include: (1) direct comparison of system performance with real vs. simulated users, (2) analysis of high-level behavioural characteristics of real vs. simulated users, (3) human evaluation of data produced by real vs. simulated users, and (4) the use of automated tests to check whether the simulator behaves as expected in specific scenarios. 

Our focus in this book is on using user simulation for automatic evaluation of systems and thus a major criterion for validating usre simulators is to what extent a user simulator can help us compare different systems or variations of the same system to identify the most effective system or system variant. 

\subsection{System Performance}

A user simulator's effectiveness is ultimately validated by directly comparing its predictions of system performance to the actual performance observed in real-world scenarios with real users.
Performance can be considered at different levels, from a specific component within the system to the overall, end-to-end performance of the entire system.

Consider query simulation as an illustrative example of component-based validation.
The effectiveness of a query simulator can be assessed by directly comparing the retrieval performance resulting from simulated queries against that of real user queries~\citep{Azzopardi:2007:SIGIR,Baskaya:2011:ECIR,Baskaya:2012:SIGIR,Keskustalo:2009:AIRS,Breuer:2022:ECIR}.  

On the other hand, end-to-end evaluation involves assessing the overall performance of the system, encompassing all its components and their interactions. This necessitates evaluating the same system(s) with both human and simulated users. The resulting performance measures can be compared in absolute terms, but often the focus is on relative comparison between different systems. In this case, it is sufficient to determine whether the relative ordering of systems in terms of performance is consistent across human and simulated users~\citep{Zhang:2020:KDD}.

\subsection{Behavioural Characteristics}

Another approach involves analyzing the statistical properties of user interaction data to assess whether simulated users exhibit behavioural patterns consistent with real users. Note that this approach focuses on examining distributions, frequencies, and aggregate measures rather than individual actions or queries.

In the context of search engines, examples include comparing query lengths \citep{Azzopardi:2007:SIGIR}, click decisions~\citep{Carterette:2015:ICTIR,Chuklin:2015:Book}, dwell times~\citep{Carterette:2015:ICTIR}, session statistics~\citep{Maxwell:2016:CIKM,Maxwell:2015:CIKM}, and gain in terms of the number of relevant documents saved~\citep{Smucker:2012:HCIR}.

For conversational agents, a collection of simulated dialogues may be compared against a human-machine dialogue corpus ``to evaluate how well the simulated dialogues match the statistical distribution of the real dialogue corpus''~\citep{Schatzmann:2006:KER}.
Various corpus-level statistics have been proposed in the literature~\citep{Schatzmann:2005:SIGDIAL,Schatzmann:2006:KER,Pietquin:2013:KER} that can be used as automatic measures:
\begin{inlinelist}
    \item \emph{High-level dialogue features}, such as average dialogue length (number of turns) and ratio of user vs. system actions.
    \item \emph{Dialogue style}, which includes the frequency or distribution of dialogue acts, user cooperativeness (proportion of slot values provided when requested).
    \item \emph{Dialogue efficiency}, including success (or task completion) rate, reward, and completion time.
\end{inlinelist}

\subsection{Human Evaluation of Simulated User Output}

While statistical measures provide valuable insights into aggregate behaviour, human evaluation offers a nuanced perspective on the quality and naturalness of the output generated by simulated users. This is particularly relevant for textual data, such as search queries in search engines and conversational utterances in dialogue systems. Human assessments can capture subtle aspects of language use. For instance, in the context of search engines, \citet{Gunther:2021:Sim4IR} investigate ``how such suggestion-based simulated sessions compare to real user sessions in the sense of coherence, realism, and representativeness of the underlying topic.''

For conversational assistants, human evaluation can be conducted at the utterance or at the dialogue level.
For \emph{utterance-level} evaluation, it is common to have human raters evaluate the generated responses along different dimensions, such as naturalness, usefulness, and grammar~\citep{Sekulic:2022:WSDM,Zhang:2022b:SIGIR,Owoicho:2023:SIGIR}.
For evaluating \emph{entire dialogues}, \citet{Zhang:2020:KDD} propose a side-by-side  human evaluation protocol, where assessors are given transcripts of two conversations, in random order, and they are told that one of the dialogues was conducted by a human and the other one by a ``bot.'' They have to guess which of the two is the human one. This follows the idea of adversarial evaluation~\citep{Li:2017:EMNLP}, inspired by the  Turing test, where the more successful a simulator is in ``fooling'' the human evaluator, the more realistic it is.

\subsection{Automated Testing}

The Tester-based framework~\citep{Labhishetty:2021:SIGIR,Labhishetty:2022:ECIR} aims to more directly evaluate a user simulator's reliability for evaluating interactive IR systems. The basic idea is to use many testers to check whether a user simulator can successfully pass those testers. A tester is composed of a pair of IR systems that are expected to show a particular pattern of performance; for example, system A is expected to perform better than system B under a certain condition (e.g., for a certain kind of queries). Given a simulated user, we can then measure the performances of the two systems by using the simulated user to interact with both of them, and see if the obtained performance pattern is consistent with the expected pattern (e.g., system A is better than system B). If it is consistent, the user simulator can be assumed to pass the tester. Using this strategy, we may compute the success or failure rate of a user simulator by using multiple testers that may also vary by using different queries or collections. The success (or failure) rate can then be treated as a quantitative measure of the reliability of a user simulator.

However, there is also a question about the reliability of the testers themselves. Intuitively, we want to trust a reliable tester more than a less reliable one, so this raises the question of how to quantify the reliability of testers. In \citep{Labhishetty:2022:ECIR}, an iterative algorithm is proposed to jointly estimate the reliability of a set of testers and the reliability of a set of user simulators. The intuition captured by the algorithm is to assume that the reliability of a user simulator would be high if the user simulator can pass multiple reliable testers with high success rates, and that the reliability of a tester would be high if the tester can be passed by many reliable user simulators. Such a mutually reinforcing heuristic naturally led to an iterative propagation algorithm to simultaneously estimate the reliabilities of all the user simulators and all the testers involved.

\citet{Sun:2023:TOIS} employ the Tester-based  framework for evaluating user simulators for conversational recommender systems. Specifically, they introduce testers based on context (varying the amount of dialogue history available), recommender (controlling retrieval capabilities), and domain (providing different amounts of training data in a particular domain).
In similar spirit, \citet{Yoon:2024:arXiv} design specific tasks to check for key properties that a user simulator should exhibit and test LLM-based simulators with task-specific prompts.

\section{User Simulation in the Evaluation Workflow}
\label{sec:practical:workflow}

It is important to emphasize that simulation is not meant to replace but to complement other evaluation methodologies. User simulations are best utilized as part of a broader evaluation workflow, providing a cost-effective and efficient way to explore a wide range of system configurations (e.g., different methods, algorithms, hyperparameter settings) before committing to expensive human evaluation (user studies or online A/B tests).

\begin{figure}[t]
    \centering
    \includegraphics[width=0.9\textwidth]{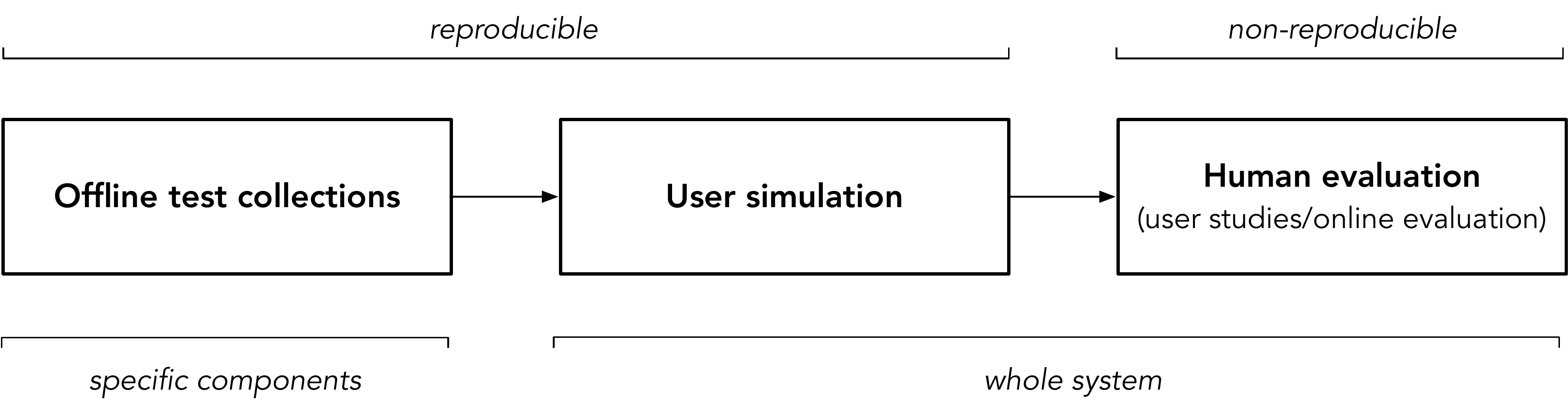}
    \caption{User simulation in the evaluation workflow.}
    \label{fig:usersim_workflow}
\end{figure}

Let us consider a modern information access system comprising multiple components, such as query suggestions, ranking algorithms, result snippets/previews, and related item recommendations.
While each of these components may have been individually evaluated offline using reusable test collections, the overall effectiveness of the system cannot be established without the involvement of human subjects, due to the complex interplay between the various components.
Existing methodologies involving human evaluation are often unsuitable for exploring a large number of system variations. 
As we have explained in Section~\ref{sec:intro:evaluation_challenges}, user studies are costly and time-consuming, while online A/B testing is constrained by the traffic volume of the service (i.e., the number of experiments that can be conducted within a reasonable timeframe). %
This is where user simulation comes into play, offering a cost-effective and efficient way to explore a large number of system variations before committing to resource-intensive user studies or online experiments. However, it is important to emphasize that simulation is not a replacement for human evaluation. It is an intermediate stage in the evaluation workflow, designed to narrow down the most promising system alternatives. Ultimately, these alternatives should be tested with real users, serving as a crucial validation step for the simulation results. See Fig.~\ref{fig:usersim_workflow} for an illustration.

The development of user simulators is often an iterative process. In the beginning, when no or limited interaction data is available, the simulator may rely on assumed user behaviour or small-scale interaction data gathered from pilot user studies. As more interaction data is collected from actual users of a service, the hyperparameters of the user simulator can be tuned, assumptions may be revisited, and the underlying models themselves can be improved, leading to more accurate and realistic simulations over time. 

It is important to note that developing and maintaining user simulators requires expertise and can be a significant investment. Modeling certain elements of human behaviour can be very challenging. Therefore, in some cases, it might be more cost-effective in the short term to run user studies, especially if the number of system variations to explore is small. However, investing in the development and refinement of a user simulator can pay off in the long run, particularly when frequent evaluations of complex systems are needed.

\section{Toolkits and Resources}
\label{sec:practical:toolkits}

In this section, we discuss available toolkits and resources for user simulation for search engines, recommender systems, and conversational assistants.

\subsection{Search Engines}

Relative to the large body of work on user simulation in IR, there is a surprisingly limited selection of toolkits that are (still) available at the time of writing.

SimIIR~\citep{Maxwell:2016:SIGIR}\footnote{\url{https://github.com/leifos/simiir}} is a framework for the simulation of interaction with search engines. A simulation is comprised of a series of topics (specified by TREC-style topic descriptions), a search interface/engine (an abstraction of a SERP with a ranked list of snippets and associated documents), an output controller (for saving data generated by the simulator), and a series of simulated searchers. The user model underlying the simulated searchers is the Complex Searcher Model (see Fig.~\ref{fig:complex_searcher_model}), and can be parameterized with a querying strategy, classifiers to determine snippet and document relevance, and a stopping strategy.

The SimIIR 2.0 framework~\citep{Zerhoudi:2022:CIKM}\footnote{\url{https://github.com/padre-lab-eu/simiir-2}} extends the original SimIIR framework and the underlying Complex Searcher Model with novel components to improve the realism of the simulated sessions.
First, an advanced query generator is added, which dynamically generates new query candidates based on the results encountered by the searcher (as opposed to generating a static pool of queries at the beginning of a session).
Second, the notion of \emph{user types} is introduced, representing groups of searchers with a specific search behaviour. For example, ```exploratory searchers' will explore a search result list more exhaustively than `lookup searchers' who will only investigate the first few results and then rephrase their queries rather quickly''~\citep{Zerhoudi:2022:CIKM}.
Third, stopping and query generation are modeled as Markov model-based decisions specific to user types.

The cwl\_eval toolkit~\citep{Azzopardi:2019:SIGIR},\footnote{\url{https://github.com/ireval/cwl}} while not a user simulator in a traditional sense, also deserves mentioning, as it implements the C/W/L framework, which reflects different beliefs about user behaviour (see Section~\ref{sec:frameworks:measures_ub}). The toolkit implements various metrics within one package and can be used to generate ``a wide range of measurements regarding the predicted user interactions with the ranked list of search results''~\citep{Azzopardi:2019:SIGIR}.

\subsection{Recommender Systems}

In the field of recommender systems research, simulation has been extensively used for measuring the long-term dynamics of recommender ecosystems, i.e., how different entities, including users, recommender algorithms, and content providers interact and influence each. Several toolkits have been created for this purpose, including RecLab~\citep{Krauth:2020:arXiv}, RecoGym~\citep{Rohde:2018:arXiv}, PyRecGym~\citep{Shi:2019:RecSys}, RecSim~\citep{Ie:2019:arXiv}, and RecSim NG~\citep{Mladenov:2021:arXiv}. However, these works operate with overly simplistic user models. We refer to \citep{Stavinova:2022:arXiv} for a detailed description and comparison of simulator toolkits for recommender systems.

\subsection{Conversational Assistants}

We discuss toolkits and resources first for traditional slot-filling-style task-oriented dialogue systems, and then those built specifically for conversational information access. 
Given the rapid pace of development in this area, a more comprehensive and up-to-date list is maintained on the companion website at \url{https://usersim.ai/toolkits}.

\subsubsection{Task-oriented Dialogue Systems}
\label{sec:sim_conv:toolkits_ds}

There are numerous open source toolkits available for developing conversational systems.
Conversational AI platforms, such as ParlAI~\citep{Miller:2018:arXiv}\footnote{\url{https://parl.ai/}} and Plato~\citep{Papangelis:2020:arXiv},\footnote{\url{https://github.com/uber-archive/plato-research-dialogue-system}} support the sharing, training, and testing of dialog models for a broad range of tasks, from open-domain chitchat to task-oriented dialogue.
Toolkits built specifically for multi-domain task-oriented dialogue include PyDial~\citep{Ultes:2017:ACL},\footnote{\url{http://pydial.org}} which follows a modular architecture, and ConvLab~\citep{Lee:2019:ACL,Zhu:2020:ACL,Zhu:2023:EMNLP},\footnote{\url{https://convlab.github.io/}} which is an end-to-end dialog system platform.
All of these toolkits include a user simulator component. %
\citet{Shi:2019:EMNLP} compare agenda-based and supervised-learning-based user simulators and release their code.\footnote{\url{https://github.com/wyshi/user-simulator}}
Another implementation of the agenda-based simulator can be found in~\citep{Li:2017:arXiv}.\footnote{\url{https://github.com/MiuLab/UserSimulator}}

Some of the most commonly used datasets for evaluating task-oriented dialogue systems are MultiWOZ~\citep{Budzianowski:2018:EMNLP}\footnote{\url{https://github.com/budzianowski/multiwoz}} and the Schema-Guided Dialogue (SGD)~\citep{Rastogi:2020:AAAI}.\footnote{\url{https://github.com/google-research-datasets/dstc8-schema-guided-dialogue}} Both cover multiple domains (including travel, events, restaurants, etc.) and come with dialogue act annotations.
For an overview of task-oriented datasets we refer to~\citep{Deriu:2021:AIR}.
\citet{Cheng:2022:EMNLP} emphasize the need for an  interactive evaluation process, i.e., user simulation, for the evaluation of TOD systems. They show that simulation-based evaluation can achieve over 98\% inform and success rates on MultiWOZ, therefore, ``TOD needs more challenging and complicated dataset and scenarios''~\citep{Cheng:2022:EMNLP}.

\subsubsection{Conversational Information Access}
\label{sec:sim_conv:toolkits_cia}

Conversational information access is an emerging research area and therefore the availability of toolkits and datasets remains limited to date.
Multiple user simulators have been proposed for answering clarifying questions in conversational search, including
CoSearcher~\citep{Salle:2021:ECIR},\footnote{\url{https://github.com/amzn/cosearcher}}
USi~\citep{Sekulic:2022:WSDM},\footnote{\url{https://github.com/isekulic/USi}} and
ConvSim~\citep{Owoicho:2023:SIGIR}.\footnote{\url{https://github.com/grill-lab/ConvSim}}
For conversational recommendation, UserSimCRS~\citep{Afzali:2023:WSDM}\footnote{\url{https://github.com/iai-group/UserSimCRS}} implements and extends the agenda-based simulator proposed in \citep{Zhang:2020:KDD}, and CFCRS~\citep{Wang:2023:KDD}\footnote{\url{https://github.com/RUCAIBox/CFCRS}} is a multi-stage recommendation dialogue simulator based on a conversation flow language model.

A series of information seeking benchmark datasets are developed within the TREC 2019--2022 Conversational Assistance track (TREC CAsT)~\citep{Dalton:2020:SIGIR}.\footnote{\url{https://www.treccast.ai/}}
Resources for clarifying questions include Qulac~\citep{Aliannejadi:2019:SIGIR} and its extension ClariQ~\citep{Aliannejadi:2020:arXiv}; both comprise (topic, facet, clarifying question, answer) tuples for single-turn interaction.
A multi-turn dataset is created and released in~\citep{Sekulic:2022:WSDM}.
There are several conversational recommendation datasets available for the movies domain, e.g., the Movie Dialogue datasets~\citep{Dodge:2016:ICLR},\footnote{\url{https://research.facebook.com/downloads/babi/}} ReDial~\citep{Li:2018:NIPS},\footnote{\url{https://redialdata.github.io/website/}} and INSPIRED~\citep{Hayati:2020:EMNLP}.\footnote{\url{https://github.com/sweetpeach/Inspired}}
DuRecDial~\citep{Liu:2021:EMNLP}\footnote{\url{https://github.com/liuzeming01/DuRecDial}} covers additional domains as well, including music, food, and restaurants.
MG-ShopDial is multi-goal dataset for conversational information access that includes conversational search, recommendation, and QA at the same time, for the e-commerce domain~\citep{Bernard:2023:SIGIR}.\footnote{\url{https://github.com/iai-group/mg-shopdial}}

\chapter{A Broader Perspective on User Simulation}
\label{ch:broader}

In this book so far, we have primarily focused on using user simulation as an evaluation tool for information access systems, such as search engines, recommender systems, and conversational assistants.
However, the applications of simulation extend to assistive AI systems in general and far beyond evaluation. Simulation techniques can be employed for data augmentation, generating synthetic user interactions to improve the training of machine learning models, or user behaviour modeling, and creating realistic simulations of how users interact with a system in general (Section~\ref{sec:broader:applications}).
As a research topic, user simulation is inherently interdisciplinary, intersecting with diverse fields both within and beyond computer science. For example, it draws upon concepts from psychology, economics, and human-computer interaction to create accurate and representative models of user behaviour (Section~\ref{sec:broader:interdisciplinary}). The recent success of large language models (LLMs) has led to their widespread use in simulation tasks across different domains and application scenarios, from generating realistic dialogue for chatbots to simulating customer interactions for business training (Section~\ref{sec:broader:llms}).
Finally, we take an even broader perspective of user simulation to discuss its connection with Artificial General Intelligence (AGI), the long-term goal of creating AI systems with human-like intelligence (Section~\ref{sec:broader:agi}).

\section{Broad Applications of User Simulation}
\label{sec:broader:applications}

Although our focus in the book is on the use of simulation for evaluation, user simulation has numerous other applications, including particularly user behaviour modeling and analysis, and data augmentation, which we briefly discuss in this section. Different applications may require or appreciate different types of simulation techniques, but there is clear synergy of those different lines of exploration.
Each application area will likely lead to the development of distinct simulators. Multiple lines of research can mutually enhance and inform each other. Eventually, these diverse lines of research may even converge as we progress towards AGI.

\subsection{Data Augmentation}

There has been much work on using user simulation for data augmentation. The generated synthetic data can then be used for training machine learning algorithms.
While interpretability is critical when we use user simulation for evaluation, it is not so important when using it for data augmentation as whether the generated data would be useful might not depend so much on the realism of a simulator, but rather whether the distribution of new data is correlated with the distribution of true user data. Such a relaxed requirement makes it easier to make progress in user simulation for data augmentation especially because of the emergence of deep learning techniques that enable end-to-end training of an algorithm for virtually any prediction problem (predicting user behaviour would enable data augmentation). The availability of LLMs further makes it feasible to simulate users to generate realistic text content, which opens up even more opportunities for using user simulation for data augmentation.

When we use a user simulator to evaluate an interactive system, the evaluation process would also naturally generate additional interaction data of the simulated users. Such data can generally be leveraged for data augmentation. When a user simulator is interpretable, the user simulator can be varied via changing meaningful user variables to simulate potentially many more users in a counterfactual way (e.g., we could make a simulated user increasingly more patient) and thus is, in general, more powerful than a non-interpretable simulator.

While there are many ways to generate synthetic data for data augmentation, a major advantage of data generated by a user simulator is that the data reflect a user's interaction with a system, which are especially useful for training an interactive intelligent system. Such data, along with evaluation measures defined on user interactions, can provide reward information for an intelligent interactive system to use for optimizing its interaction policy, thus enabling end-to-end training of a whole interactive agent.

However, it is also worth pointing out that although simulating a whole user is desirable, for data augmentation, even simulating a single user behaviour (e.g., clicking in search) would still be useful for obtaining additional training data. Focusing on specific actions would also simplify making progress in data augmentation. Since data augmentation benefits various machine learning models, using user simulation for this purpose further expands the scope of research on user simulation into potentially many other areas of AI.

\subsection{User Behaviour Analysis}

Analysis of user behaviour is important for many reasons. For example,
it enables understanding users in general, which further enables understanding our society and and answering many social science research questions. Also, formal user models to capture user behaviour can be used by an intelligent agent to personalize its service to individual users.

User simulation can be regarded as taking a computational approach to studying users' behaviour. Specifically, once we build a user simulator, the simulator can be regarded as a formal (mathematical) model of a user, where all the hypotheses about the user's behaviour are explicitly articulated in some form, thus enabling us to test  those hypotheses by comparing the behaviour of a simulated user with that of a real user. Such a comparison can directly facilitate validation of a user simulator and provide insights about how to further improve a simulator. For example, we may discover the simulator is unable to simulate a particular behaviour of a user well, and thus consider this weakness when using the user simulator for evaluating systems. Such insights can also guide further research on improving user simulation.

Analysis of user behaviour has often been done by doing controlled user studies. Such work often reveals interesting observations or insights about behaviours of real users that can be leveraged directly for building user simulators. In this sense, such user studies can be viewed part of the general research agenda of building realistic user simulators, only that they may focus on simulating one aspect of user behaviour in a particular user study. From another perspective, user simulation enables testing any hypothesis that might be formulated based on any real user study.

Analysis of user behaviour has also been done by analyzing log data generated by real users. Such work can be regarded as directly contributing to research on user simulation, especially when we can also use such log data for instantiating or seeding a user simulation (see more discussion of this issue in Section~\ref{sec:practical:instantiating}). Indeed, it is natural to view all work on analyzing log data of users as part of study of user simulation, thus the scope of research in user simulation generally covers analysis of all kinds of user interaction logs in any system, which does not necessarily have to be an AI system.

An especially interesting direction of using user simulation for user modeling is to conduct simulation experiments to model a community of users interacting with an interactive system over a long time period as has been done in, e.g., \citep{Hazrati:2024:UMUAI,Yao:2023:ICML}. Such simulation experiments would enable counterfactual analysis of impact of any system to be deployed as well as understanding user behaviour and preferences in a simulated application environment such as recommender systems~\citep{Zhang:2020:ISR}.

\section{User Simulation as an Interdisciplinary Research Field}
\label{sec:broader:interdisciplinary}

The broader discussion of user simulation in the previous section also suggests that the research of user simulation is a highly interdisciplinary and touches many different topic areas, including, e.g.,
information science, information access,  machine learning, natural language processing, knowledge representation, human-computer interaction, and psychology.
It is thus desirable to integrate research in all these different areas in the future.
In the rest of this section, we will discuss some specific connections between user simulation with multiple other areas.

\subsection{User Simulation and Intelligent Agents}

A sophisticated user simulator can be regarded as an intelligent agent that operates in an environment of a system by interacting with the system. The MDP framework that we discussed earlier provides a foundation for formally defining the behaviour of such an agent. As such, research on user simulation requires studying how to build an intelligent agent in general and many techniques used for building an intelligent agent are also useful for building a user simulator. We thus anticipate increasingly more such techniques to be used for user simulation. From another perspective, research on user simulation enables intelligent agents to interact with simulated users for training their interaction policy, thus enabling improvement of interactive intelligent agents.

Considering human-AI collaboration,  we can expect  sophisticated user simulators to not only simulate human-like behaviour but also simulate/model human's ability of problem solving since in the future, an intelligent agent may (proacively) ask a human user for help, in which case, the user simulator would need to simulate what kind of help the user may potentially provide to an AI system. Since it is also desirable to make an intelligent (task) agent human-like, building user simulation agents and building intelligent task agents may be viewed as approaching the same goal of building an intelligent human-like agent from different angles, both moving toward artificial general intelligence as we will further discuss in Section~\ref{sec:broader:agi}

\subsection{User Simulation and Machine Learning}

User simulation provides many interesting opportunities for applying a broad variety of ML techniques, including unsupervised learning, supervised learning, imitation learning, and reinforcement learning (RL). Particularly relevant is inverse reinforcement learning~\citep{Arora:2021:AI}, as it is especially designed to recover the reward function and the policy just from the analysis of the observed behaviour. %
This is unlike RL where access to the reward function is assumed, which is often difficult to define. Generative probabilistic models may be applied to learn patterns of user actions in an unsupervised way. The observed user actions can be used as labels for use with supervised learning or imitation learning to build a user simulator. Reinforcement learning is necessary for user simulation since it enables a user simulator to optimize sequential decision making (policy) by learning from its interactions with a system.  User simulation can also be used to generate synthetic user (interaction) data which can be useful for training machine learning algorithms, especially complex algorithms, such as reinforcement learning, that often requires either massive amounts of user interaction data or an interactive environment for training. User simulation techniques would enable the training of such algorithms that are currently infeasible.

\subsection{User Simulation and Knowledge Representation and Natural Language Processing}

A realistic user simulator must include a model of a user's knowledge state, including specifically a representation of a user's background knowledge (about relevant and non-relevant content, how an information access system works, what content is available in the collection for search, etc.). A user's actions such as formulation of a specific query or clicking on a particular search result are often a consequence of a particular knowledge state. Thus, user simulation is closely related to the study of knowledge representation, and involves reasoning over knowledge (to decide which action to take). Existing work has so far used very simple knowledge representations, such as unigram language models; in the future, it would be interesting to apply more sophisticated knowledge representation techniques (e.g., knowledge graphs) to build more realistic user simulators.

All such knowledge also needs to be updated to reflect learning by a search user during the search process~\citep{Vakkari:2016:JIS}. Therefore, it is necessary to study the cognitive processes of searchers. This makes the study of user simulation closely related to intelligent tutoring systems, where simulated learners have also been used for evaluating intelligent education systems~\citep{Dorcca:2015:AIED}.  In general, how to represent the knowledge that a user possesses, how to model the reasoning a user makes over the knowledge, and how to capture a user's knowledge gain during a search process are all important yet challenging open questions that may require deeper understanding of a fundamental question in AI---how knowledge should be represented formally.

Since user simulation requires advanced natural language understanding (e.g., understanding search results and assessing relevance) and natural language generation capabilities (e.g., formulating queries and conversational utterances), there is a clear connection to the field of natural language processing (NLP) as well. NLP techniques can be leveraged to simulate users' cognitive processes more accurately.

\subsection{User Simulation and HCI and Psychology}

Psychology and HCI research play a vital role in establishing the theoretical foundations for simulating users, making them not only closely linked to user simulation but also essential for constructing realistic simulators. Looking ahead, we anticipate a growing integration of knowledge and theories from psychology and HCI into the development of user simulators. Leveraging any insights gained about users has the potential to enhance the accuracy and fidelity of user simulation.
Starting points might include \citeauthor{Fu:2007:HCI}'s work on SNIF-ACT, well-motivated and explicit computational cognitive model that explains navigation behaviour on the Web~\citep{Fu:2007:HCI}. Another excellent example of a user simulator modeling user surfing behaviour can be found in~\citep{Chi:2003:CHI}.

At the same time, an interpretable user simulator can be instrumental in analyzing user interaction data to gain a deeper understanding of user behaviour and preferences. Existing research has already employed various models to analyze search logs for insights into search users. In the future, more advanced user simulators can offer opportunities to more deeply understand users. These simulators can also serve as hypothesized ``computational models'' of users, enabling the formulation of testable hypotheses. By conducting user studies specifically designed to test such hypotheses, we can confirm and establish theoretical knowledge about user behaviour.

\subsection{User Simulation and Specific User Tasks}

Domain-specific or system-specific user simulation would require collaborative research between user simulation researchers and domain experts. The incorporation of domain-specific knowledge and assumptions would be necessary. Since information access systems are widely used by many people in all kinds of application domains, there will be widespread opportunities for collaboration with domain experts, leading to many specialized user simulators developed for specific domains. For example, in the case of simulating searches in the medical domain, we likely would need to incorporate specific knowledge about medical search tasks, whereas in e-commmerce search, it would be necessary to incorporate other kinds of domain knowledge, depending on the specific type of products to be purchased by the user. Simulation of learners in education is another example, which has attracted much attention recently~\citep{Kaser:2023:AIED}. Since learners often rely on an information access system to learn, simulation of learners is very closely related to simulation of users of information access systems, and similar issues such as validity of a simulator have also been studied and discussed in the education context~\citep{Kaser:2023:AIED}.

\subsection{User Simulation and Software Systems}
\vspace*{-0.25\baselineskip}

Finally, in contrast to the static test collections currently used for evaluation, user simulators need to function as operational software platforms to facilitate evaluation. Consequently, an interesting relationship emerges between user simulation and software systems, as the challenge lies in constructing ``high-performance'' user simulators capable of simultaneously interacting with multiple search systems (to enable their evaluation).
Given that a sophisticated user simulator will likely have to use advanced algorithms and data structures (e.g., for representation of knowledge) and support large-scale experiments (that possibly explore various configurations), there will be significant new challenges in
building
efficient user simulators.

\section{Large Language Models and User Simulation}
\label{sec:broader:llms}
\vspace*{-0.25\baselineskip}

LLMs have shown great promise in many AI tasks. With multimodality, the  most sophisticated foundation LLMs are now able to simulate many aspects of human intelligence, notably natural language understanding and generation, computer vision, and software coding. While those LLMs are still lacking in their capacity of reasoning and planning and they are also unable to regulate their hallucination behaviours, it is reasonable to assume that they will become increasingly capable and human-like in the future. As such, they provide a good foundation for building both intelligent task agents and user simulation agents. In this section, we briefly discuss some recent trends in using LLMs for user simulation and suggest some promising future directions.

First, since LLMs are able to generate fluent realistic natural language text, they directly enable simulation of all kinds of user actions where the user would produce text, including, e.g., generation of queries for search or the generation of natural language conversational utterances. While simulation of keyword queries has so far been mostly based on simple language models (e.g., n-gram language models) as we discussed in Section~\ref{sec:sim_search:queries}, LLMs have recently been exploited to simulate query generation~\citep{Abolghasemi:2023:CIKM,Kiesel:2024:ECIR}.
In the case of conversational systems, traditional simulation methods were limited in their capability of generating natural language conversations that a human could produce. LLMs overcame this limitation and can be leveraged to simulate human conversations in various contexts, including, e.g.,
conversational AI~\citep{Meyer:2022:CUI},
task-oriented dialogue~\citep{Hu:2023:CIKM,Davidson:2023:arXiv},
conversational search~\citep{Sekulic:2022:WSDM,Owoicho:2023:SIGIR},
conversational recommendation~\citep{Wang:2023:EMNLP,Friedman:2023:arXiv}, and
conversational QA~\citep{Abbasiantaeb:2024:WSDM}.

In addition, LLMs have also been used for simulation in recommender systems~\citep{Zhang:2023:arXiv,Wang:2024:arXiv}
and for generating relevance label as a way to replace humans in evaluating IR systems~\citep{Faggioli:2023:ICTIR}.
The latter may be seen as simulating human assessments of utility of a document in search results (using a black box model).

As the uses of LLMs for various information access tasks become more widespread, it is also crucial to acknowledge their limitations.
As discussed in Chapter~\ref{ch:sim_conv}, LLM-based responses can be uncontrollable, leading to unrealistic or incoherent behaviours, and a lack of the natural variation found in human interactions~\citep{Terragni:2023:arXiv,Davidson:2023:arXiv,Wang:2024:WWW,Yoon:2024:arXiv}. Moreover, LLMs often exhibit greater knowledge than average users and generate overly ``perfect'' responses, potentially leading to the simulation of unrealistic ``superusers''~\citep{Wang:2023:EMNLP,Wang:2024:WWW}. While prompting can guide LLM behaviour, ensuring strict adherence to instructions remains a challenge~\citep{Kiesel:2024:ECIR,Wang:2024:WWW}. The reproducibility of LLM simulations is also questionable, as model updates may alter behaviour.

LLMs offer great potential for simulating diverse user actions, opening up exciting avenues for future research. For example, LLMs have already been used to build generative agents, i.e., computational software agents that simulate believable human behaviour, enabling various simulations of social environments (e.g., simulating a small town of twenty-five generative agents using natural language~\citep{Park:2023:UIST}).
For comprehensive surveys of work combining agent-based modeling and LLMs across various fields (social science, natural science, and engineering) and domains (e.g., cyber, physical, social environment) we refer readers to~\citep{Gao:2023:arXiv} and \citep{Wang:2024:FCS}.

\section{User Simulation as a Step toward AGI}
\label{sec:broader:agi}

The overarching goal of developing a realistic user simulator is, in many ways, aligned with the general goal of developing intelligent agents with human-like intelligence, i.e., the goal of achieving Artificial General Intelligence (AGI).
A sophisticated user simulator, behaving like a human user, would respond to system interactions and choose appropriate actions to complete a task. As an intelligent agent, such a simulator can be effectively modeled using the MDP framework discussed in this book.
Similarly, an intelligent agent designed to support a user in completing a task would also respond to user actions and select appropriate actions (i.e., choose appropriate system services to serve users), and can be generally modeled with an MDP framework as well.
Thus, both user simulation agents and intelligent interactive task agents share a common modeling approach, allowing for optimization through reinforcement learning algorithms.
The difference lies in the definition of states and actions as well as the reward functions.
For user simulation, the primary goal (reward) is to exhibit behaviour similar to real users, while for task agents, the reward is often tied to providing user-friendly service to facilitate task completion. Additionally, the actions differ, with user simulation focusing on user actions and intelligent agents on system responses.

Due to this similarity, the technical challenges in building intelligent user simulators may mirror those faced in developing intelligent task agents. We therefore anticipate broader connections between research in user simulation and many other areas of AI.

We may also view intelligent user simulation agents and intelligent task agents as approaching the same (eventual) goal of AGI from the two ends of a spectrum with variable trade-off between ``human-like'' and ``task support.'' At one end of the spectrum, we have intelligent task agents that may emphasize only task support without being human like; at the other end, we have intelligent user simulators that emphasize only human-like without providing task support. Future research can be expected to converge to the middle of the spectrum as those task agents become increasingly human like, while the user simulators become increasingly capable of supporting tasks.
The emergence of LLMs may accelerate the integration and synergy discussed above since the foundation LLMs may very well serve as the foundation for building both kinds of agents (eventually).

AI systems can be broadly categorized into two types: \emph{autonomous AI}, where machines aim to perform tasks normally performed by humans (e.g., self-driving cars), and \emph{assistive AI}, where the goal is not to replace humans, but rather aid them to fulfill their tasks more efficiently and effectively, often by augmenting human intelligence.  A main challenge unique to assistive AI systems is how to collaborate with a human user effectively so as to maximize the combined intelligence of humans and machines. Effective human-AI collaboration would require  user simulation agents to be tightly integrated with task agents so that the task agent would be able to leverage a user simulation agent to obtain reward and guide its optimization of interaction policy. Similarly the user simulation agent would also need to adapt to model how new AI systems would behave over time. Thus the dependency between research on intelligent task agents and that on user simulation is inherent and might not disappear until we achieve AGI.

\chapter{Conclusion and Future Challenges}
\label{ch:concl}

Accurate evaluation of information access systems---such as search engines, recommender systems, and conversational assistants---is critical both for confident real-world deployment and for making progress in scientific research. However, the task of effectively evaluating these systems has long remained a challenging endeavor.
At least one major reason for the difficulty lies in the fact that the utility of such a system generally has to be assessed by its users, and different users (or even the same user in a different use context) often perceive the utility and behave differently. This also reflects the general challenge of evaluating any interactive intelligent system whose utility must be assessed by its user in a context-specific way.

As one promising way to address this challenge, user simulation has been proposed and studied in the context of information access systems, with much progress made in the last two decades.
We started the book with a broad overview of user simulation and its many applications with its use for evaluation as our focus.
We showed that current evaluation methods and measures can generally be viewed as naive user simulators. Thus, user simulation naturally serves as a unified framework for developing better evaluation methods in the future. Indeed, it is our belief that user simulation may be the only way to support reproducible experiments on evaluating any interactive intelligent system whose utility must be assessed by its users empirically.
We then systematically reviewed the frameworks, models, and techniques for constructing user simulators by using an MDP framework to tie multiple lines of scattered research work together in a unified formal framework. %
We now conclude this survey by highlighting open issues and providing several potential research directions. We end with some recommendations of actions for the broader research communities.

\section{Embracing Simulation-Based Evaluation}
\label{sec:concl:embracing}

There has been significant progress in the development of user simulation techniques, and we have seen numerous instances of successful application of simulation-based approaches for evaluating search engines and conversational assistants (though the use of simulation for evaluating recommender systems appears to be less common). 
The benefits of using user simulation for system evaluation are numerous. Simulation allows researchers and developers to test their systems under various scenarios and conditions, which may be difficult or impossible to achieve in a real-world setting. Additionally, user simulation can help identify potential flaws or weaknesses in a system before it is deployed, leading to more robust and effective systems~\citep{Bernard:2024:CUI}.

Despite its clear benefits, simulation-based evaluation has not been widely adopted in the information retrieval and recommender systems communities.
This may be due to several factors, including the complexity of creating realistic simulations, the lack of consensus on simulation-based evaluation methodology, open questions regarding the validity of simulations, and the resources required to develop and run simulations. Naturally, in the future, we need to do more research to address these challenges. Meanwhile, we also argue that it is time to recognize the value and benefits of user simulation and that it should be considered as a valuable addition to the evaluation toolbox of researchers and developers, alongside with the methodologies of test collection-based evaluation, user studies, and online evaluation. Indeed, by viewing an existing evaluation measure using the user simulation evaluation framework, we can often obtain a more meaningful interpretation of the evaluation measure and assess any unrealistic assumption about a user implicitly made by the evaluation measure.\footnote{One specific example is the natural justification in the user simulation framework for the use of variable $k$ when using NDCG@k to aggregate over a set of queries~\citep{Karmaker:2020:CIKM}.}

To promote user simulation for evaluation, as a first step, we should leverage the large number of existing test collections, notably those created at TREC, CLEF, and NCTIR, to turn those into user simulators that are made publicly available to the community. For example, we may use a detailed topic description from a TREC collection as assumed information about a target user to be simulated, based on which we can then build a simulator to attempt to simulate that user's behaviour. 
Search log data can also be used as a basis. For example, in the case of product search, we may attempt to simulate a user who is known to have bought a particular product (based on information available in a search log) to simulate the user's query formulation behaviour. 
Even though present-day user simulators are far from perfect or accurate, compared with the current evaluation methods used (where unrealistic or extremely naive user behaviour has been assumed), utilizing user simulation for evaluation is still desirable, let alone the opportunity to evaluate interactive systems that we currently cannot easily do without relying on user simulation~\citep{Zhang:2017:ICTIR}.\footnote{Note that despite the incompleteness of relevance judgments, relative comparison of systems using such imperfect relevance judgments is still meaningful, so in this sense, the community is already accepting this kind of ``imperfect evaluation.''}

User simulators can be released as toolkits, such as SimIIR~\citep{Maxwell:2016:SIGIR,Zerhoudi:2022:CIKM} or UserSimCRS~\citep{Afzali:2023:WSDM}, or a web-based API service as has been done in living lab-style evaluation~\citep{Hopfgartner:2019:bookchapter,Jagerman:2018:JDIQ}.
These relevant existing efforts have already demonstrated the feasibility of using more interactive evaluation approaches.  
To promote the use of user simulation for evaluation, it is also necessary to organize evaluation activities regularly (e.g., by establishing a user simulation track at TREC,  CLEF, or NTCIR) to facilitate research and sharing of knowledge.

\section{Fostering Industry-Academia Collaboration}

Accurate evaluation of information access systems (or interactive AI systems in general) is a shared interest of both industry and academia.
Thus, user simulation is a technology that can help to foster collaboration between academia and industry. From an academic perspective, access to realistic datasets for evaluation is always a major challenge (barrier), which often hinders research progress.
Due to privacy concerns, industry labs are generally unable to release their datasets publicly. Interestingly, using user simulation for evaluation is not only theoretically desirable, but also a natural way for removing this barrier in data access, since releasing user simulators trained/estimated using commercial search log data should have much less privacy concerns than releasing any log data (directly). This means that if academic researchers can develop algorithms and models for user simulation, and make them available as open source, then commercial service providers would be able to train and validate user simulators against their logs, and publish the trained user simulators without having to share any actual user data. Those published simulators would then enable academic researchers to develop and validate new search and recommendation algorithms, as well as new simulation and evaluation technologies, forming a positively reinforced and potentially self-sustainable innovation ecosystem. From an industry perspective, such an ecosystem would enable them to have timely access to the most advanced algorithms developed by (external) researchers, thereby essentially outsourcing research that can directly benefit their information services.

\section{Extending Current User Simulation Technologies for Evaluation}

Although the current work has already yielded valuable user simulation technologies applicable for immediate use in evaluation and fostering collaboration between academia and industry, numerous limitations and opportunities for future research remain. We briefly discuss some of them here. \\

\parheading{User simulation for recommender systems.}  
Even though we intended to cover both search engines and recommender systems in the book, most of the content in the previous chapters has been conducted in the context of search engines; user simulation for the evaluation of recommender systems has only been studied recently~\citep{Rahdari:2022:IntRS,Hazrati:2024:UMUAI,Ghanem:2022:ECRA,Rahdari:2024:TORS}. We took the stance that search engines and recommender systems are ``two sides of the same coin'' and there are indeed several shared elements, e.g., when a user interacts with a ranked list of recommended information items the similarities. This is also reflected in the fact IR evaluation methodology has been adopted and extended for evaluating recommender systems~\citep{Cremonesi:2010:RecSys,Parapar:2021:RecSys}. However, there is undoubtedly a much better understanding of user behaviour when it comes to interacting with search engines than in the case of recommender systems, which is an obvious future opportunity to address. \\

\parheading{Holistic user simulation.} As can be seen easily from Chapter~\ref{ch:sim_search}, most existing work on in the context of search and recommender systems has focused on a simulating a specific user action (e.g., formulating a query or clicking on a document), without much work on integrating the modeling of multiple actions into a coherent and holistic user simulator.
The few exceptions with a complete solution only attempted to create a basic user simulator. Thus, an important direction for future work is to study how to build more sophisticated holistic user simulators by integrating the various component-level simulation techniques. \\

\parheading{Modeling user simulation as an MDP.} Another limitation of existing work is that it focuses on developing and evaluating specific techniques, while there is very little work on developing general formal models for user simulation. Earlier we attempted to use an MDP as a general formal framework to bring together separate lines of research in user simulation. There is clearly a need for doing more research along this direction. Specifically, future research may use MDPs as a formal framework in which specific simulators may be instantiated by specifying states, actions, rewards, and policy. \\

\parheading{Infrastructures for user simulation-based evaluation.}
Evaluation using user simulation requires infrastructures that go beyond traditional test collections. As we briefly discussed in Section~\ref{sec:practical:toolkits}, the current work in this area is still quite limited, which may also be a reason why user simulation has not yet been widely used for evaluation.  Thus, there is a need for further research and development of scalable and sustainable infrastructures for supporting user simulation-based evaluation. This includes, for example, releasing toolkits with implementations of user simulators, establishing web-based services that researchers can readily use to evaluate their interactive systems, maintaining a repository of experimental results on user simulators, and developing open-source infrastructures to facilitate large-scale collaboration among researchers in contributing new user simulators. Such infrastructures would also naturally facilitate collaboration between industry and academia, which is required for sustaining the innovation and delivering broader impact in this area.

\section{Long-Term Challenges and Future Research}

Beyond the short-term research problems and advancement of user simulation for evaluation, there are also many long-term broader difficult challenges that need to be addressed in this general topic area. In this last section, we briefly discuss some of the major ones and how we might tackle them, particularly by encouraging interdisciplinary research work involving multiple research communities.

\subsection{Realism}
\label{sec:concl_future:realisticity}

The first general challenge lies in a more rigorous definition of the problem of user simulation itself. While informally we can easily understand what it means by simulating a user computationally, which naturally implies that an ideal user simulator should be as similar to a real user as possible, how to rigorously and mathematically define the problem remains a major open challenge. In particular, the question ``How to make simulators more realistic?'' needs to be formally specified as some kind of optimization problem. Viewed in the MDP framework, this general challenge has to do with the overall formulation of the problem, i.e., how to define an agent with all the necessary components in the MDP, including the definition of states, actions, rewards, and the policy to be simulated.

The importance of this problem was already recognized by the community as one of the open questions identified at the Sim4IR workshop~\citep{Balog:2022:SIGIRForum} concerned the realism of simulators.
The following quote from the workshop report explains well this important foundational question about user simulation:
\begin{quote}
    ``It remains an open question as to how realistic (i.e. human-like) simulators can be, or indeed should be. It is important to note that simulators do not need to be perfect mirrors of human behaviour, but instead simply need to be “good enough.” By this, we mean that output from simulations should correlate well with human assessments on a given task with respect to some evaluation metric. The main requirement is reproducibility.''
\end{quote}
This touches on the issue of formulating the problem of user simulation and evaluating user simulation itself. Depending on the purpose and the anticipated use of a user simulator, we may want to frame and evaluate the problem of user simulation differently. There are inevitably many trade-offs that have to be made, such as interpretability vs. prediction accuracy, task-specific simulation of users vs. general simulation, individual user simulation vs. user group simulation~\citep{Nguyen:2019:UMUAI}.
Practical factors, such as data availability and the possibility of overfitting models to specific observed users in a dataset, are also important to consider.

For example, for the purpose of data augmentation, the user simulators do not have to be interpretable and thus do not have to be built based on realistic user models; instead, a user simulator would be acceptable if the generated data are realistic. This, however, raises the question about how we should quantify the similarity of the generated data with real user data. Is it acceptable as long as it is very close to one user's data or does it have to be close to the data from a group of users? For analyzing user behaviour, a realistic user simulator would have to be based on realistic user models. In such a case, interpretability is important, whereas how well the generated data match empirical data observed from a real user may be secondary. This of course raises another open question, which is how to rigorously determine whether a model is realistic.

The trade-off between interpretability and prediction accuracy is reflected in the wide spectrum of possible computational models and algorithms for user simulation already developed, which included both highly interpretable, but overly simplistic models (e.g., the cascade model for scanning and clicking) and purely data-driven models that can fit empirical user rating/clicking behaviour well (e.g., deep neural networks for predicting clicks in recommender systems), with many models in between. A major variable here is the nature of parameters used in the model. When all the parameters are interpretable, we would have not just one single user simulator but a whole family of user simulators that can vary meaningfully to model individual differences of users in the family by varying some of the parameter values. However, due to the constraints on the interpretability of parameters, the model's predictive power would also be inevitably limited (by the parameters allowed), thus their empirical performance in predicting user behaviour would generally be not as high as that of pure data-driven models, where the empirical accuracy of prediction is an explicitly defined objective function to be optimized. Ideally, in the future, we should integrate these two kinds of models by using increasingly more interpretable parameters, which may be viewed as deeper understanding of the users as well, since such interpretable user simulators can be regarded as theoretical claims about users. The data-driven models could be used to understand the gap between a theoretical model and empirical observations, i.e., reveal what kind of patterns the theoretical models fail to explain. Those can then become a research problem to be further studied, leading to new research results that can eventually help bridge the gap with additional menaningful parameters introduced to the simulator or additional relations captured between those latent user variables.

Another dimension of variation is a user's task. A different task generally implies a different simulator, since a full specification of a user simulation would include the user, the system, the user's task, and the general context and environment where the user attempts to complete the task. Thus, when we attempt to define a user simulator, we would need to fully characterize these other factors in addition to the user. The user's task is particularly important, as a user's behaviour varies significantly depending on the task, even though there is clear dependency on the environment or context as well. The incorporation of a user's task in simulation formulation is challenging because of the lack of formal ways to differentiate user tasks. To address this problem, it is necessary to develop an ontology of user tasks (or otherwise some kind of formal language to express different kinds of user tasks). Once the task is specified, related variables in the context or environment can also be identified to allow for a more complete specification of the user simulation problem. %

Finally, regardless how we formulate the user simulation problem, eventually we would need to address the common challenge of how to measure the similarity of simulated and real users. There are many dimensions along which the two may be compared; even for just one particular dimension, queries, it is still unclear what the best way of measuring the similarity is. The main challenge lies in that different deviations may have variable cost. For example, we might use word overlap to measure the similarity between a simulated and a real query, but how should we weigh the cost of missing one query word vs. addition of an extra word not in the original query? It may be desirable to parameterize the similarity function so that the parameters can be optimized by using machine learning (e.g., metric learning~\citep{Kaya:2019:Symmetry}).

\subsection{Evolving System Functions and User Interfaces}

The second general challenge is that a full specification of a user simulator must include a full characterization of the system functions and interfaces that a user would interact with. However, both the functions and user interfaces vary across systems and they also evolve over time. For example, while early-generation search systems used simple list-based interfaces, modern recommender system organize items into slates and SERPs have non-linear layouts. There is also the question of conversational UIs, which support multiple modalities. Since the actions available for a user to take generally depend on the interfaces and functions of a system, this variation creates a major challenge in developing general user simulation techniques, since the most effective simulation techniques for a user action may depend on the whole user interface or other functions that the user has access to.

In the MDP framework, this general challenge is reflected in the complexity of the definition of the states, the definition of the space of actions, and the transition of states due to a particular action being taken. Multi-modal interactions in conversational assistants would clearly further increase the complexity since the number of potential actions that are available for a user to take might even be infinite. However, note that in the case of simulating a user's query, we are already facing a situation when the number of possible actions is infinite, but a user can be assumed to use queries of only a limited length, which helps make the simulation more tractable. In the case of a conversational assistant that allows a user to express any request in natural language, it would be inappropriate to add a constraint on the length of a sentence that a user would use. A user could still be potentially modeled as solving an optimization problem here in that the user would attempt to find a natural language sentence that would optimize some kind of criteria. In this sense, some ideas developed for query simulation (e.g., the PRE optimization framework~\citep{Labhishetty:2022:ICTIR}) might be applicable to other user simulation scenarios.

A particularly interesting general question here is to what extent we can develop general policies that might be invariant to the variations of user interfaces. This is presumably feasible for simulating some user actions, since there may be some general psychological mechanism that all/most users follow. For example, despite many variations of how a list of search results would be shown exactly to a user, from a user's perspective, perhaps a sequential browsing (linear traversing) model would often be used by the user to prioritize the scanning of results; the difference would be more reflected in the clicking decisions, which in turn might be affected more by how exactly a snippet is shown. The challenge here is how to decompose a complex policy in user simulation into multiple potentially generalizable component policies. In some sense, the current work has already made such a decomposition when query simulation and clicking simulation are studied separately.
However, it is unclear how such a decomposition is to be performed in principled way and how common components and workflows from multiple systems are to be extracted. One possibility may be to analyze the interaction data collected from multiple systems and use data mining approaches, such as frequent pattern mining, or latent variable models, such as hidden Markov models, to discover empirically meaningful common interface components. This way, we may create some basic user simulators designed based on some common interfaces or interface components shared by many systems, and use them as a prior for adaptation to a specific system for construction of a more system-specific simulator.  A major challenge here is determining what kind of parameters would need to be introduced to enable customization and how to set the values of those parameters.

With more complex interfaces, a user's action may have to be modeled at a lower level, including particularly the modeling of cognitive effort involved in recognizing possible actions that the user could potentially consider. This may require eye-tracking studies. Just as a clicking decision generally follows the scanning of results, and the two steps can be modeled separately, it would be interesting to study how to decompose other actions in a similar way. In general, it appears that we could identify three steps that a user possibly follows: (1) recognize a new interface and try to make sense of it, by recognizing the various functions the interface supports,
(2) assess the potential utility of taking each action as well as the cost associated with taking it, and
(3) make a choice between those candidate actions and take a specific action.
How to model each step in a general way is an interesting future challenge---one that certainly requires interdiscplinary research intersecting with HCI, information science, and decision theory. The interface card model~\citep{Zhang:2015:SIGIR} may be a relevant framework for such a modeling task.

As new systems and new interfaces are deployed, they also create new opportunities in the sense of the log data they can collect.
It would likely contain traces of many interesting new actions, or interaction patterns, that were not available in old systems.
The new log data could be compared with the old data collected from a previous version of the system/interface to understand users' behaviour in adapting to the new system/interface. Such knowledge about users can be very useful for building user simulators that can adapt automatically to future interfaces.
From another perspective, simulations can also inform the design of the search interface, especially a new search interface; for example, how much information to show and how to arrange various content representations~\citep{White:2006:IPM}.

\subsection{Modeling Variable Latent States of Users: Cognitive and Emotional States}

The third general challenge lies in the modeling of latent states of users, including, e.g., cognitive states and emotional states. A series of studies by Kuhlthau have examined information retrieval from both cognitive and affective aspects~\citep{Kuhlthau:1988:RQ,Kuhlthau:1991:JASIST}. To model a user holistically, it is important to model both their cognitive and emotional states, as these are interconnected. Both states can be latent, making them challenging to model directly. However, observable user behaviours often result from (or are determined by) specific latent states. Thus, modeling these latent states is crucial for accurately predicting a user's observable behaviour.

While there has been some work on modeling users' knowledge and cognitive state, and simulating query formulation and clicking based on these, the current work has often used very simple models (e.g., unigram language models).
Also, the knowledge state is generally limited to a user's familiarity with relevant and non-relevant content. In the future, more research is needed in this area. %

First, for the simulation of any user interacting with a information access system, there is a need to model the user's information need. The ASK theory~\citep{Belkin:1982:JD} provides some basis for modeling the case when the user has a real need for some information that they are unfamiliar with; it can be used to conveniently build a comprehensive model of information need by leveraging knowledge graphs that are widely available nowadays. According to ASK, an information need reflects an anomalous state of knowledge (ASK), and thus we might assume that part of a knowledge graph is ``unavailable'' to a user and the user's goal (task) is to obtain relevant information to the missing part, thus filling the ``hole'' in the user's knowledge graph. However, in a broader context of search applications, such as product search, we also often have users search for known items (e.g., purchasing a consumed product regularly). In that scenario, ASK might not be the right framework as the user might have excellent knowledge about relevant content. There, the information need is more about identifying the exact information item(s) in a collection that matches well with what the user has already had in mind. (We might still consider there is some ``hole'' in the user's knowledge graph, which, e.g., is the exact URL to the product that the user wanted to purchase.)
In other cases, a user's information need may be quite complex (e.g., in medical search), and the user may also face a significant vocabulary gap. Then, modeling how a user learns to bridge the vocabulary gap during the search process would be required in order to accurately simulate the user's behaviour. In general, for a complex information need, we will need complex variables to characterize a user's information need accurately, along with ways to capture the associated (and inevitable) uncertainty.
Moreover, a user's information need likely evolves during the search process as the user is being exposed to more information items, making it necessary to model how the information need would be updated over time.

Second, a user's search behaviour depends much on the user's knowledge background in general. Thus, another essential element in any user simulator is a model of the user's background knowledge. Of particular importance is to model the user's familiarity with relevant and non-relevant content. This would allow user simulation to be employed for both expert and novice users, as well as any users in between. To make such a background knowledge model meaningful, there should be some additional interpretable categorical variables associated with users. For example, elementary school students might be modeled using a similar background knowledge model, which would clearly be different from that used for graduate student users. To make the background knowledge models realistic, such models can be estimated based on the real content produced by a certain group of users. In the extreme case, if we can collect content generated by an individual user, we may use all the observed content from the user to build a background model for them. As in the case of an information need, a user's knowledge would also evolve over time during the search process, from which the user can be expected to gain new knowledge. This requires modeling the cognitive process of a user during search and updating the user's knowledge model based on the new information the user has been exposed to. An interesting connection may be made here between user simulation and work on ``search as learning''~\citep{Vakkari:2016:JIS,Collins:2017:Report}.

Third, a user's search behaviour depends on the user's past experiences with the system.
For example, if the system tends to perform better with keyword queries as opposed to natural language questions, then users will learn to issue the former. Another important component of a user simulator, therefore, is a model of a user's knowledge about how the system works. In some sense, the PRE framework for query formulation~\citep{Labhishetty:2022:ICTIR} is based on the assumption that a user would imagine what would happen to a candidate query if that query were to be executed by the system, and then try to assess the expected Recall and Precision of the candidate query (i.e., assessing the optimality). However, there is no explicit model about how a system works proposed in the PRE framework so far. In the future, modeling and updating a user's knowledge about the search system would be necessary. How to develop such models and incorporate them into user simulation is a highly interesting challenge for future research.

Finally, a user's emotional state has a significant impact on the user's behaviour.  Hence, a realistic user simulator must also include models of emotional states and how they evolve during the search process. Perhaps the most obvious one is to model whether a user is frustrated (e.g., with the difficulty in finding any relevant information to a query) since this particular emotion has much to do whether the user would quit the search process or would have the patience to browse more deeply in the search result list. Clearly, this state is also related to a user's knowledge about how the search system works; the user would be more likely frustrated if they had an unrealistic expectation about the capabilities of the system.

In general, the cognitive and emotional states of a user may interact with each other in a complex way. An important research direction would be to fully understand their interactions and develop formal causal models to capture their dependency. Moreover, it is also necessary to study how they are related to the observable behaviours of users; once again, the development of causal models to capture how internal latent states may cause specific behaviours of a user when interacting with a system would be necessary. As the latent variables are not directly observable, collecting information from users in the form of surveys or questionnaires may be needed for such studies.

\subsection{Evaluation of User Simulators: Validity and Reliability}

As already touched upon in the discussion of realism earlier, how to evaluate user simulators is yet another major difficult challenge.
In theory, utilizing user simulation for evaluation seems logical; however, in practice, the reliability of this evaluation methodology is called into question if the user simulators themselves lack accuracy.
Given that the current measures are using essentially quite naive user simulators, one might assume that imperfect user simulators would also be acceptable. However, the real challenge here is to quantify the reliability of a user simulator, so that we at least would know the limitations when we use a certain kind of user simulators for evaluation. Compared with the work on developing user simulation techniques, there has been little research on how to evaluate the user simulators themselves. The Tester-based framework~\citep{Labhishetty:2021:SIGIR} provides a basis for evaluating user simulators using a set of necessary requirements, but those requirements are indeed far from sufficient. The reliability estimation algorithm proposed in~\citep{Labhishetty:2022:ECIR} would allow for assigning reliability scores to a set of Testers and user simulators simultaneously. This may provide us with an initial step toward quantifying the validity and reliability of user simulators.
However, how to develop more Testers is also a new challenge that must be tackled before we can have a sufficiently large set of Testers for evaluating user simulators.

To ensure realistic user modeling, there is no way to escape from the measure of similarity between a user simulator and an actual user. Here we may distinguish different perspectives of similarity. First, we might consider model similarity. In such a case, we would have to have user models available for us to compare again. Realistically, we may only have available a component model for a specific action or behaviour of a user though, as the current understanding of users is still quite limited. With the availability of any component user model, we can then compare the model used in a simulator with the component user model. %
Second, we might consider data similarity. In such a case, we would compare the simulator-generated data with the data observed from real users. The comparison can be done from different perspectives. A natural decomposition to compute the overall similarity score based on aggregation over similarity scores on each action to be simulated; see our discussion earlier under Realism (Section~\ref{sec:concl_future:realisticity}.)

In sum, there are many challenges associated with the evaluation of user simulators. While human involvement for validation is inevitable, the question of how it might be minimized is an important one for obvious reasons.

\chapter*{Acknowledgements}
\addcontentsline{toc}{chapter}{Acknowledgements}
\markboth{Acknowledgements}{Acknowledgements}

We are grateful to the editors-in-chief, Yiqun Liu and, especially, Pablo Castells, for handling our manuscript.  We are also extremely grateful to our two anonymous reviewers: their valuable comments and suggestions helped us improve the book.

The authors thank Nolwenn Bernard for her invaluable assistance in setting up and maintaining the annotated bibliography on the companion website.
The authors further want to thank Craig Macdonald, Maria Vlachou, To Eun Kim, Ryen White, Yaxiong Wu, and Yong Li for their suggestions on literature to include.

Krisztian Balog was partially supported by the Norwegian Research Center for AI Innovation, NorwAI (Research Council of Norway, project number 309834). ChengXiang Zhai was partially supported by the US National Science
Foundation and the Institute of Education Sciences through Award \# 2229612 (National AI Institute for Inclusive Intelligent Technologies for Education).

\backmatter  %

\printbibliography

\end{document}